\documentclass[]{pasj02} 
\usepackage[switch,mathlines]{lineno} 
\usepackage{ dsfont }

\usepackage{graphicx}
\usepackage{caption}
\jyear{2026}
\Received{}%{yyyy/mm/dd}
\Accepted{}%{yyyy/mm/dd}
\usepackage{float}
\usepackage{adjustbox}
\usepackage{threeparttable}
\usepackage{ upgreek }
\usepackage{stfloats}
\usepackage{lscape}

\usepackage{fix-cm}

\begin{document} 

\title{Investigating \textit{r}-Process Diversity in Mildly Metal-Poor Stars Using the Th~{\sc ii} 5989\,\AA\ Line}

\author{
 Kurumi \textsc{Furutsuka},\altaffilmark{1,2}\altemailmark\orcid{0009-0008-0515-7492} \email{furutsuka@nhao.jp} 
 Satoshi \textsc{Honda},\altaffilmark{2}\altemailmark\orcid{0000-0001-6653-8741}\email{honda@nhao.jp}
}
\altaffiltext{1}{Graduate School of Science, University of Hyogo, 3-2-1 Kouto, Kamigori-cho, Ako-gun, Hyogo, 678-1297, Japan}
\altaffiltext{2}{Nishi-Harima Astronomical Observatory, Center for Astronomy, University of Hyogo, 407-2 Nishigaichi, Sayo-cho, Hyogo, 679-5313, Japan}


\maketitle

\begin{abstract}
The origin of the heaviest elements remains a fundamental mystery, with the "actinide-boost" (Thorium overabundance) serving as a critical diagnostic of \textit{r}-process nucleosynthesis conditions. 
While this diversity is well documented in extremely metal-poor (EMP) stars, its frequency in the mildly metal-poor (MMP) regime ($-2.0 < $ [Fe/H] $< -0.5$) is poorly constrained due to blending issues in the commonly used 4019 \AA~ line. 
We present a systematic abundance analysis of mildly metal-poor stars, exploiting the relatively unblended Th~{\sc ii} 5989\, \AA~ line.
High-dispersion spectra were obtained using the 1.5m telescope at Gunma Astronomical Observatory with GAOES, complemented by archival data from the Subaru Telescope/HDS retrieved via SMOKA.
We successfully detected Th in 32 out of 42 sample stars. Our analysis demonstrates that the 5989 \AA~ line is a useful diagnostic for Th abundance measurements in cool red giants.
The mean [Th/Eu] ratio remains nearly constant over the MMP metallicity range, suggesting that progressive homogenization of the interstellar medium during Galactic chemical evolution would reduce abundance variations inherited from individual \textit{r}-process events, or the dominant \textit{r}-process events may share similar initial Th/Eu production ratios.
While the [Th/Eu] ratio is nearly constant overall in the MMP regime, its dispersion appears to increase toward the metal-rich end of this regime and reaches approximately 1 dex near solar metallicity. 
This behavior may reflect the combined effects of variations in the initial Th/Eu production ratios of the progenitor objects, the mixing of \textit{r}-process products in the interstellar medium, and the radioactive decay of Th over stellar lifetimes.
\end{abstract}

\section{Introduction}
The origin and nucleosynthetic sites of \textit{r}-process elements remain among the most significant mysteries in nuclear astrophysics. Detailed analysis of \textit{r}-process enhanced metal-poor stars revealed an elemental abundance pattern for heavy elements with Z > 56 that matches the solar pattern, suggesting an origin governed by "\textit{r}-process universality" -— a process in which heavy elements are synthesized in the same proportions every time (e.g. \cite{1996ApJ...467..819S}, \cite{2002A&A...387..560H}, \cite{2021RvMP...93a5002C}). While neutron star mergers are confirmed sites (e.g., \cite{2017PASJ...69..102T}), the "actinide-boost" phenomenon --a significant overabundance of Thorium (Th) relative to Europium (Eu)-- suggests high diversity in \textit{r}-process ejecta conditions (c.f. \cite{2002A&A...387..560H}; \cite{2018ApJ...859L..24H}).
\par Actinide-boost stars are defined by comparing the initial Th/Eu ratio of the star. $^{232} ${Th} is a radioactive element whose half-life is 14.05 Gyr. Therefore, if Th and Eu were produced with the same ratio in each progenitor, Th/Eu (ratio with stable lanthanide elements) will decrease as the age increases.  
Approximately 30\% of \textit{r}-process enhanced (RPE) stars in the very metal-poor regime (VMP; [Fe/H] $< -2.0$) show log $\upvarepsilon$ (Th/Eu) $> -0.35$ ([Th/Eu] $> +0.15$)  (\cite{2014A&A...569A..43M}; \cite{2018ApJ...868..110S}; \cite{2019ApJ...881....5H}). 
These variations likely reflect intrinsic differences in the progenitor
systems, such as the electron fraction ($Y_e$) or entropy (e.g., \cite{2023ApJ...942...39F}).
However, it is unknown whether this diversity persists as the Galaxy matures. Therefore, establishing the frequency of the actinide-boost in the mildly metal-poor regime (MMP; $-2.0 < $[Fe/H]$ < -0.5$) is crucial to understanding the evolution of heavy elements.  
This allows us to test whether \textit{r}-process conditions remain universal or shift as the enrichment source transitions from early halo progenitors to the disk-forming era.
\par The formation and chemical evolution of the Milky Way are recorded in the chemical fingerprints and kinematic properties of its stars. 
Our understanding of the Galaxy has advanced significantly at the two extremes of metallicity.
In the EMP regime, stars primarily located in the halo preserve the nucleosynthetic footprints of the first stars and early core-collapse supernovae.
Conversely, in the metal-rich regime ([Fe/H] $> -0.5$), the thin disk population reflects the integrated, averaged state of the modern Galaxy.
The MMP regime represents the most turbulent transition era in Galactic history. Recently, observational research on neutron-capture elements in the MMP regime has been actively advancing, as exemplified by the MINCE (Measuring at Intermediate metallicity Neutron-Capture Elements) project (\cite{2022A&A...668A.168C}), which targets this previously overlooked metallicity range.
This metallicity range, corresponding to 1/100 to 1/3 of the solar abundance, tracks the merger of massive satellite systems --most notably the Gaia-Sausage--Enceladus (GSE)-- followed by intense in-situ star formation and the assembly of the thick disk (e.g. \cite{2018Natur.563...85H}, \cite{2022A&A...661A.103M}). 
Chemically, this period marks the critical onset of contributions from intermediate-mass AGB stars and Type Ia supernovae. 
Resolving the detailed chemical patterns --specifically the \textit{r}-process elements-- during this phase is the key to deciphering the assembly and chemical maturation of the Milky Way.
\par There is a systematic observational bias that has hindered our understanding of \textit{r}-process evolution. To date, Th detections in very metal-poor stars ([Fe/H] $<-2.0$) have relied exclusively on the 4019 \AA~ line in giant stars. While detectable at low metallicity, this line becomes increasingly unusable as metallicity rises due to severe blending. Furthermore, even in the relatively metal-rich regime ([Fe/H] $>-1.0$), most previous Th measurements --primarily targeting solar-neighborhood dwarf stars-- have continued to use the 4019 \AA~ line (e.g., \cite{2022MNRAS.516.3786M}). However, at such high metallicities, the 4019 \AA~ line is plagued by extreme line crowding, rendering the resulting Th abundances highly uncertain and less reliable. To overcome this "metallicity barrier", the 5989 \AA~ line has recently emerged as a crucial alternative. The utility and reliability of this line for high-precision abundance determination have been explicitly demonstrated in recent literature (\cite{2025A&A...699A.276A}). By targeting MMP giants and utilizing the 5989 \AA~ line, we can bypass the reliability issues inherent in the 4019 \AA~ line studies.
Giants allow us to probe a vast volume of the Galaxy beyond the immediate solar neighborhood, and the 5989 \AA~ line provides a cleaner, highly reliable diagnostic that has never been systematically applied to giants in this crucial transition metallicity range. \par This paper is outlined as follows. In Section 2, we describe our observations and the data reduction process. Section 3 presents spectrum analysis, including the estimation of atmospheric parameters. Section 4 presents measured Thorium and Europium abundances and a comprehensive error analysis. In Section 5, we show results of Th and Eu abundances and discuss the implications of our findings in the context of Galactic chemical evolution. Finally, our conclusions are summarized in Section 6.
\section{Observations and reduction}\label{sec:2}
\subsection{Data and sample selection}
For the purpose of this study, we selected bright giant stars for which atmospheric parameters and metallicities had been previously estimated in the literature. 
Our selection criteria targeted stars with $-2 \le \mathrm{[Fe/H]} \le 0$, effective temperatures $T_{\mathrm{eff}} \le 5000$\,K, and surface gravities $\log g \le 3$. 
The sample was selected to contain not only $r$-process enhanced stars ($\mathrm{[Eu/Fe]} > +0.3$) but also $r$-process normal stars ($\mathrm{[Eu/Fe]} \le +0.3$). 
This covers a wide metallicity range, primarily focusing on the MMP regime but also extending to metal-rich stars up to and beyond solar metallicity. 
Such extensive coverage allows for a comprehensive investigation of $r$-process variations from early Galactic evolution through the enrichment phases reaching super-solar metallicities.
\par To compile this comprehensive sample, our target selection and data collection were executed in two sequential phases:
First, we conducted a dedicated high-dispersion spectroscopic observation program targeting observable bright giants. 
Candidate stars were searched and selected via the SIMBAD database (\cite{2000A&AS..143....9W}). 
The spectroscopic observations for these targets were performed using the Gunma Astronomical Observatory Echelle Spectrograph (GAOES) mounted on the 150~cm telescope at the Gunma Astronomical Observatory (GAO).
Second, to significantly expand our sample size and cover a wider metallicity range, we subsequently compiled high-quality archival data. 
We utilized the SAGA database (\cite{2008PASJ...60.1159S}) to screen candidate stars with appropriate mildly metal-poor parameters. 
Their high-dispersion spectra were then retrieved from the SMOKA science archive (\cite{2002ASPC..281..298B}), which provides public access to the data obtained with the High Dispersion Spectrograph (HDS : \cite{2002PASJ...54..855N}) on the Subaru Telescope. In total, 16 objects were observed by GAOES, and 27 with HDS; accounting for the overlap of one star (HD124897, observed with both instruments), our final merged sample of 42 objects is analyzed in this work.

\subsection{Observations and data reduction}\label{ssec:21}
\par High-dispersion spectroscopic observations and data collection were conducted using multiple facilities as described above. 
For our new observations at GAO, we utilized GAOES, which provides a high nominal spectral resolving power of $R (= \lambda / \Delta \lambda) \sim 80,000$. 
For the complementary archival data, the spectra obtained with the Subaru/HDS spectrograph feature a comparable high resolving power of $R \sim 80,000$.
The obtained spectra cover a wide wavelength range spanning approximately 5000--7000,\AA, which critically encompasses the target lines for our chemical abundance analysis. 
Across the final sample of 42 objects, a high signal-to-noise ratio (S/N) of 92--590 was achieved at 6000,\AA. 
The complete list of target stars and their corresponding observation logs, including observation dates, exposure times, and specific S/N ratios, is summarized in Table~1.
The data reduction for all spectra was carried out using standard procedures.
For the obtained GAOES data, as well as the archival frames from SMOKA where applicable, the processing included overscan reduction, cosmic-ray removal, flat-fielding, scattered light subtraction, wavelength calibration, and continuum normalization.
These reduction steps were performed using the IRAF (Image Reduction and Analysis Facility) software package\footnote{Software package developed by NOAO (National Optical Astronomy Observatory) for astronomical image analysis and data reduction (https://iraf.net/).}.

\begin{table}[H]
\centering
\caption{Summary of observations for the 42 target stars.}
\tiny
\begin{adjustbox}{center}
\begin{tabular}{lrrr}
\hline
Object      & Obs.Date   & Exp.Time (s) & S/N (6000 \AA) \\
\hline
GAOES       &            &          &              \\
\hline
HD 10761     & 2008/11/14 & 1200     & 94           \\
HD 118055    & 2008/3/22  & 9000     & 112          \\
HD 124897    & 2008/3/22  & 60       & 207          \\
HD 14770     & 2008/11/14 & 2400     & 92           \\
HD 18970     & 2008/11/14 & 1200     & 143          \\
HD 35369     & 2008/11/14 & 1200     & 162          \\
HD 37160     & 2008/11/14 & 1200     & 141          \\
HD 41597     & 2009/1/8   & 4800     & 200          \\
HD 54131     & 2008/11/14 & 2400     & 140          \\
HD 54810     & 2008/11/14 & 1800     & 101          \\
HD 58367     & 2008/11/14 & 1800     & 182          \\
HD 6186      & 2008/11/14 & 1298     & 153          \\
HD 67447     & 2008/11/14 & 2400     & 166          \\
HD 76294     & 2008/11/14 & 600      & 227          \\
HD 77912     & 2008/11/14 & 1200     & 267          \\
HD 85503     & 2009/1/6   & 2400     & 244          \\
\hline
HDS         &            &          &              \\
\hline
BD +01 3070 & 2003/2/20  & 900      & 200          \\
BD +30 2611 & 2006/3/12  & 2190     & 332          \\
BD --01 1792 & 2013/3/22  & 300     & 112           \\
BD +05 4314 & 2010/6/18  & 1600     & 316           \\
HD 111721    & 2008/7/27  & 600      & 590          \\
HD 112126    & 2010/5/26  & 400      & 412          \\
HD 124897    & 2003/2/20  & 1        & 447          \\
HD 141531    & 2004/6/2   & 600      & 278          \\
HD 171496    & 2010/5/26  & 400      & 316          \\
HD 206739    & 2007/9/21  & 900      & 274          \\
HD 210295    & 2010/5/26  & 900      & 342          \\
HD 220838    & 2006/9/14  & 600      & 322          \\
HD 221170    & 2005/6/19  & 180      & 392          \\
HD 37828     & 2007/1/29  & 20       & 171          \\
HD 6833      & 2017/2/19  & 60       & 116          \\
KIC 10096113 & 2018/7/12  & 4800     & 114          \\
KIC 10737052 & 2018/7/12  & 5200     & 130          \\
KIC 11802968 & 2015/7/4   & 500      & 126          \\
KIC 1726211  & 2014/9/9   & 600      & 132          \\
KIC 4351319  & 2014/9/9   & 300      & 184          \\
KIC 5184073  & 2018/7/12  & 5400     & 126          \\
KIC 5530598  & 2014/9/9   & 100      & 184          \\
KIC 5698156  & 2018/7/12  & 600      & 153          \\
KIC 6611219  & 2018/7/12  & 1800     & 126          \\
KIC 7205067  & 2014/9/9   & 300      & 114          \\
KIC 8350894  & 2018/7/12  & 3600     & 133          \\
KIC 9583607  & 2018/7/11  & 2400     & 122          \\
\hline
\end{tabular}
   \end{adjustbox}
\end{table}

\begin{table}[H]
\centering
\caption{Derived atmospheric parameters of the program stars.}
\tiny 
\begin{adjustbox}{center}
\begin{tabular}{lrrrrrrr}
\hline
Object  & $T_{\rm eff}$ & $\log g$  & $v_{\rm t}$  & [Fe/H]  & [Fe/H]\_err \\
 & (K) & (cgs) & (km/s) & (dex) & (dex)\\
\hline
GAOES       &      &      &      &            &             \\
\hline
HD 10761     & 5045 & 2.55 & 1.46 & 0.03       & 0.06 \\
HD 118055$^\ast$    & 4400 & 1.80 & 2.55 & $-$1.80      & 0.04 \\
HD 124897    & 4259 & 1.49 & 1.48 & $-$0.62      & 0.05 \\
HD 14770     & 4978 & 2.52 & 1.50 & 0.01       & 0.06 \\
HD 18970     & 4794 & 2.52 & 1.40 & $-$0.05      & 0.06 \\
HD 35369     & 4959 & 2.62 & 1.28 & $-$0.12      & 0.05 \\
HD 37160     & 4763 & 2.68 & 1.16 & $-$0.56      & 0.04 \\
HD 41597     & 4516 & 1.76 & 1.45 & $-$0.59      & 0.05 \\
HD 54131     & 4675 & 2.34 & 1.36 & $-$0.18      & 0.05 \\
HD 54810     & 4709 & 2.51 & 1.18 & $-$0.26      & 0.05 \\
HD 58367     & 4902 & 1.78 & 2.11 & $-$0.13      & 0.09 \\
HD 6186      & 4837 & 2.40 & 1.38 & $-$0.30      & 0.05 \\
HD 67447     & 5040 & 2.24 & 2.11 & 0.00       & 0.08 \\
HD 76294     & 4833 & 2.39 & 1.47 & $-$0.12      & 0.09 \\
HD 77912     & 4995 & 2.15 & 2.08 & $-$0.07      & 0.08 \\
HD 85503     & 4594 & 2.62 & 1.16 & 0.44       & 0.12 \\
\hline
HDS         &      &      &      &            &      \\
\hline
BD +01 3070$^\ast$ & 5404 & 3.65 & 1.18 & $-$1.36      & 0.14 \\
BD +30 2611 & 4330 & 1.06 & 1.66 & $-$1.41      & 0.08 \\
BD --01 1792 & 4961 & 3.12 & 0.93 & $-$0.93      & 0.06 \\
BD +05 4314     & 5180 & 3.28 & 1.30 & $-$1.36      & 0.07 \\
HD 111721    & 4942 & 2.57 & 1.35 & $-$1.40      & 0.07 \\
HD 112126    & 4159 & 1.09 & 1.71 & $-$1.39      & 0.05 \\
HD 124897    & 4258 & 1.59 & 1.40 & $-$0.59      & 0.05 \\
HD 141531    & 4350 & 1.03 & 1.79 & $-$1.83      & 0.10 \\
HD 171496    & 4954 & 2.21 & 1.48 & $-$0.67      & 0.04 \\
HD 206739    & 4694 & 1.80 & 1.59 & $-$1.61      & 0.07 \\
HD 210295    & 4729 & 2.06 & 1.37 & $-$1.37      & 0.05 \\
HD 220838    & 4370 & 1.21 & 1.93 & $-$1.95      & 0.12 \\
HD 221170$^\ast$    & 4510 & 1.00 & 1.80 & $-$2.19      & 0.12 \\
HD 37828     & 4394 & 1.33 & 1.65 & $-$1.49      & 0.09 \\
HD 6833      & 4516 & 1.76 & 1.34 & $-$0.69      & 0.05 \\
KIC 10096113 & 4883 & 2.41 & 1.39 & $-$0.74      & 0.06 \\
KIC 10737052 & 5076 & 2.66 & 2.19 & $-$1.22      & 0.10 \\
KIC 11802968 & 4904 & 3.66 & 0.74 & $-$0.11      & 0.06 \\
KIC 1726211  & 4920 & 2.35 & 1.38 & $-$0.66      & 0.04 \\
KIC 4351319  & 4915 & 3.50 & 0.95 & 0.29       & 0.08 \\
KIC 5184073$^\ast$  & 4864 & 1.87 & 1.58 & $-$1.42      & 0.02 \\
KIC 5530598  & 4652 & 3.04 & 1.20 & 0.34       & 0.10 \\
KIC 5698156  & 4594 & 1.88 & 1.40 & $-$1.37      & 0.05 \\
KIC 6611219  & 4552 & 1.58 & 1.42 & $-$1.30      & 0.07 \\
KIC 7205067  & 5055 & 2.66 & 1.45 & $-$0.02      & 0.06 \\
KIC 8350894  & 4594 & 1.57 & 1.27 & $-$1.09      & 0.05 \\
KIC 9583607  & 4950 & 2.00 & 1.53 & $-$0.80      & 0.05  \\
\hline

\end{tabular}
\end{adjustbox}
\begin{tabnote}
For almost all objects, we derived atmospheric parameters by using TGVIT.\\
$^\ast$ Value adopted from the literature due to the lack of high-quality spectra in our observation: \citet{2000AJ....120.1841F} for HD 118055, \citet{2012ApJ...753...64I} for BD +01 3070, \citet{2021ApJ...912...72M} for KIC 5184073, \citet{2006ApJ...645..613I} for HD 221170.
\end{tabnote}

\end{table}

\section{Determination of stellar parameters}\label{ssec:3}
\par Although stellar atmospheric parameters for our targets have been estimated in previous studies, we re-determined them using a consistent method to maintain a uniform analysis across the entire sample.
\par The atmospheric parameters, effective temperature ($T_{\rm eff}$), surface gravity ($\log g$), microturbulent velocity ($v_{\rm t}$), and metallicity ([Fe/H]), were determined under the assumption of Local Thermodynamic Equilibrium (LTE). We used TGVIT program (\cite{2005PASJ...57...27T}), which is based on Kurucz's ATLAS9 model atmosphere grids (\cite{1993KurCD..13.....K}). 
First, we measured the equivalent widths (EWs) of Fe~{\sc i} and Fe~{\sc ii} absorption lines by fitting  Gaussian profiles using SPTOOL (\cite{2002PASJ...54..451T}).
SPTOOL is also based on the ATLAS9 model and designed for comprehensive spectral synthesis and analysis.
Typically, 50--100 Fe~{\sc i} lines were used for each target.
$T_{\rm eff}$ was iteratively adjusted until the trend of Fe~{\sc i} abundance with excitation potential was minimized. 
Log g was derived from the ionisation balance of abundances from Fe~{\sc i} and Fe~{\sc ii} lines. $v_{\rm t}$ was determined by minimizing the Fe slope of abundance versus EW. 
\par To verify the validity and consistency of our derived atmospheric parameters, the excitation and ionization equilibria for the representative star HD~54810 ($T_{\mathrm{eff}}$; 4709, $\log g$; 2.51,$v_t$; 1.18, [Fe/H]; $-$0.26) are illustrated in Figure~1. 
As shown in the left panel of Figure~1, the iron abundances derived from individual $\mathrm{Fe~\textsc{i}}$ lines exhibit no significant dependence on the excitation potential, confirming the accuracy of our effective temperature ($T_{\mathrm{eff}}$). 
Similarly, the right panel demonstrates that the derived abundances are independent of the equivalent widths, justifying our choice of microturbulent velocity ($v_t$). 
Furthermore, the systematic agreement between the mean abundances of $\mathrm{Fe~\textsc{i}}$ and $\mathrm{Fe~\textsc{ii}}$ lines ensures the reliability of our spectroscopic gravity ($\log g$).
We adopted [Fe~{\sc i}/H] for [Fe/H] because Fe~{\sc ii} lines can only be detected in fewer than 10 lines, in contrast to Fe~{\sc i} lines, which can be detected in nearby 100 lines. 
In TGVIT, calculations are performed by simultaneously varying $T_{\mathrm{eff}}$, $\log g$, $v_t$, and Fe abundance (\cite{2002PASJ...54..451T}, section 3.5). The error is evaluated based on the variability in Fe abundances obtained from Fe lines. On the other hand, \citet{2008PASJ...60..781T} demonstrated, through a comparison of atmospheric parameters obtained by TGVIT with values from previous studies, that the differences are greater than the formal error calculated by TGVIT. Therefore, relying solely on the formal error calculated by TGVIT may underestimate the actual uncertainty. Then, for the estimation of elemental abundances, we adopted an error estimate that takes this external comparison into account (see also subsection 4.3.2).
For four stars, the number of measurable Fe~{\sc i} lines fell below 50 because many features became too weak to be reliably detected or suffered from relatively low S/N. 
In these stars, we adopted atmospheric parameters from previous studies (Table 2). 
\par To check for systematic differences between GAOES and HDS, we analyzed the HD 124897 observed by both spectrographs, and confirmed that the derived atmospheric parameters are consistent within the measurement errors.
We also compared atmospheric parameters with previous studies. For details, please refer to Figures \ref{fig:compTeff} -- \ref{fig:compVt} in Appendix 1.

\begin{figure}[h]
  \centering
    \includegraphics[width=8cm]
    {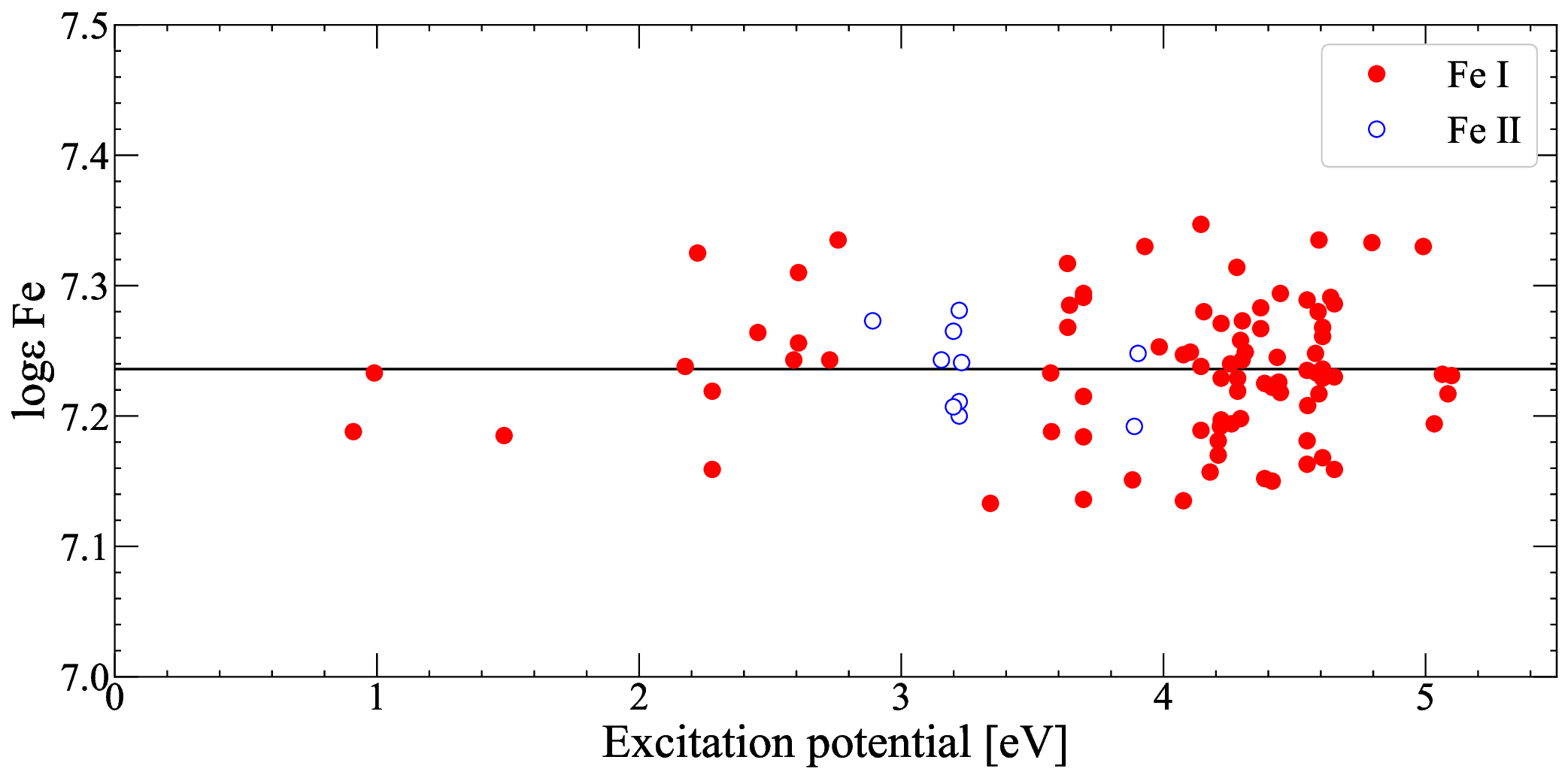}
   \includegraphics[width=8cm]
   {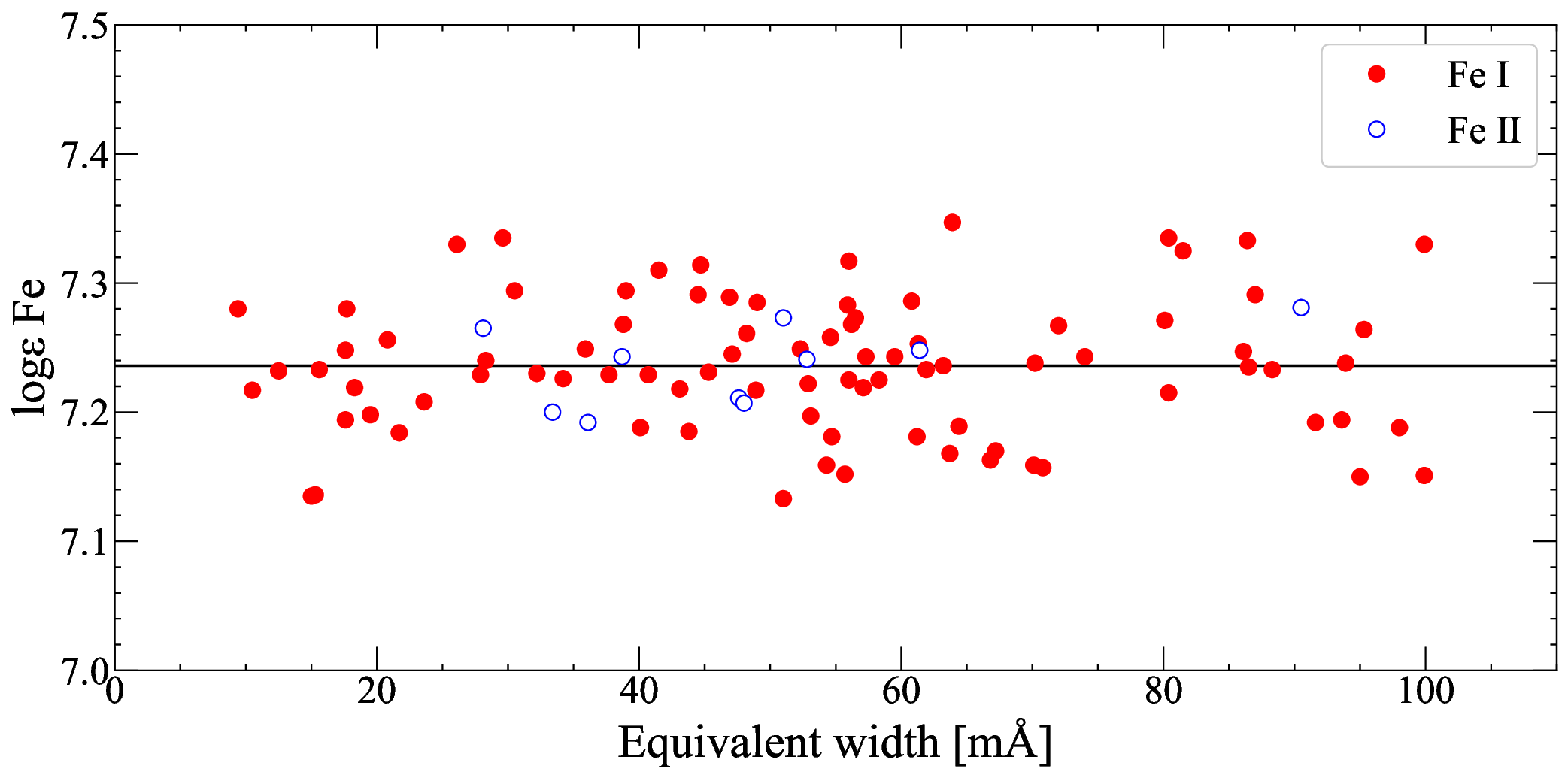}
    \caption{Top: Excitation potential of each Fe line vs Fe abundance, which was derived from each Fe line in the HD 54810 spectrum. Bottom: Equivalent width of each Fe line vs Fe abundance, which was derived from each Fe line in the HD 54810 spectrum. Red points show Fe~{\sc i}, and blue circles show Fe~{\sc ii}. The black solid line shows the estimated Fe abundance.
    {Alt text: Determination of Stellar atmospheric Parameters. }
    }
    \label{fig:FeabundancevsEW_EP}

\end{figure}

\begin{figure*}[b]
  \centering
   \includegraphics[width=0.3\linewidth]{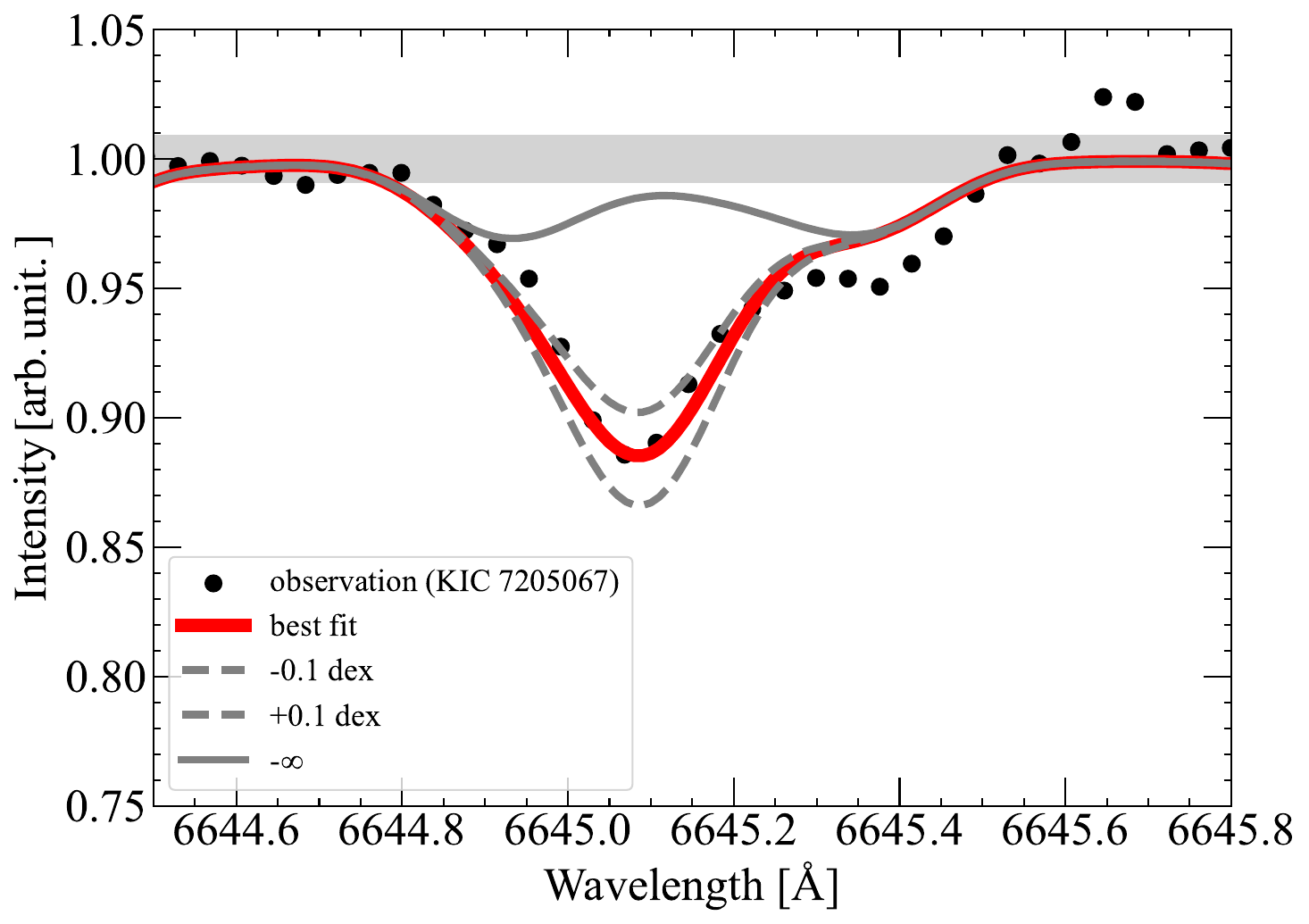}
   \includegraphics[width=0.3\linewidth]{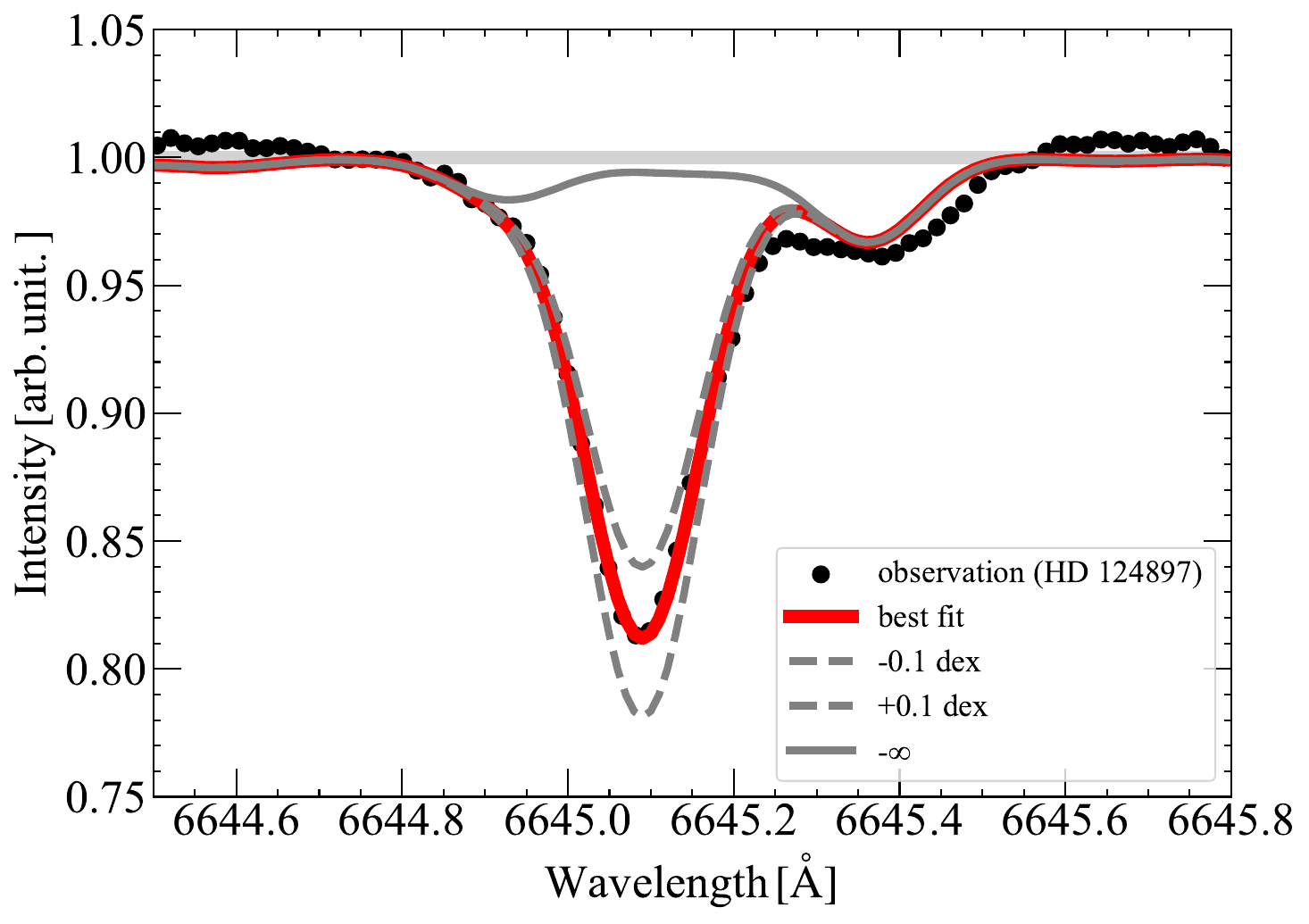}
      \includegraphics[width=0.3\linewidth]{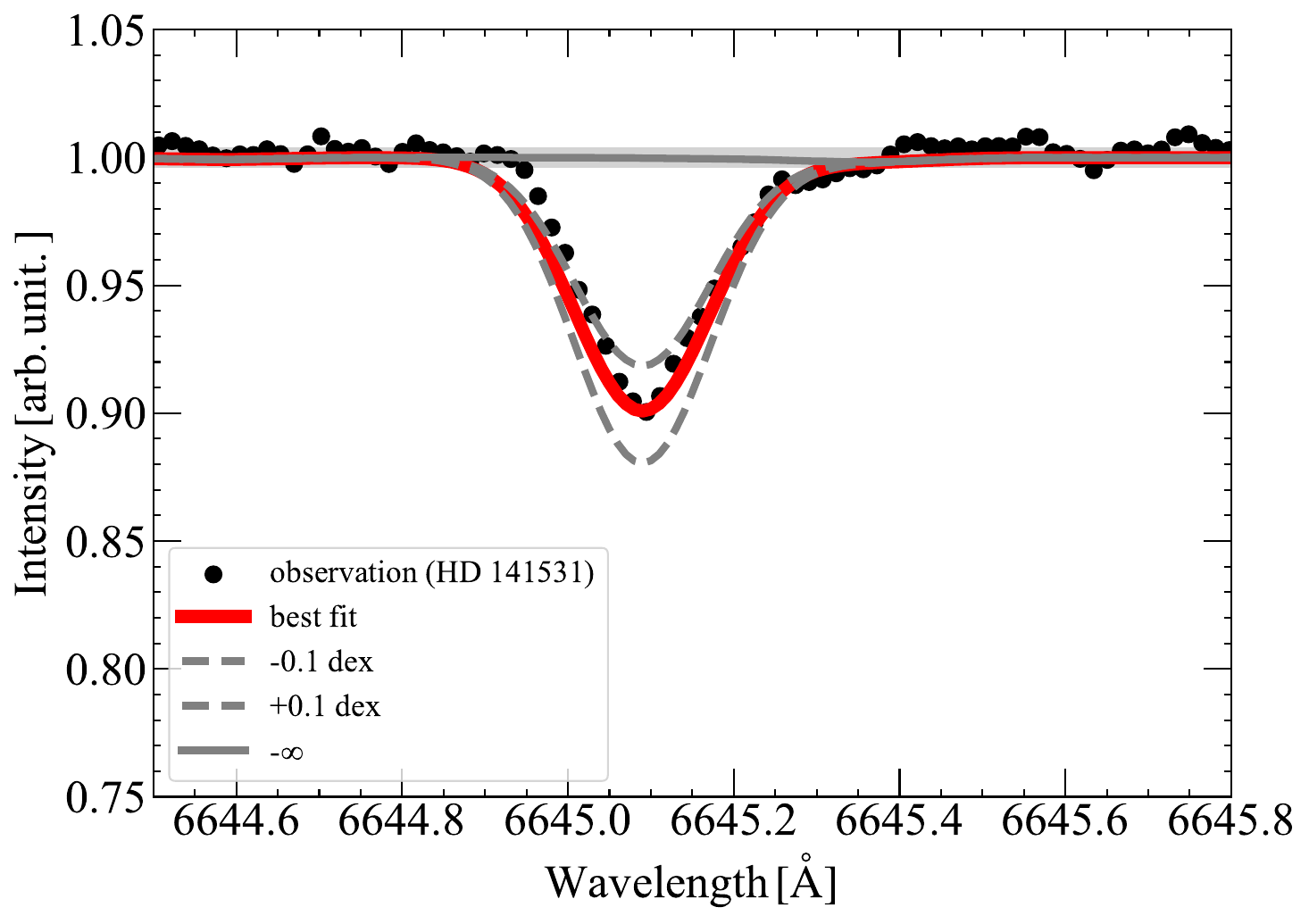}
  \caption{Absorption line of Eu 6645 \AA~ in (left) KIC 7205067; [Fe/H] = --0.01, (center) HD 124897 ; [Fe/H] = --0.58, (right) HD 141531; [Fe/H] = --1.83. Black points show the observed spectrum, and the red solid line shows the best-fit synthetic spectrum, the gray dashed lines show synthetic spectra calculated with the Eu abundance changed by 0.1 dex, and the gray solid line shows the synthetic spectrum calculated without Eu. The gray shaded region shows 1$\upvarsigma$ noise level from S/N.
  {Alt text: Absorption line of Eu 6645 angstrom of three stars.}
  }
    \label{fig:absEu6645}
\end{figure*}

\begin{figure*}[b]
  \centering

   \includegraphics[width=0.3\linewidth]{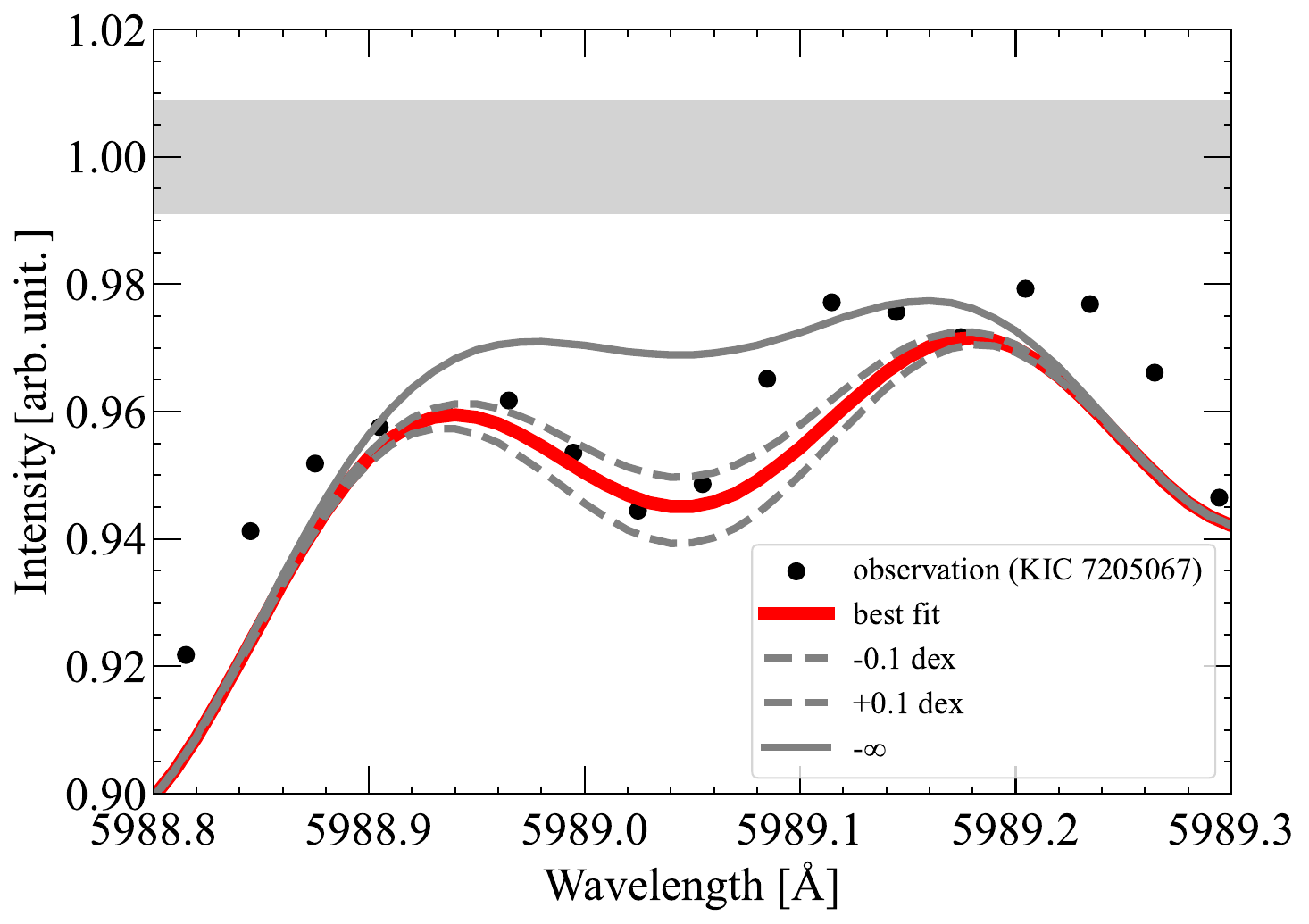}
   \includegraphics[width=0.3\linewidth]{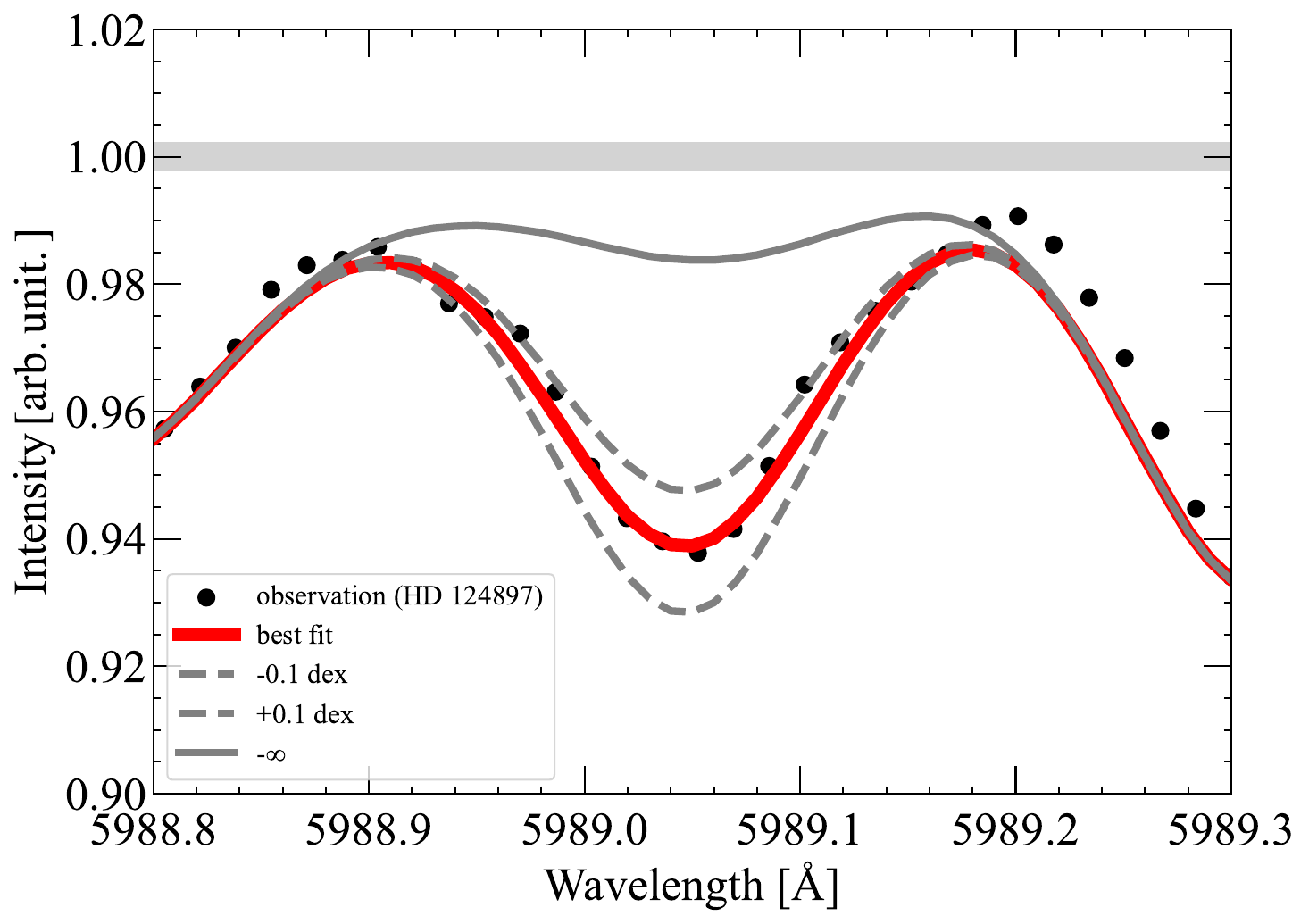}
    \includegraphics[width=0.3\linewidth]{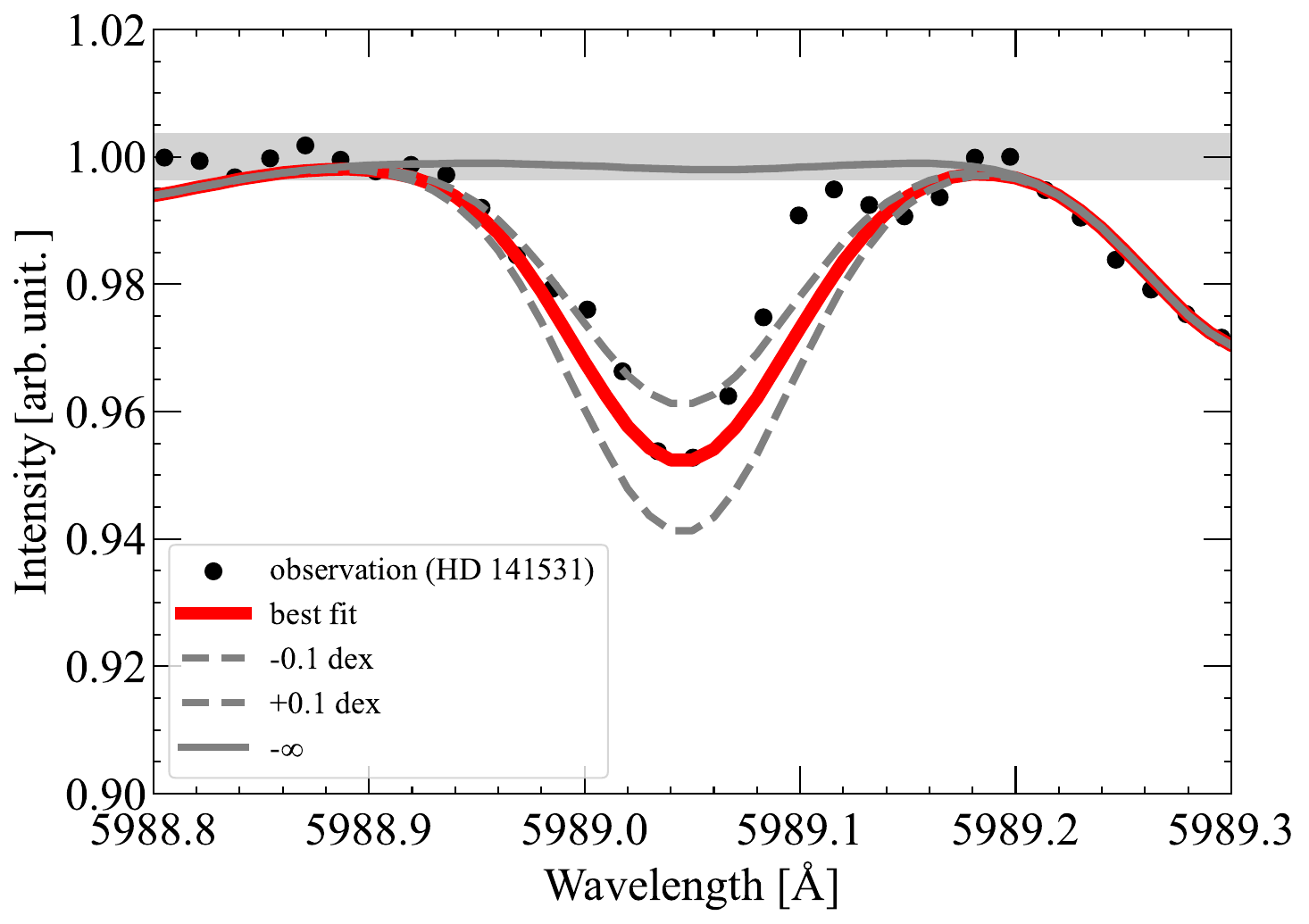}
  \caption{Absorption line of Th 5989 \AA~ in (left) KIC 7205067; [Fe/H] = --0.01, (center) HD 124897; [Fe/H] = --0.58, (right) HD 141531; [Fe/H] = --1.83. Black points show the observed spectrum, and the red solid line shows the best-fit synthetic spectrum, the gray dashed lines show synthetic spectra calculated with the Th abundance changed by 0.1 dex, the gray solid line shows the synthetic spectrum calculated without Th. Gray shading part shows 1$\upvarsigma$ noise level from S/N.
  {Alt text: Absorption line of Th 5989 angstrom of three stars.}
  }
    \label{fig:absTh5989}
\end{figure*}

\section{Determination of Thorium and Europium Abundances}\label{sec:4}
The abundances of Eu and Th were derived under the assumption of Local Thermodynamic Equilibrium (LTE) using the SPTOOL software (\cite{2002PASJ...54..451T}). The [X/Fe] values were normalized by using the solar abundance from \citet{2009ARA&A..47..481A}. 
Elemental abundances were determined by comparing the observed spectra with synthetic spectra while keeping the atmospheric parameters fixed. The atomic line data are summarized in Table \ref{tab:linelist} in Appendix~4.

\subsection{Eu abundances}\label{ssec:41}
In the spectral synthesis of the 6645.1 \AA~ line, we accounted for the blending Fe lines by adopting the derived [Fe/H] of each target star. Furthermore, it is necessary to account for line splitting due to hyperfine structure and isotopic shifts. Following \citet{2019A&A...631A.113F}, we estimated the Eu abundances using the line list shown in Appendix~4.
Furthermore, four molecular CN absorption lines are located around 6644.3 \AA~(see also Appendix~2, 4). For relatively metal-rich stars ([Fe/H] $>-1$), we took these CN features into account by assuming an estimated N abundance, as detailed in subsection 4.2. Examples of the resulting synthetic spectrum fits for the Eu line are displayed in Figure \ref{fig:absEu6645} for three representative stars covering different metallicity regimes. All spectra of program stars are shown in Figure \ref{fig:Eu_allspectra_1} in Appendix~2.
\subsection{Th abundances}\label{ssec:42} For the 5989 \AA~region, several neighboring elemental lines must be considered. We first estimated the individual abundances of $\mathrm{Si~\textsc{i}}$ (from 5988.7 \AA), $\mathrm{Ti~\textsc{i}}$ (5988.5 \AA), and $\mathrm{Nd~\textsc{ii}}$ (5989.3 \AA), which are essentially independent of the Th feature. Next, we took into account the possibility of blending CN molecule absorption lines in relatively metal-rich stars ([Fe/H] $>-1$). We assumed $\mathrm{[C/Fe]} = 0.0$ and derived the N abundance by synthesizing the CN absorption feature at 5973\,\AA. Finally, fixing $\mathrm{[C/Fe]} = 0.0$ and adopting this estimated N abundance, we varied the Th abundance to fit the 5989\,\AA\ profile \citep{2025A&A...699A.276A}. 
The final best-fit synthetic spectra for the Th line are illustrated in Figure \ref{fig:absTh5989}. All spectra of program stars are shown in Figures \ref{fig:allspe_Thdetct1}, \ref{fig:allspe_Thuplim} in Appendix~3. 
The uncertainty of N abundance estimation is estimated by [Fe/H] error and atmospheric parameters. The average uncertainty in the N abundance is 0.2 dex. It was found that this blend overestimates the Th abundance by 0.59 dex at maximum (Table \ref{table:CNestim} in Appendix~5). This tendency is particularly pronounced at high metallicity. 
\subsection{Uncertainties of abundances}\label{ssec:43}
The total uncertainty in our abundance measurements arises from three main sources: random errors due to spectral noise, systematic errors due to uncertainties in atmospheric parameters, and potential systematic shifts caused by non-local thermodynamic equilibrium (NLTE) effects.
In the following subsections, we describe each of these error sources in detail. 
We estimate the total abundance uncertainty by adding the random and systematic uncertainties in quadrature. For Th determination, it also includes the accuracy of the CN abundance estimates obtained using the same method.
\subsubsection{Random error from spectral noise}\label{ssec:431}
The uncertainty in elemental abundance due to random errors caused by spectral noise can be estimated using the standard deviation of the abundance obtained from multiple absorption lines of the same element. In this study, since the abundance of Th and Eu was estimated from a single absorption line, we instead used the standard deviation of the Fe~{\sc i} abundance, which was calculated from approximately 100 absorption lines.

\begin{table*}[t]
\tiny
 \caption{Abundance variation from changing atmospheric parameters for representative objects.}
 \label{Tab:atmpara_representive_error}
\begin{tabular}{llrrrrrrrrrr}
\hline
Object  & Element &  $\Delta T_{\rm eff}$ (K) & $\Delta T_{\rm eff}$ (K) & $\Delta$log$g$ (dex) & $\Delta$log$g$ (dex) & $\Delta v_t$ (km/s) & $\Delta v_t$ (km/s) & $\Delta${[}Fe/H{]} (dex) & $\Delta${[}Fe/H{]} (dex) & Total (dex)&\\
($T_{\rm eff}$, log$g$, $v_t$, [Fe/H]) & & +100  &$-$100  & +0.2 &$-$0.2 & +0.1  & $-$0.1  & +0.1  & $-$0.1  & &\\ 
\hline
KIC 7205067&Eu & 0.00 & 0.00 & 0.09 & $-$0.09  & 0.00 & 0.00 & 0.02  & $-$0.02  &  0.13 &\\
 (5055, 2.66, 1.45, $-$0.02)  &Th & 0.05 & $-$0.05 & 0.10 & $-$0.08  & 0.00 & 0.00 & 0.02  & $-$0.02 &  0.15&\\ 
\hline
HD 54810  & Eu & 0.00 & $-$0.23 & 0.10 & $-$0.08 & 0.00 & 0.02 & 0.04 & $-$0.02 & 0.27&\\
(4709, 2.51, 1.18, $-$0.26) & Th & 0.05 & $-$0.05 & 0.10 & $-$0.12 & 0.00 & 0.00 & $-$0.02 & 0.00 &0.17 &\\
\hline
HD 124897 (HDS)& Eu & $-$0.02 & 0.00 & 0.08 & $-$0.09 & $-$0.01 & 0.00 & 0.03 & $-$0.04 & 0.13&\\
 (4258, 1.58, 1.40, $-$0.58)  & Th & 0.05 & $-$0.05 & 0.11 & $-$0.10 & 0.00 & $-$0.01 & 0.00 & 0.02 & 0.17&\\
\hline
HD 112126 & Eu & $-$0.01 & 0.01 & 0.08 & $-$0.08 & 0.00 & 0.00 & 0.03 & $-$0.03 & 0.12&\\
 (4159, 1.09, 1.71, $-$1.39) & Th & 0.05 & $-$0.05 & 0.09 & $-$0.09 & 0.00 & 0.00 & 0.02 & $-$0.03 & 0.14&\\
\hline
KIC 8350894 & Eu & 0.01 & 0.00 & 0.08 & $-$0.09 & $-$0.02 & 0.01 & 0.02 & $-$0.03 & 0.13&\\
(4594, 1.57, 1.27, $-$1.09)  & Th & 0.05 & $-$0.05 & 0.08 & $-$0.08 & 0.00 & 0.00 & 0.03 & $-$0.02 & 0.14&\\
\hline

\end{tabular}
\end{table*}

\begin{table*}[t]
\tiny
\caption{Th and Eu abundance of samples. }
\label{Table:final_abundance}
\begin{tabular}{lrrrcrcrrrrrrc}
\hline
Object           & {[}Fe/H{]} & Fe\_err & log$\upvarepsilon$Th           & Th\_err & {[}Th/Fe{]}     & {[}Th/Fe{]}\_err & log$\upvarepsilon$Eu & Eu\_err     & {[}Eu/Fe{]} & {[}Eu/Fe{]}\_err & log$\upvarepsilon$Th/Eu        & {[}Th/Eu{]}     & {[}Th/Eu{]}\_err \\
\hline
BD +30 2611       & $-$1.41      & 0.08    & $-$0.75            & 0.16    & 0.64            & 0.18             & $-$0.31  & 0.15        & 0.58        & 0.17             & $-$0.44            & 0.06            & 0.22\\
HD 10761          & 0.03       & 0.06    & 0.20             & 0.26    & 0.15            & 0.27             & 0.57   & 0.28        & 0.02        & 0.29             & $-$0.37            & 0.13            & 0.38\\
HD 112126         & $-$1.39      & 0.05    & $-$0.92            & 0.15    & 0.45            & 0.16             & $-$0.34  & 0.13        & 0.53        & 0.14             & $-$0.58            & $-$0.08           & 0.20\\
HD 118055         & $-$1.80      & 0.04    & $-$0.78            & 0.07    & 1.00            & 0.08             & $-$0.51  & 0.06        & 0.77        & 0.07             & $-$0.27            & 0.23            & 0.09\\
HD 124897 (GAOES) & $-$0.62      & 0.05    & $-$0.38            & 0.23    & 0.22            & 0.24             & 0.22   & 0.14        & 0.32        & 0.15             & $-$0.60            & $-$0.10           & 0.27\\
HD 124897 (HDS)   & $-$0.59      & 0.05    & $-$0.37            & 0.23    & 0.20            & 0.24             & 0.18   & 0.14        & 0.25        & 0.15             & $-$0.55            & $-$0.05           & 0.27\\
HD 141531         & $-$1.83      & 0.10    & $-$1.08            & 0.17    & 0.73            & 0.20             & $-$0.72  & 0.16        & 0.59        & 0.19             & $-$0.36            & 0.14            & 0.23\\
HD 14770          & 0.01       & 0.06    & 0.23             & 0.22    & 0.20            & 0.23             & 0.74   & 0.14        & 0.21        & 0.16             & $-$0.51            & $-$0.01           & 0.26\\
HD 171496         & $-$0.67      & 0.04    & $-$0.35            & 0.21    & 0.30            & 0.21             & 0.12   & 0.14        & 0.27        & 0.14             & $-$0.47            & 0.03            & 0.25\\
HD 18970          & $-$0.06      & 0.06    & 0.50             & 0.24    & 0.53            & 0.25             & 0.68   & 0.27        & 0.21        & 0.28             & $-$0.18            & 0.32            & 0.36\\
HD 206739         & $-$1.61      & 0.07    & $-$0.90            & 0.15    & 0.69            & 0.17             & $-$0.58  & 0.14        & 0.51        & 0.16             & $-$0.32            & 0.18            & 0.21\\
HD 210295         & $-$1.37      & 0.05    & $-$0.80            & 0.17    & 0.55            & 0.18             & $-$0.49  & 0.14        & 0.36        & 0.15             & $-$0.31            & 0.19            & 0.22\\
HD 220838         & $-$1.95      & 0.12    & $-$1.26            & 0.18    & 0.67            & 0.22             & $-$0.91  & 0.17        & 0.52        & 0.21             & $-$0.35            & 0.15            & 0.25\\
HD 221170         & $-$2.19      & 0.12    & $-$1.49            & 0.02    & 0.68            & 0.12             & $-$0.90  & 0.02        & 0.77        & 0.12             & $-$0.59            & $-$0.09           & 0.03\\
HD 37828          & $-$1.49      & 0.09    & $-$1.16            & 0.16    & 0.35            & 0.18             & $-$0.66  & 0.15        & 0.35        & 0.18             & $-$0.50            & 0.00            & 0.22\\
HD 41597          & $-$0.59      & 0.05    & $-$0.34            & 0.23    & 0.23            & 0.24             & 0.13   & 0.14        & 0.20        & 0.15             & $-$0.47            & 0.03            & 0.27\\
HD 54131          & $-$0.18      & 0.05    & 0.04             & 0.23    & 0.20            & 0.24             & 0.48   & 0.27        & 0.14        & 0.28             & $-$0.44            & 0.06            & 0.36\\
HD 54810          & $-$0.26      & 0.05    & 0.09             & 0.23    & 0.33            & 0.24             & 0.50   & 0.27        & 0.24        & 0.28             & $-$0.41            & 0.09            & 0.36\\
HD 6186           & $-$0.30      & 0.05    & 0.04             & 0.27    & 0.32            & 0.27             & 0.44   & 0.27        & 0.22        & 0.28             & $-$0.40            & 0.10            & 0.38\\
HD 67447          & 0.00       & 0.08    & 0.71             & 0.28    & 0.69            & 0.29             & 0.77   & 0.15        & 0.25        & 0.17             & $-$0.06            & 0.44            & 0.32\\
HD 6833           & $-$0.69      & 0.05    & $-$0.18            & 0.23    & 0.49            & 0.24             & 0.29   & 0.14        & 0.46        & 0.15             & $-$0.47            & 0.03            & 0.27\\
HD 76294          & $-$0.12      & 0.09    & 0.15             & 0.23    & 0.25            & 0.25             & 0.52   & 0.16        & 0.12        & 0.18             & $-$0.37            & 0.13            & 0.28\\
HD 77912          & $-$0.07      & 0.08    & 0.12             & 0.25    & 0.17            & 0.26             & 0.71   & 0.15        & 0.26        & 0.17             & $-$0.59            & $-$0.09           & 0.29\\
HD 85503          & 0.44       & 0.12    & 0.41             & 0.28    & $-$0.05           & 0.30             & 0.87   & 0.29        & $-$0.09       & 0.32             & $-$0.46            & 0.04            & 0.41\\
KIC 10096113      & $-$0.74      & 0.06    & $-$0.52            & 0.23    & 0.20            & 0.24             & 0.22   & 0.15        & 0.44        & 0.16             & $-$0.74            & $-$0.24           & 0.27\\
KIC 11802968      & $-$0.11      & 0.06    & 0.01             & 0.24    & 0.10            & 0.25             & 0.74   & 0.27        & 0.33        & 0.28             & $-$0.73            & $-$0.23           & 0.36\\
KIC 1726211       & $-$0.66      & 0.04    & $-$0.49            & 0.23    & 0.15            & 0.23             & 0.30   & 0.27        & 0.44        & 0.27             & $-$0.79            & $-$0.29           & 0.36\\
KIC 4351319       & 0.29       & 0.08    & 0.51             & 0.26    & 0.20            & 0.27             & 0.88   & 0.28        & 0.07        & 0.29             & $-$0.37            & 0.13            & 0.38\\
KIC 5530598       & 0.34       & 0.10    & 0.54             & 0.27    & 0.18            & 0.29             & 0.95   & 0.29        & 0.09        & 0.30             & $-$0.41            & 0.09            & 0.39\\
KIC 5698156       & $-$1.37      & 0.05    & $-$1.12            & 0.17    & 0.23            & 0.18             & $-$0.55  & 0.14        & 0.30        & 0.15             & $-$0.57            & $-$0.07           & 0.22\\
KIC 6611219       & $-$1.30      & 0.07    & $-$0.52            & 0.15    & 0.76            & 0.17             & $-$0.04  & 0.15        & 0.74        & 0.16             & $-$0.48            & 0.02            & 0.21\\
KIC 7205067       & $-$0.02      & 0.06    & 0.40             & 0.22    & 0.40            & 0.23             & 0.59   & 0.14        & 0.09        & 0.16             & $-$0.19            & 0.31            & 0.26\\
KIC 8350894       & $-$1.09      & 0.05    & $-$0.47            & 0.15    & 0.75            & 0.16             & $-$0.13  & 0.14        & 0.59        & 0.15             & $-$0.34            & 0.16            & 0.20\\
BD +01 3070       & $-$1.36      & 0.14    & \textless{}~$-$0.26 &   --      & \textless{}~1.10 &       --          & $-$0.12  & 0.19 & 0.74        & 0.24             & \textless{}~$-$0.14 & \textless{}~0.36 & --\\
BD +05 4314       & $-$1.36      & 0.07    & \textless{}~$-$0.24 &   --     & \textless{}~1.10 &         --        & $-$0.35  & 0.15 & 0.49        & 0.16             & \textless{}~0.11  & \textless{}~0.61 & --\\
BD --01 1792       & $-$0.93      & 0.06    & \textless{}~$-$0.31 &   --     & \textless{}~0.60 &         --        & 0.08   & 0.14 & 0.49        & 0.15             & \textless{}~$-$0.39 & \textless{}~0.11 &--\\
HD 111721         & $-$1.40       & 0.07    & \textless{}~$-$0.67 &  --      & \textless{}~0.70 &        --         & $-$0.62  & 0.15 & 0.25        & 0.16             & \textless{}~$-$0.05 & \textless{}~0.45 &--\\
HD 35369          & $-$0.12      & 0.05    & \textless{}~0.61  &   --     & \textless{}~0.70 &       --          & 0.63   & 0.14   & 0.22        & 0.15             & \textless{}~$-$0.02 & \textless{}~0.48 &--\\
HD 37160          & $-$0.56      & 0.04    & \textless{}~0.26  &   --     & \textless{}~0.80 &        --         & 0.32   & 0.27 & 0.36        & 0.27             & \textless{}~$-$0.06 & \textless{}~0.44 &--\\
HD 58367          & $-$0.13      & 0.09    & \textless{}~0.20  &   --     & \textless{}~0.30 &        --         & 0.55   & 0.16 & 0.15        & 0.18             & \textless{}~$-$0.35 & \textless{}~0.15 &--\\
KIC 10737052      & $-$1.22      & 0.10     & \textless{}~$-$0.11 &   --     & \textless{}~1.10 &       --          & $-$0.1   & 0.16 & 0.61        & 0.19             & \textless{}~$-$0.01 & \textless{}~0.49 &--\\
KIC 5184073       & $-$1.42      & 0.02    & \textless{}~$-$0.50 &    --    & \textless{}~0.90 &        --         & $-$0.46  & 0.13 & 0.44        & 0.13             & \textless{}~$-$0.04 & \textless{}~0.46 &--\\
KIC 9583607       & $-$0.80       & 0.05    & \textless{}~$-$0.04 &   --     & \textless{}~0.74 &       --          & $-$0.06  & 0.14   & 0.22        & 0.15             & \textless{}~0.02  & \textless{}~0.52 &--\\
\hline
\end{tabular}
\begin{tabnote}
\raggedright
The units for all abundance-related columns are dex.
\end{tabnote}
\end{table*}

\subsubsection{Systematic errors from atmospheric parameters}\label{ssec:432}
\par For error estimation, we adopt the typical uncertainties in the atmospheric parameters 
derived from the TGVIT program (cf. \cite{2008PASJ...60..781T}): 100~K for $T_{\mathrm{eff}}$, 
0.2~dex for $\log g$, 0.1~dex for $\mathrm{[Fe/H]}$, and 0.1~km/s for $v_t$.
At first, we selected representative objects by effective temperature to cover all temperature range. Second, we changed each parameter by the typical error and re-estimated abundances for five stars (Table \ref{Tab:atmpara_representive_error}). For the remaining objects, we adopted the uncertainty estimated for the representative star with the most similar atmospheric parameters.
To decide the reference object for the remaining object, we normalized the atmospheric parameters using the mean and standard deviation of all samples and selected the one with the smallest RMS distance from the five representative values. The total uncertainty from atmospheric parameters for Th and Eu abundance is 0.15 dex and 0.16 dex on average. 
\par For objects that adopted previous studies' atmospheric parameters, we changed each parameter by the previous studies' error and reestimated abundances for each object.
Total uncertainty is adopted as the RMS of the random and systematic errors mentioned earlier. Total uncertainty of Th is 0.02 -- 0.28, Eu is 0.02 -- 0.29, and [Th/Eu] is 0.03 -- 0.41.

\subsubsection{Non-local thermodynamic equilibrium effects for abundances}\label{ssec:433}
According to 3D NLTE calculation, Eu abundance from 6645 \AA~ is less than 1D LTE calculation by 0.05 dex in [Fe/H] < $-$1 (\cite{2025MNRAS.538.3284S}).
On the other hand, the difference is no more than 0.1 dex in cool giants ($T_{\rm eff}$ \textdblhyphenchar~ 4500 K, log$g$ \textdblhyphenchar~ 1.5) at [Fe/H] \textgreater~--2 (\cite{2025A&A...693A.211G}). For Th abundance from 5989 \AA, there is no calculation of 3D NLTE effect; the uncertainty of NLTE effect is not included in this work. For reference, in the case of Th abundance from 4019 \AA, Th abundance is at most 0.21 dex higher than the 1D LTE calculation in a cool metal-poor giant (\cite{2012A&A...540A..98M}). 
The present analysis assumes LTE for both Th and Eu abundance determinations.
Since NLTE calculations for the 5989 \AA~ line are not yet available, a systematic uncertainty of several tenths of a dex cannot be completely ruled out.
Future theoretical calculations of the 5989 \AA~ are highly desirable.
\par Final Th and Eu abundances and total error are shown in Table \ref{Table:final_abundance}.
\subsubsection{Comparison with previous studies}\label{ssec:434}
We compared our derived Th and Eu abundances with previous studies. First, we verified the validity of our analysis by comparing our results with those of \citet{2025A&A...699A.276A}, which analyzed high-metal-content objects using the same method. For the four objects in common, we summarized the differences between our results and those of the previous study in Table 5. Differences of 0.1 dex or more were observed for [Eu/H] in KIC 5530598, KIC 4351319, and HD 124897, and for [Th/H] in KIC 7205067. Possible causes include uncertainties in atmospheric parameters, in the level of continuous emission, and in the broadening of absorption lines. Although systematic errors of approximately 0.1 -- 0.2 dex are expected, the analysis method was applied consistently, and the precision is sufficient to discuss the overall distribution and metal abundance dependence.
Furthermore, for the \textit{r}-process enhanced metal-poor star (HD 221170), we confirmed that the 5989 \AA~ line is detectable, consistent with previous studies. While confirmation is difficult on the high-metallicity side due to the large uncertainty at 4019 \AA, we consider that systematic errors due to the lines do not need to be taken into account, at least based on the results from the metal-deficient region.
Therefore, in the following discussion, we combine our results with those of previous studies in the following section.

\begin{table*}[]
\caption{Differences of Th and Eu abundances with \citet{2025A&A...699A.276A}.}
\centering
\begin{tabular}{lrrrrrr}
\hline
object         & $\Delta${[}Fe/H{]} (dex) & $\Delta T_{\rm eff}$ (K)& $\Delta$log$g$ (dex)& $\Delta v_{\rm t}$ (km/s) & $\Delta${[}Eu/H{]} (dex) & $\Delta${[}Th/H{]} (dex) \\
\hline
KIC 5530598     & $-$0.03 & 53    & 0.19  & 0.16 & 0.18        & 0.00           \\
KIC 4351319     & 0.00     & 39    & 0.18  & 0.04 & 0.21        & 0.08        \\
KIC 7205067     & $-$0.04 & $-$9    & 0.08  & 0.13 & 0.09        & 0.18        \\
HD 124897 (HDS) & $-$0.06 & $-$28   & $-$0.07 & 0.08 & $-$0.12       & $-$0.04       \\
\hline
\end{tabular}
\end{table*}

\section{Result and discussion}\label{sec:5}
 \subsection{Detection of the 5989 \AA~ line and determination of Th abundances}\label{ssec:51}
Th was detected in 32 objects, and an upper limit was obtained in 10 objects in $-2.2 < $ [Fe/H] $< +0.4$. This represents a new determination of Th abundances for 27 objects. For HD 221170, we confirmed the same [Th/Fe] obtained mainly by using the 4019 \AA~ line (\cite{2006ApJ...645..613I}). Therefore, it indicates that there are few differences in abundance between 4019 \AA~ and 5989 \AA. In our analysis, Th abundance was detected in stars whose atmospheric parameters were 4159 -- 5055 K for $T_{\rm eff}$, 1.00 -- 3.66 for log $g$, 0.74 -- 2.55 km/s for $v_{\rm t}$ with the 5989 \AA~ line. The 5989 Å line is therefore useful for stars within this parameter range.

\begin{figure}[H]
 \begin{center}
  \includegraphics[width=8cm]{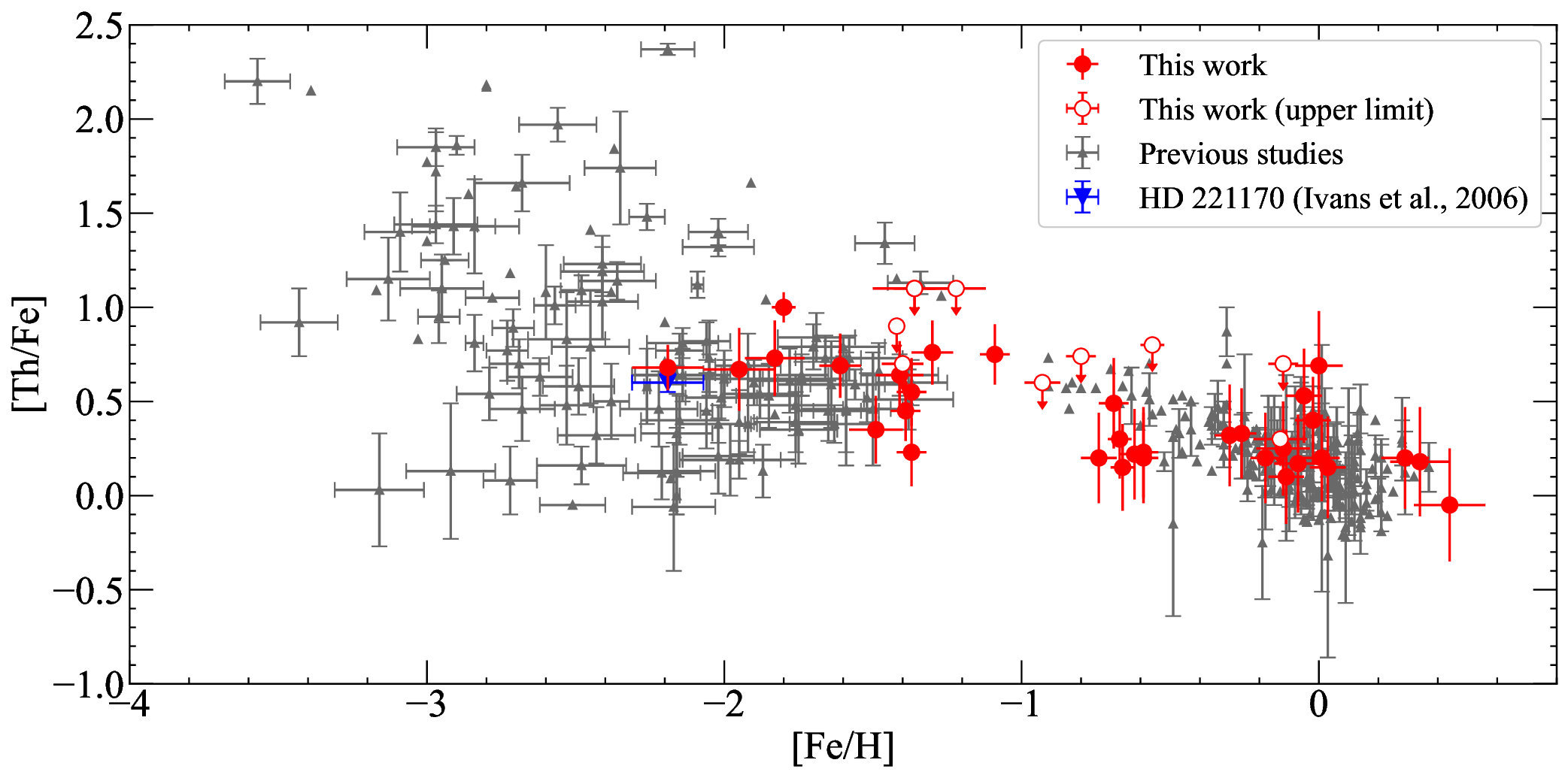}  
\caption{Observed [Th/Fe] ratios as a function of metallicity [Fe/H]. 
The red-filled circles represent our program stars. Upper limits are indicated by downward arrows.
Previous study values are obtained from \citet{2026arXiv260412892S}, \citet{2022MNRAS.516.3786M}, \citet{2025A&A...699A.276A}, and the SAGA database (\cite{2008PASJ...60.1159S}), 
are shown in black dots. The blue point shows HD 221170 from \citet{2006ApJ...645..613I}. For HD 124897, the HDS data is displayed as representative data.
{Alt text: Result of Th and Eu abundances.}
}
 \label{fig:Th/Fe_result}
 \end{center}
\end{figure}

\begin{figure}[H]
 \begin{center}
    \includegraphics[width=8cm]{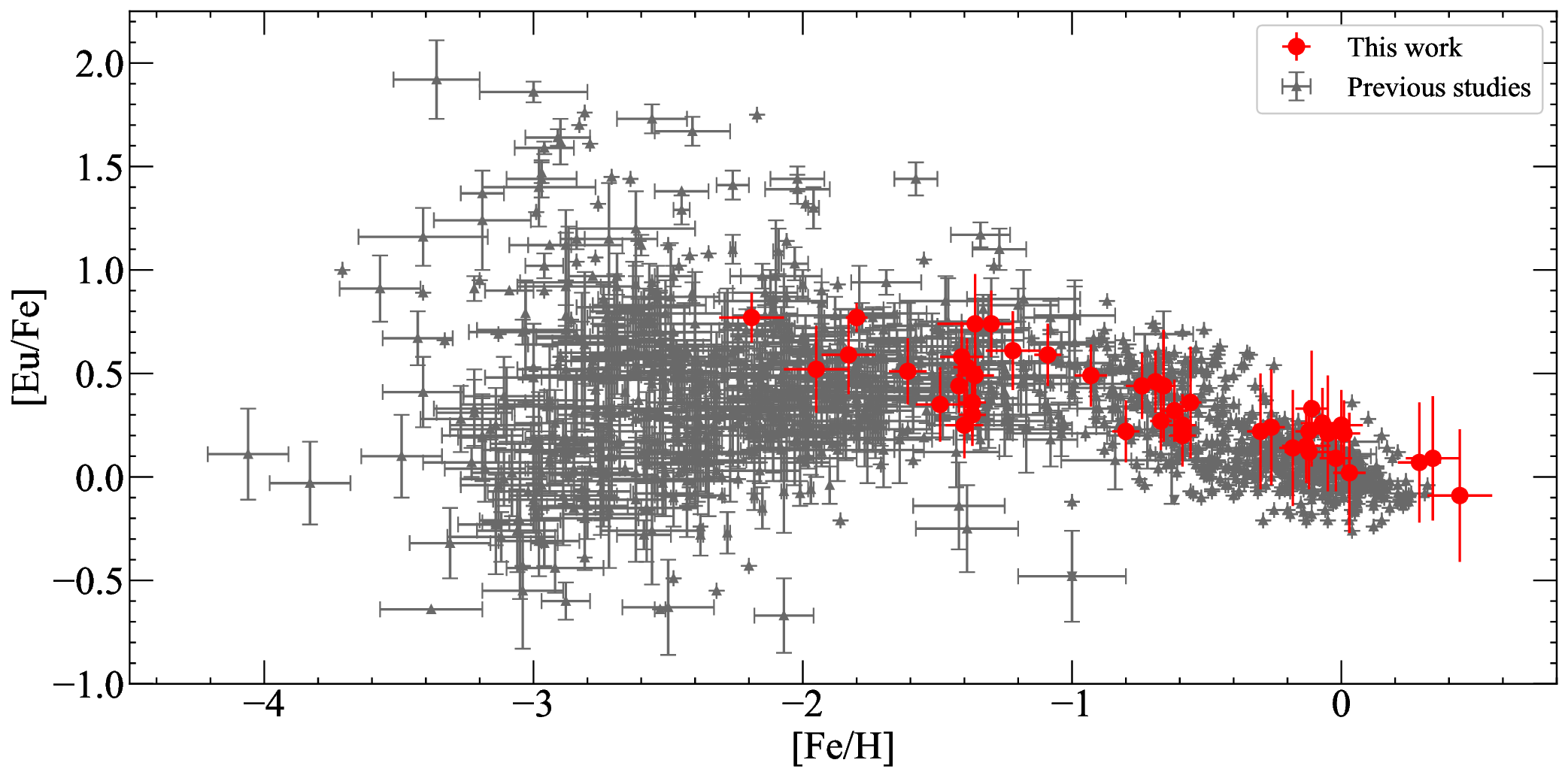} 
\caption{Observed [Eu/Fe] ratios as a function of metallicity [Fe/H]. 
The red-filled circles represent our program stars. 
Previous study values are obtained from the SAGA database (\cite{2008PASJ...60.1159S}).
{Alt text: Result of Eu abundances.}
}
 \label{fig:Eu/Fe_result}
 \end{center}
\end{figure}

 \subsection{Abundance trends of Th and Eu with metallicity}\label{ssec:52}
Figure \ref{fig:Th/Fe_result} shows the abundance trends of Th, whereas Figure \ref{fig:Eu/Fe_result} shows that of Eu.
The [Th/Fe] ratio decreases with increasing metallicity, whereas the [Eu/Fe] ratio follows the well-established trend reported in previous studies.
These behaviors are broadly consistent with the Galactic chemical evolution of neutron-capture elements and indicate that the present abundance analysis is reliable.
In contrast to [Th/Fe] and [Eu/Fe], the [Th/Eu] ratio exhibits little dependence on metallicity throughout the mildly metal-poor regime (shown in Figure \ref{fig:Th/Eu_result}). 
Although the measurement uncertainties remain non-negligible, no systematic trend with [Fe/H] is evident within the observed metallicity range.
Another important observational result is that no clear actinide-boost star was identified in the present sample.
Previous studies of extremely metal-poor stars have reported that actinide-boost stars constitute a significant fraction of \textit{r}-process-enhanced objects, whereas we found no such object among the mildly metal-poor stars analyzed here.
Consequently, the star-to-star scatter of [Th/Eu] appears considerably smaller than that observed among extremely metal-poor stars. 
These observational results suggest that the diversity of [Th/Eu] ratios is already substantially reduced in the mildly metal-poor regime. 
The possible astrophysical implications of this behavior are discussed in the following section.

\subsection{Implications for galactic chemical evolution and \textit{r}-process enrichment}\label{ssec:53}
In $-2 <$ [Fe/H] $< -1.4$, both [Th/Fe] and [Eu/Fe] decrease with increasing metallicity. The similar metallicity dependence of [Th/Fe] and [Eu/Fe] indicates that Th and Eu broadly follow similar chemical-evolutionary trends in this metallicity range. 
Such behavior is broadly consistent with Galactic chemical evolution, in which the relative contributions of \textit{r}-process enrichment and Fe production evolve with metallicity (e.g. \cite{2019ApJ...875..106C}).
One possible explanation for this downward trend is the delayed contribution of Fe from SNe Ia. Since SNe Ia supply Fe but do not significantly produce \textit{r}-process elements, an increase in their contribution as Galactic chemical evolution progresses would tend to lower the abundance ratios of Th and Eu relative to Fe.
However, the relative contribution of SNe Ia to Fe enrichment depends on the stellar population, and normal core-collapse supernovae may still provide a substantial fraction of Fe at these metallicities (e.g. \cite{2021MNRAS.506.5410I}). Therefore, the contribution of SNe Ia should be regarded as one possible factor in the observed decrease rather than its sole explanation.
\par
\begin{figure}[H]
 \begin{center}
  \includegraphics[width=8cm]{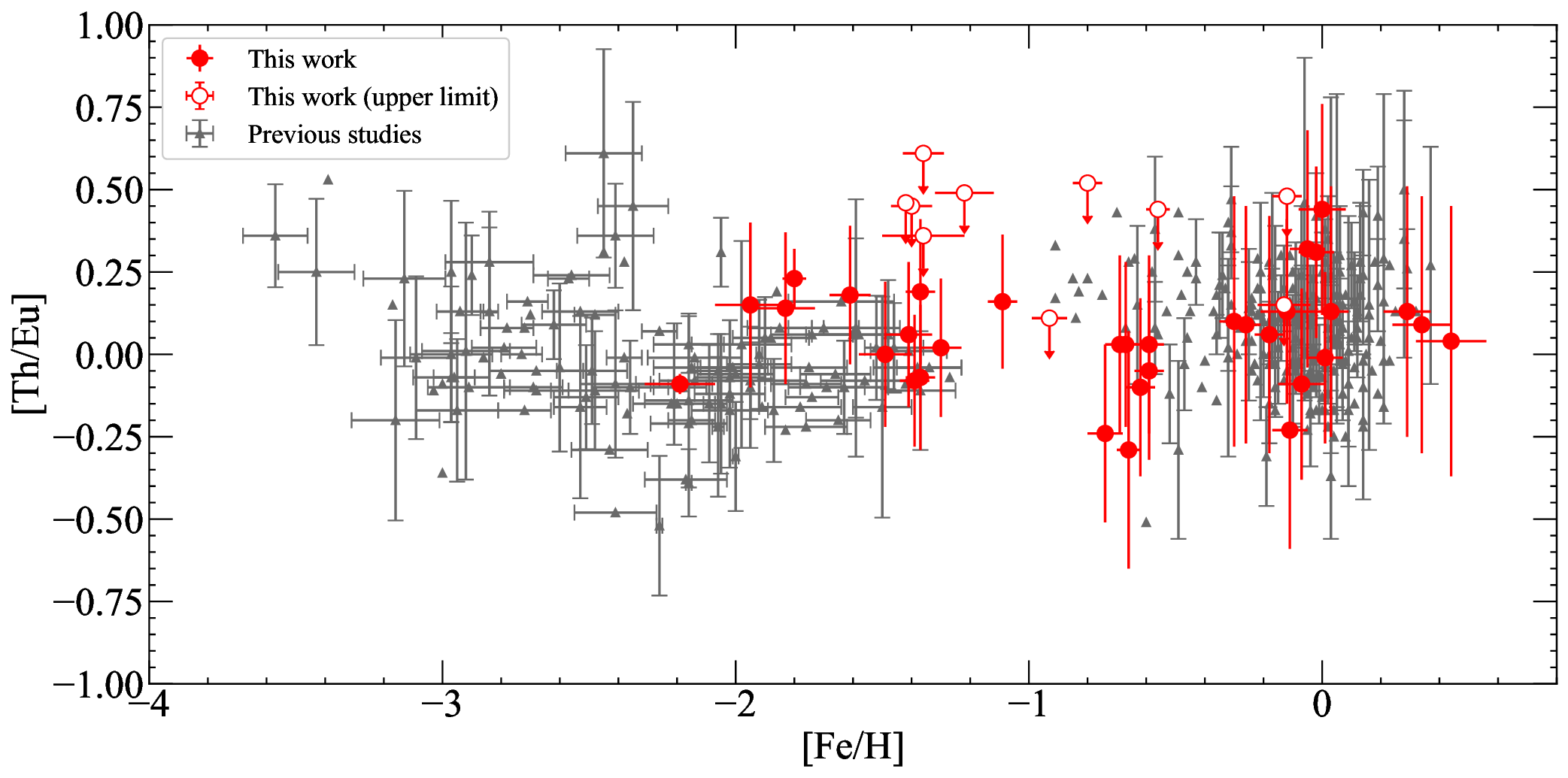} 
 \end{center}
\caption{Observations of [Th/Eu] as a function of metallicity; [Fe/H]. Red points show
this analysis, and red circles show upper limits of Th abundance. Gray points show previous studies (\cite{2026arXiv260412892S}, \cite{2022MNRAS.516.3786M}, \cite{2025A&A...699A.276A}, and the SAGA database ; \cite{2008PASJ...60.1159S}).
{Alt text: Result of [Thorium over Europium] ratio.}
}\label{fig:Th/Eu_result}
\end{figure}

\begin{figure}[H]
 \begin{center}
  \includegraphics[width=8cm]{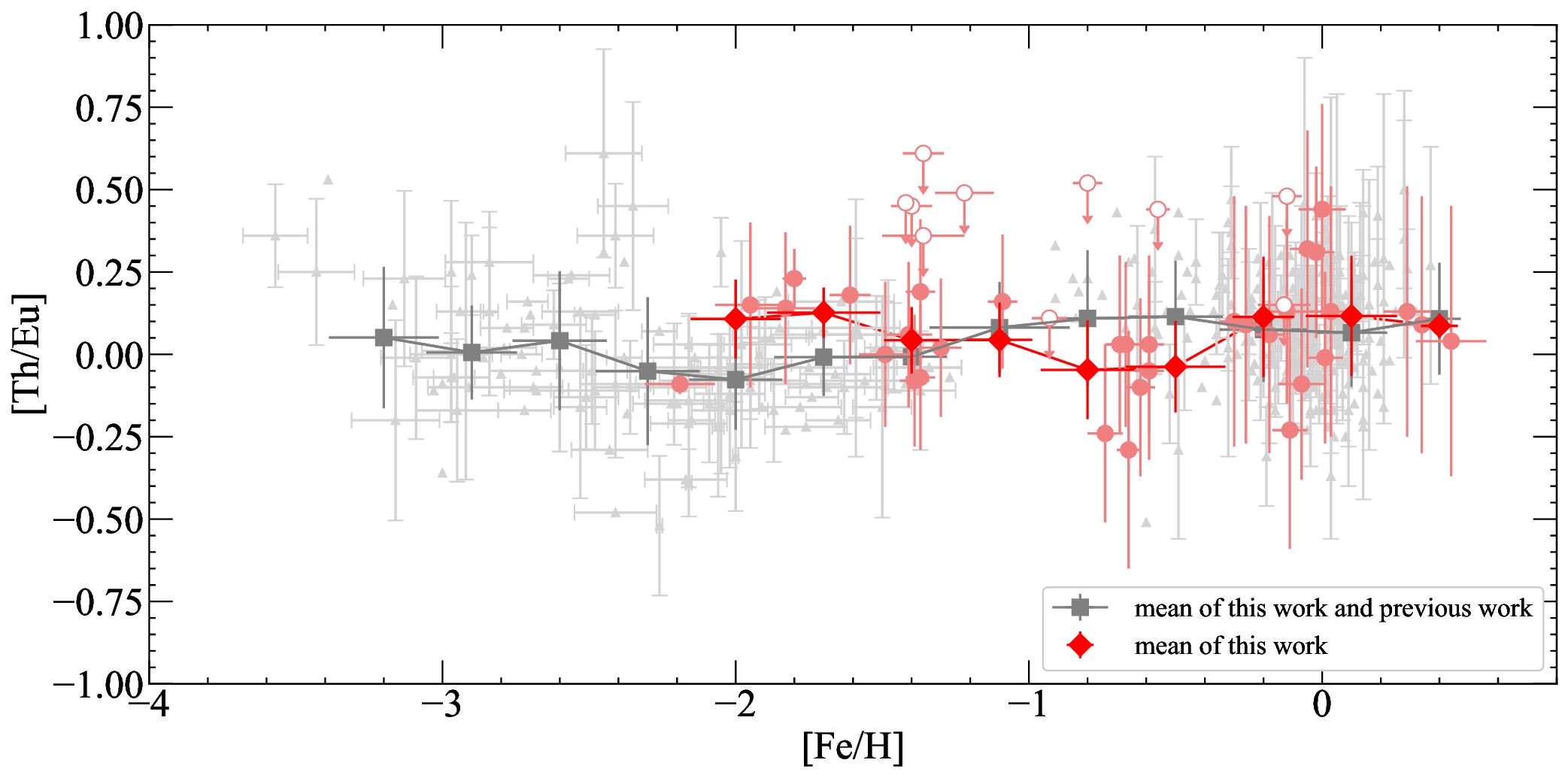} 
 \end{center}
\caption{Trend of [Th/Eu] as a function of metallicity; [Fe/H]. Red points and solid line show the average of stellar [Th/Eu] from this work at each 0.6 dex interval. 
Likewise, gray points and the solid line show the value from this work and previous work. Note that for adjacent intervals, we allowed a 0.3 dex overlap only on the low [Fe/H] side. {Alt text: Trend of [Thorium over Europium] ratio. }
}\label{fig:Th/Eu_trend}
\end{figure}

 Figure \ref{fig:Th/Eu_trend} summarizes the mean abundance ratios in each metallicity bin and highlights the overall trends that are less apparent in the individual stellar measurements shown in Figure \ref{fig:Th/Eu_result}.
In contrast to [Th/Fe] and [Eu/Fe], the [Th/Eu] ratio exhibits little dependence on metallicity throughout the mildly metal-poor regime, and shows that the mean values remain nearly constant within the uncertainties. 
Moreover, no clear actinide-boost star was identified in the present sample. 
Previous studies of \textit{r}-process-enhanced extremely metal-poor stars have shown that actinide-boost stars constitute a significant fraction of the population and produce a large star-to-star scatter in [Th/Eu].
Taken together, the absence of actinide-boost stars and the nearly constant [Th/Eu] ratios indicate that the pronounced abundance diversity characteristic of extremely metal-poor stars is no longer evident in the mildly metal-poor regime. 
This transition provides an important clue to understanding how \textit{r}-process enrichment evolved during the early chemical evolution of the Galaxy.
\par Several mechanisms may account for this behavior. 
One possible explanation for this transition is the progressive homogenization of the interstellar medium (ISM) as Galactic chemical evolution proceeded.
As the number of enrichment events increased, the chemical composition of newly formed stars would have reflected the cumulative contributions from many independent \textit{r}-process events, thereby reducing the abundance scatter inherited from individual nucleosynthetic events. 
Another possibility is that the dominant \textit{r}-process events contributing to the mildly metal-poor population shared a similar initial production ratio of Th relative to Eu. 
In this case, the nearly constant [Th/Eu] ratio would be consistent with a relatively uniform initial Th/Eu production ratio among the dominant \textit{r}-process sources.
Although the mean [Th/Eu] ratio is approximately constant, Figure \ref{fig:Th/Eu_trend} may also suggest a slight decrease toward higher metallicity. 
Since $^{232} ${Th} is radioactive with a half-life of 14.05 Gyr, this tendency could partly reflect radioactive decay associated with stellar age rather than variations in the nucleosynthetic production ratio itself. 
Older stars are expected to have experienced greater radioactive decay, leading to lower present-day [Th/Eu] ratios. 
However, stellar ages have not been determined for the present sample, and therefore the relative importance of radioactive decay, Galactic chemical evolution, and possible variations in the initial production ratio cannot yet be distinguished.
Future investigations combining homogeneous Th abundance measurements with precise stellar ages derived from Gaia and asteroseismology, together with Galactic chemical evolution models, will be important for clarifying the origin of the remarkably uniform [Th/Eu] ratios observed in MMP stars. 
Such studies will help determine whether the actinide-boost phenomenon is confined to the earliest stages of Galactic evolution or whether other physical mechanisms are responsible for the transition from the large abundance scatter observed in extremely metal-poor stars to the chemically more homogeneous population studied here.

\section{Summary}\label{sec:6}
In this study, we determined Thorium abundances for 42 mildly metal-poor (MMP) stars using the $\mathrm{Th~\textsc{ii}}$ 5989 \AA~ line, substantially increasing the number of stars with reliable Th abundance measurements in this metallicity regime. 
The abundance trends of [Th/Fe] and [Eu/Fe] are consistent with previous studies of Galactic chemical evolution, demonstrating that the present abundance analysis provides reliable Th and Eu abundances for MMP stars.
\par One of the principal results of this study is that the [Th/Eu] ratio exhibits little dependence on metallicity throughout the MMP regime. 
In addition, no clear actinide-boost star was identified in our sample. 
In contrast to the large star-to-star scatter in [Th/Eu] reported for \textit{r}-process-enhanced extremely metal-poor (EMP) stars, the MMP stars analyzed here show a much smaller dispersion. 
These observational results suggest that the diversity of Th/Eu abundance ratios observed in the earliest stellar populations becomes considerably reduced in the mildly metal-poor regime.
Several mechanisms may contribute to this behavior. 
Progressive homogenization of the interstellar medium during Galactic chemical evolution would reduce abundance variations inherited from individual \textit{r}-process events, while a relatively uniform initial production ratio of Th relative to Eu among the dominant \textit{r}-process sites could also account for the observed constancy of [Th/Eu]. 
In addition, the slight decrease in the mean [Th/Eu] ratio toward higher metallicity may partly reflect the effects of radioactive decay associated with stellar age. 
Although the present data do not allow these possibilities to be distinguished, they provide new observational constraints on the chemical evolution of \textit{r}-process elements in the early Galaxy.
\par Future studies combining homogeneous Th abundance measurements with precise stellar ages derived from Gaia and asteroseismology, together with Galactic chemical evolution models, will help clarify the origin of the remarkably uniform [Th/Eu] ratios observed in mildly metal-poor stars. 
The present results are consistent with a scenario in which the large star-to-star diversity in \textit{r}-process enrichment observed among EMP stars becomes progressively reduced as Galactic chemical evolution proceeds, although further observational and theoretical studies are required to establish the underlying physical mechanisms.

\begin{ack}
We thank the staff of the Nishi-Harima Astronomical Observatory and Gunma Astronomical Observatory for their support in the observations and data analysis, and the Optical and Infrared Synergetic Telescopes for Education and Research (OISTER) program funded by the MEXT of Japan.
Based in part on data collected at Subaru Telescope and obtained from the SMOKA, which is operated by the Astronomy Data Center, National Astronomical Observatory of Japan.
This research has made use of the SIMBAD database, operated at CDS, Strasbourg, France.
We are grateful to Wako Aoki for his helpful support during the early stages of this work. We thank the anonymous referee for the careful reading of the manuscript and constructive comments that helped improve the manuscript.
\end{ack}
\section*{Funding}
 This research was supported by the JST SPRING program (Grant Number JPMJSP2175) and JSPS KAKENHI (Grant Numbers 21H04499, 25K01046).

\appendix 

    \section{Comparison with previous atmospheric-parameter determinations}
\begin{figure*}[]
 \begin{center}

  \includegraphics[width=16cm]{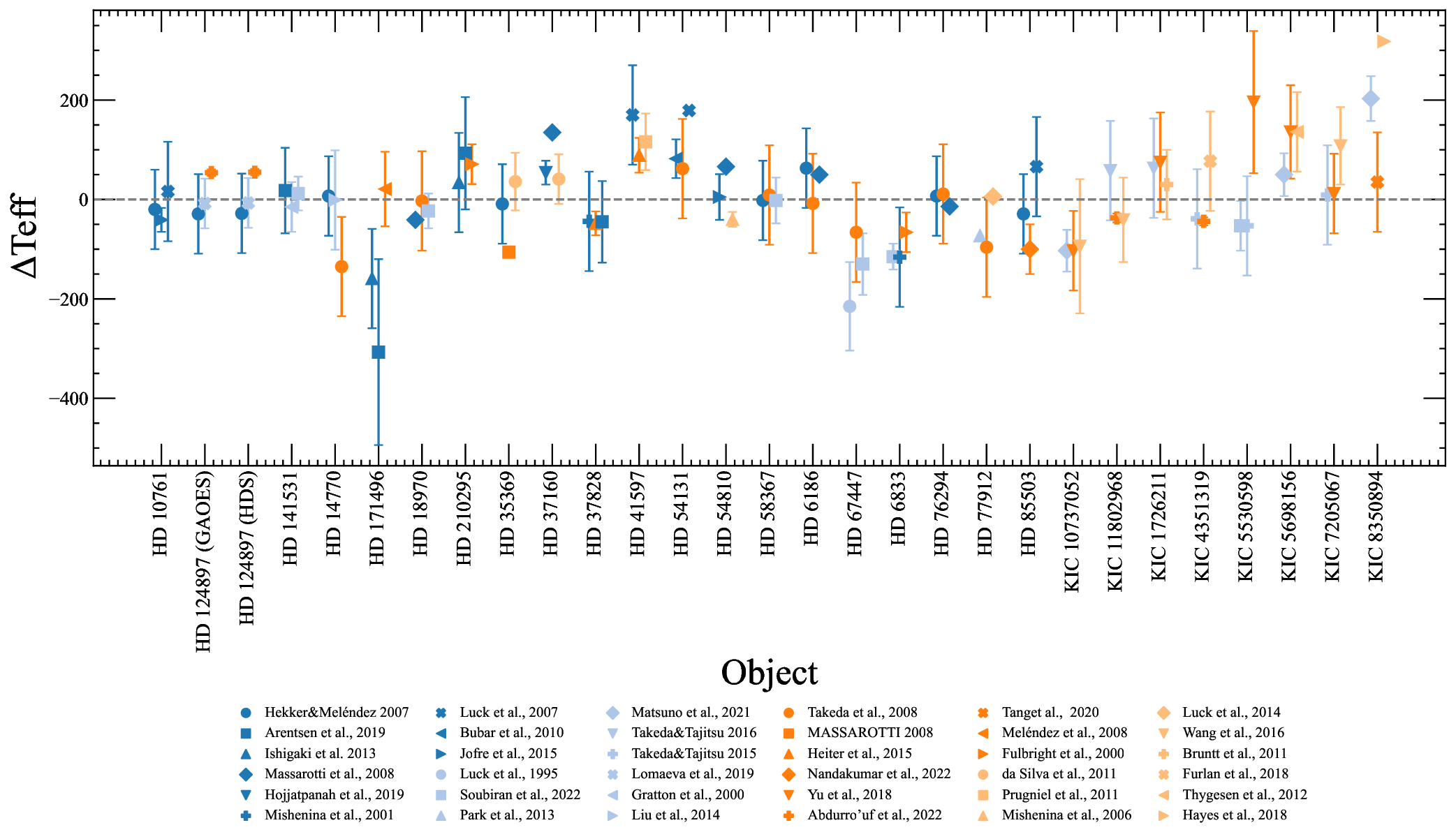} 
 \end{center}
\caption{Effective temperature of each star compared with three previous studies.
{Alt text: Comparison of effective temperature with previous studies. } 
}
\label{fig:compTeff}
\end{figure*}

\begin{figure*}[]
 \begin{center}
  \includegraphics[width=16cm]{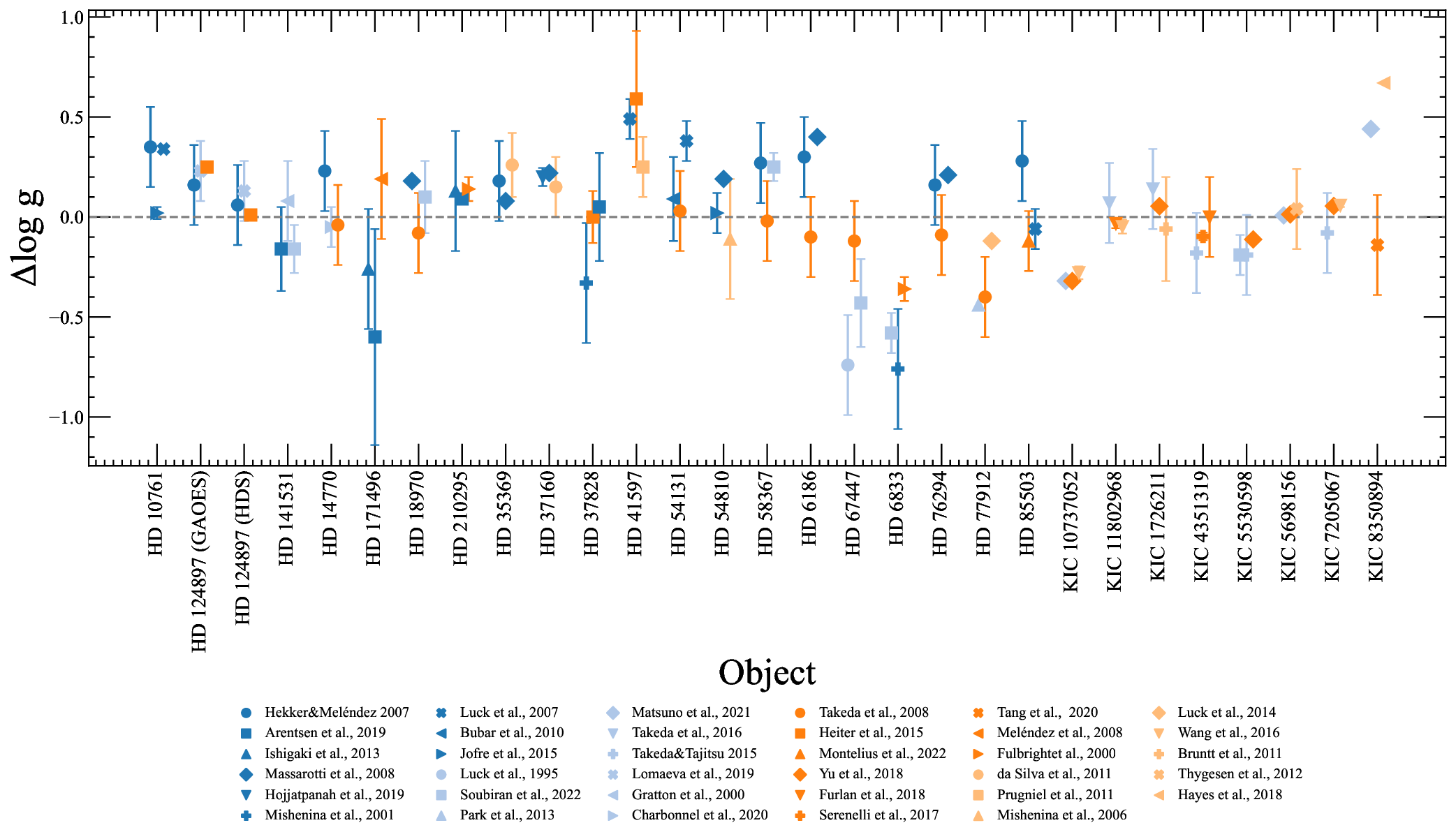} 
 \end{center}
\caption{Surface gravity of each star compared with three previous studies.
{Alt text: Comparison of surface gravity with previous studies. } 
}  
\label{fig:complogg}
\end{figure*}

\begin{figure*}[]
 \begin{center}
  \includegraphics[width=16cm]{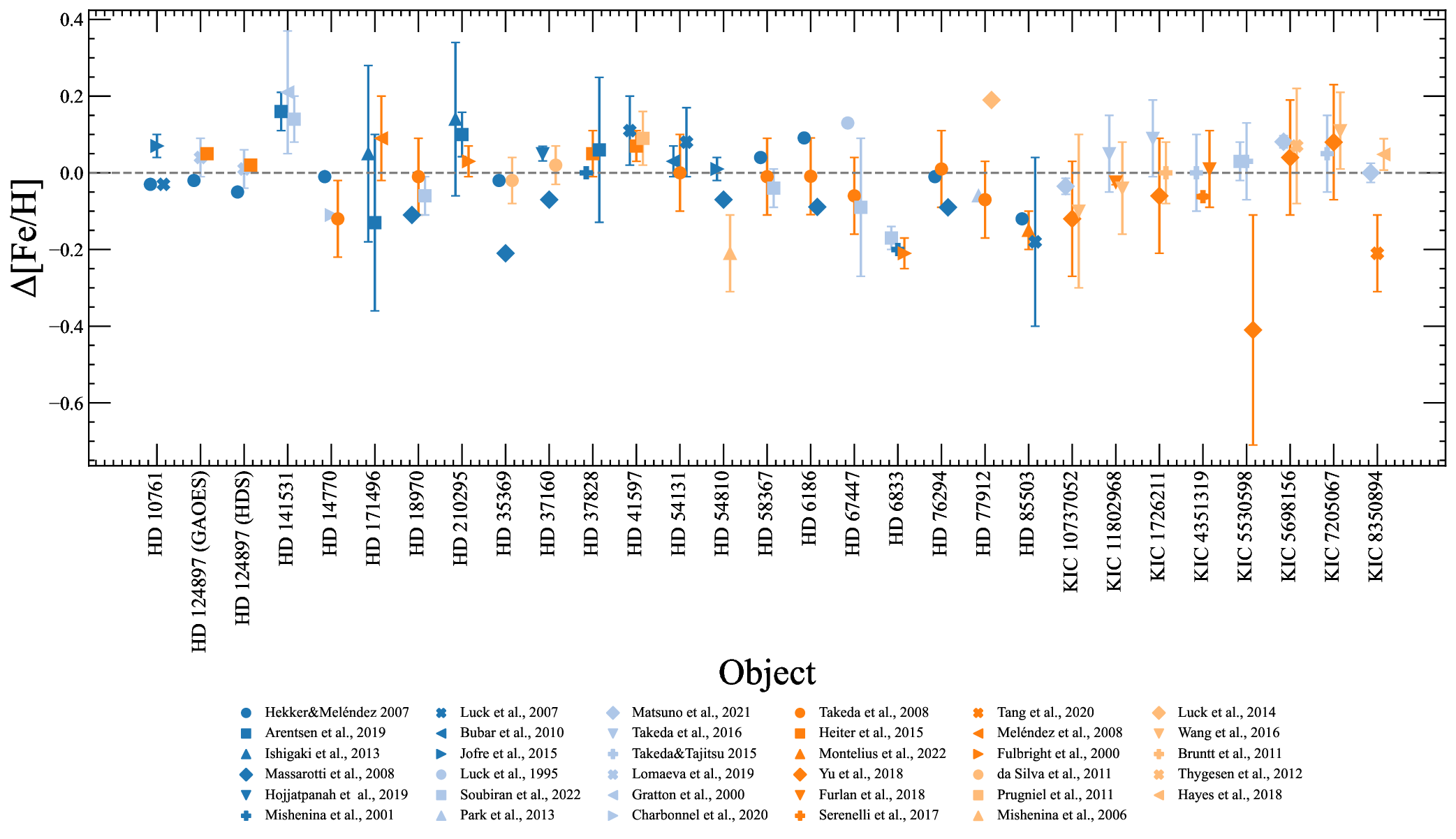}

 \end{center}
 \caption{Metallicity ([Fe/H]) of each star compared with three previous studies.
 {Alt text: Comparison of Metallicity ([Fe over H]) with previous studies. } 
 }  
\label{fig:comp[Fe/H]}
\end{figure*}

\begin{figure*}[]
 \begin{center}
  \includegraphics[width=16cm]{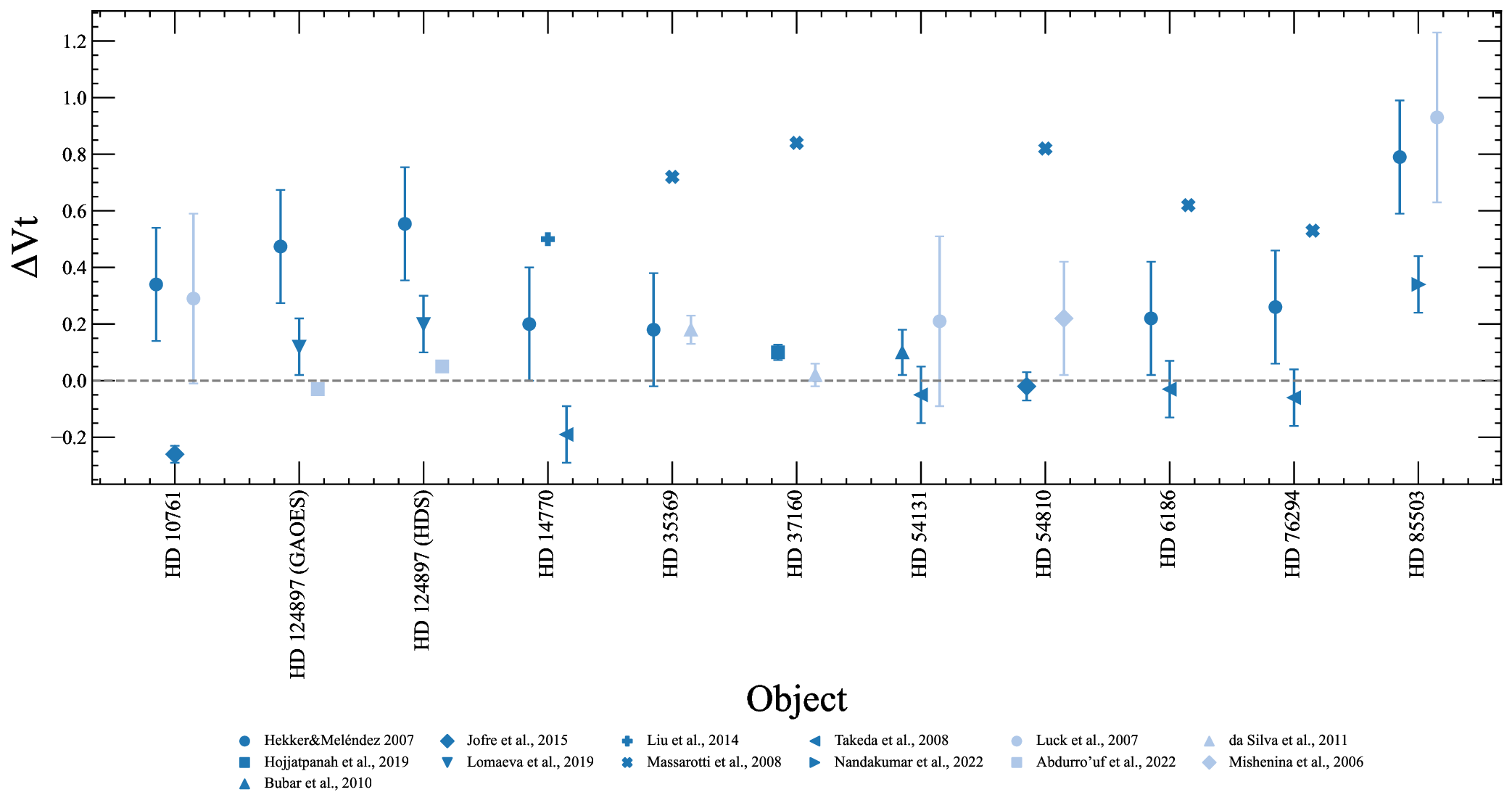} 
 \end{center}
 \caption{Microturbulent velocity of each star compared with three previous studies.  \citet{2008PASJ...60..781T} determined the atmospheric parameters using the same method, and our results are consistent with theirs. Although there appears to be significant variation in $v_{\rm t}$, this is because of the method to determine it. \citet{2008AJ....135..209M} assumed  $v_t$=2 km/s for all objects. Although HD85503 is less than the previous study by about 0.9 km/s, it makes no abundance change for Th and only 0.06 dex for Eu.
 {Alt text: Comparison of microturbulent velocity with previous studies. } 
 } 
\label{fig:compVt}
\end{figure*}

    We compared $T_{\rm eff}$ (Figure \ref{fig:compTeff}), $\log g$ (Figure \ref{fig:complogg}), and [Fe/H] (Figure \ref{fig:comp[Fe/H]}) with three previous studies for 28 stars and $v_{\rm t}$ (Figure \ref{fig:compVt}) for 11 stars. Almost all objects are consistent with previous studies, this estimation is reliable, and we can discuss uniformly.

\clearpage

 \textbf{\section{All spectra; Europium}}
    Synthetic spectra fits for the Eu line for all samples (Figure \ref{fig:Eu_allspectra_1}).

 \begin{figure*}
\centering

\includegraphics[width=0.24\textwidth]{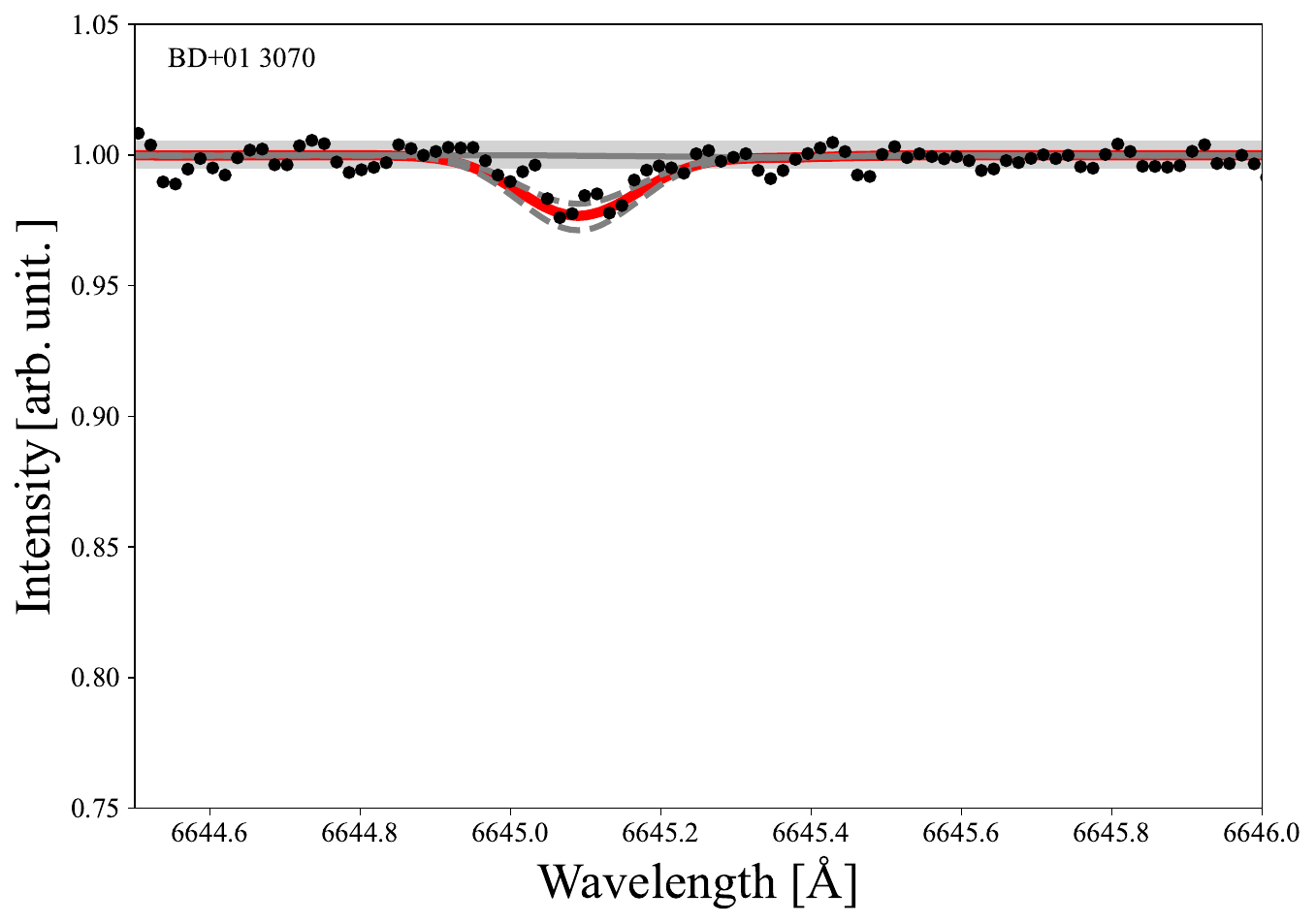}
\includegraphics[width=0.24\textwidth]{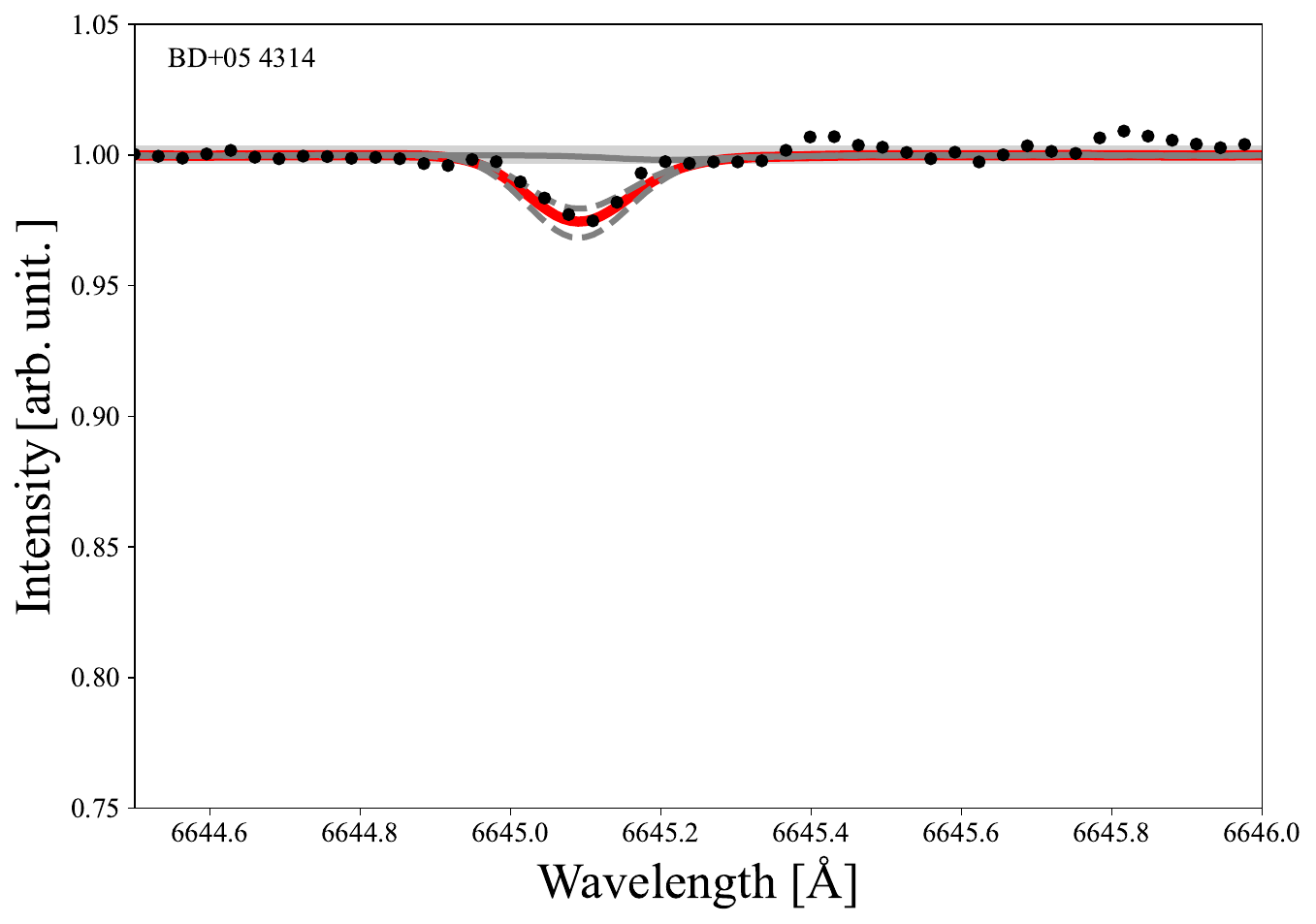}
\includegraphics[width=0.24\textwidth]{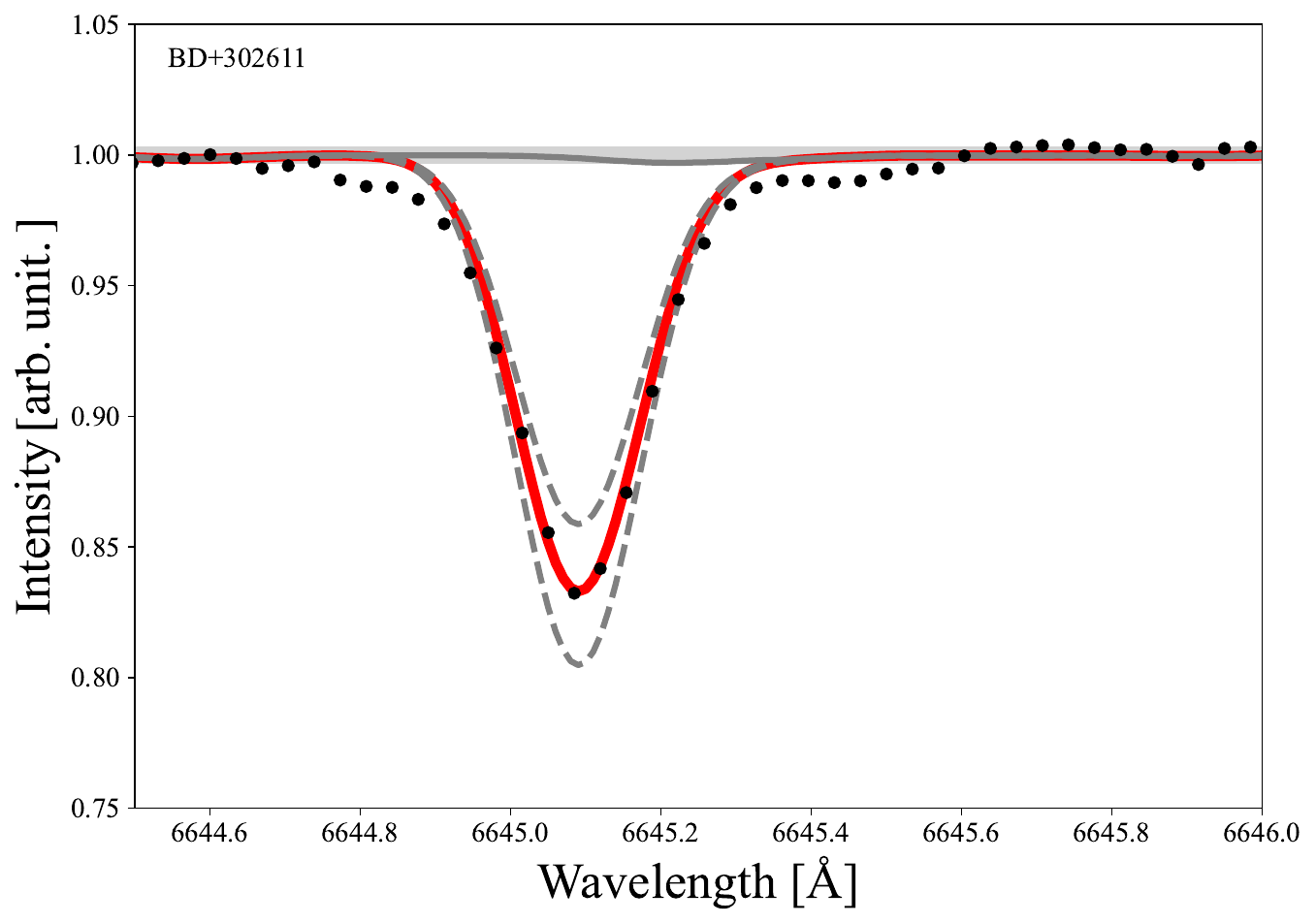}
\includegraphics[width=0.24\textwidth]{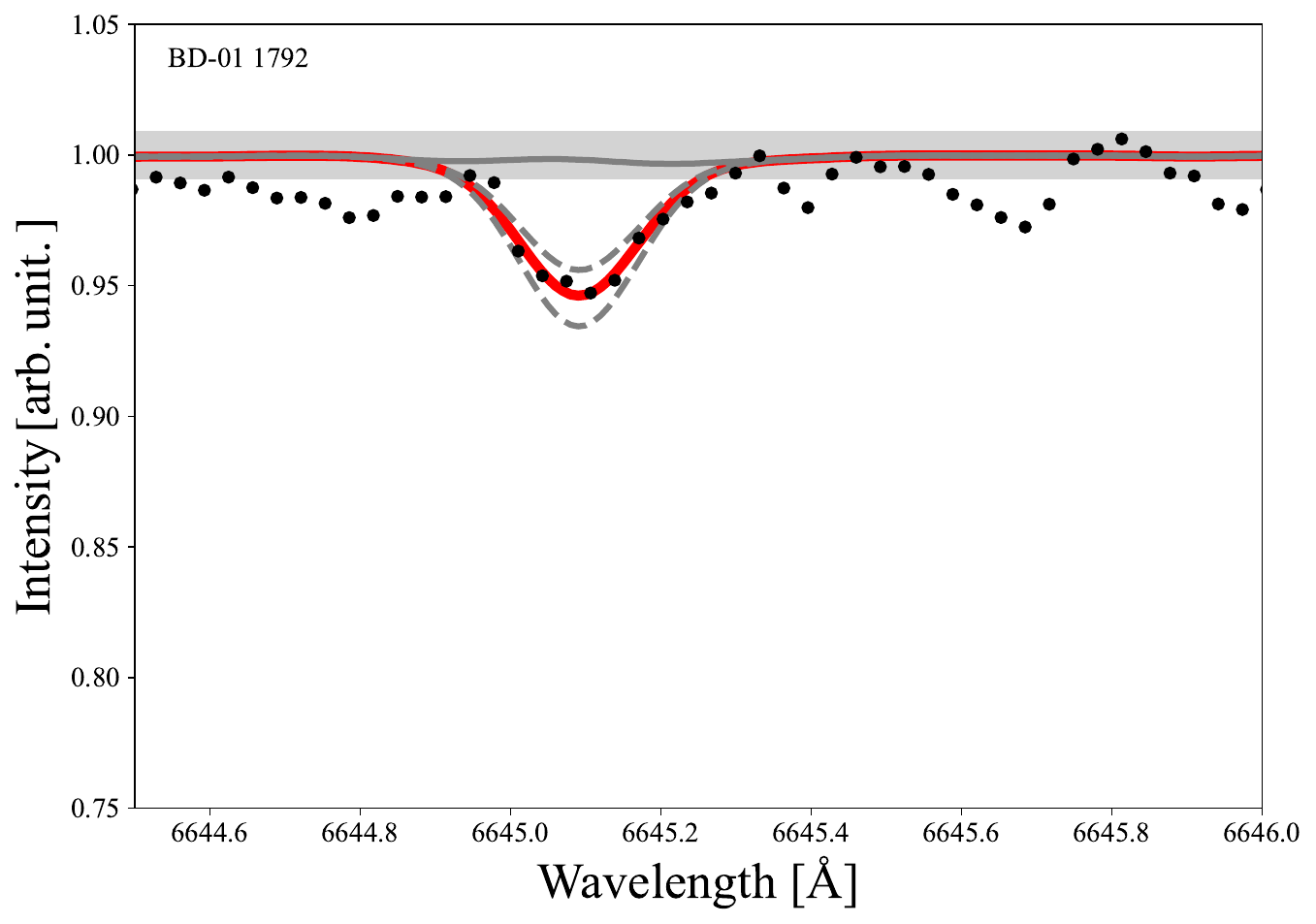}
\\[2mm]

\includegraphics[width=0.24\textwidth]{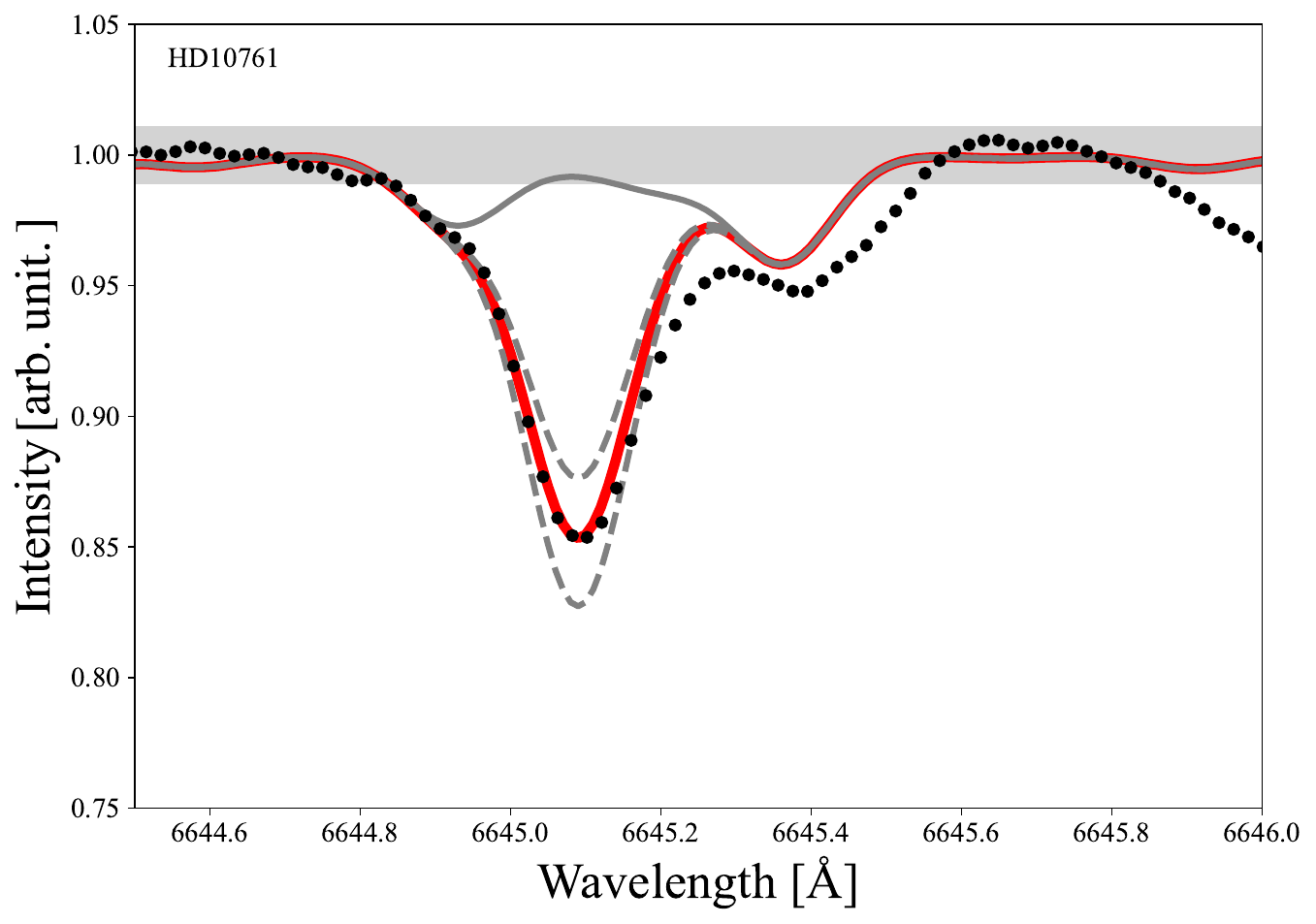}
\includegraphics[width=0.24\textwidth]{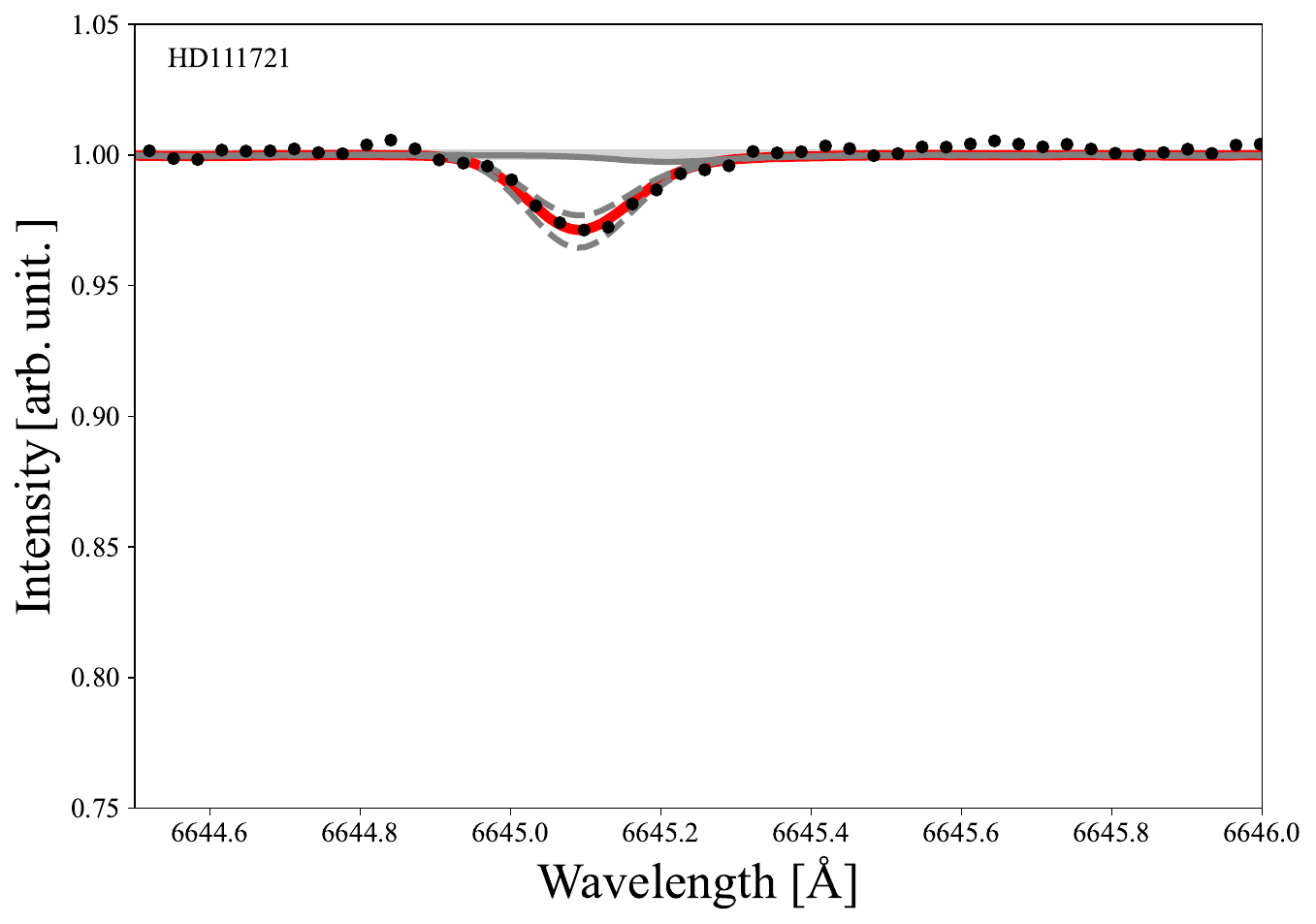}
\includegraphics[width=0.24\textwidth]{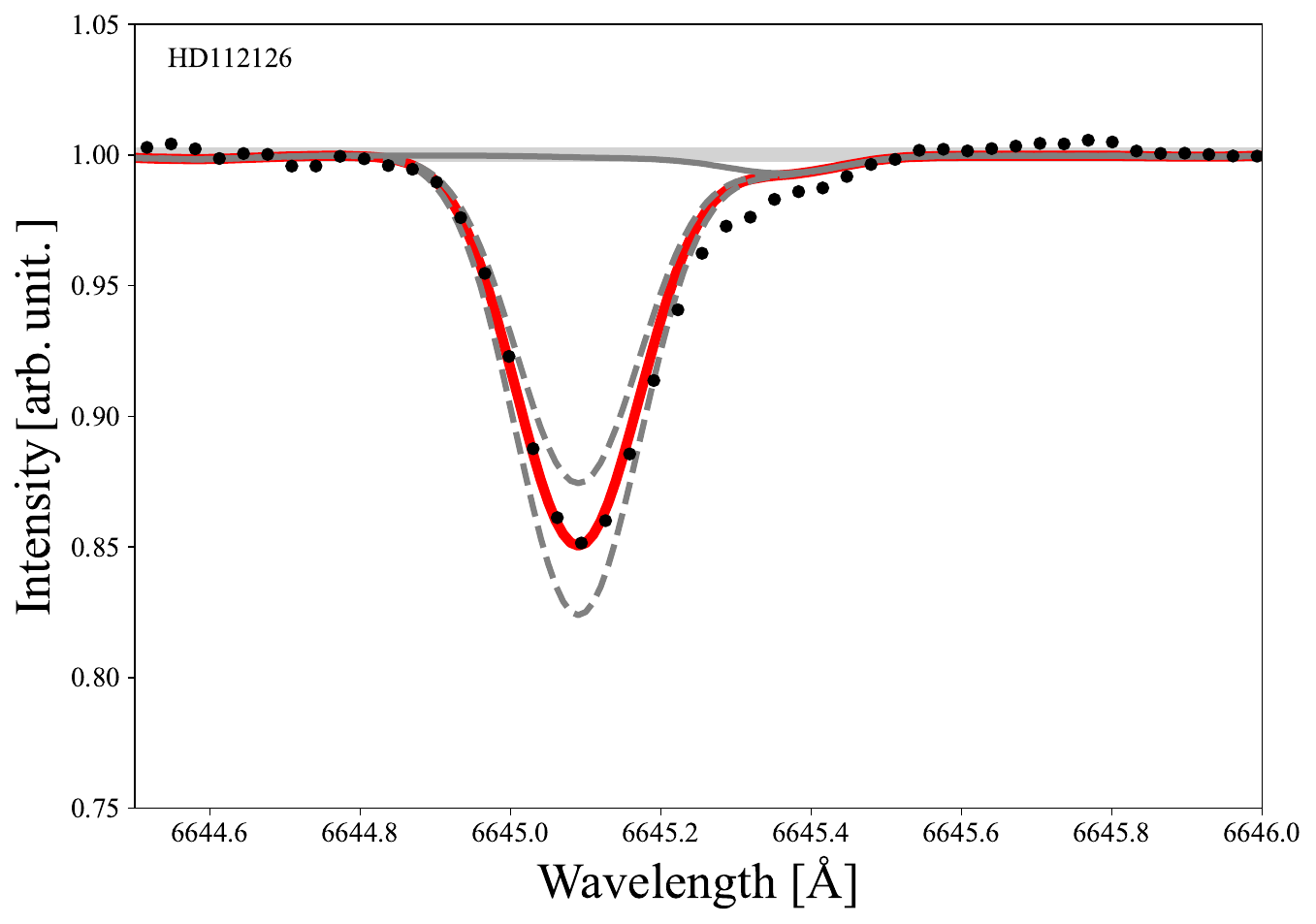}
\includegraphics[width=0.24\textwidth]{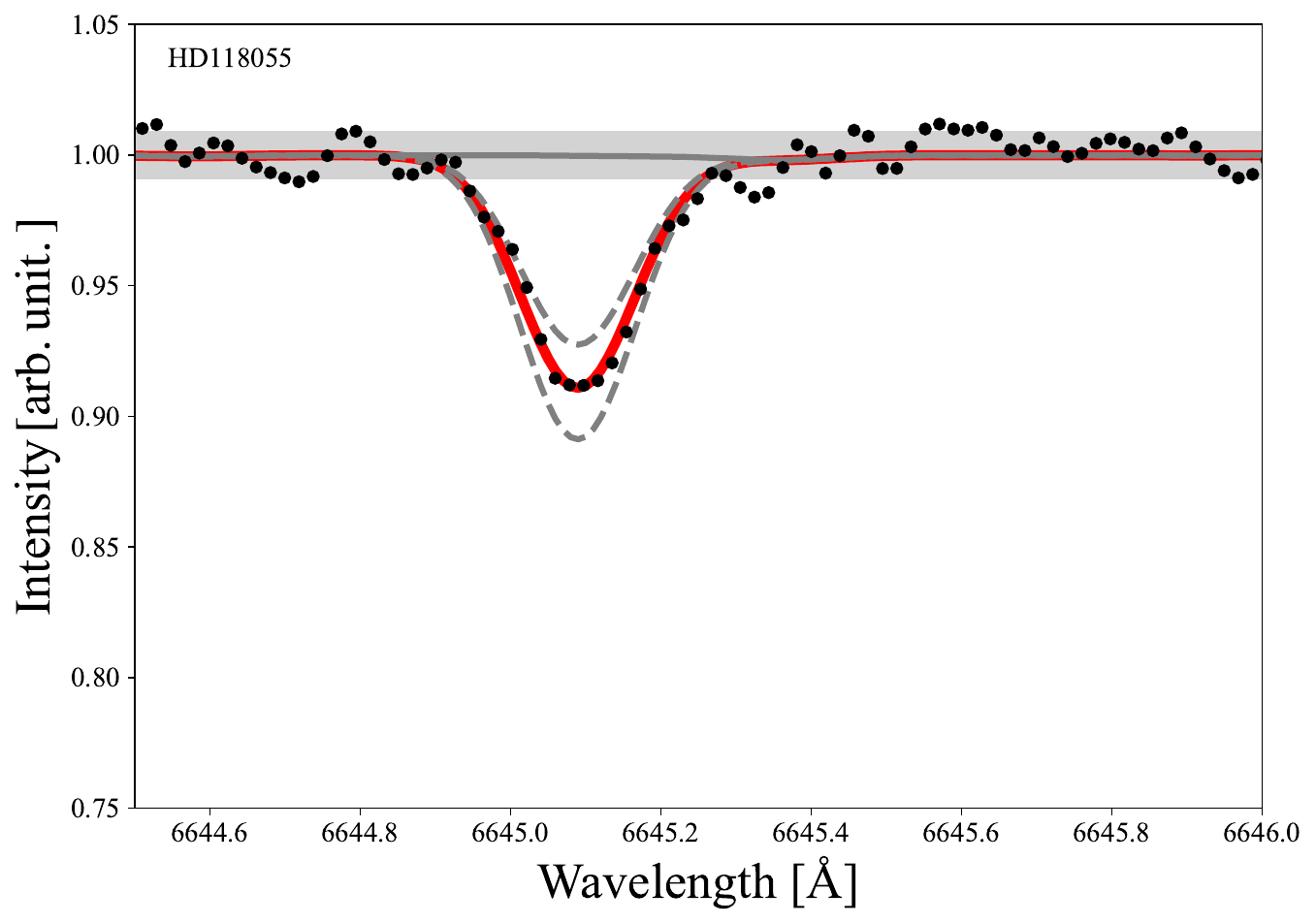}
\\[2mm]

\includegraphics[width=0.24\textwidth]{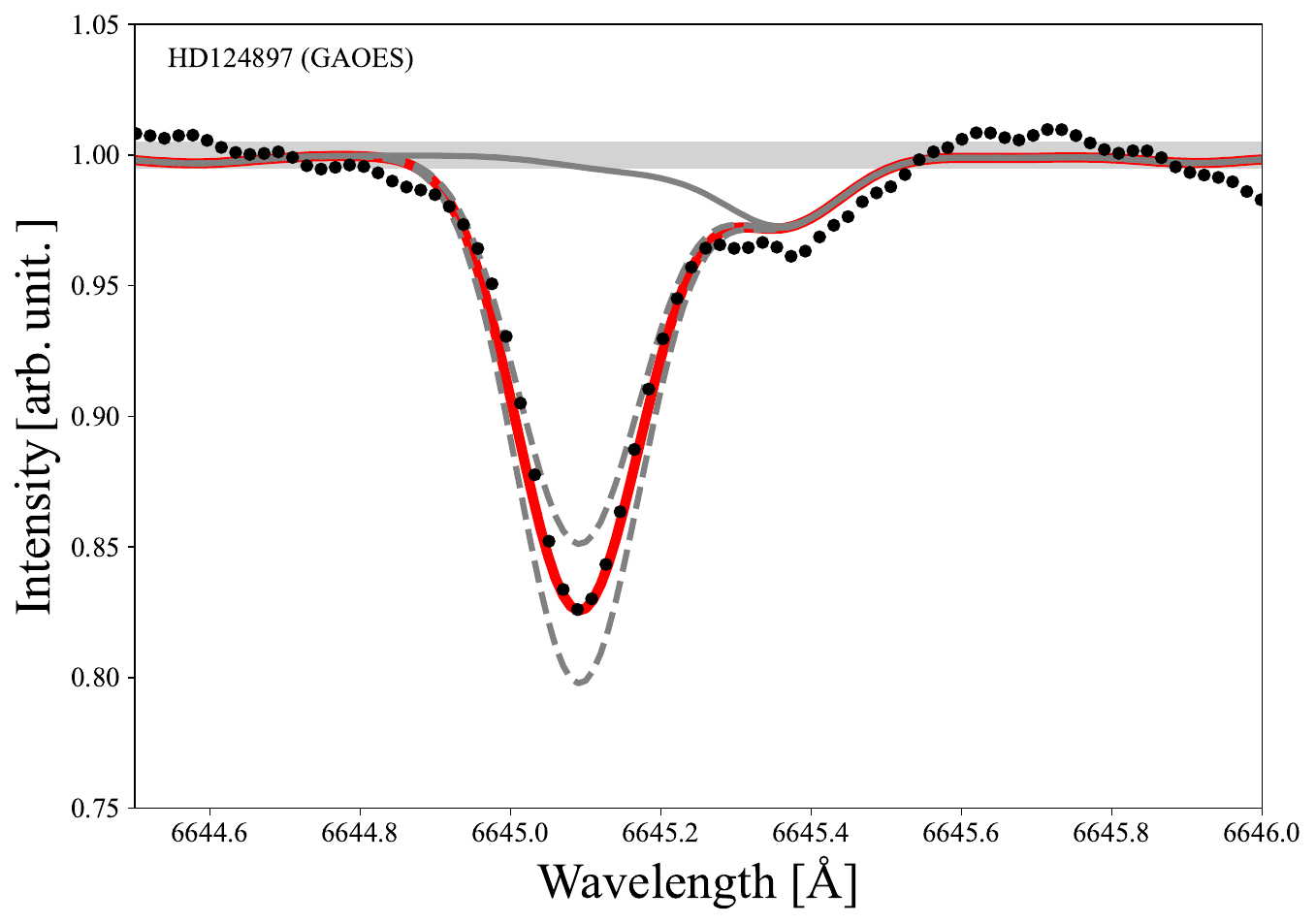}
\includegraphics[width=0.24\textwidth]{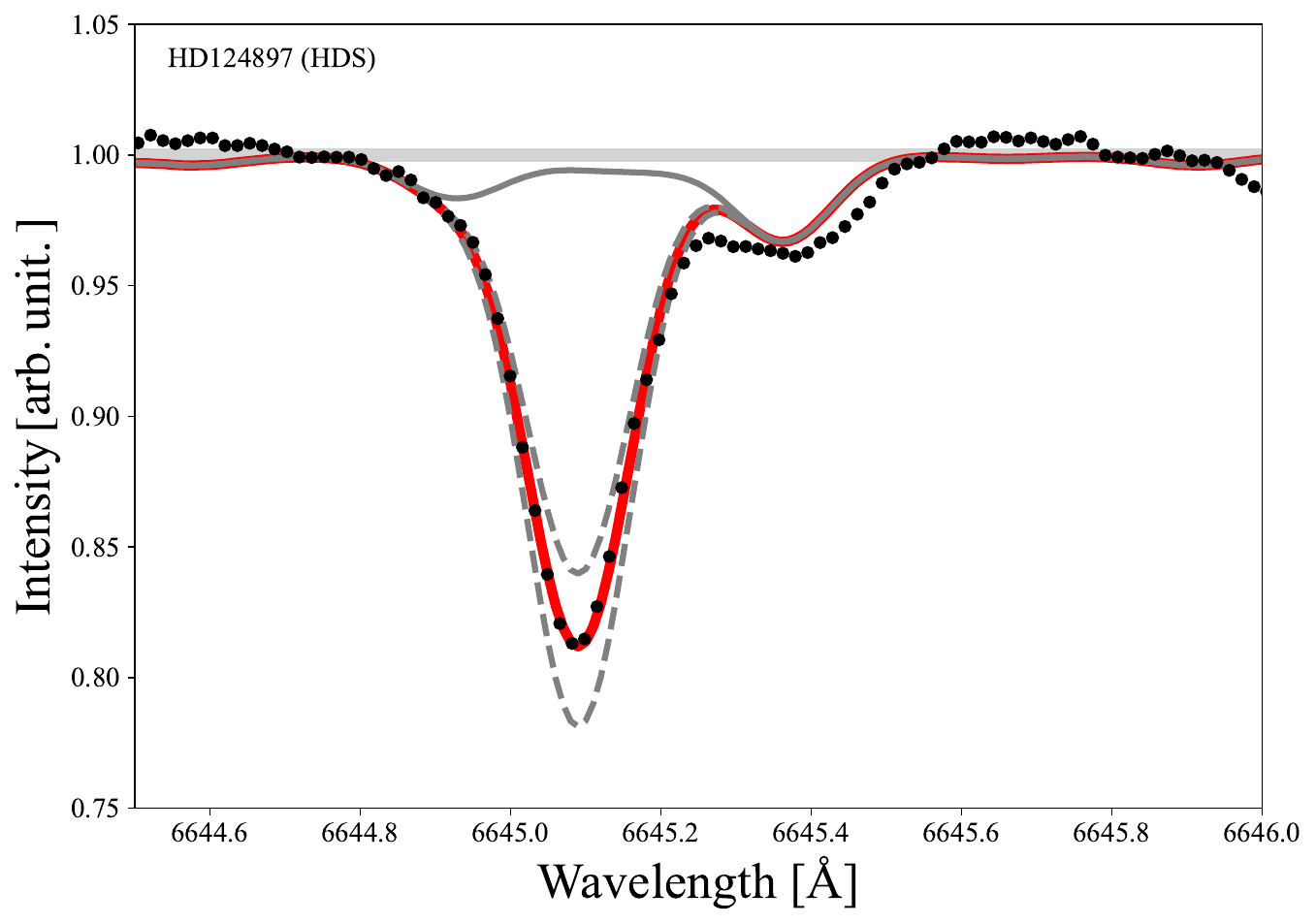}
\includegraphics[width=0.24\textwidth]{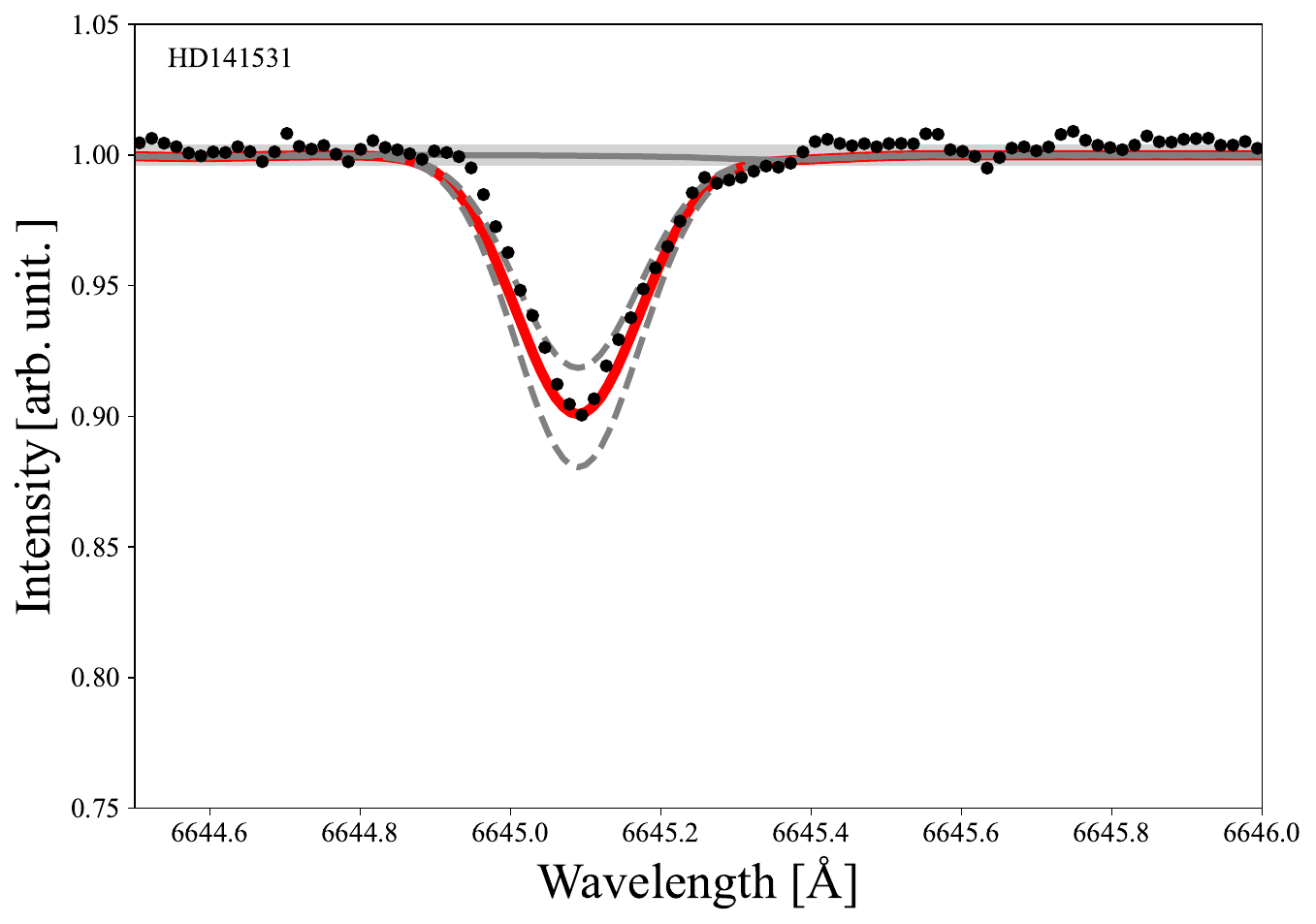}
\includegraphics[width=0.24\textwidth]{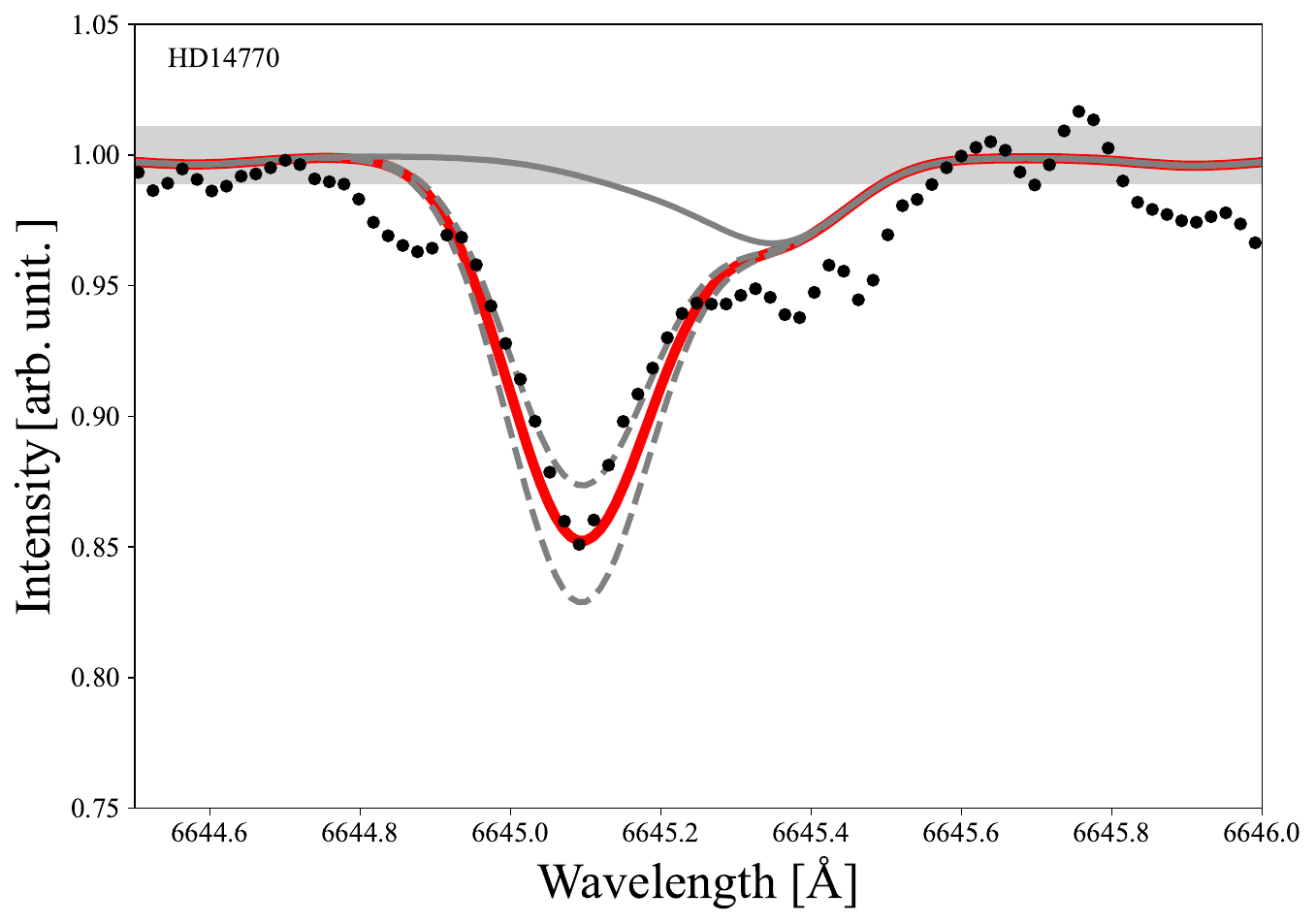}
\\[2mm]

\includegraphics[width=0.24\textwidth]{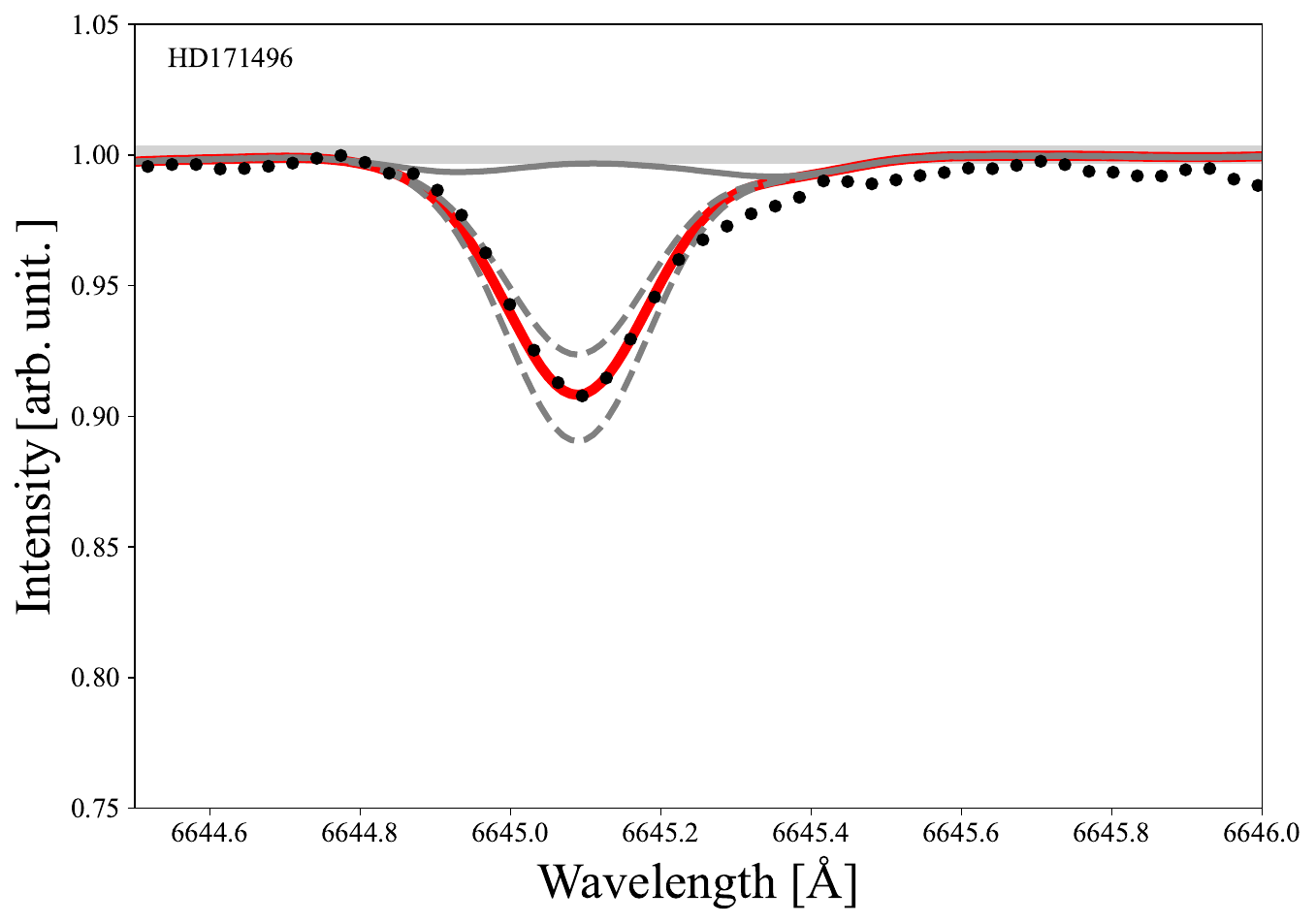}
\includegraphics[width=0.24\textwidth]{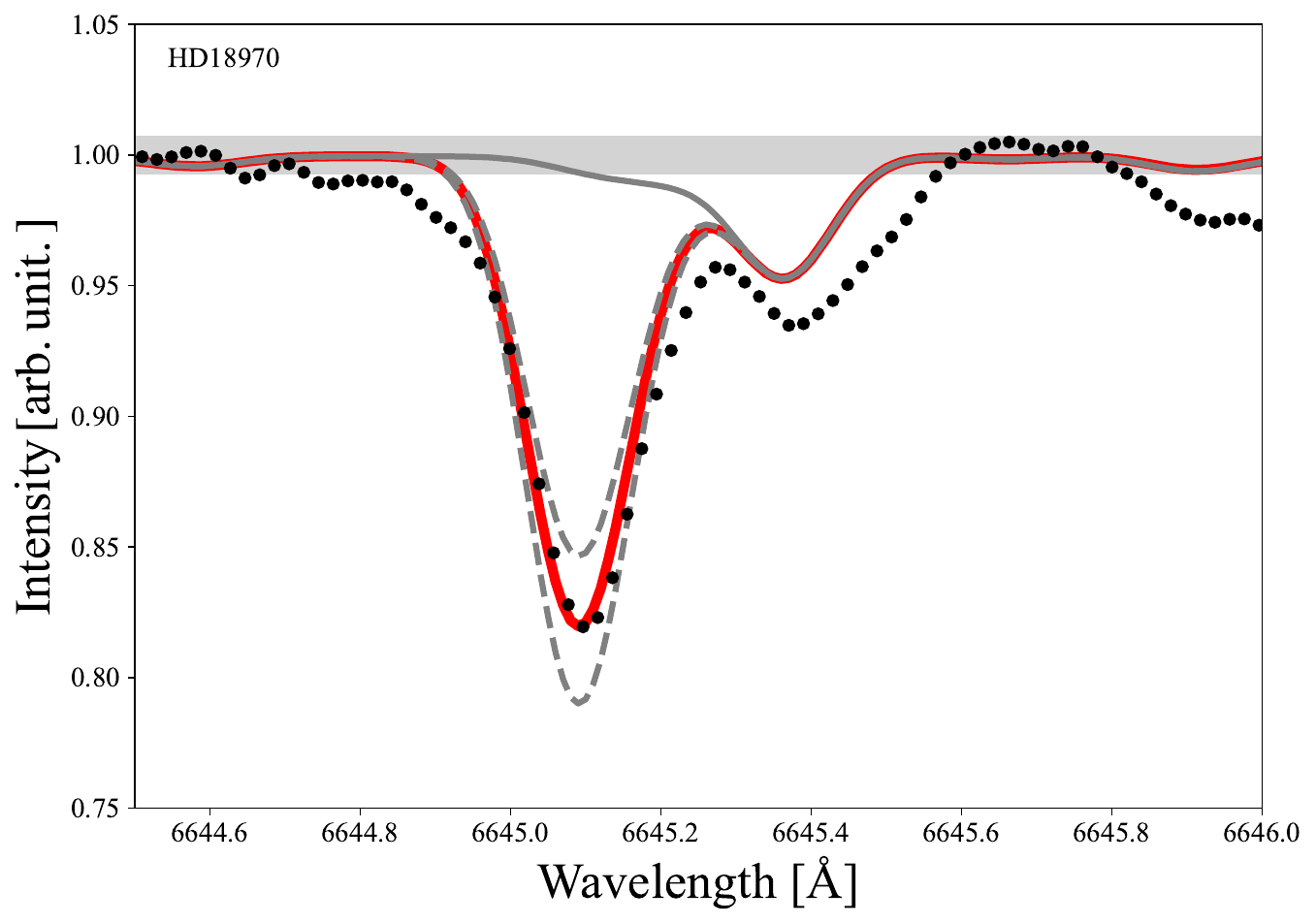}
\includegraphics[width=0.24\textwidth]{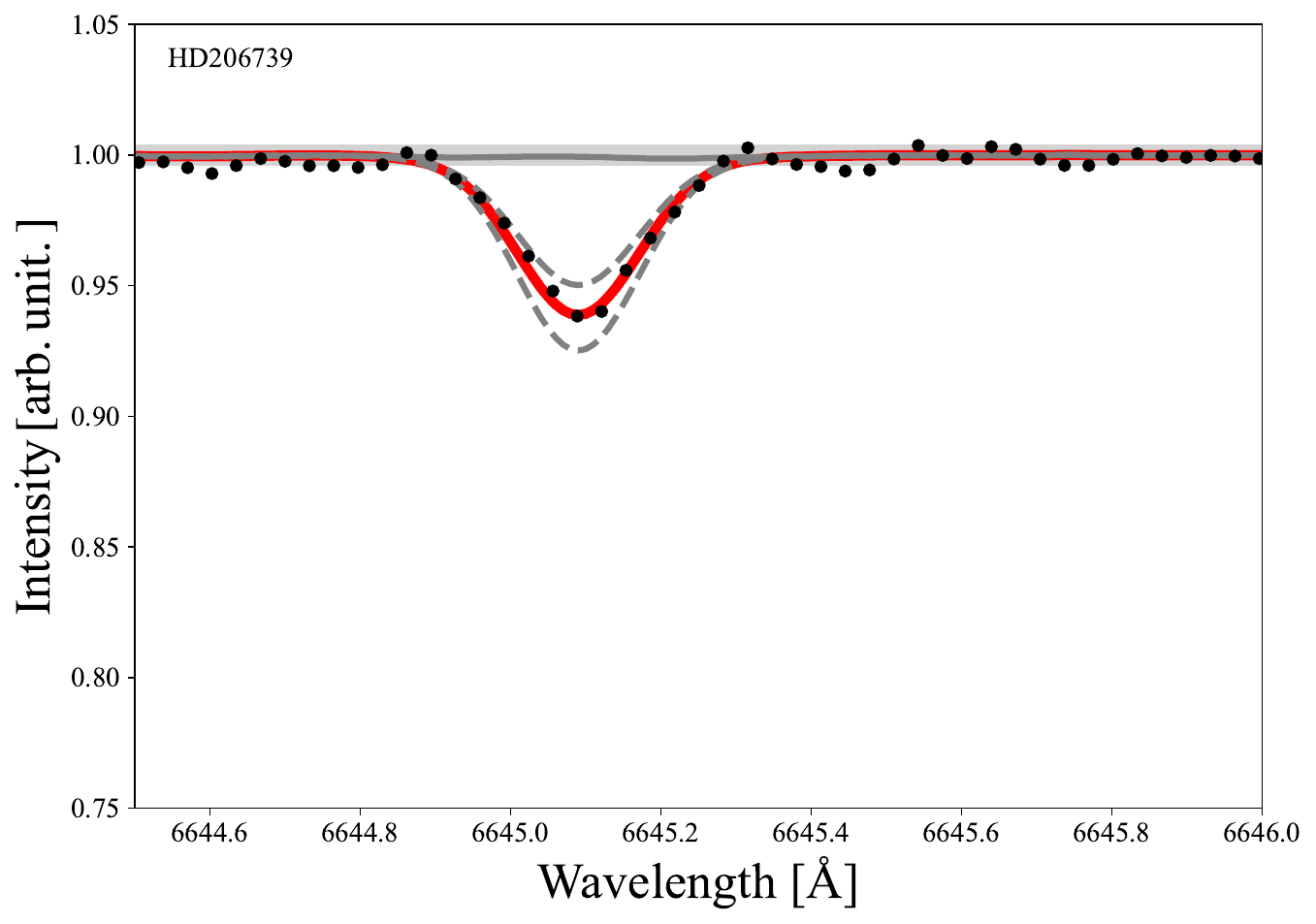}
\includegraphics[width=0.24\textwidth]{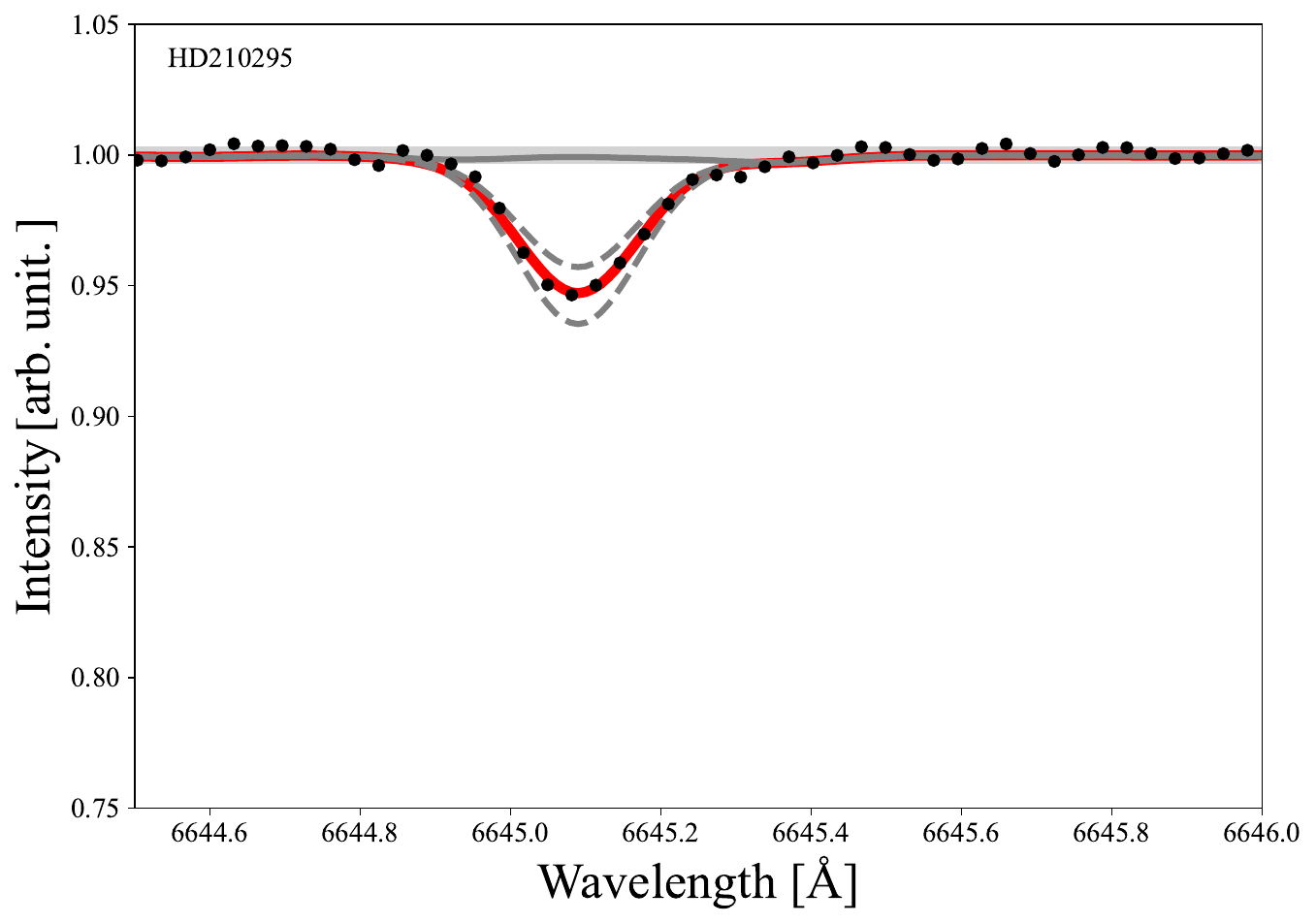}
\\[2mm]

\includegraphics[width=0.24\textwidth]{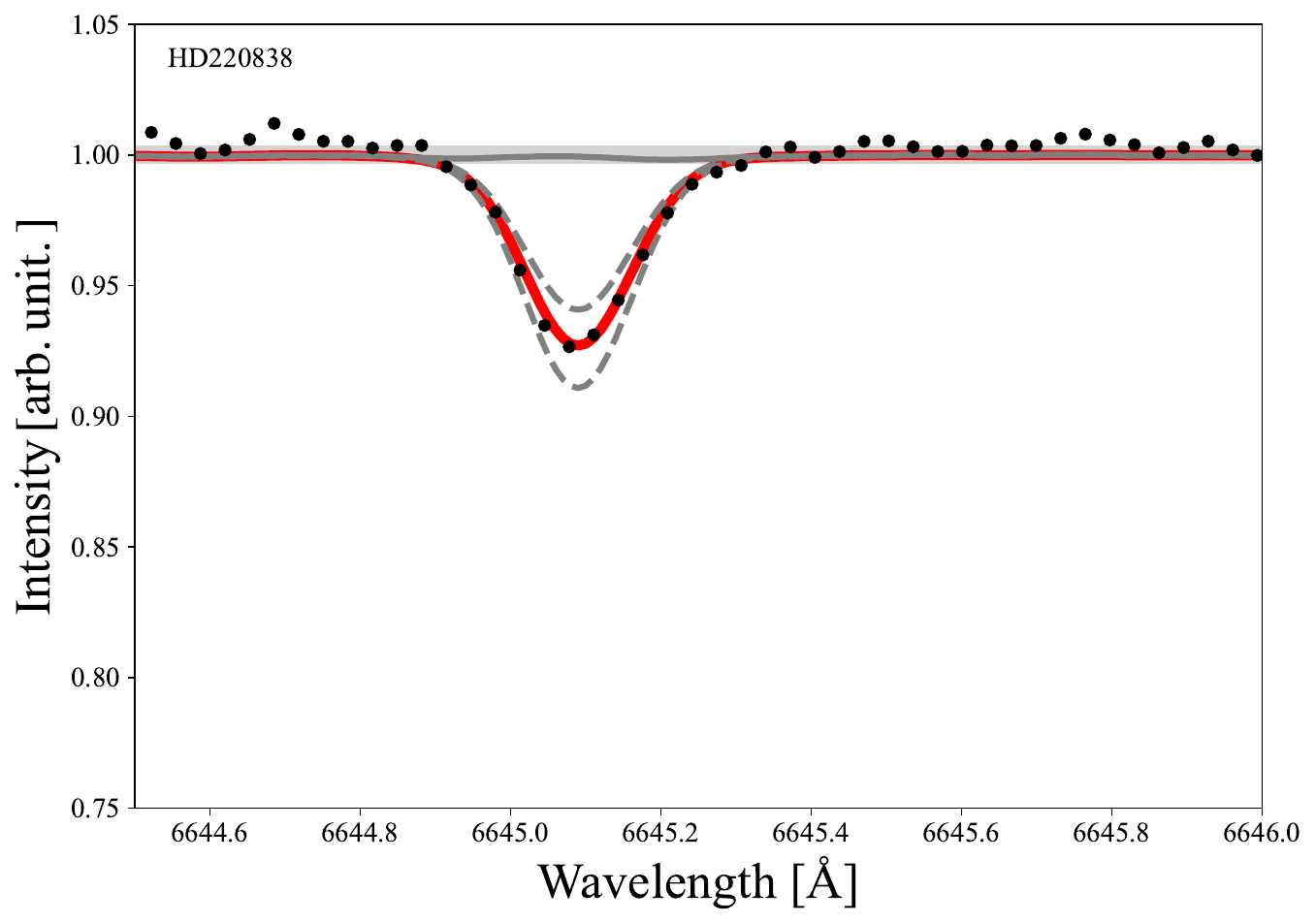}
\includegraphics[width=0.24\textwidth]{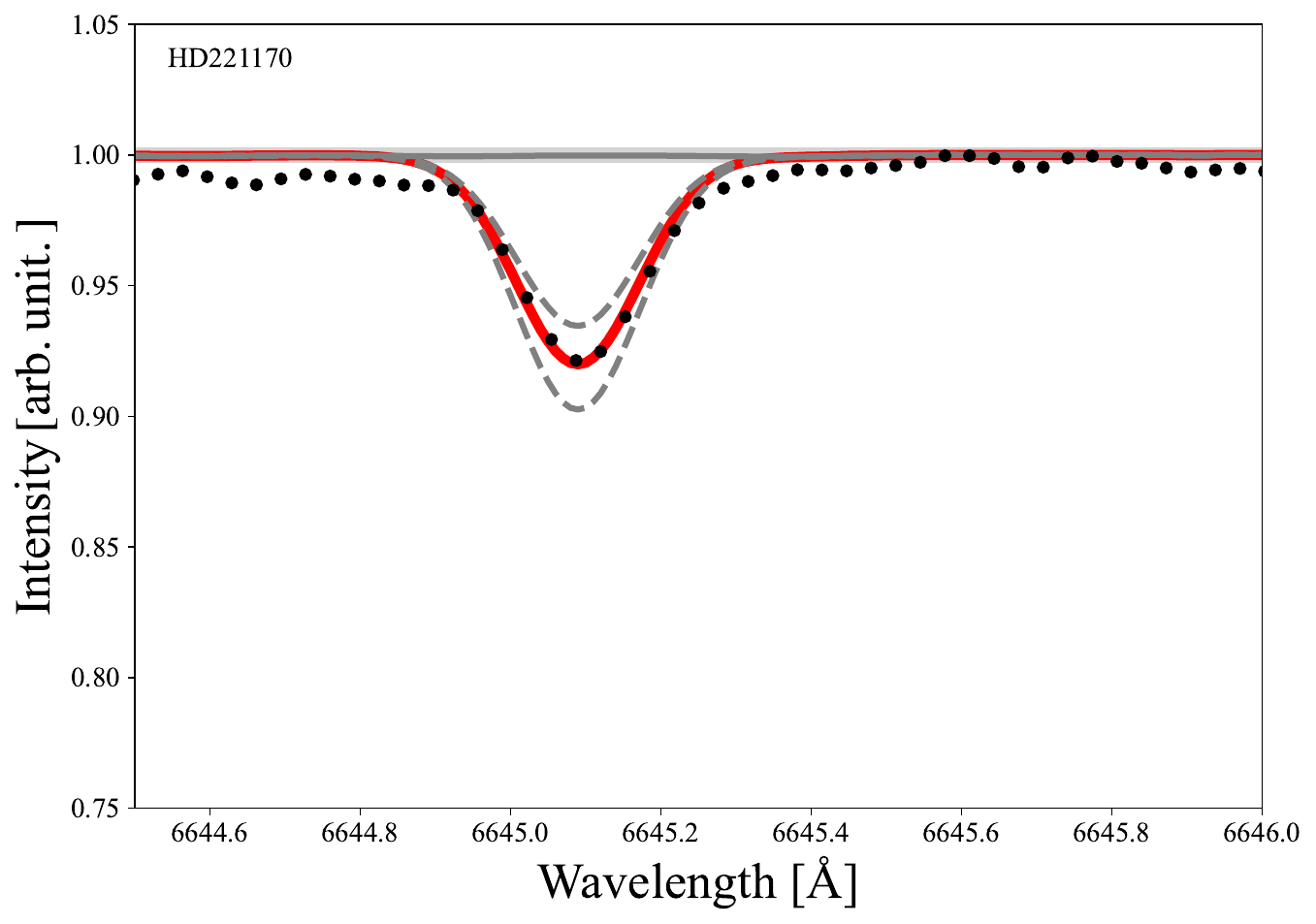}
\includegraphics[width=0.24\textwidth]{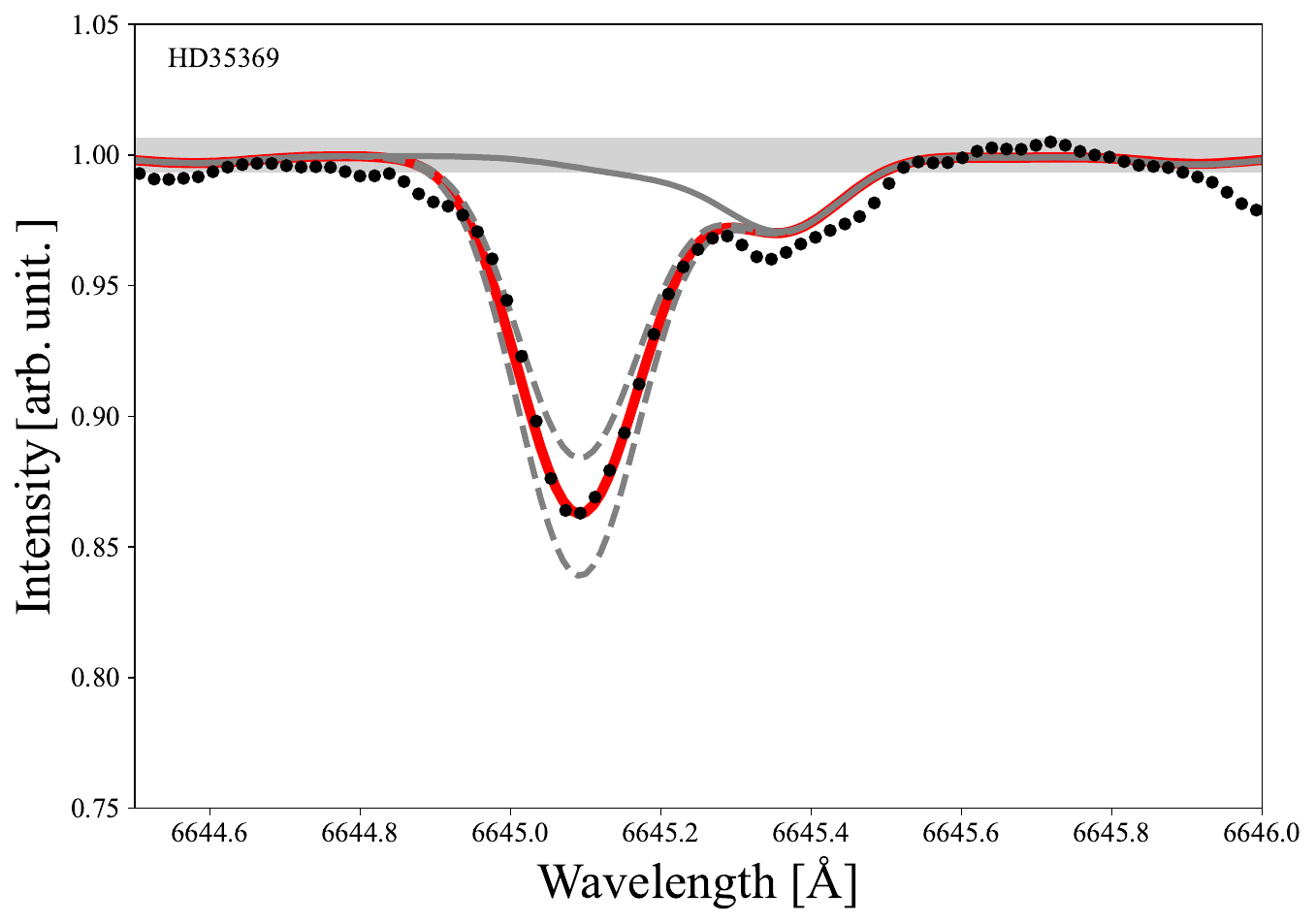}
\includegraphics[width=0.24\textwidth]{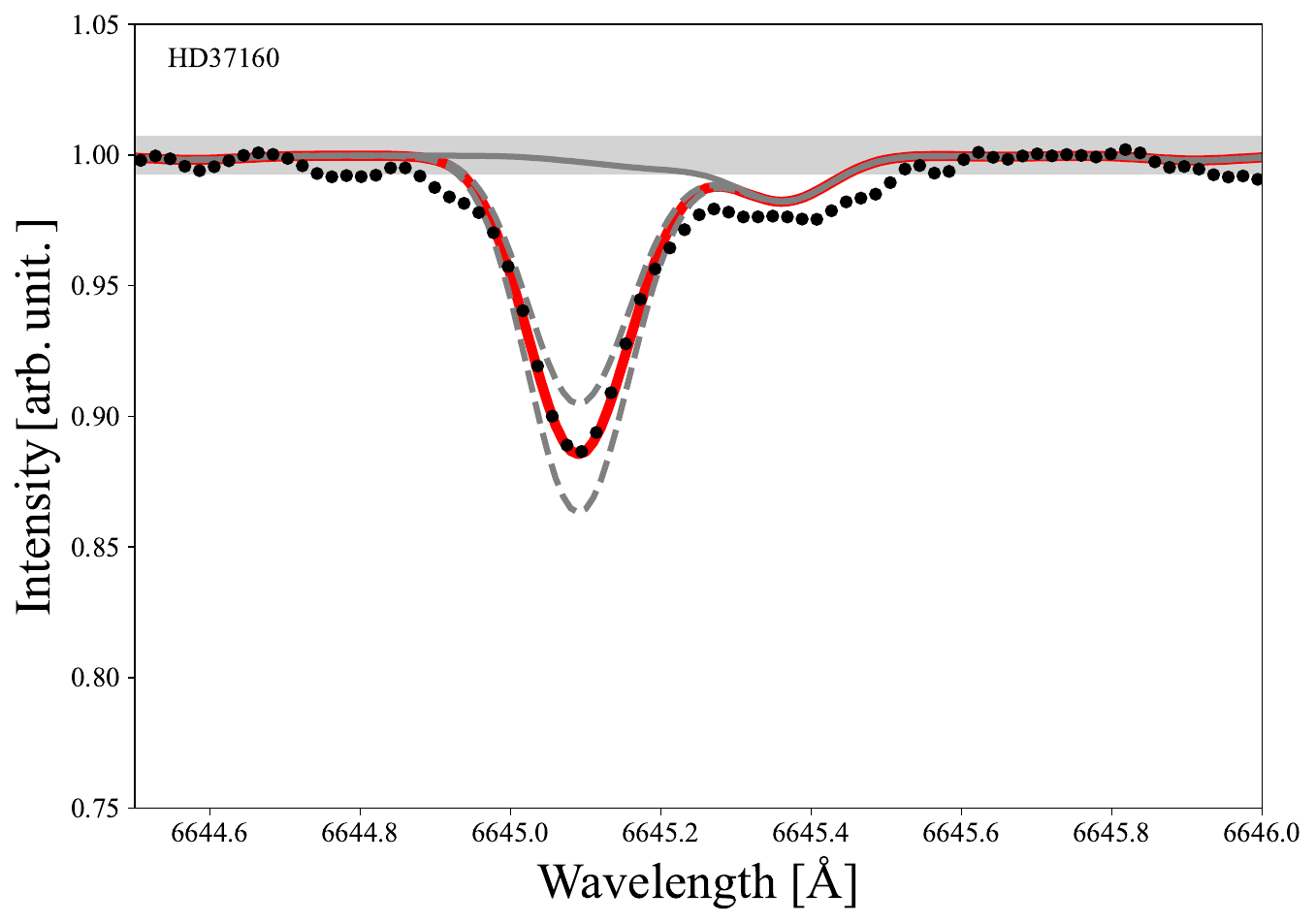}
\\[2mm]

\includegraphics[width=0.24\textwidth]{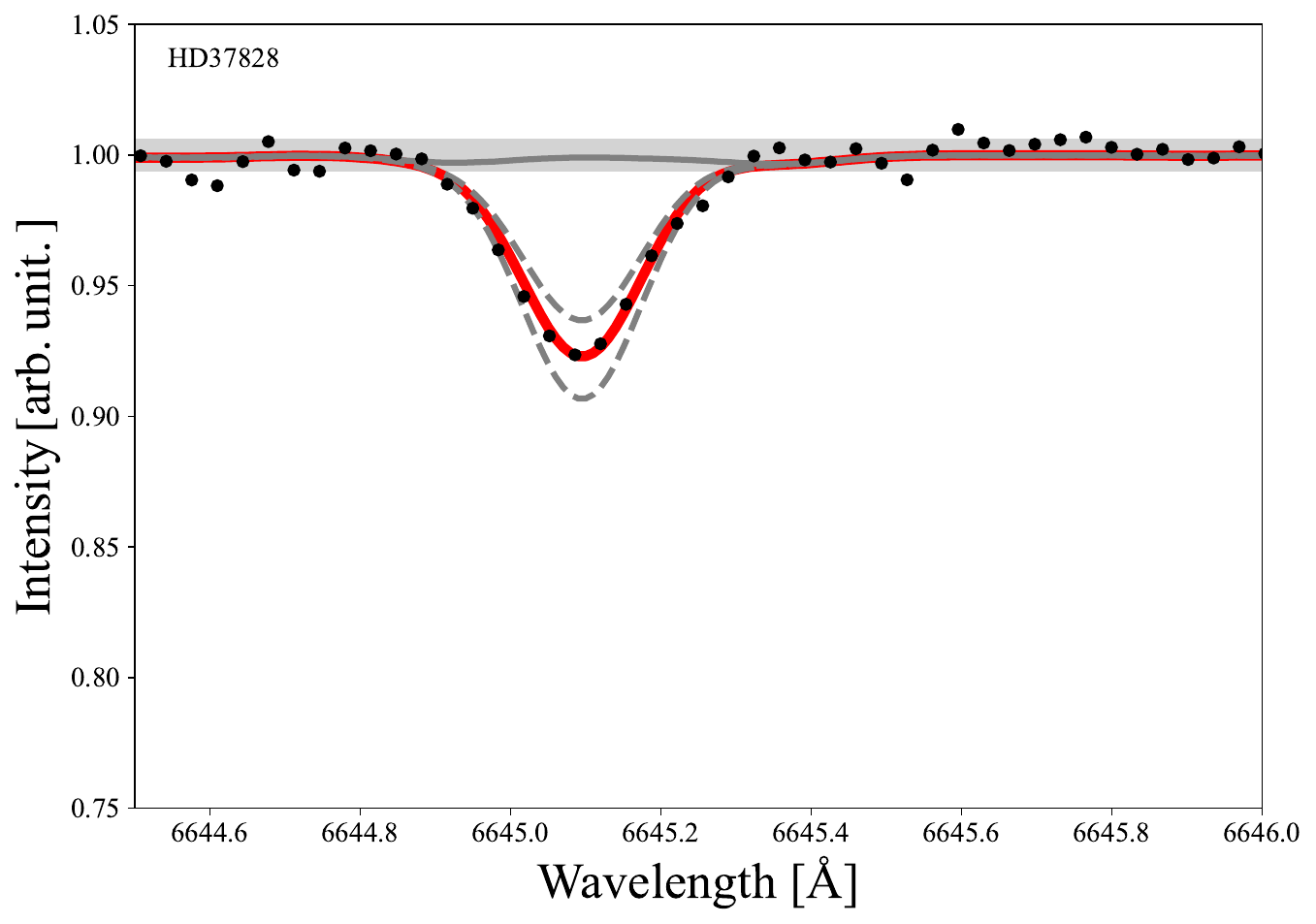}
\includegraphics[width=0.24\textwidth]{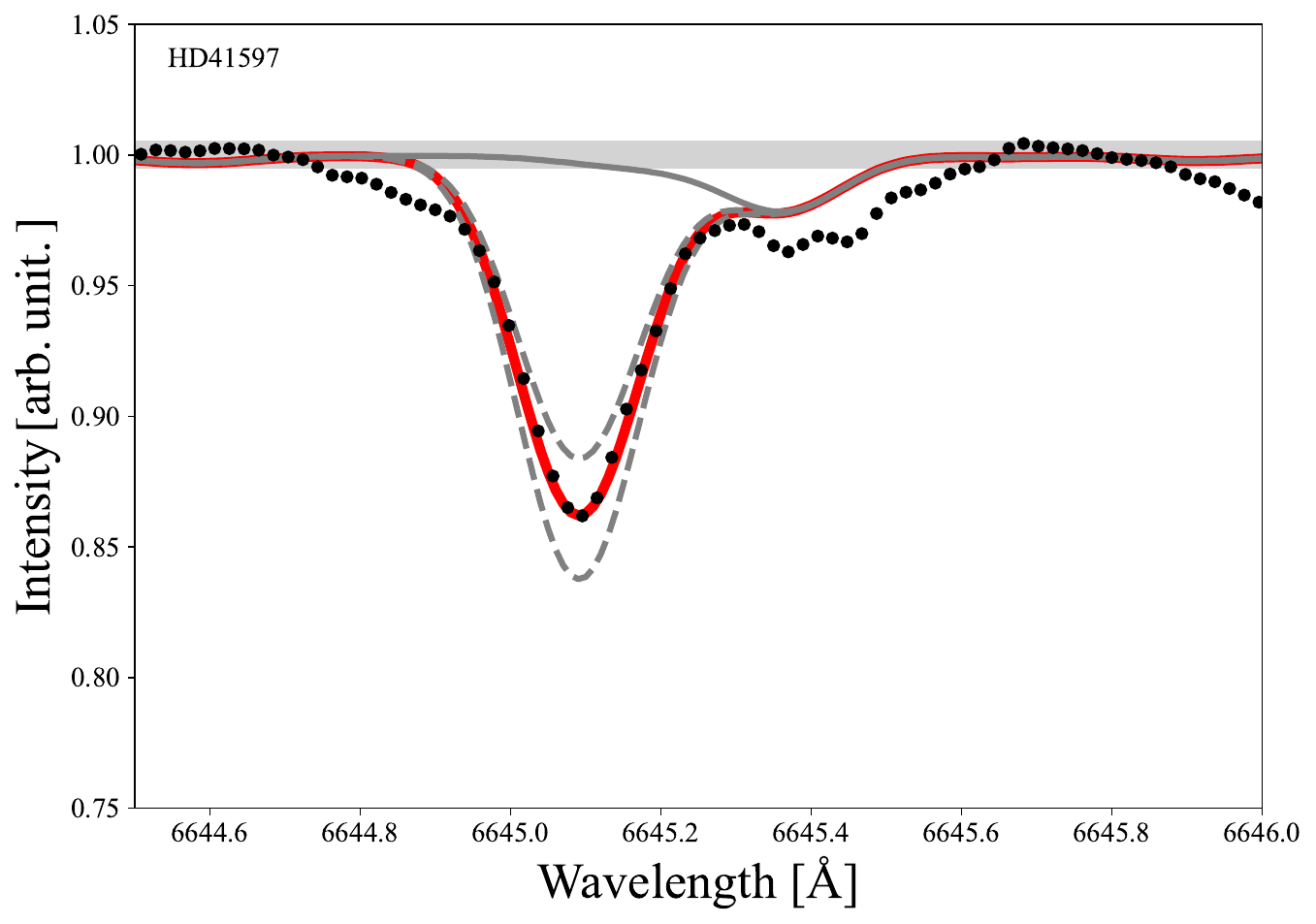}
\includegraphics[width=0.24\textwidth]{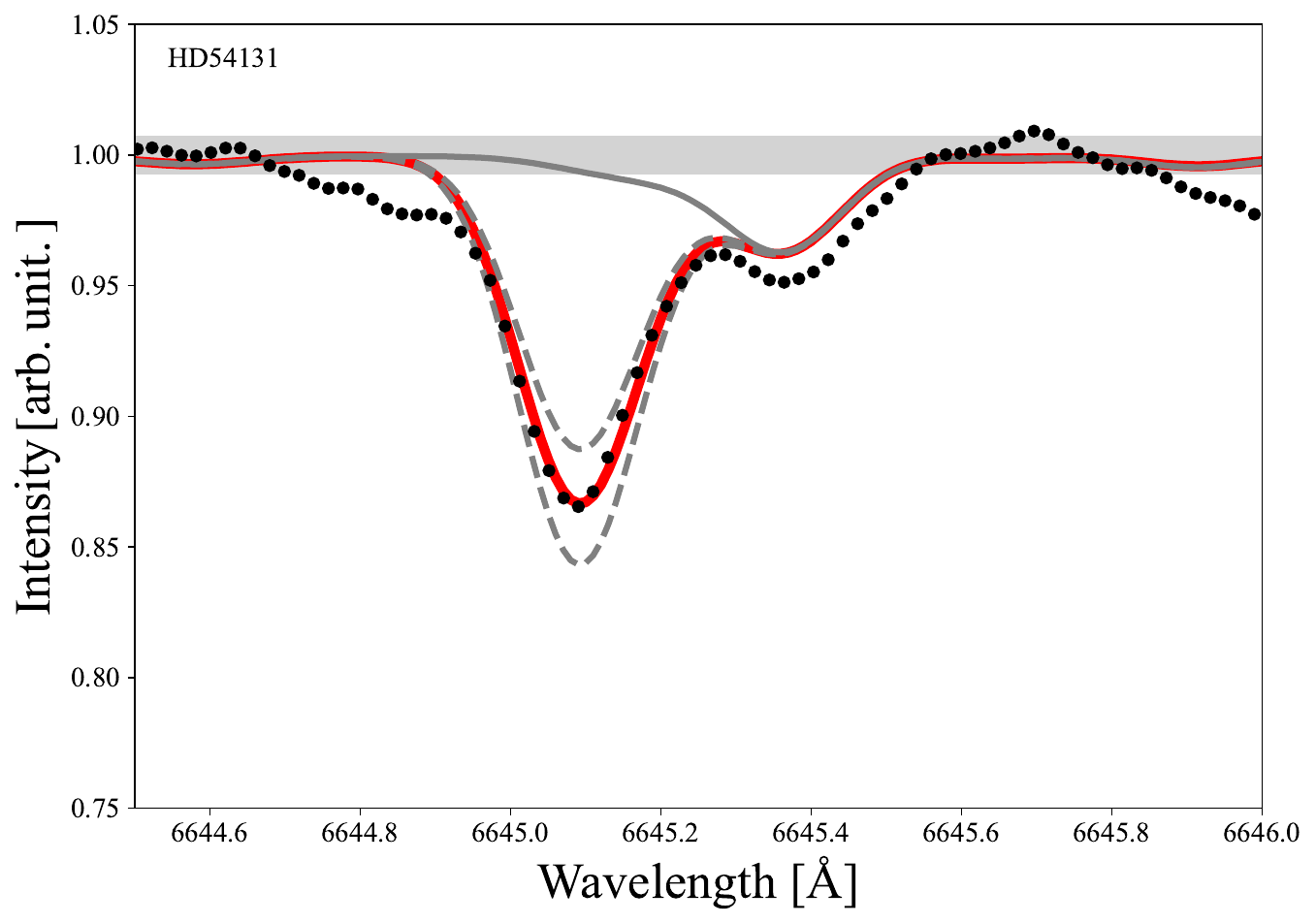}
\includegraphics[width=0.24\textwidth]{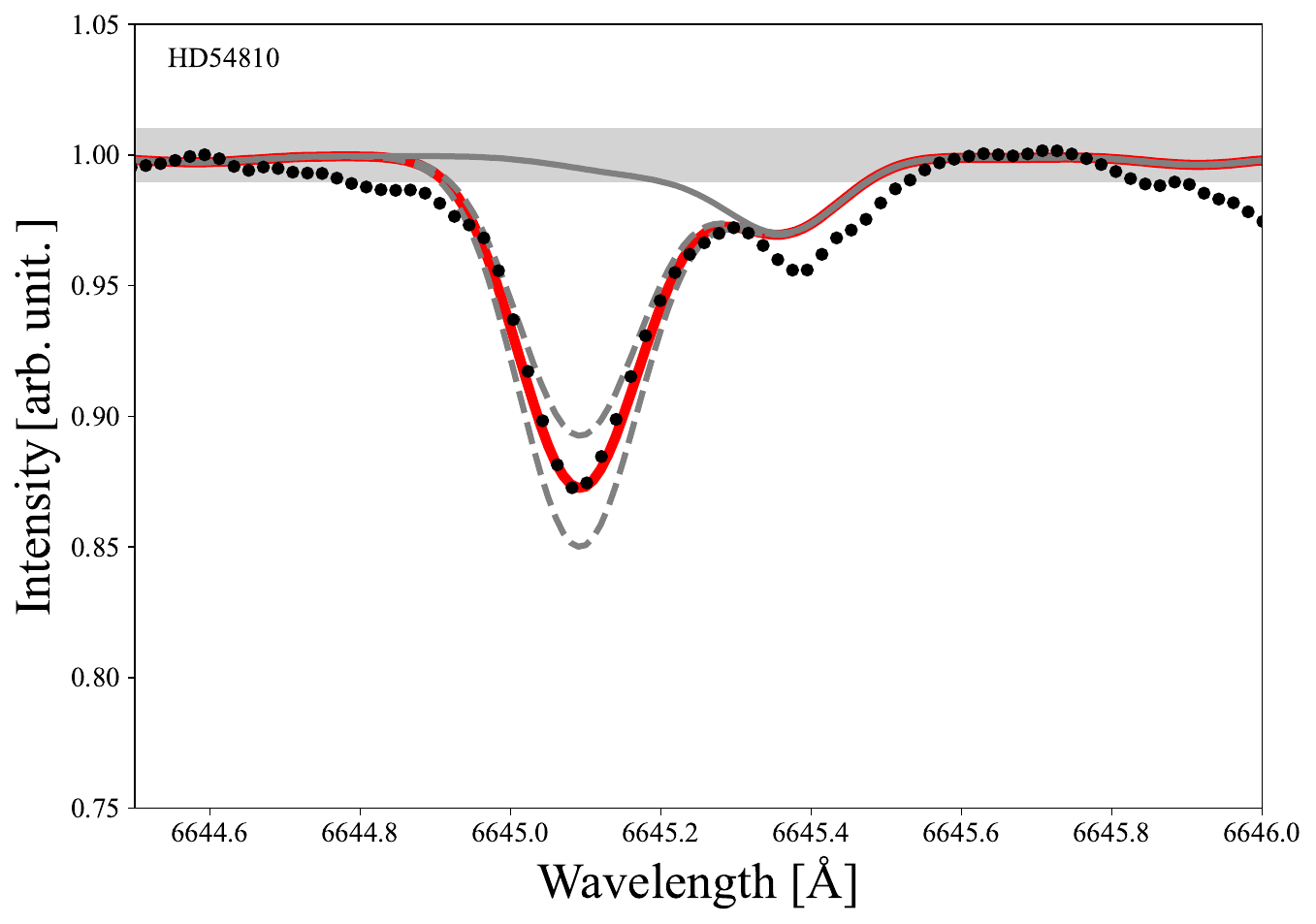}
\\[2mm]

 \caption{All spectra of Eu 6645 \AA~line. Black points show observations, red solid lines show the best-fit synthetic spectra, dashed lines show Eu abundance changed by 0.1 dex, and gray solid line shows no Eu. Gray shade shows 1 $\upvarsigma$ error from S/N noise.
  {Alt text: All spectra of Europium 6645 angstrom line. } 
 }
 \label{fig:Eu_allspectra_1}
 \end{figure*}

\addtocounter{figure}{-1}

 \begin{figure*}
\includegraphics[width=0.24\textwidth]{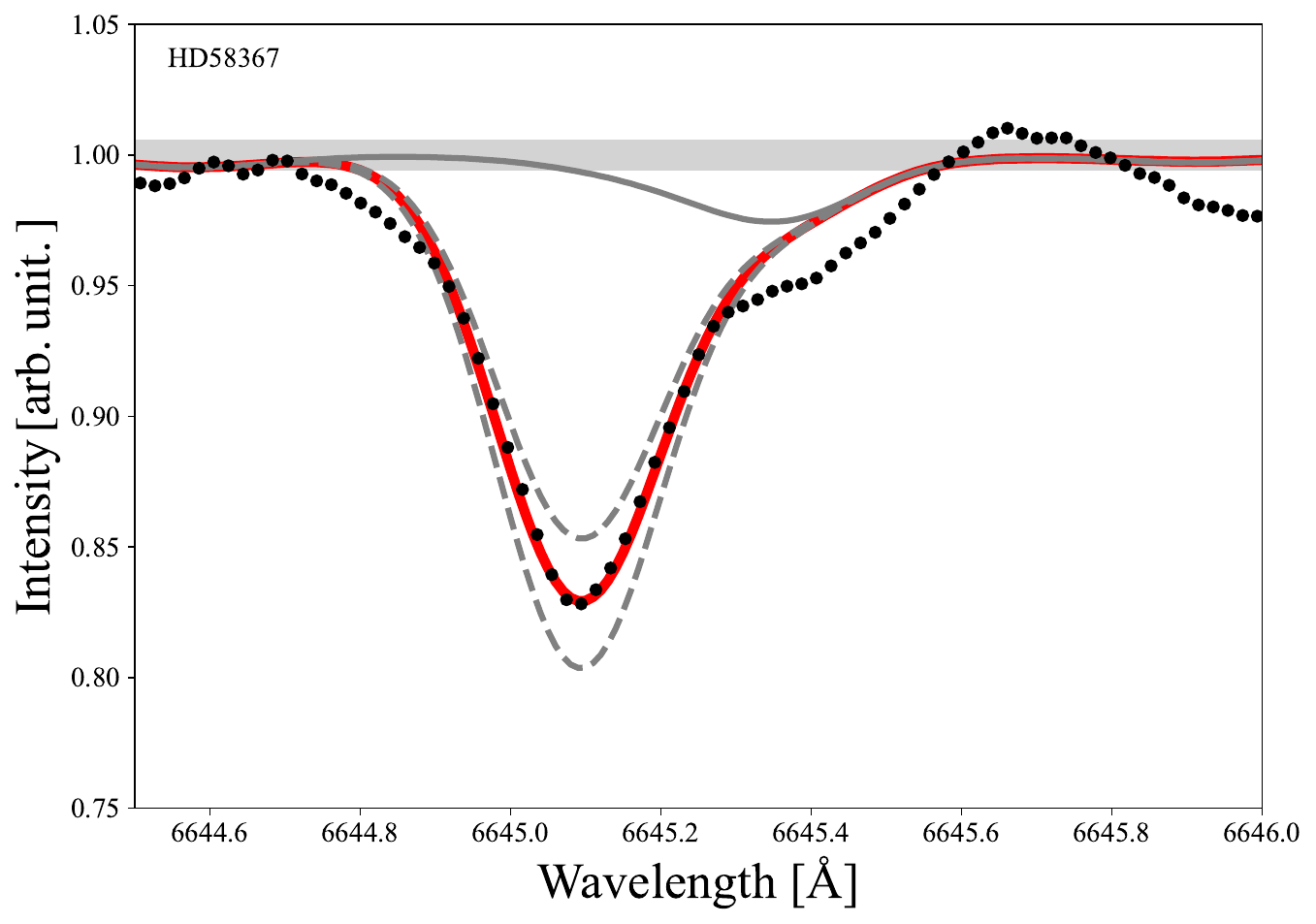}
\includegraphics[width=0.24\textwidth]{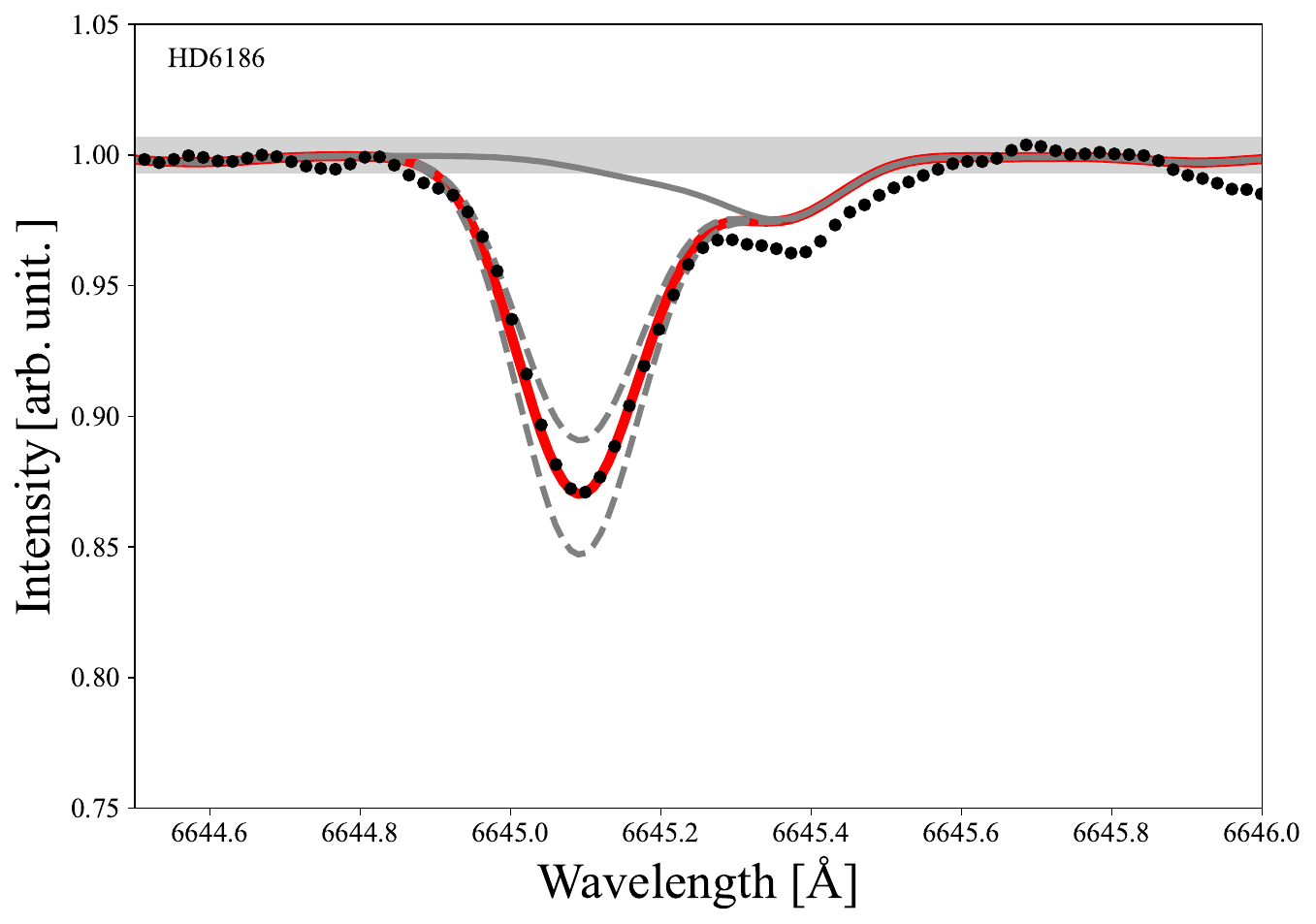}
\includegraphics[width=0.24\textwidth]{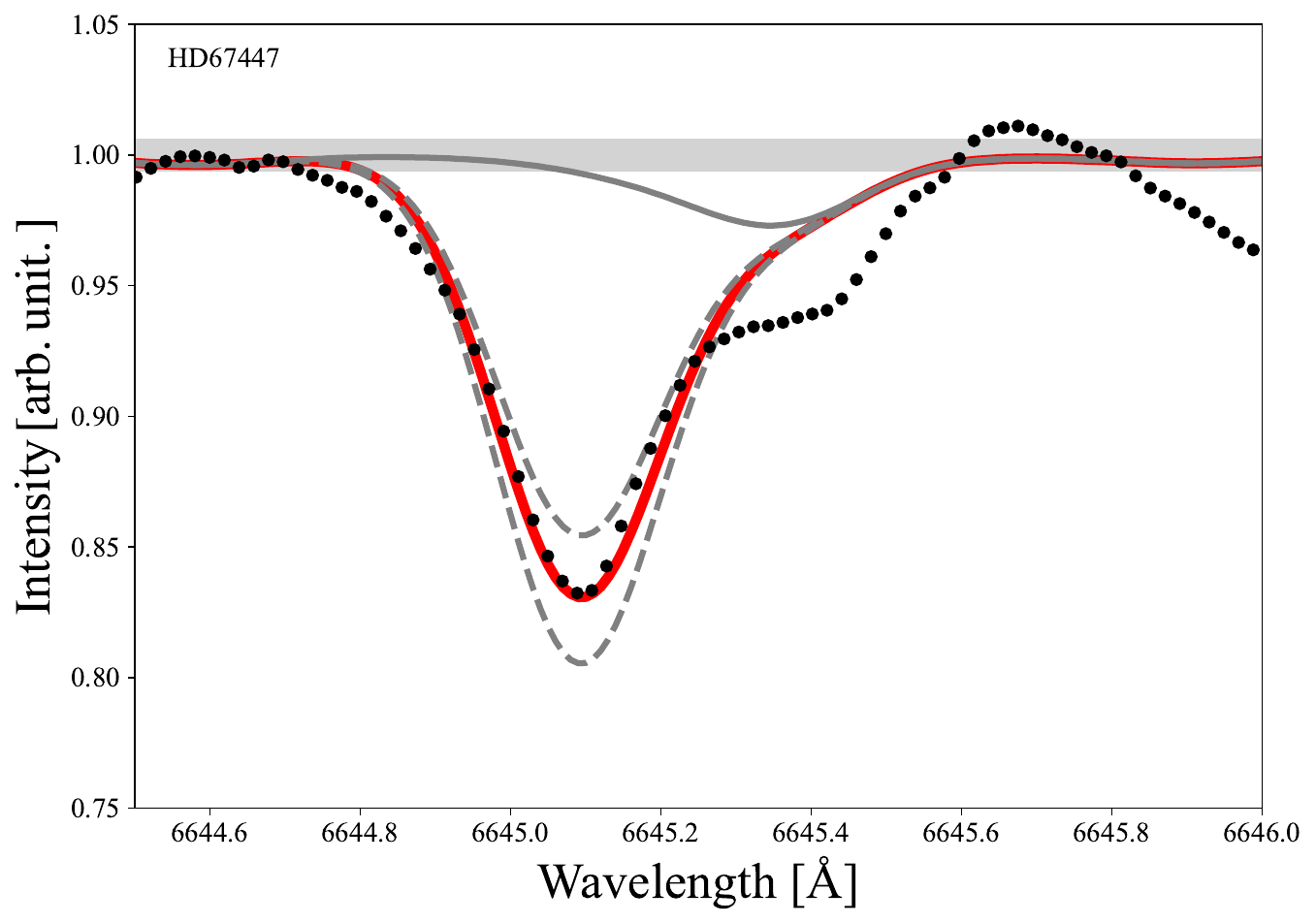}
\includegraphics[width=0.24\textwidth]{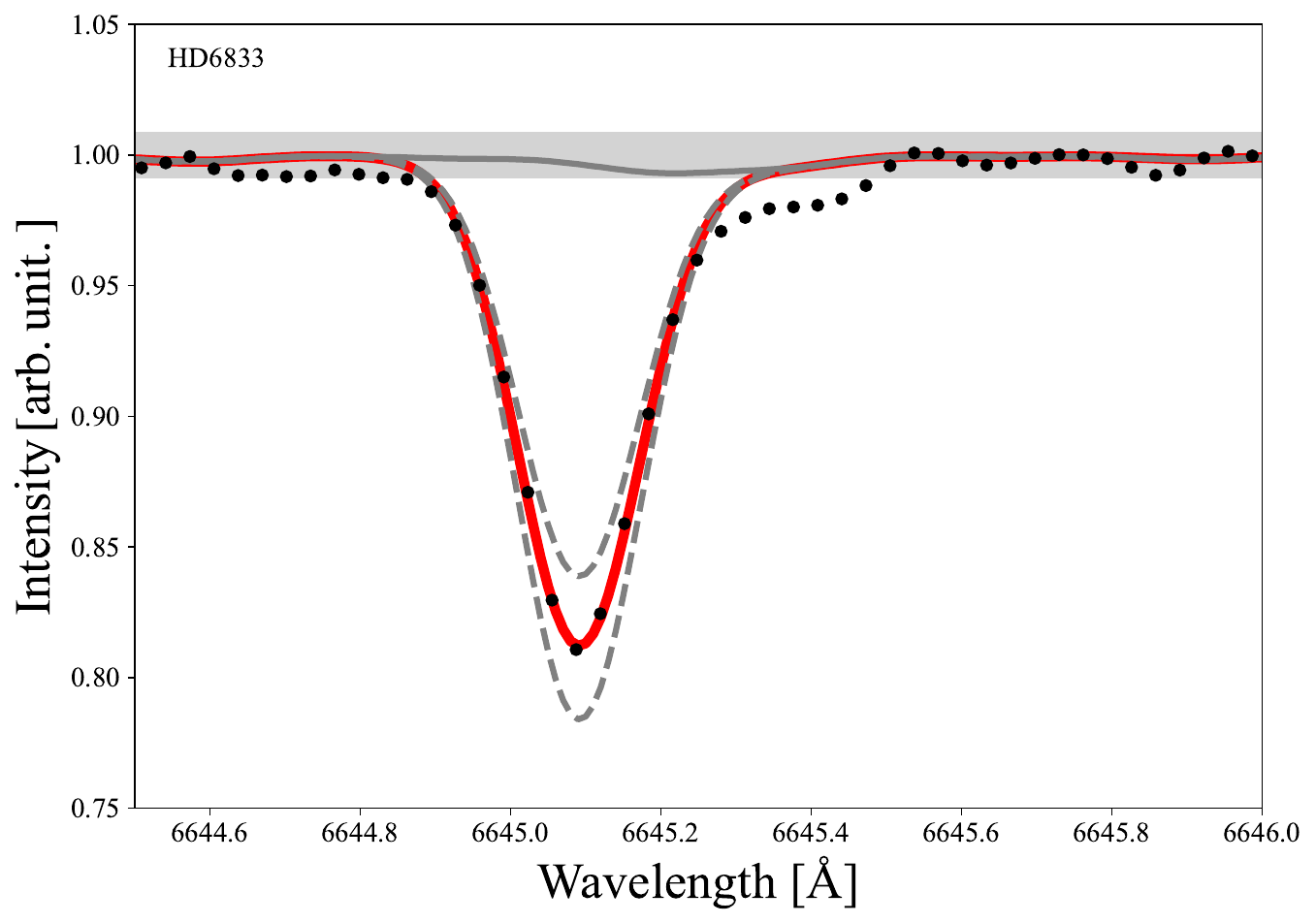}
\\[2mm]

\centering
\includegraphics[width=0.24\textwidth]{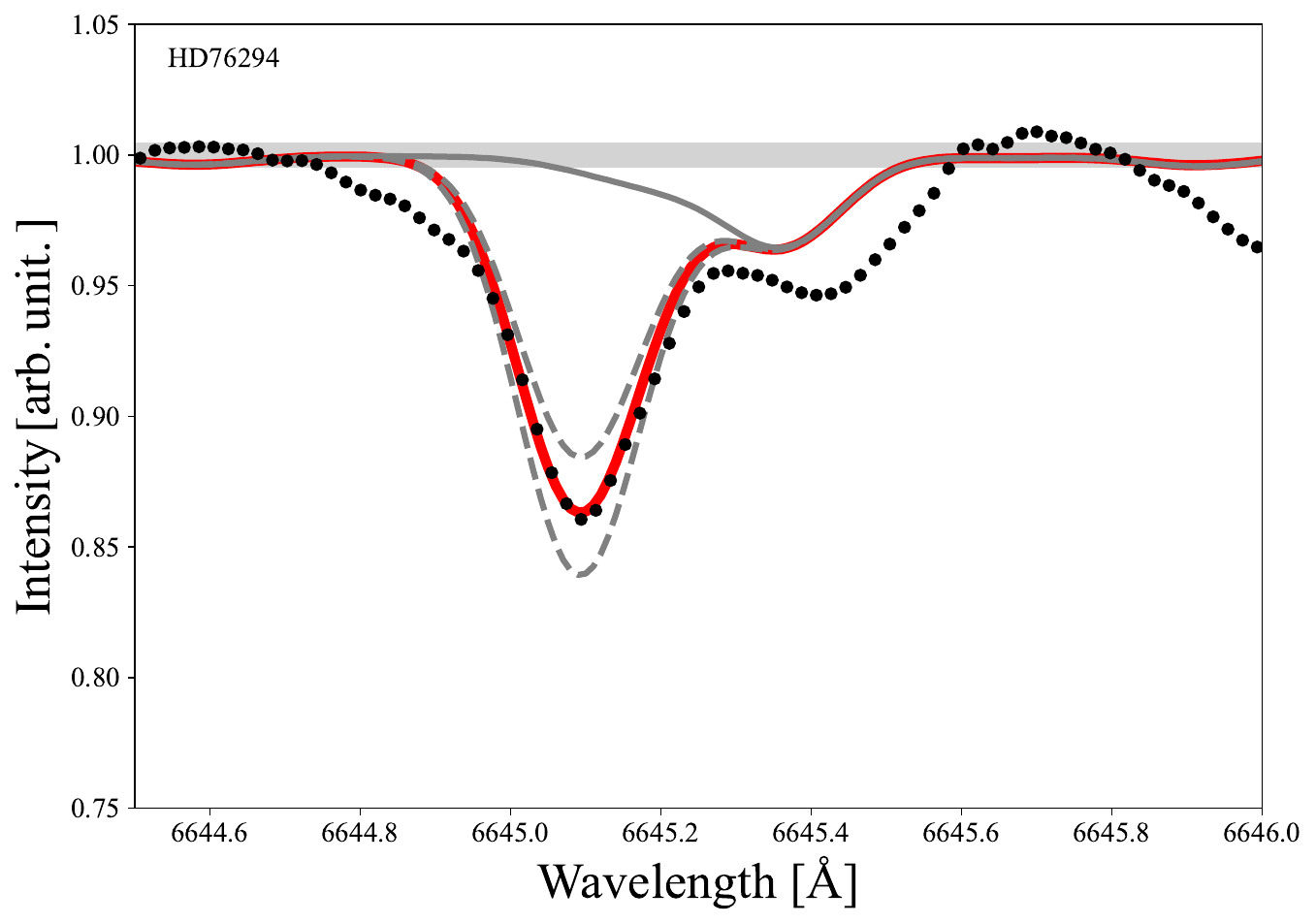}
\includegraphics[width=0.24\textwidth]{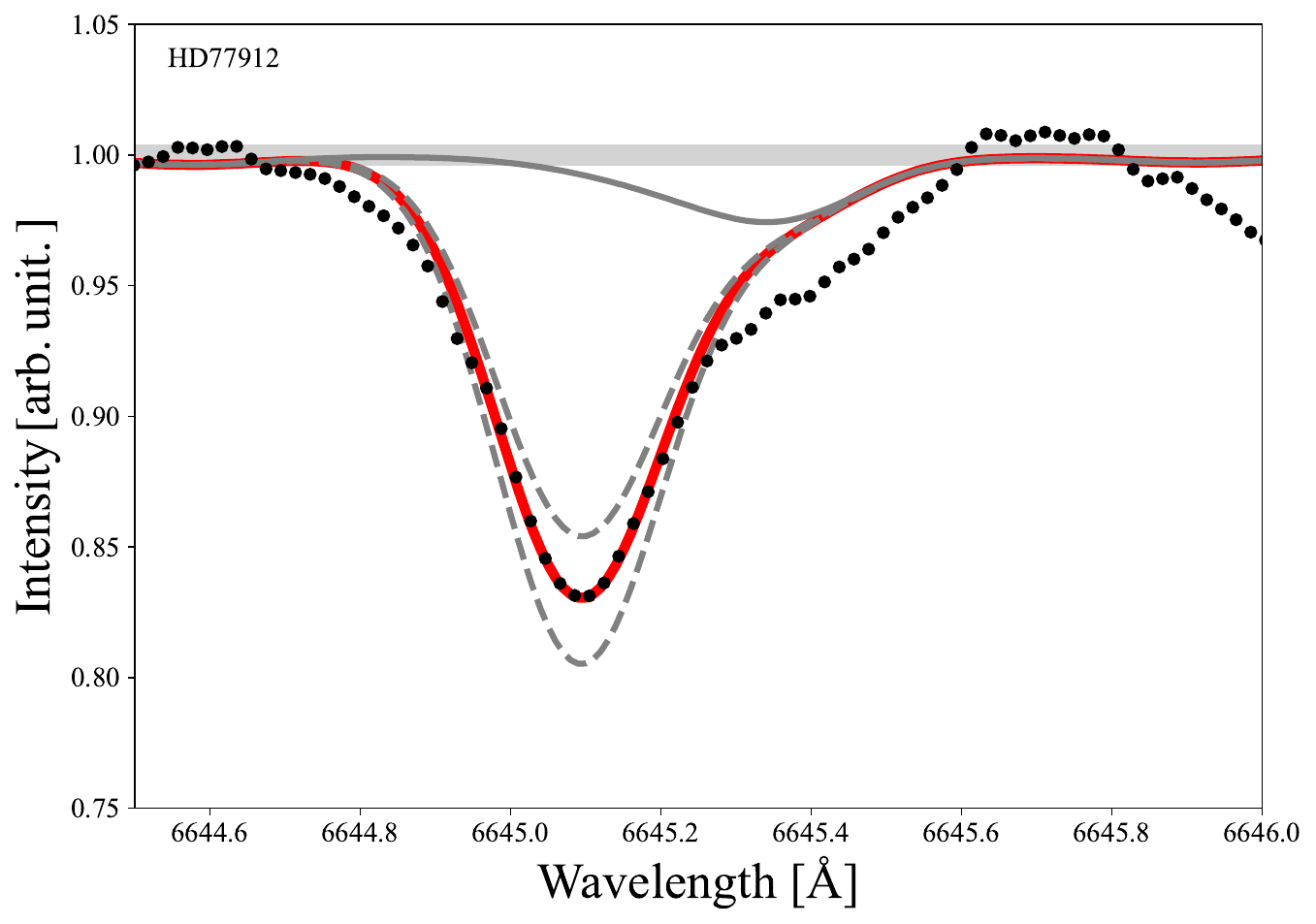}
\includegraphics[width=0.24\textwidth]{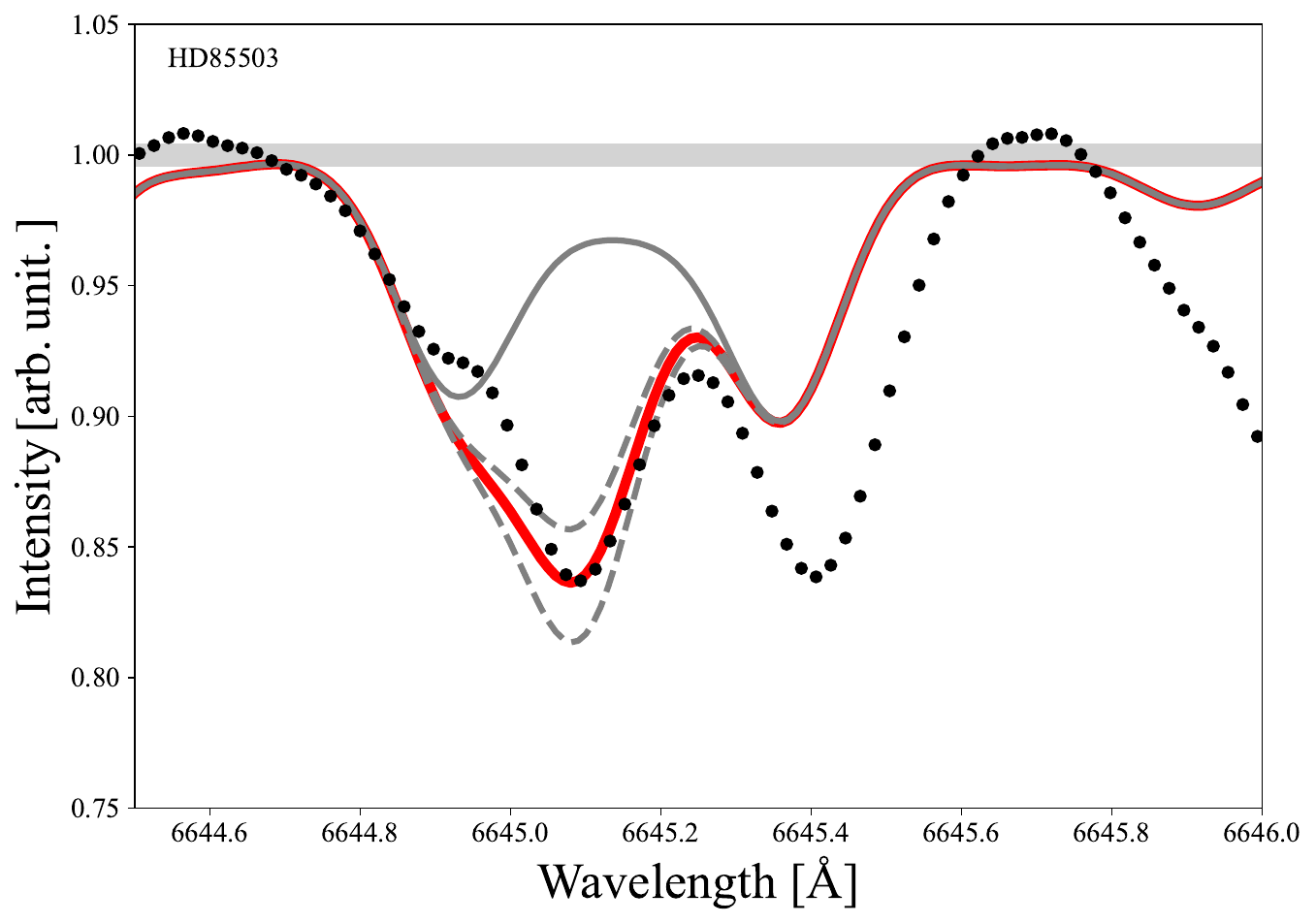}
\includegraphics[width=0.24\textwidth]{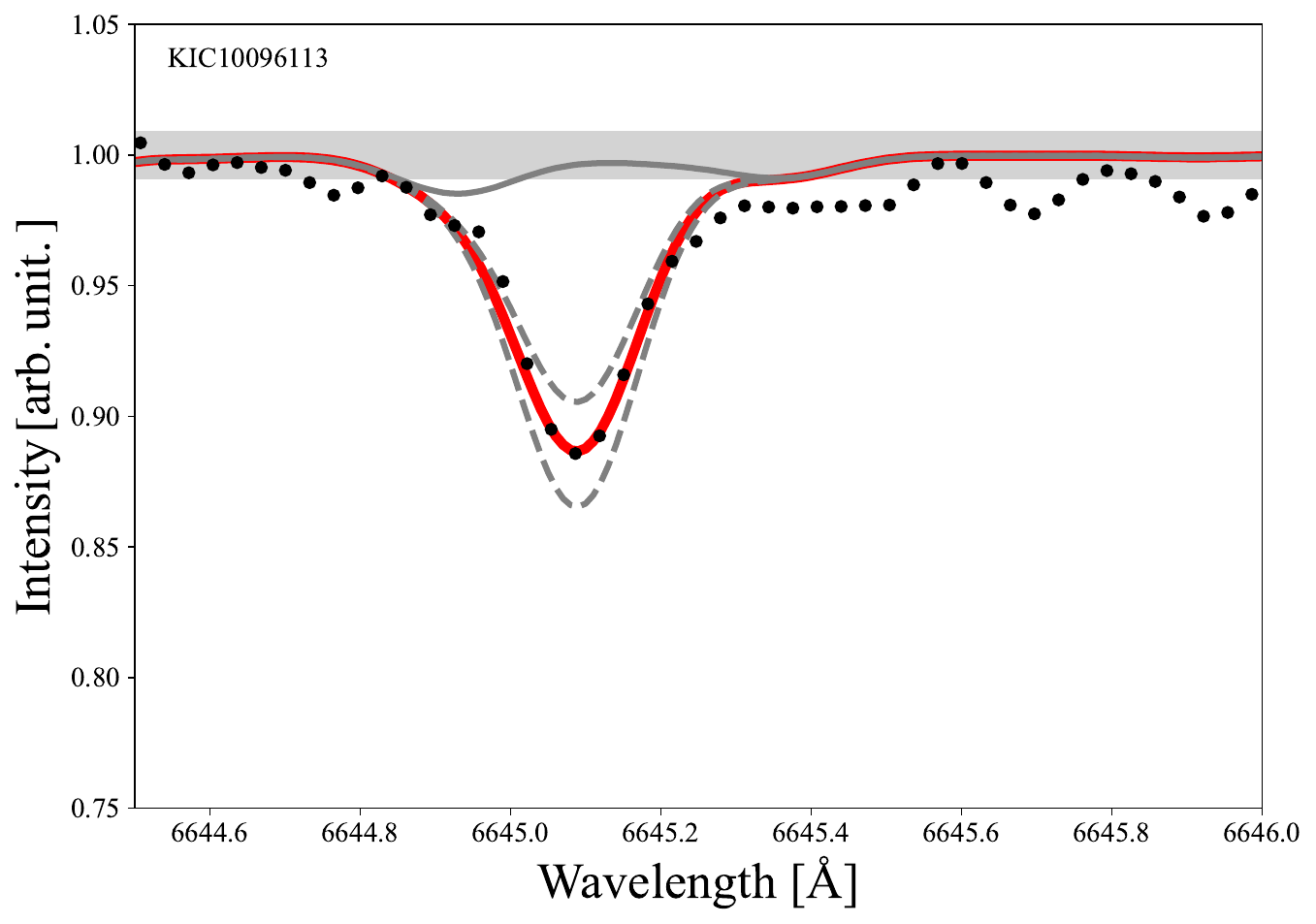}
\\[2mm]

\includegraphics[width=0.24\textwidth]{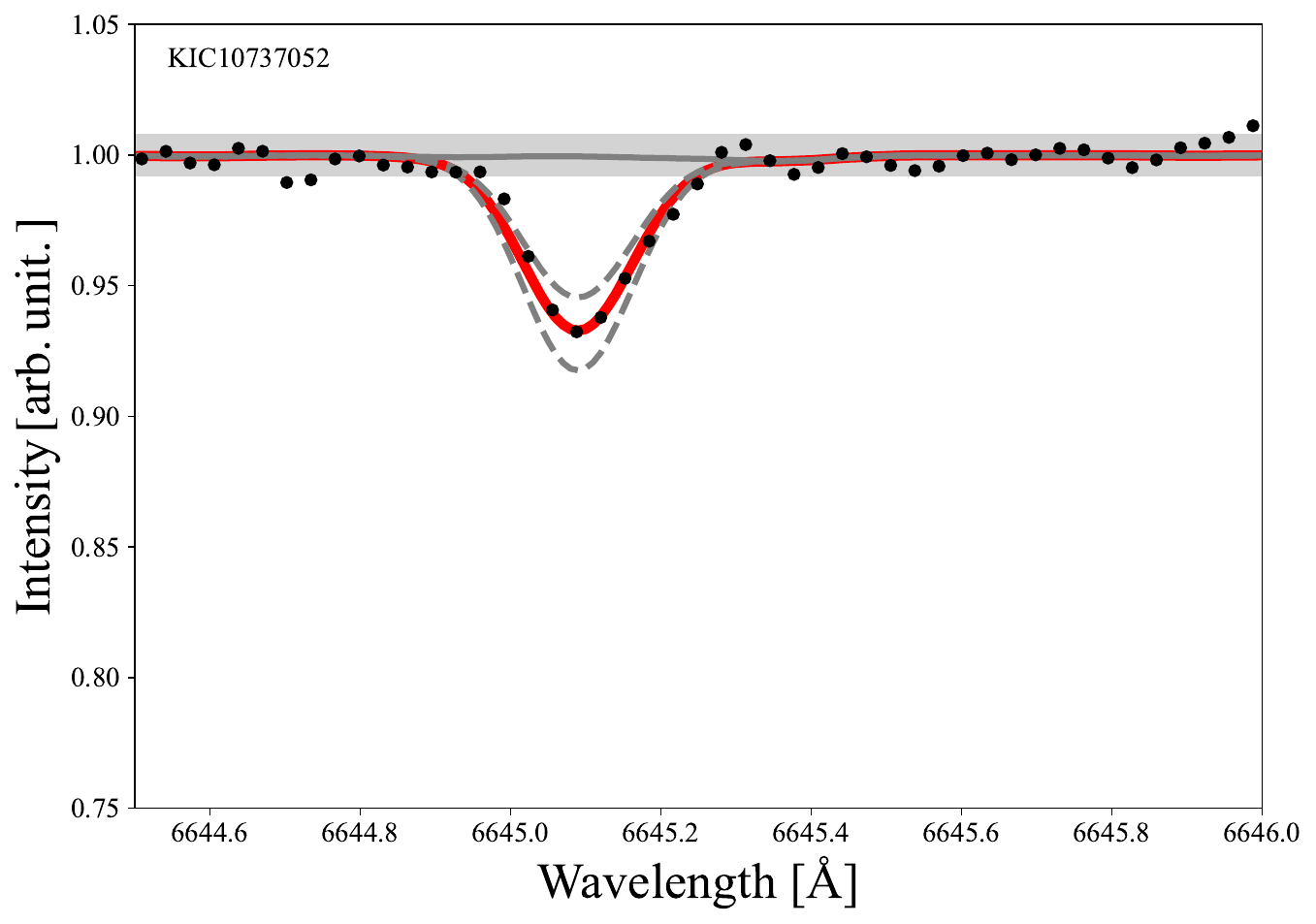}
\includegraphics[width=0.24\textwidth]{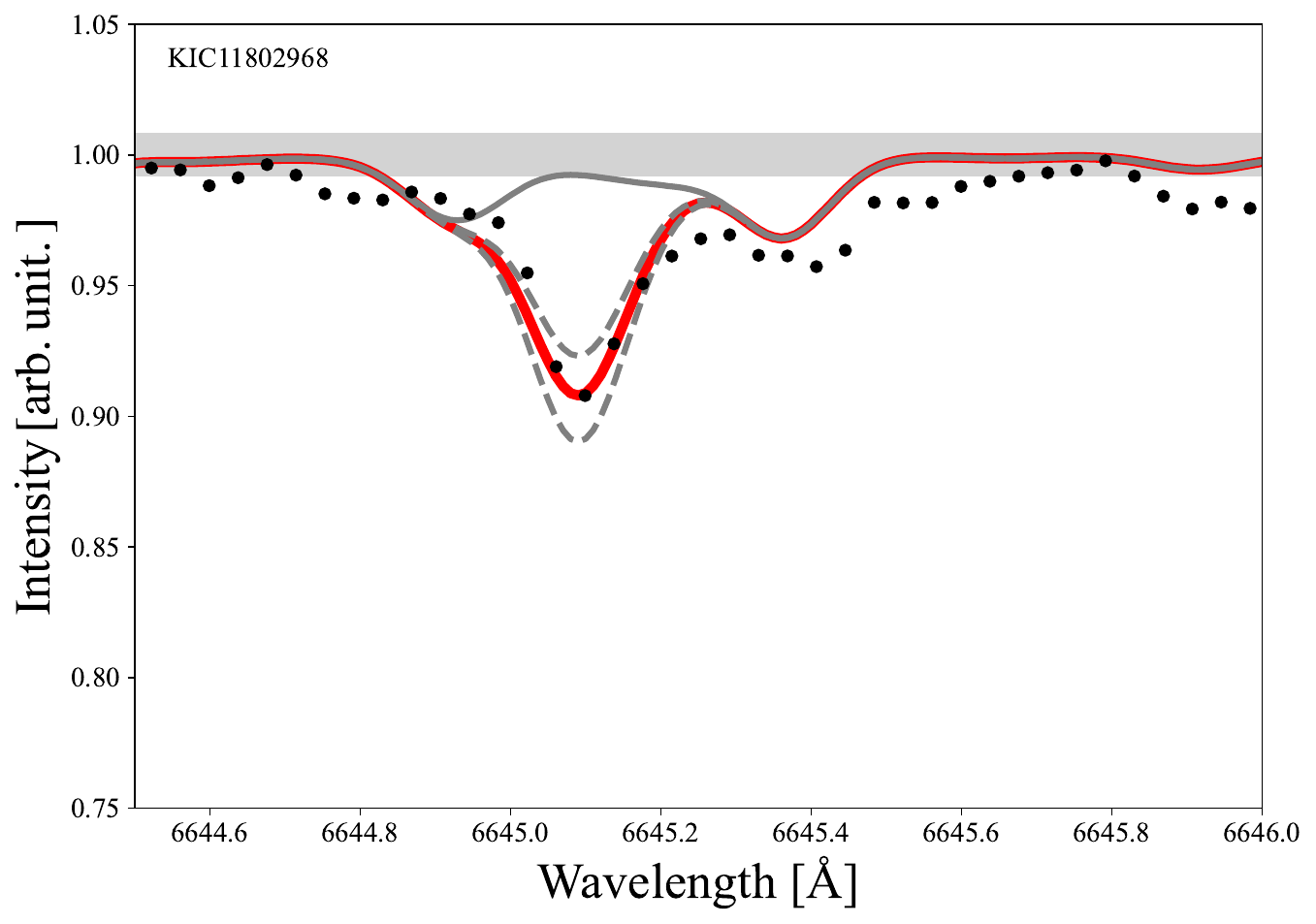}
\includegraphics[width=0.24\textwidth]{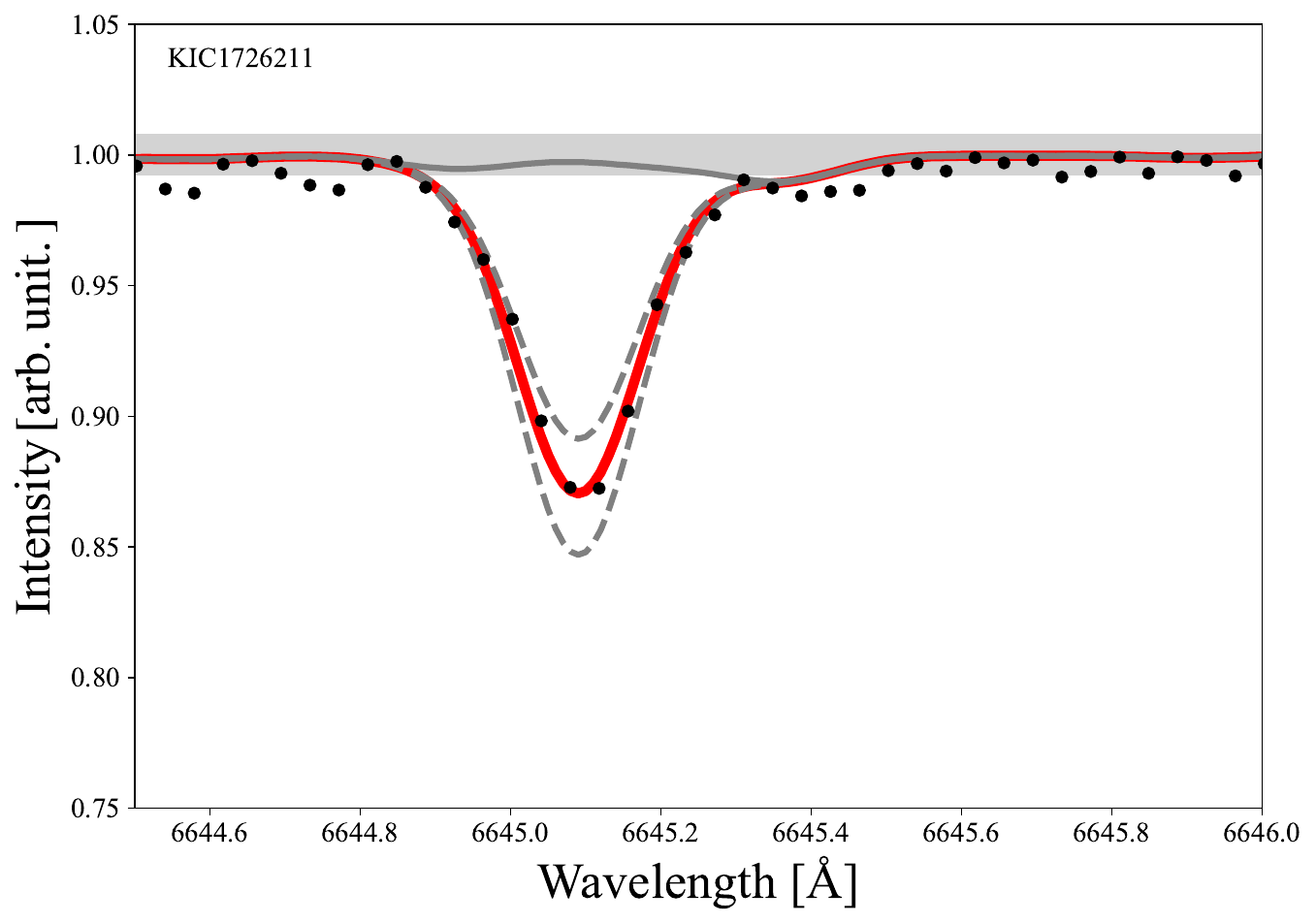}
\includegraphics[width=0.24\textwidth]{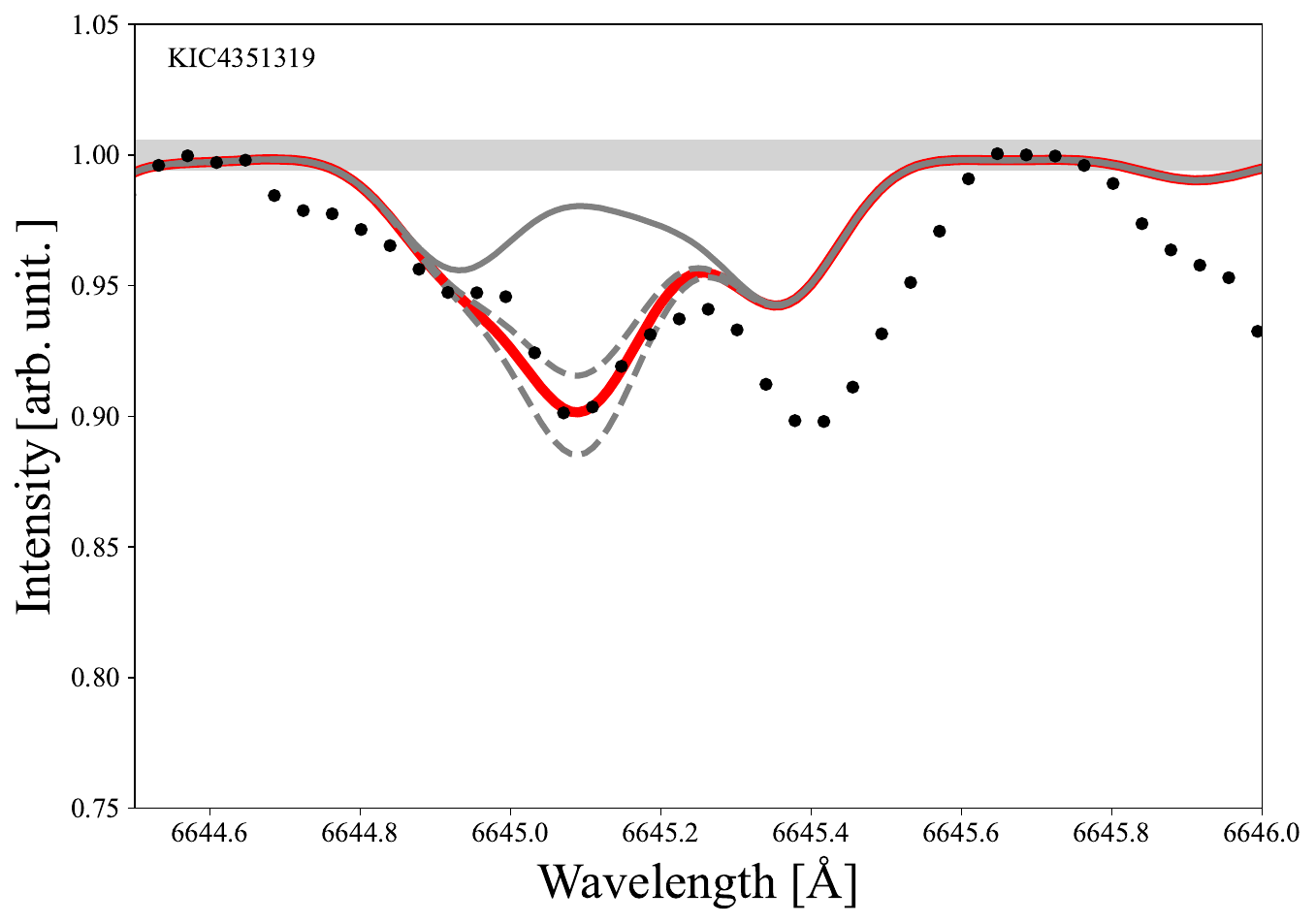}
\\[2mm]

\includegraphics[width=0.24\textwidth]{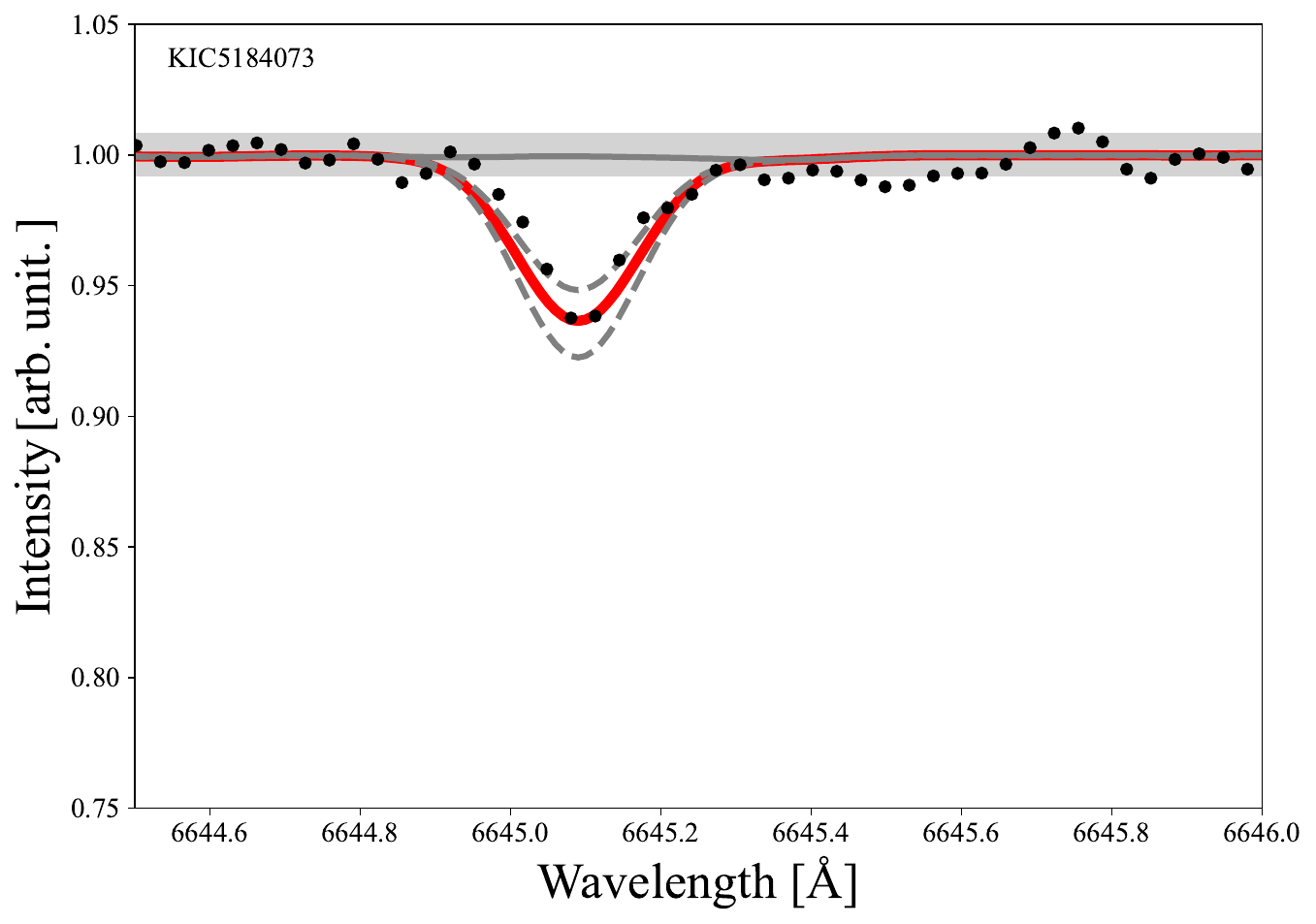}
\includegraphics[width=0.24\textwidth]{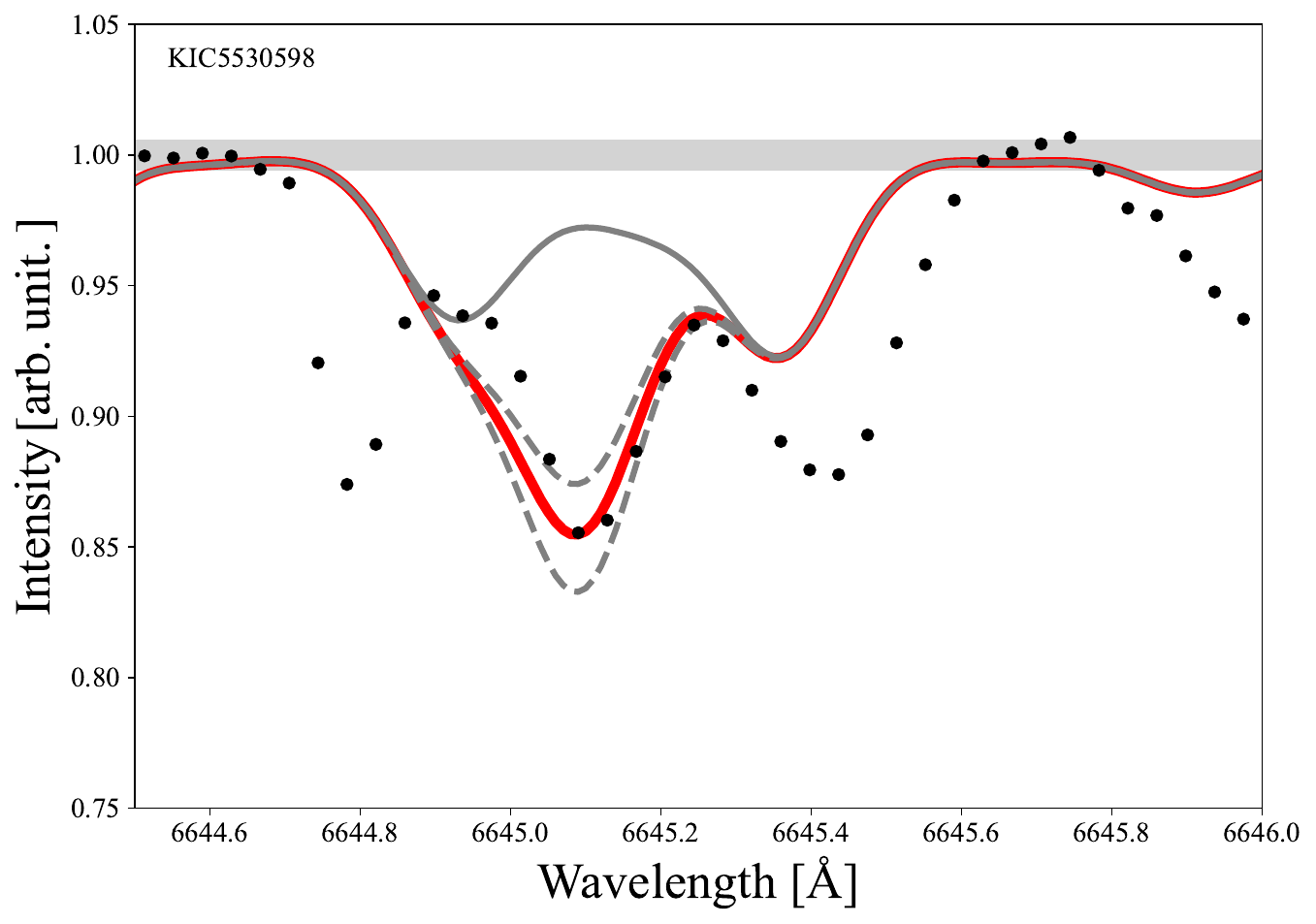}
\includegraphics[width=0.24\textwidth]{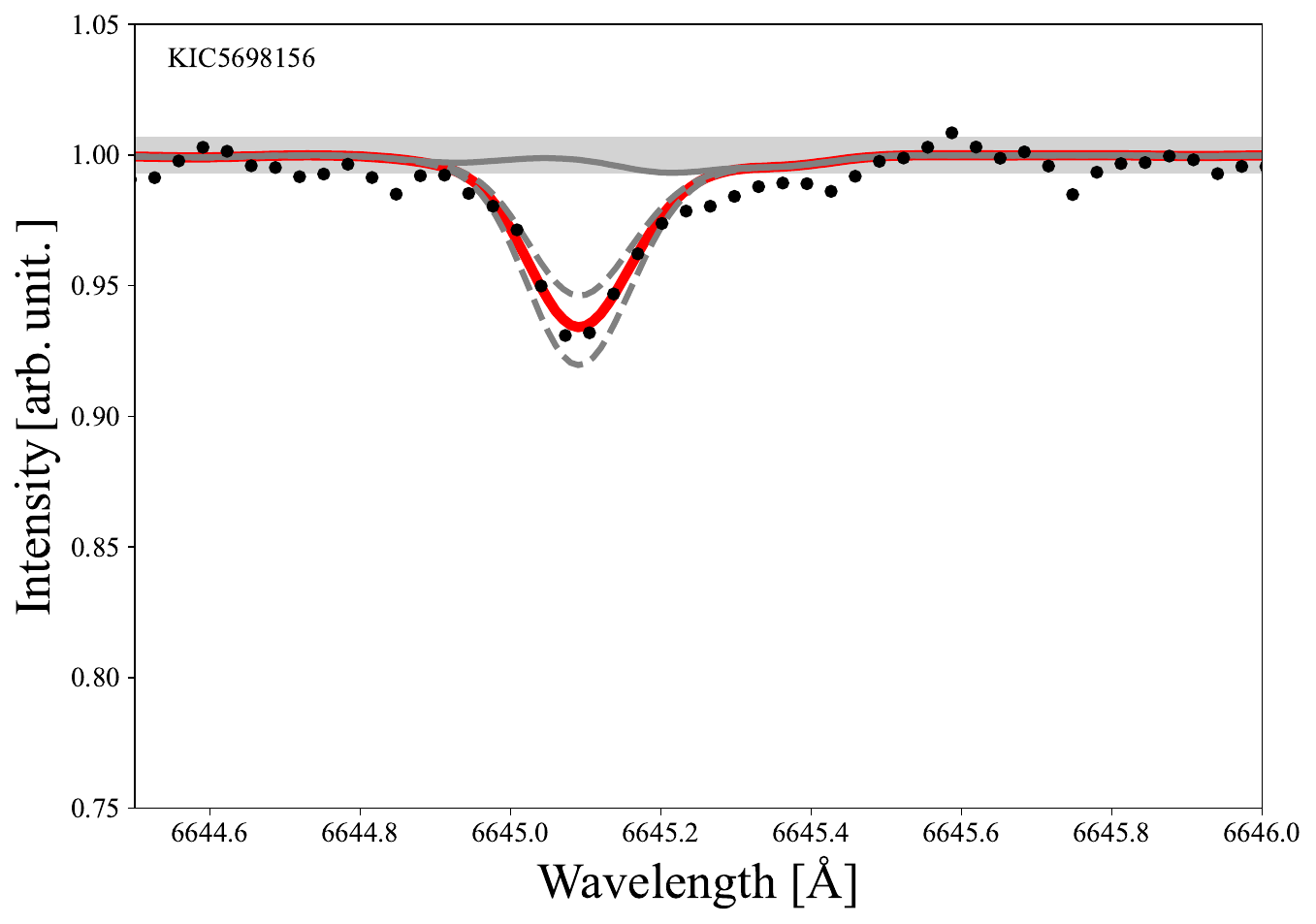}
\includegraphics[width=0.24\textwidth]{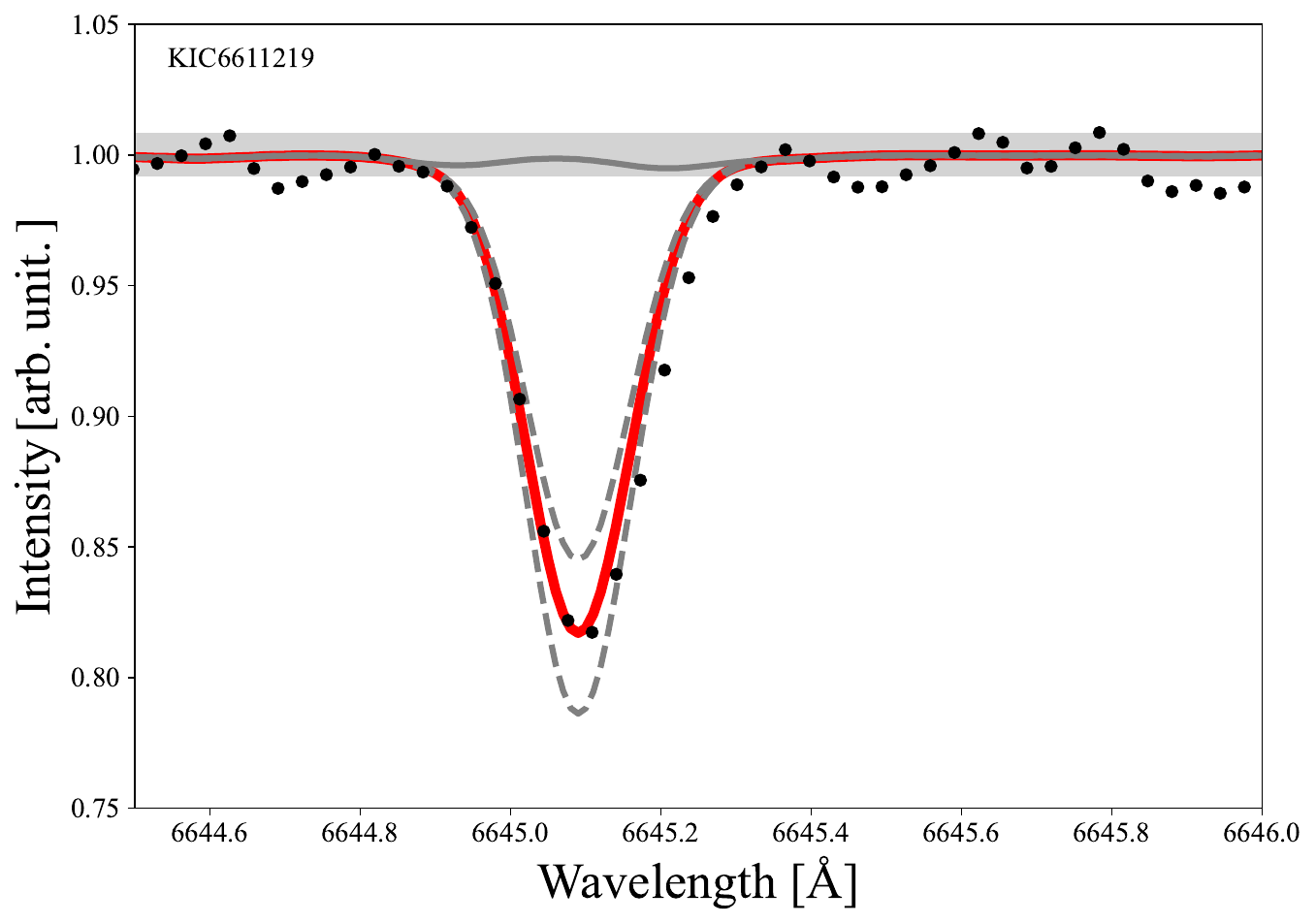}
\\[2mm]

\includegraphics[width=0.24\textwidth]{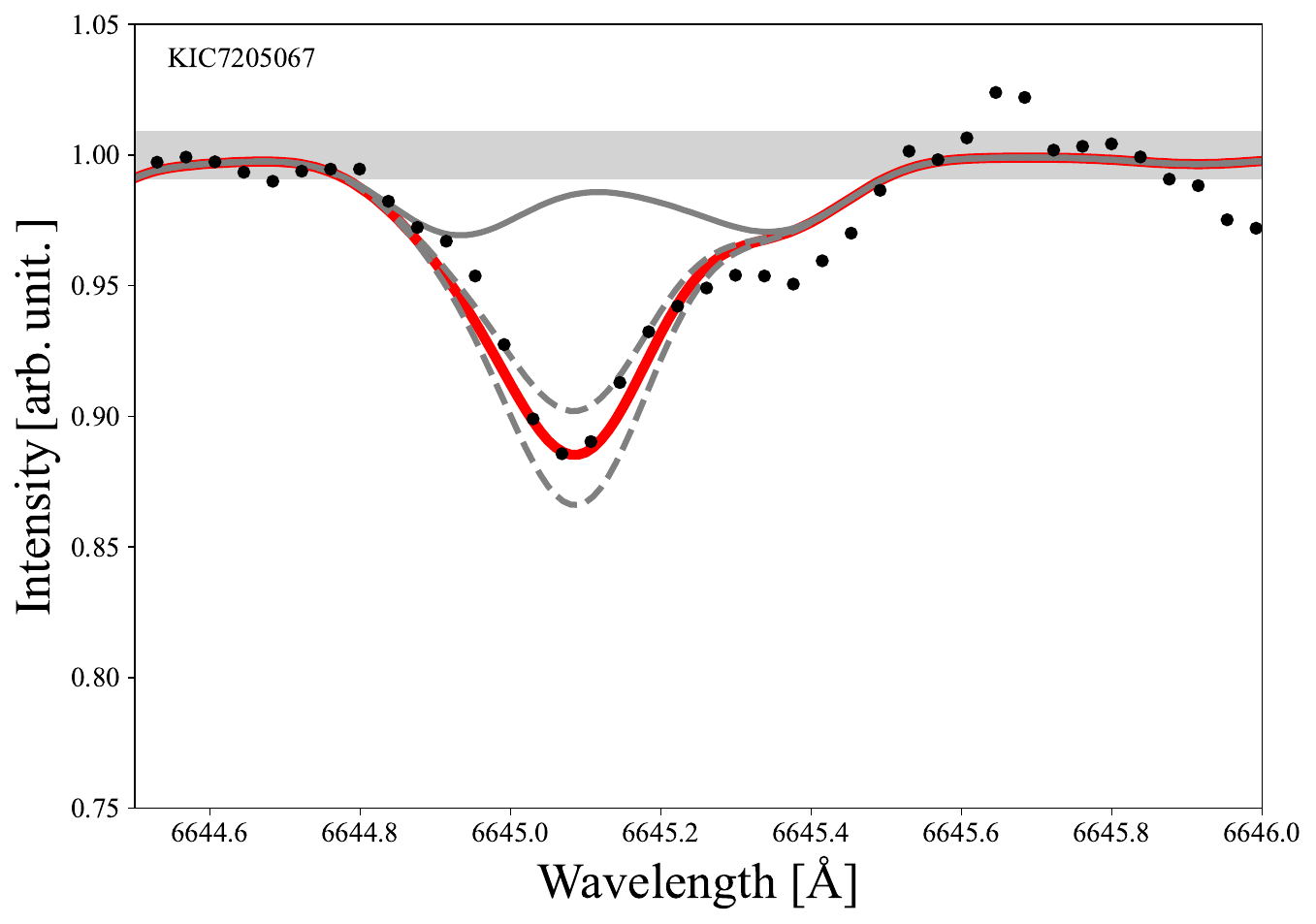}
\includegraphics[width=0.24\textwidth]{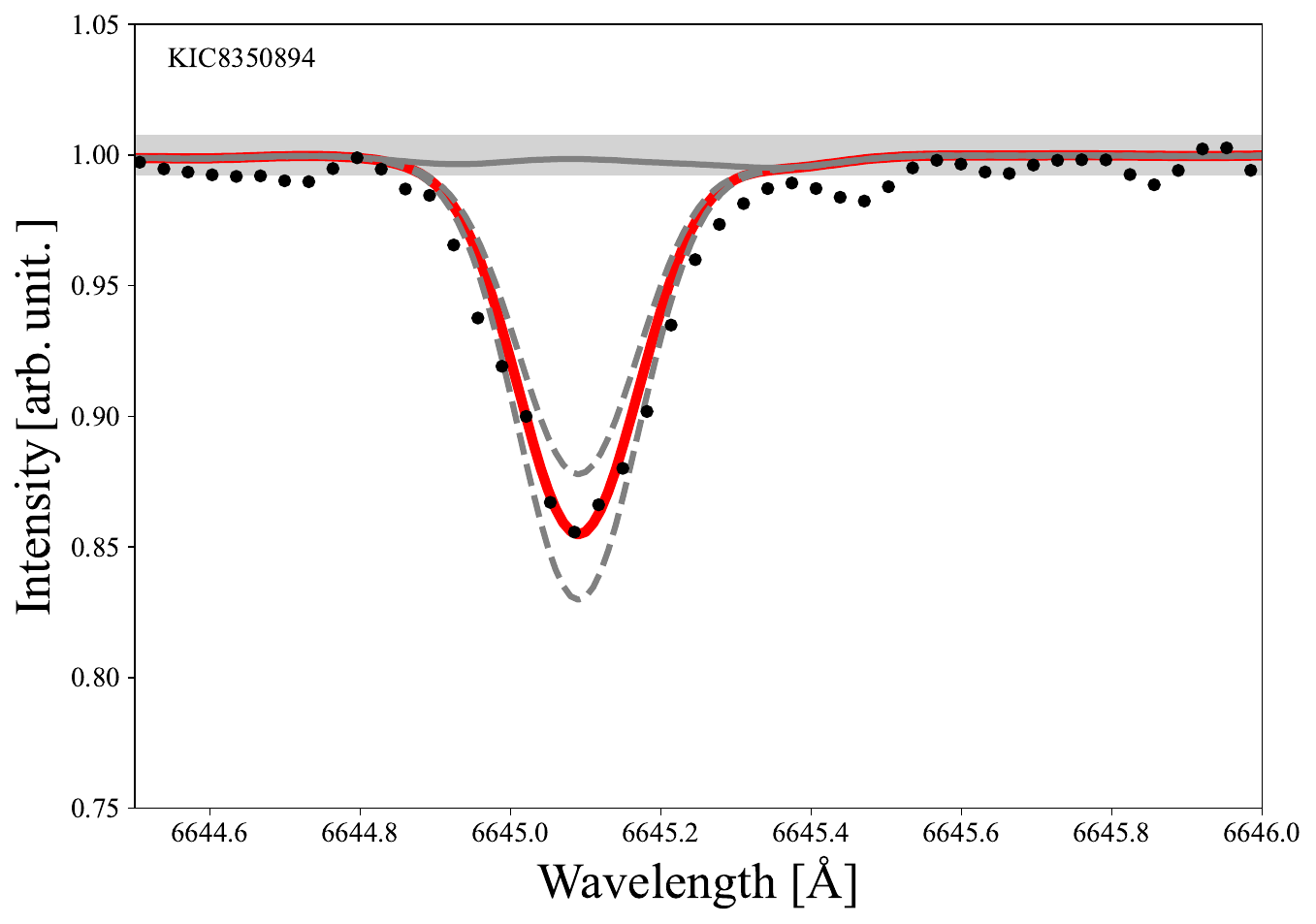}
\includegraphics[width=0.24\textwidth]{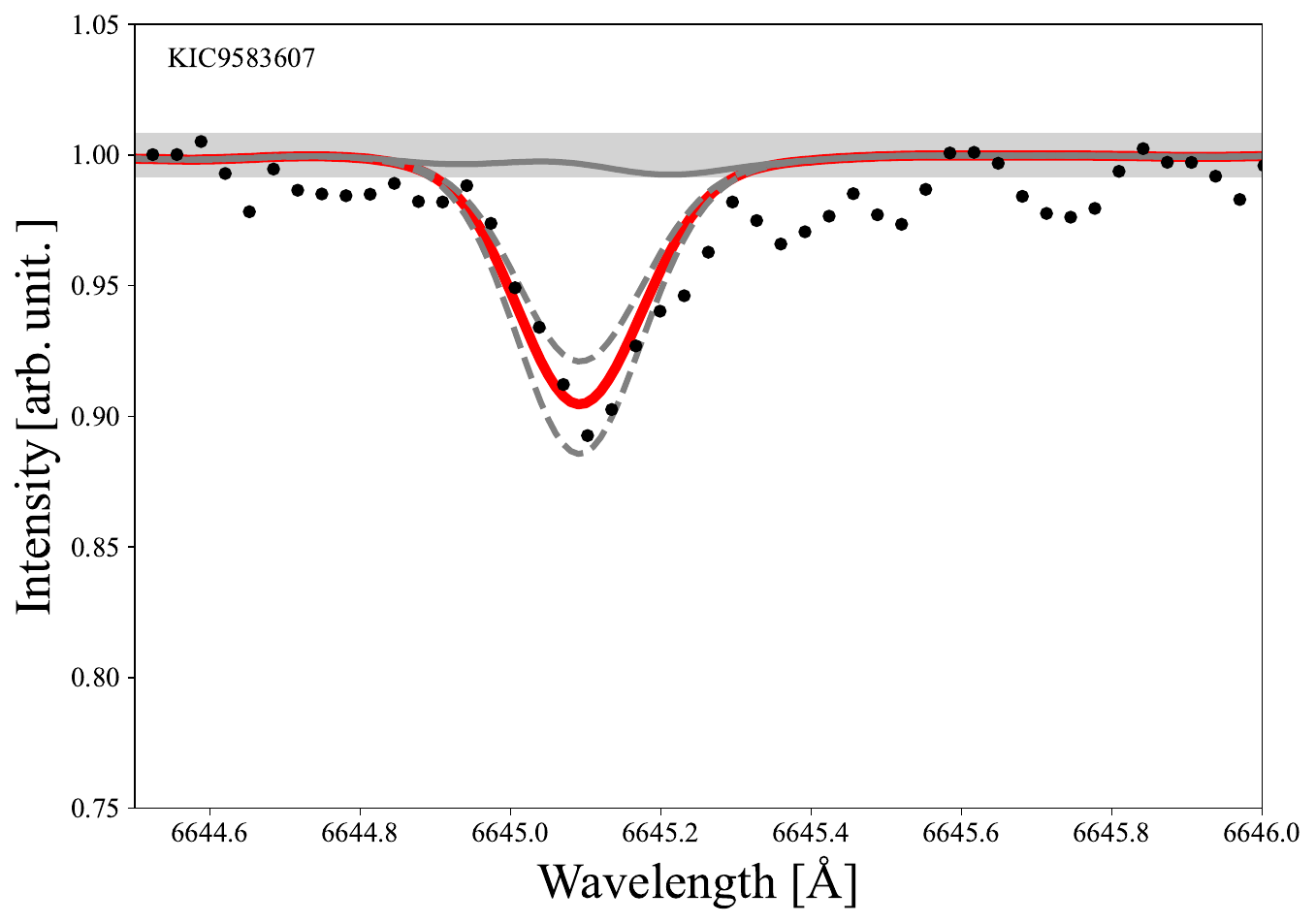}

\caption{(Continued)
 {Alt text: All spectra of Europium 6645 angstrom line (Continued). } 
 }
 \label{fig:Eu_allspectra_2}
 \end{figure*}

 \textbf{\section{All spectra; Thorium}}
    Synthetic spectra fits for the Th line for all samples. Th detected objects are shown in Figure \ref{fig:allspe_Thdetct1}, Th upper limit objects are shown in Figure \ref{fig:allspe_Thuplim}.

\begin{figure*}[t]
\centering

\includegraphics[width=0.24\textwidth]{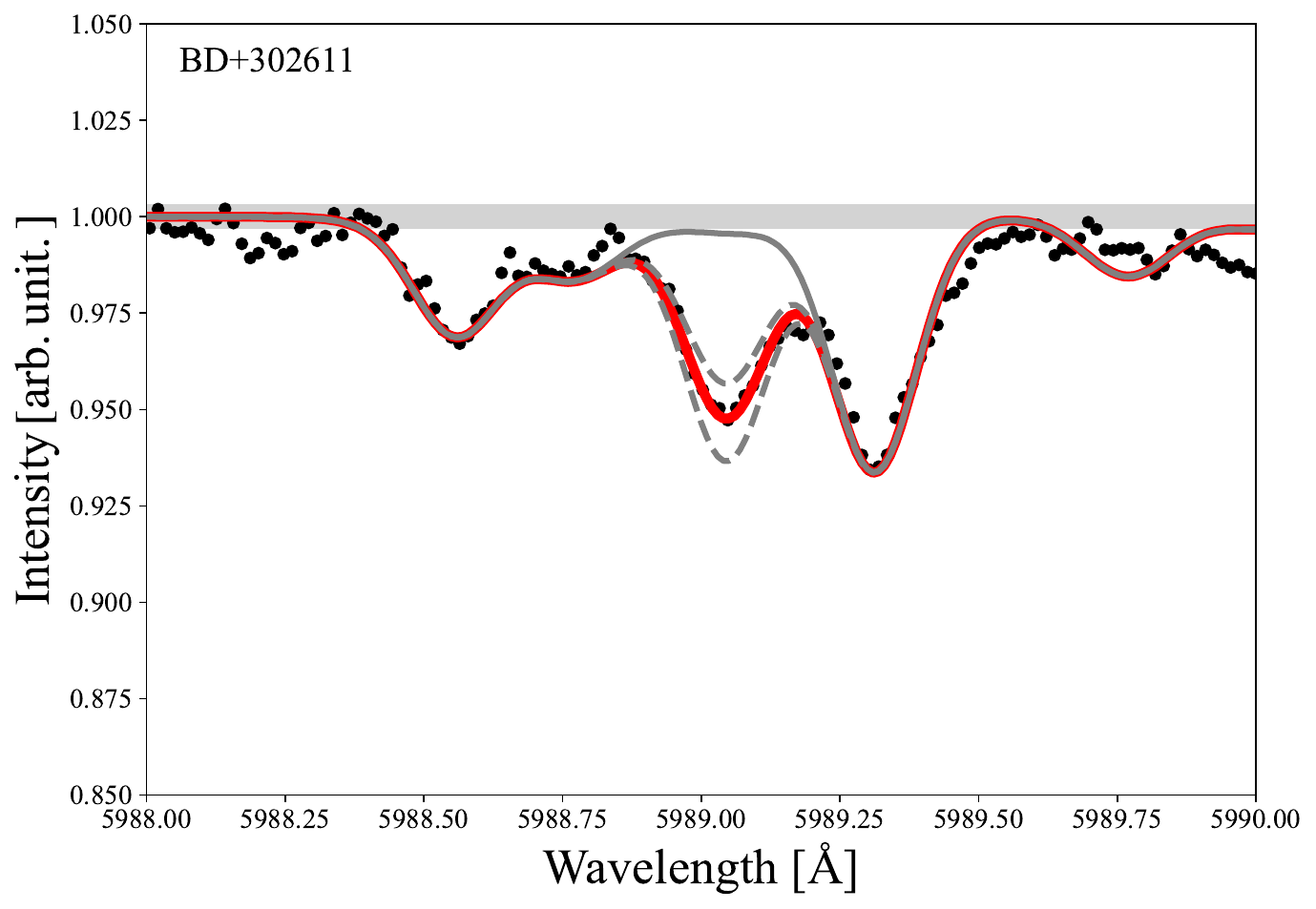}
\includegraphics[width=0.24\textwidth]{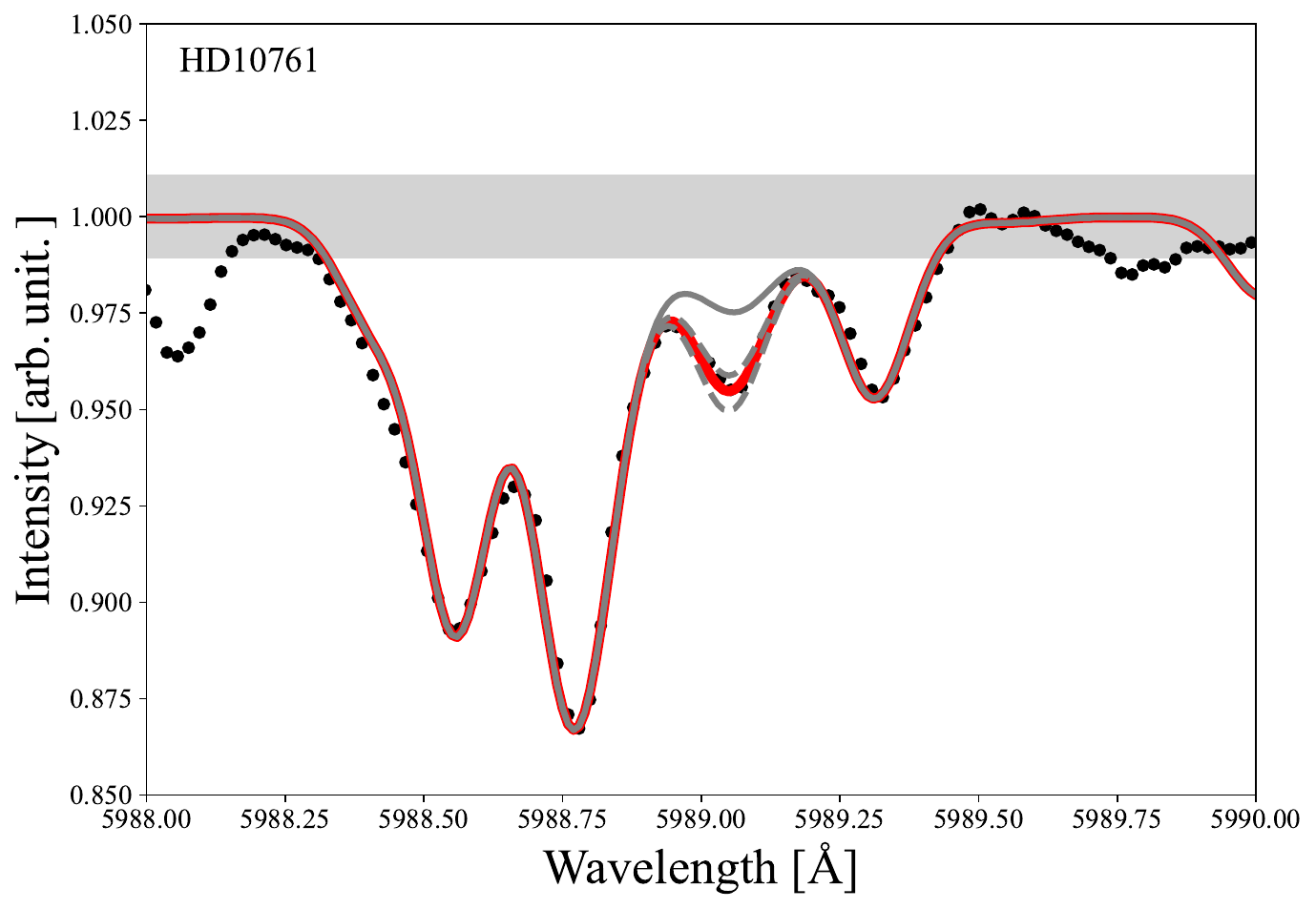}
\includegraphics[width=0.24\textwidth]{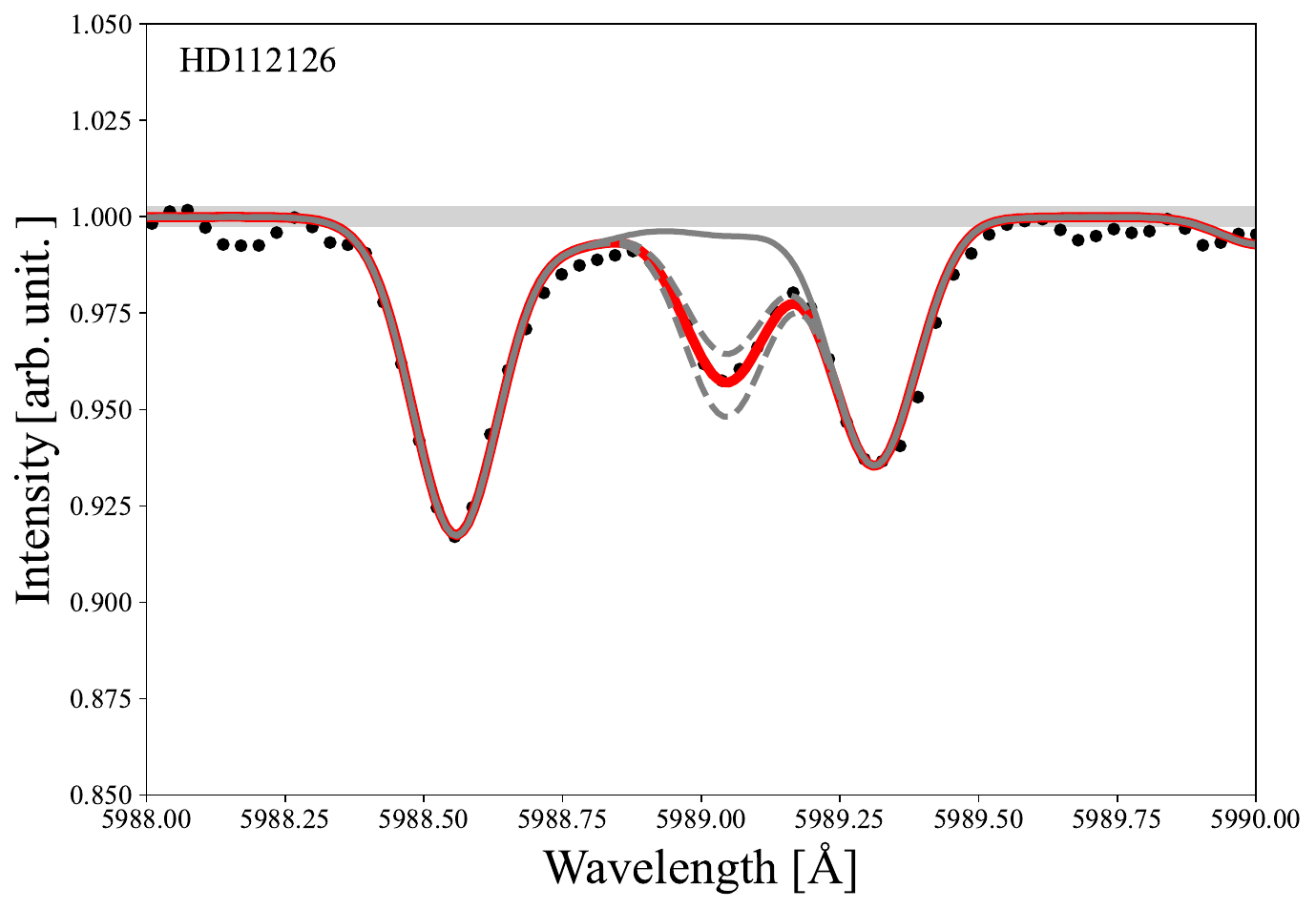}
\includegraphics[width=0.24\textwidth]{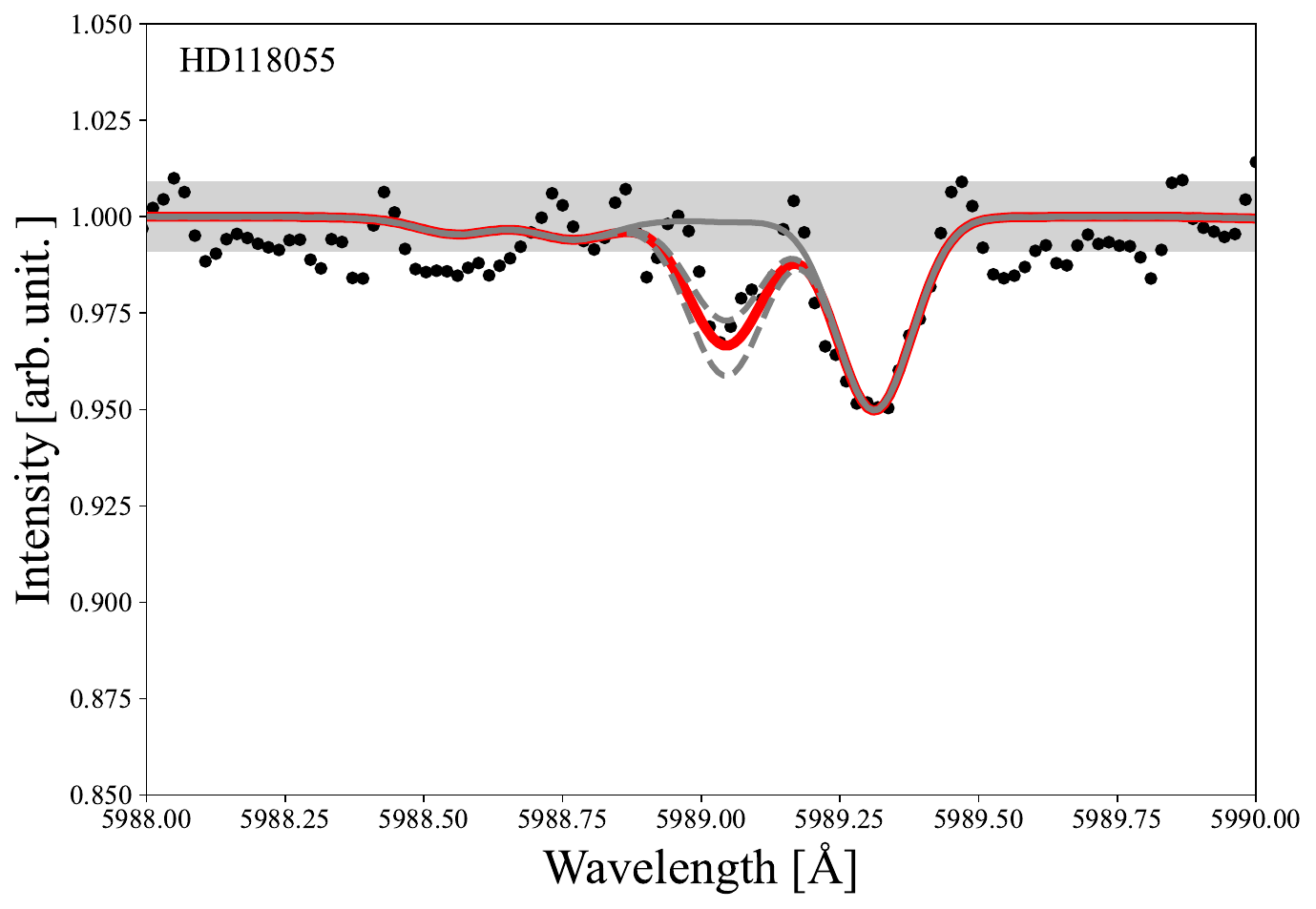}
\\[2mm]

\includegraphics[width=0.24\textwidth]{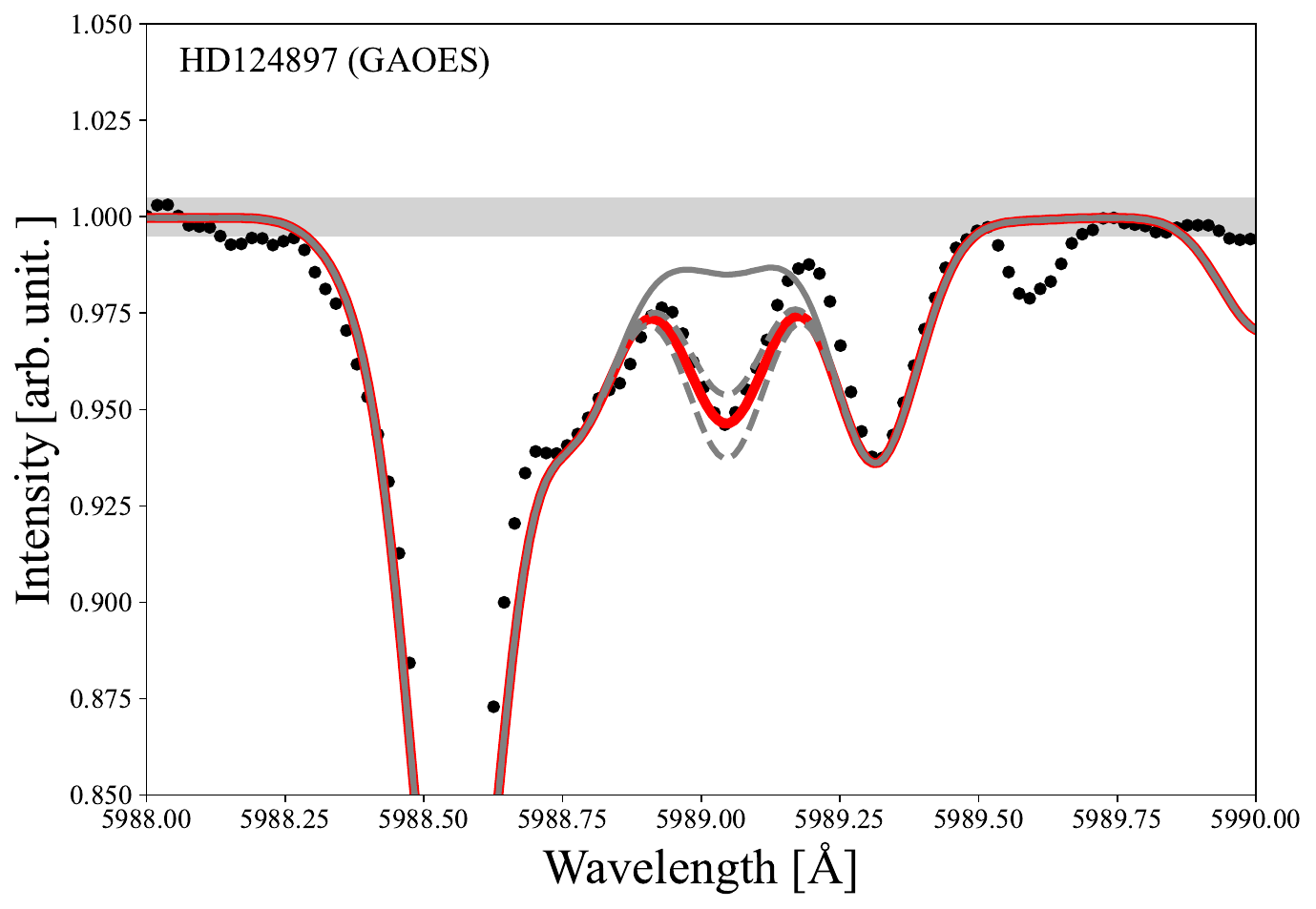}
\includegraphics[width=0.24\textwidth]{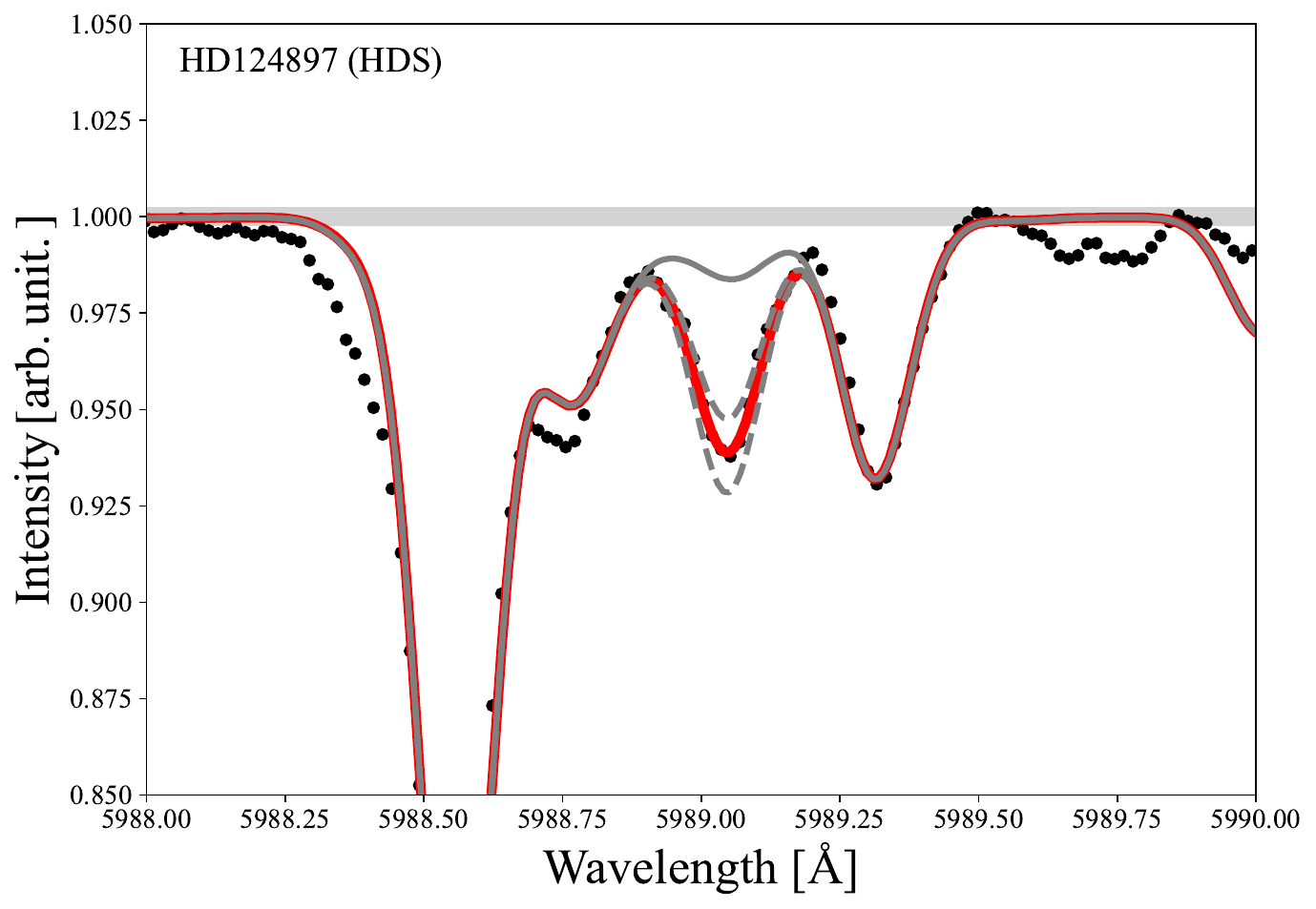}
\includegraphics[width=0.24\textwidth]{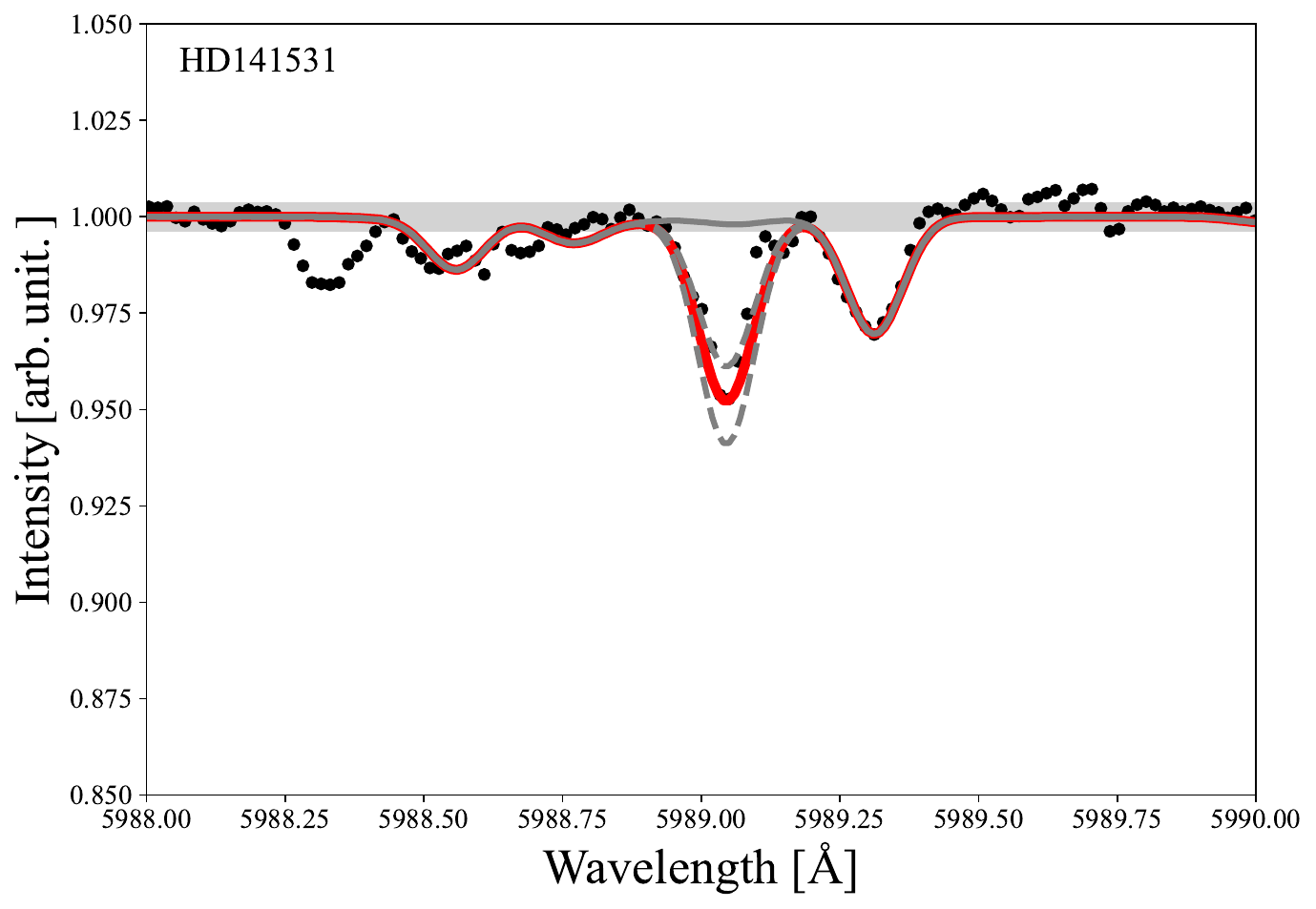}
\includegraphics[width=0.24\textwidth]{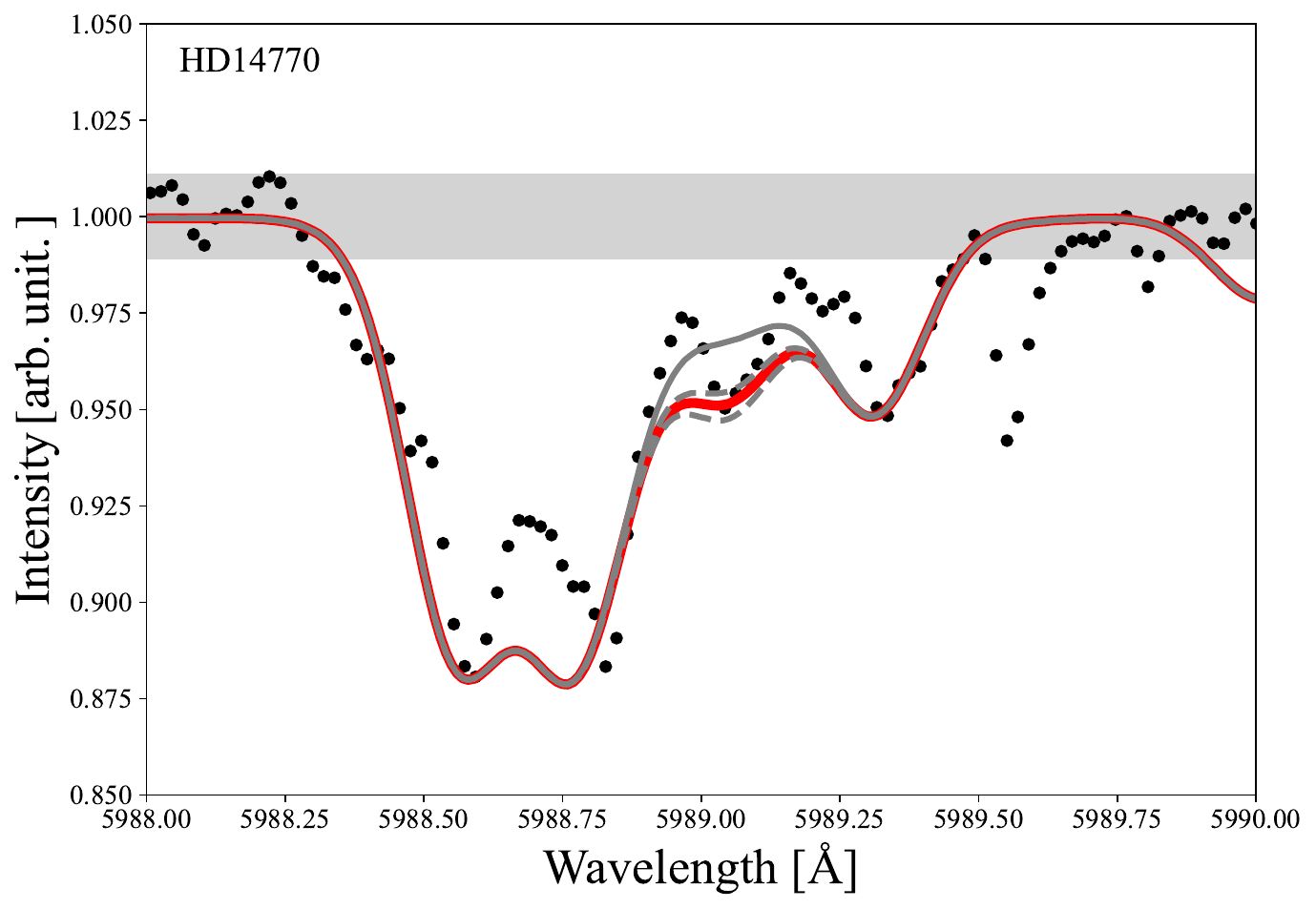}
\\[2mm]

\includegraphics[width=0.24\textwidth]{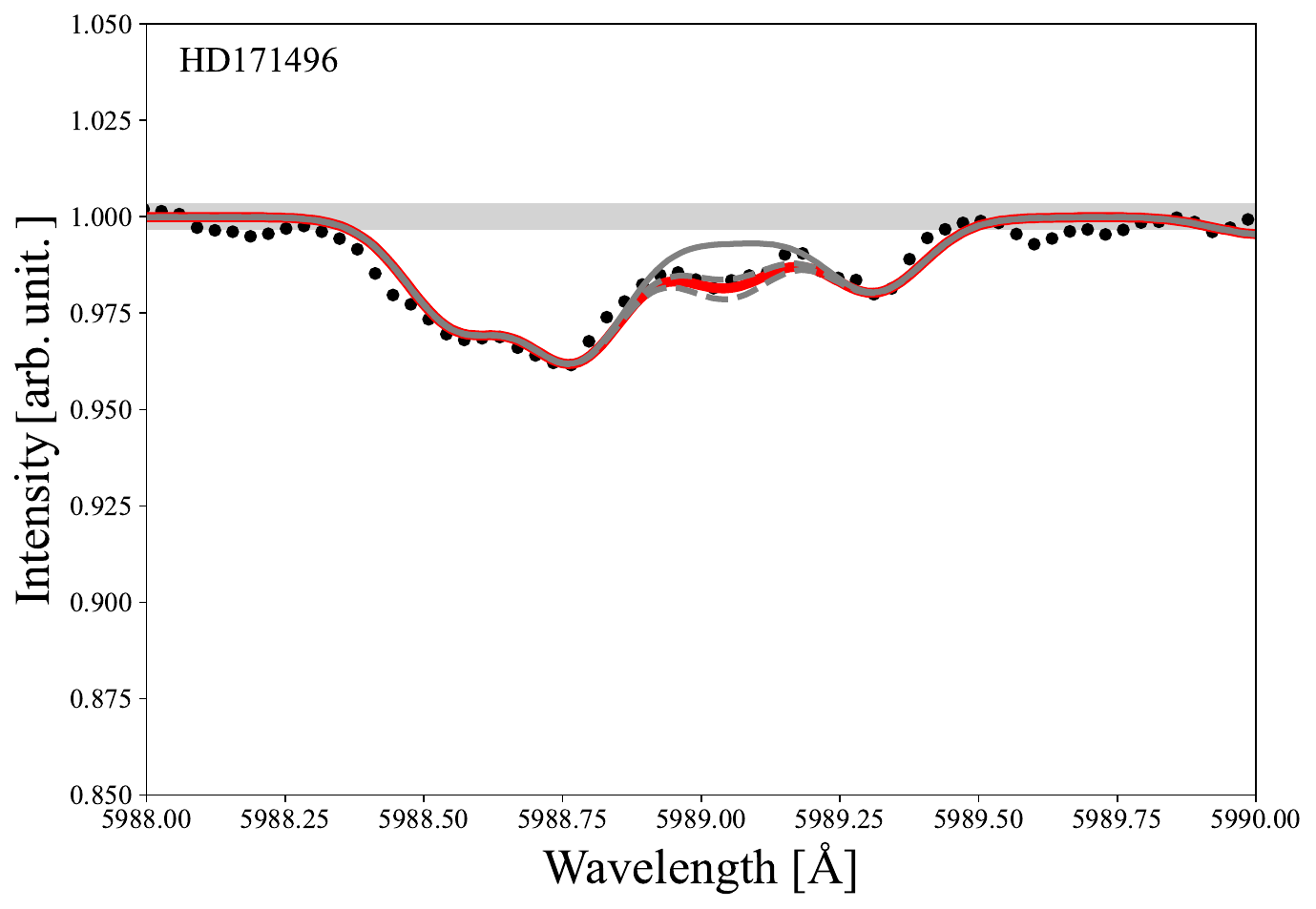}
\includegraphics[width=0.24\textwidth]{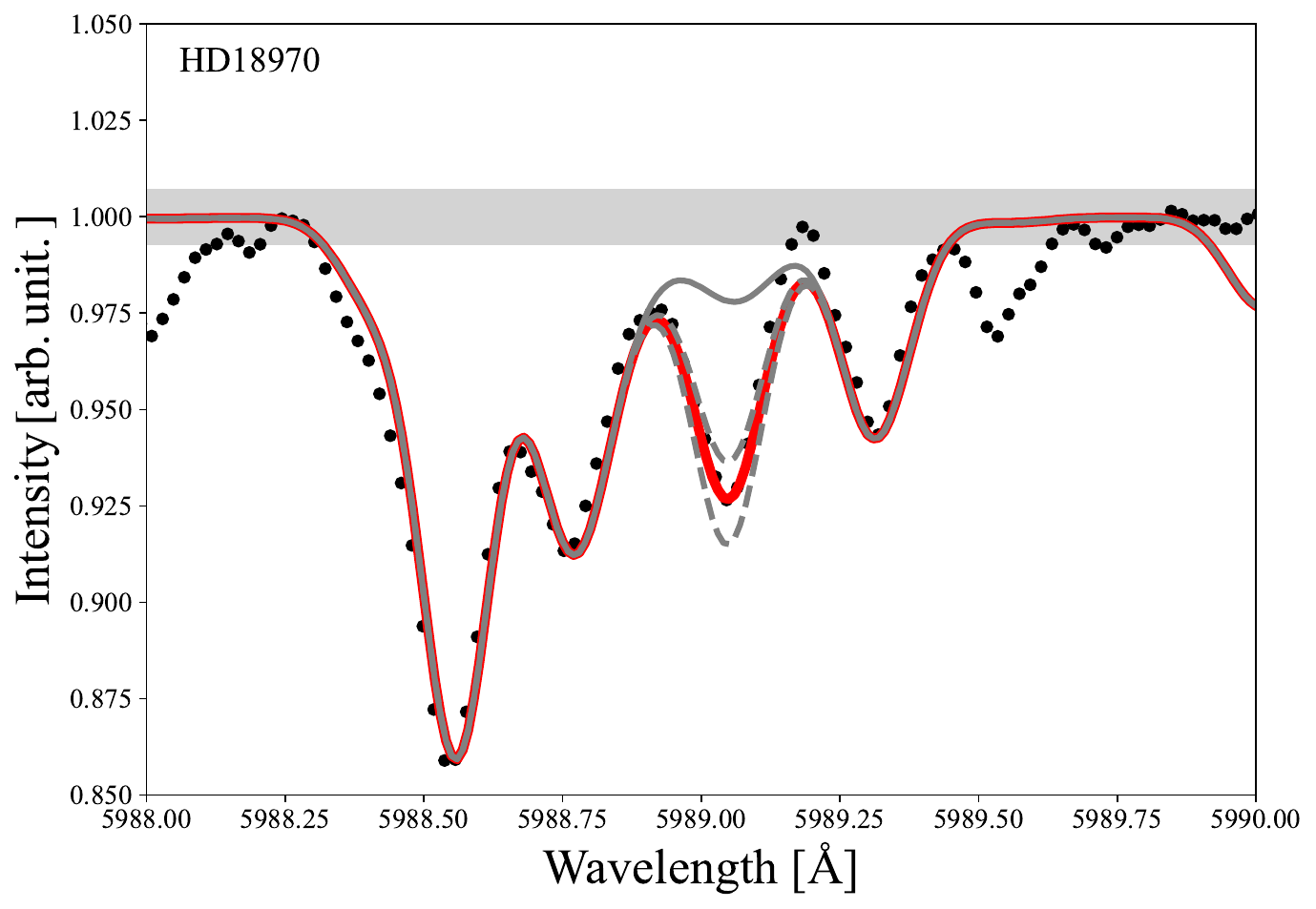}
\includegraphics[width=0.24\textwidth]{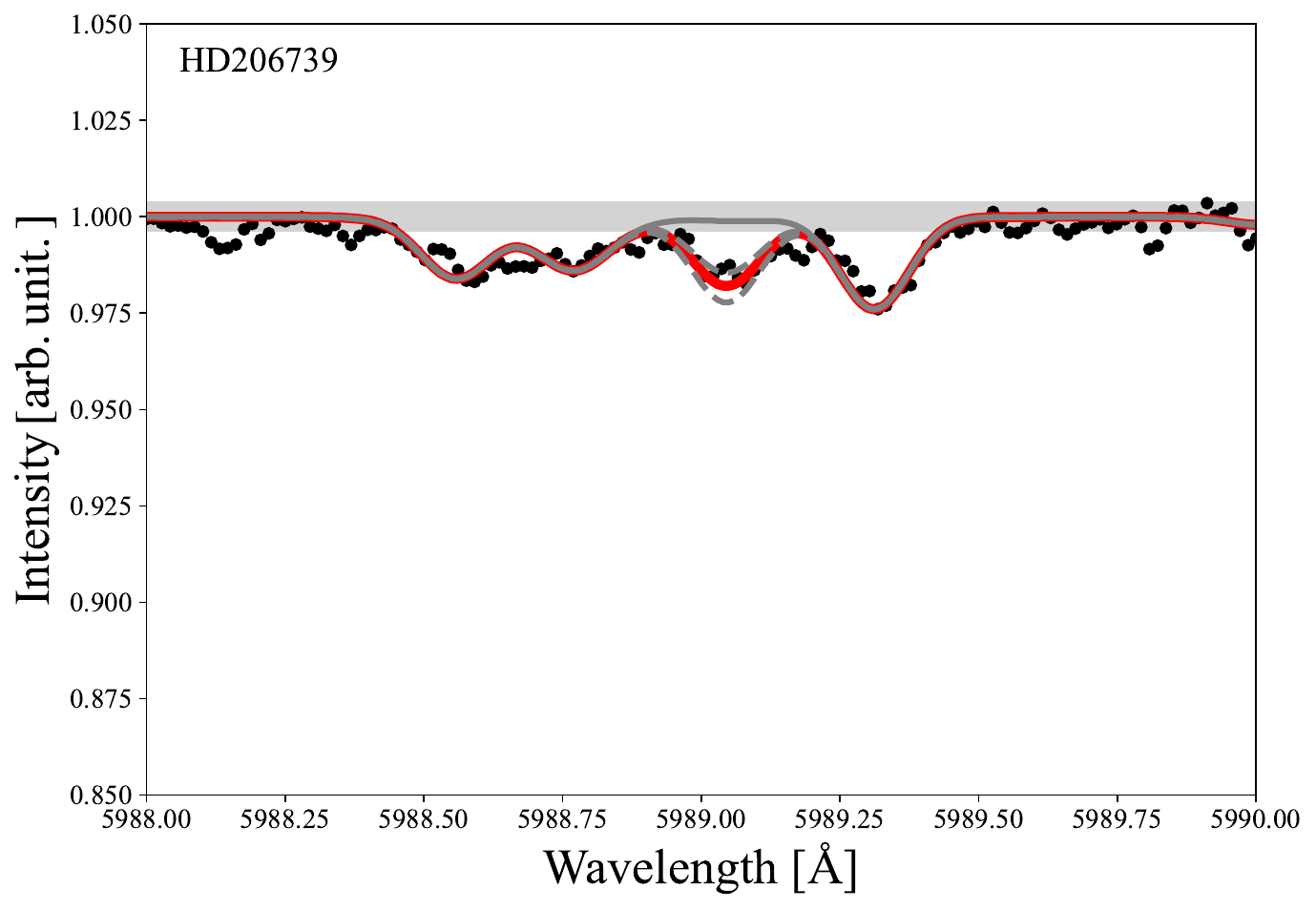}
\includegraphics[width=0.24\textwidth]{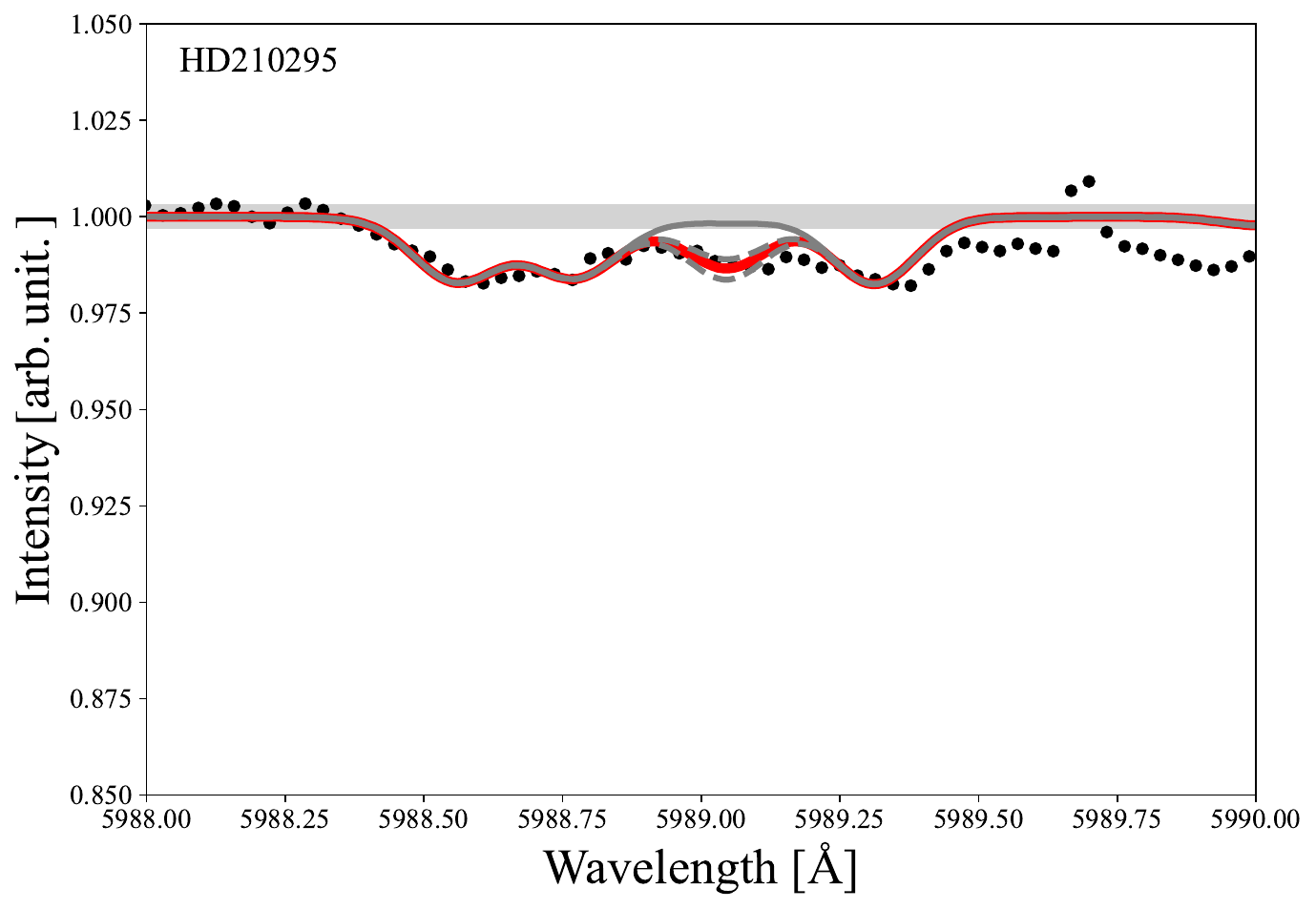}
\\[2mm]

\includegraphics[width=0.24\textwidth]{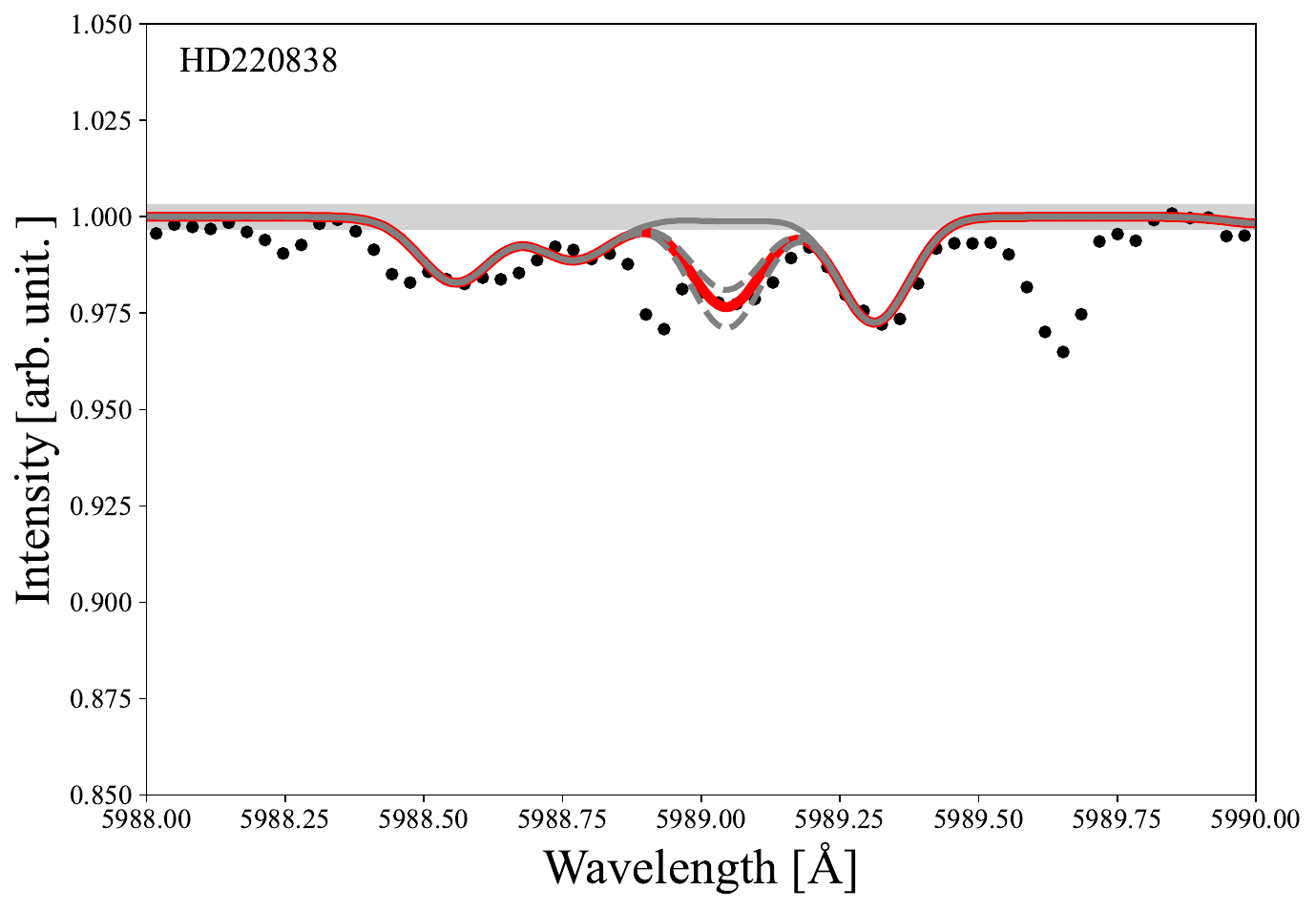}
\includegraphics[width=0.24\textwidth]{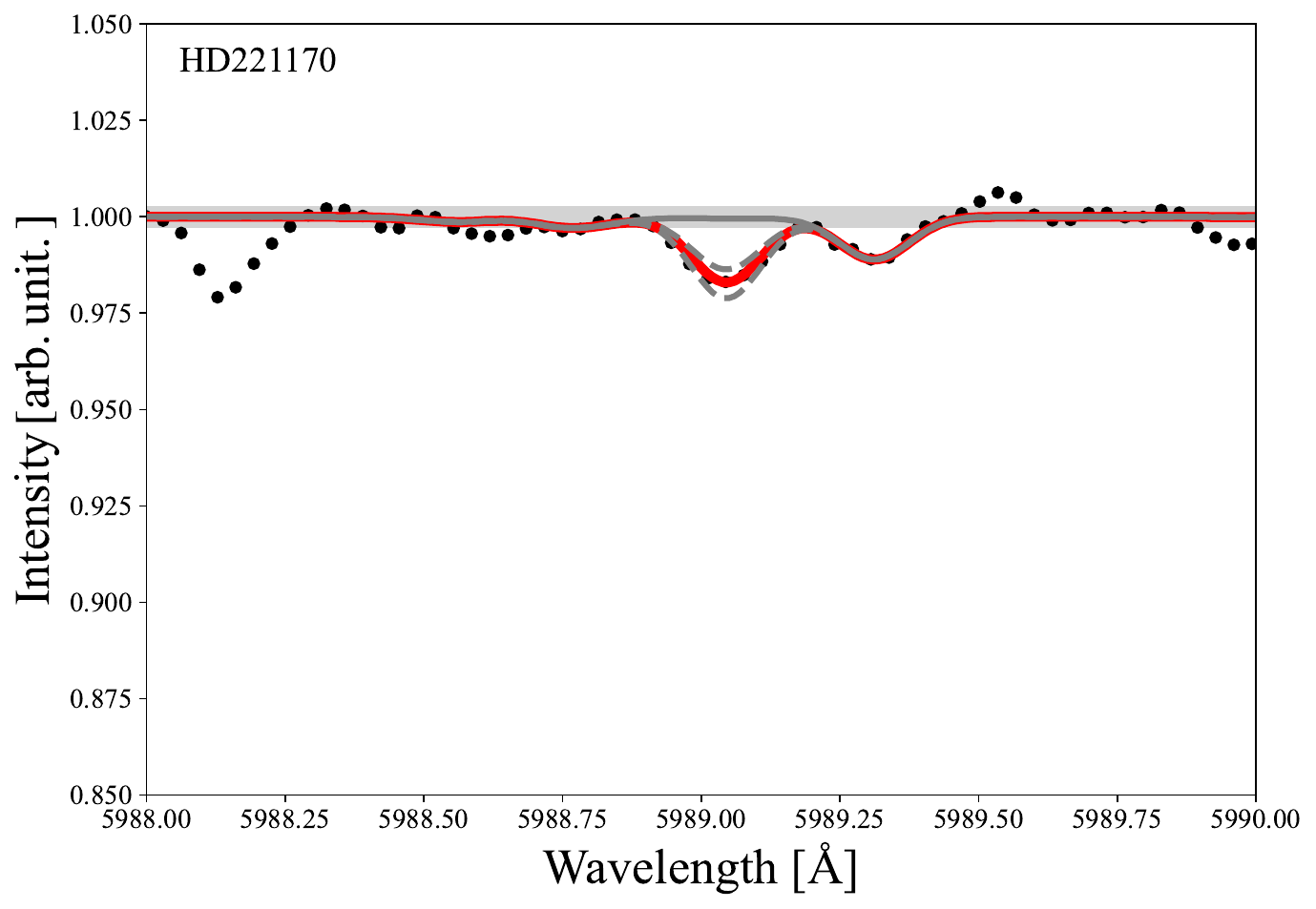}
\includegraphics[width=0.24\textwidth]{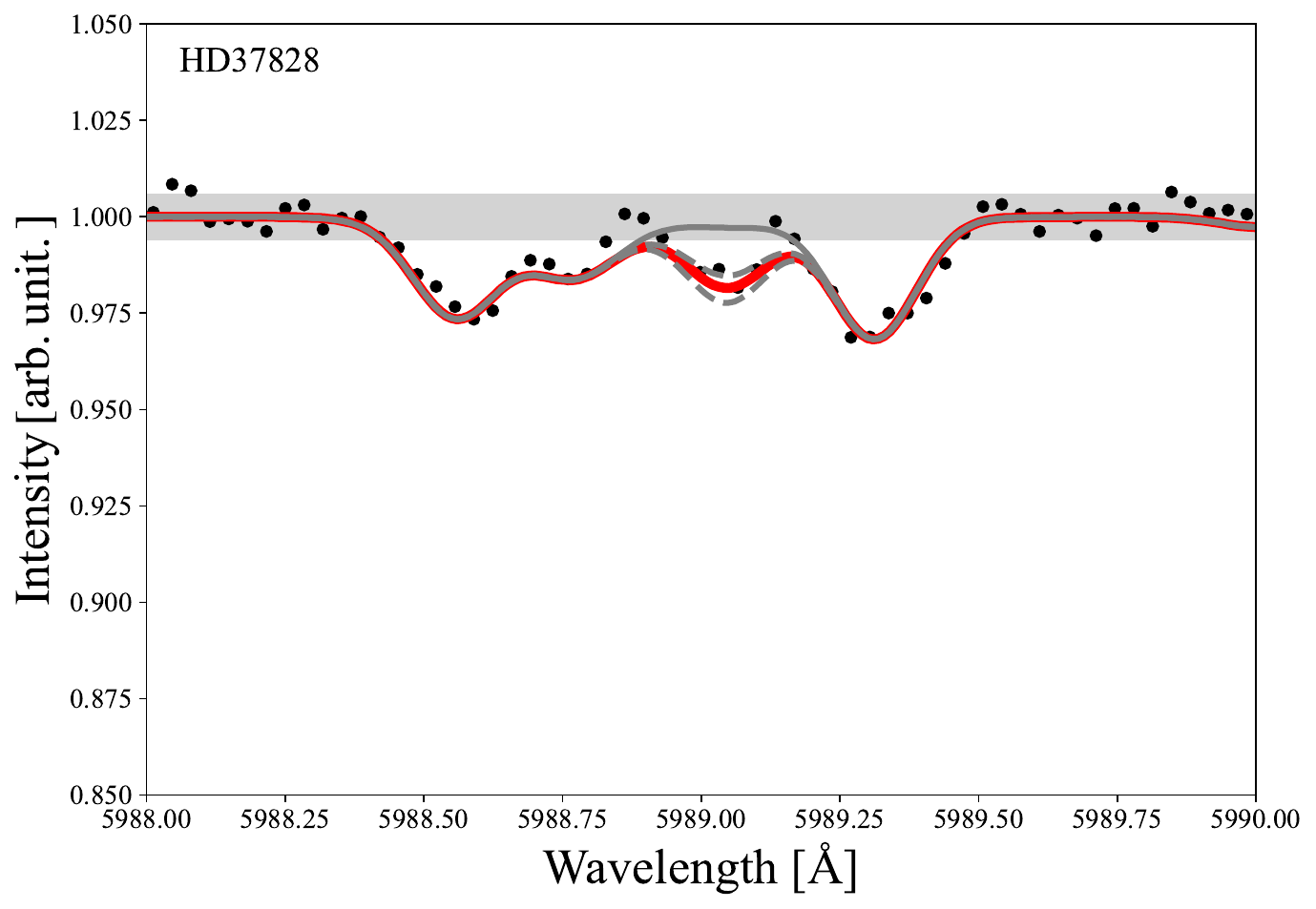}
\includegraphics[width=0.24\textwidth]{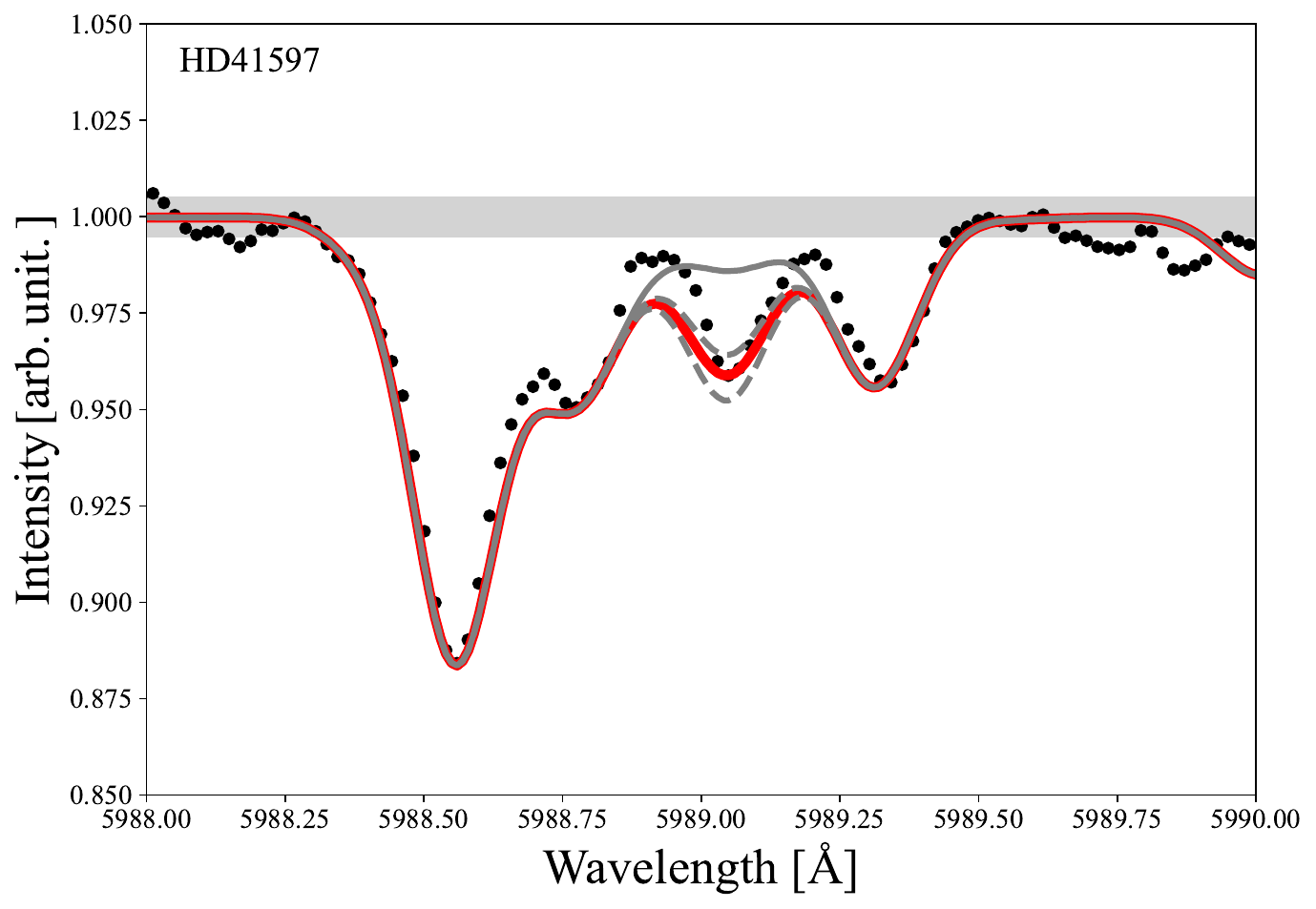}
\\[2mm]

\includegraphics[width=0.24\textwidth]{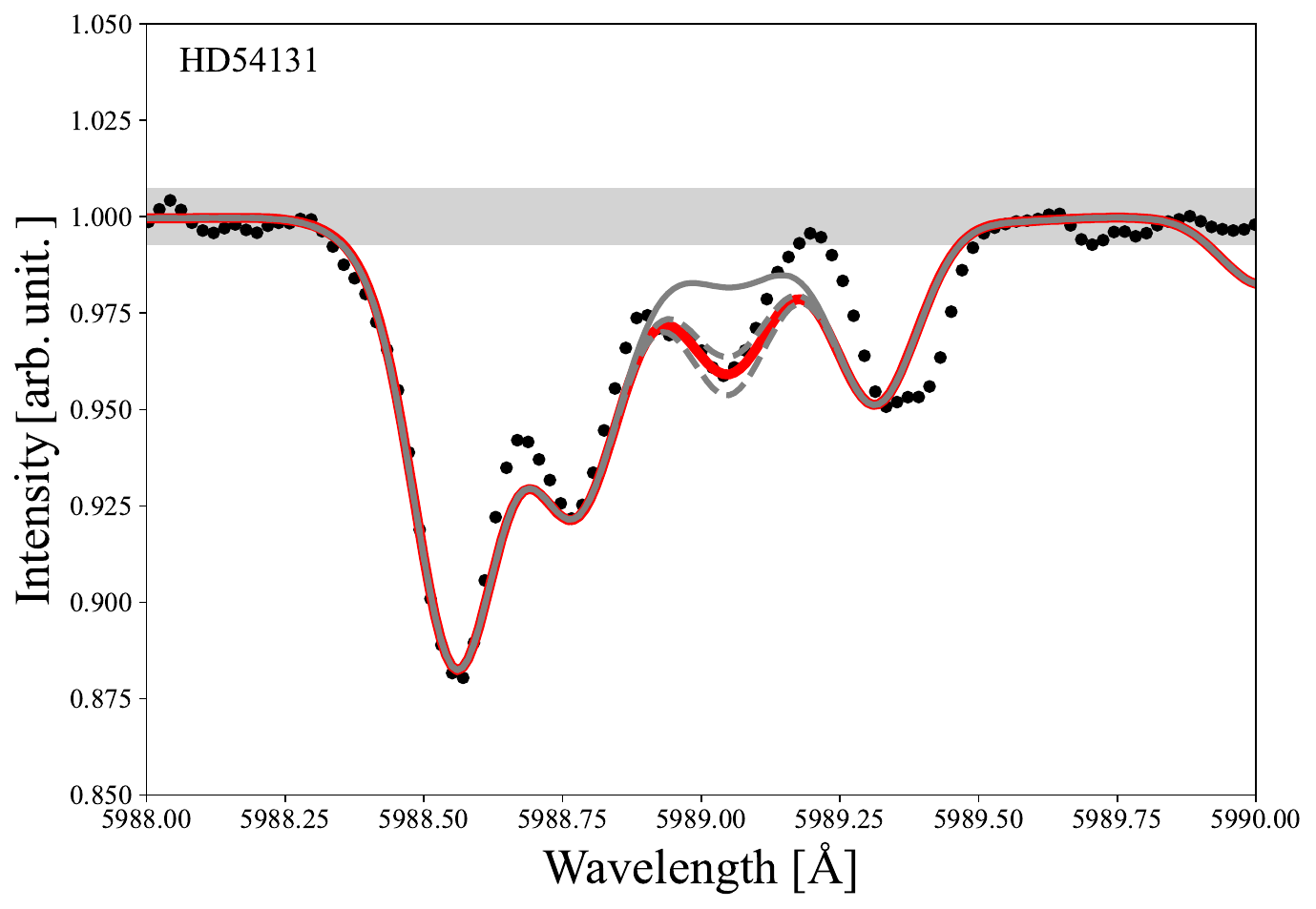}
\includegraphics[width=0.24\textwidth]{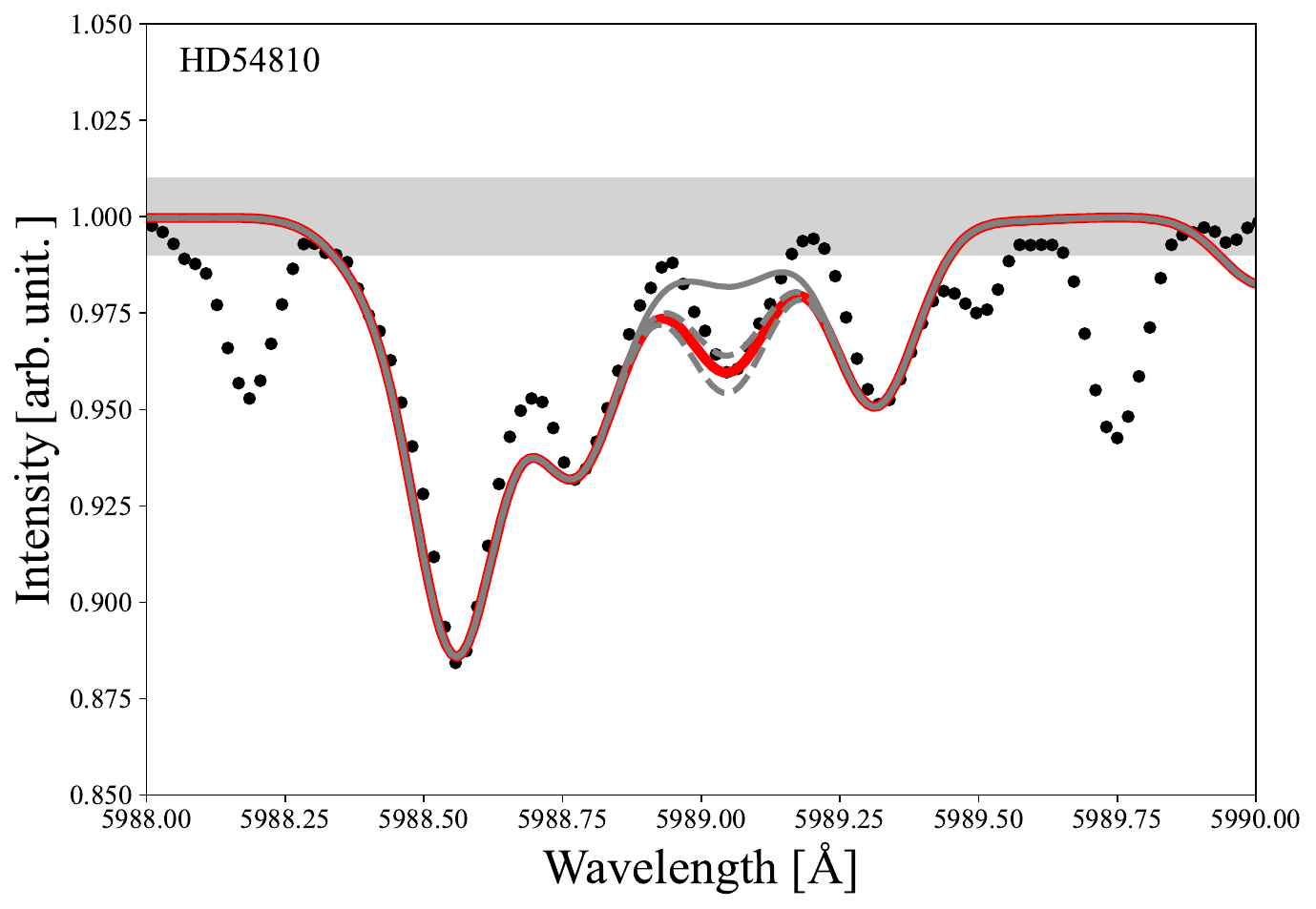}
\includegraphics[width=0.24\textwidth]{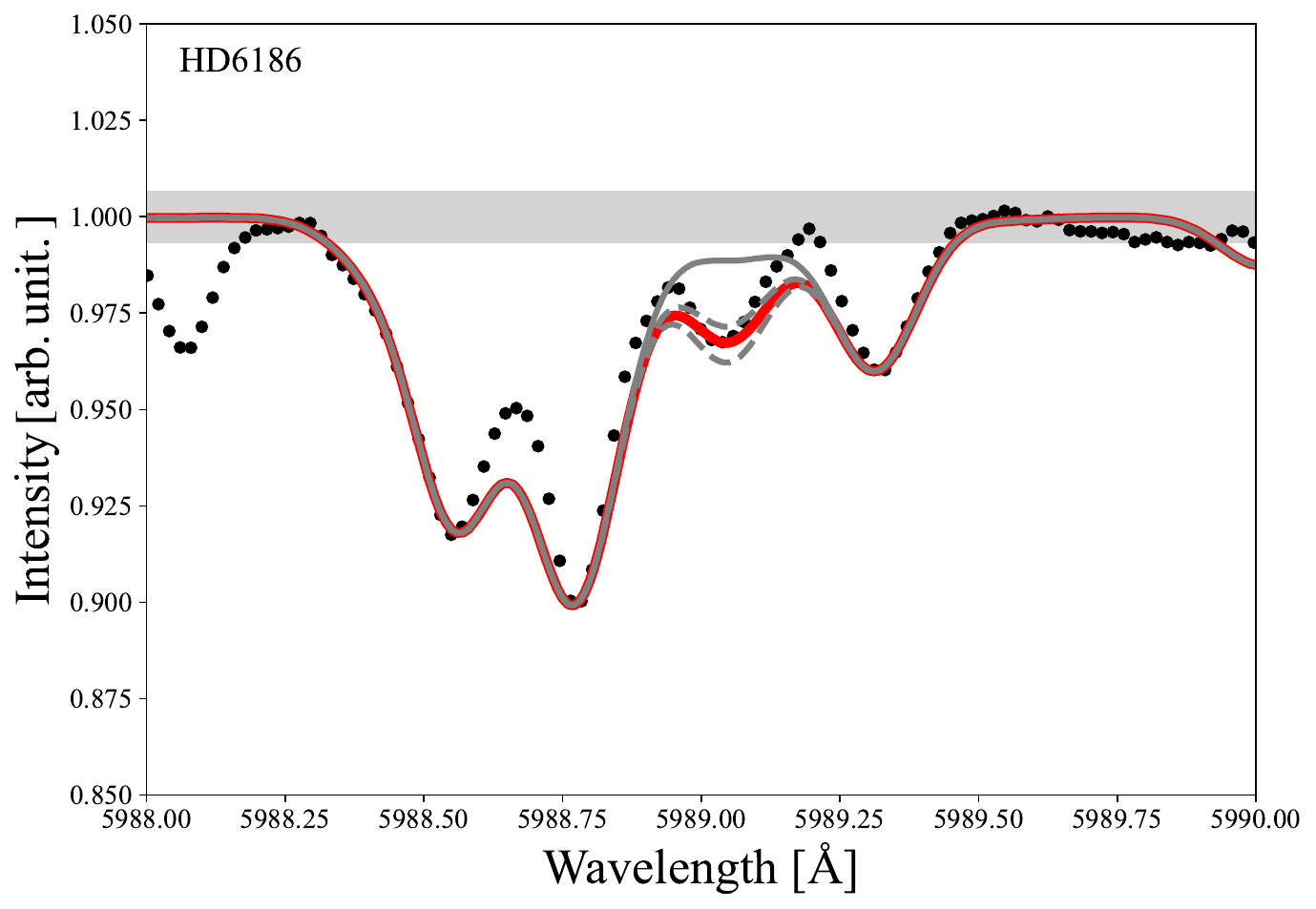}
\includegraphics[width=0.24\textwidth]{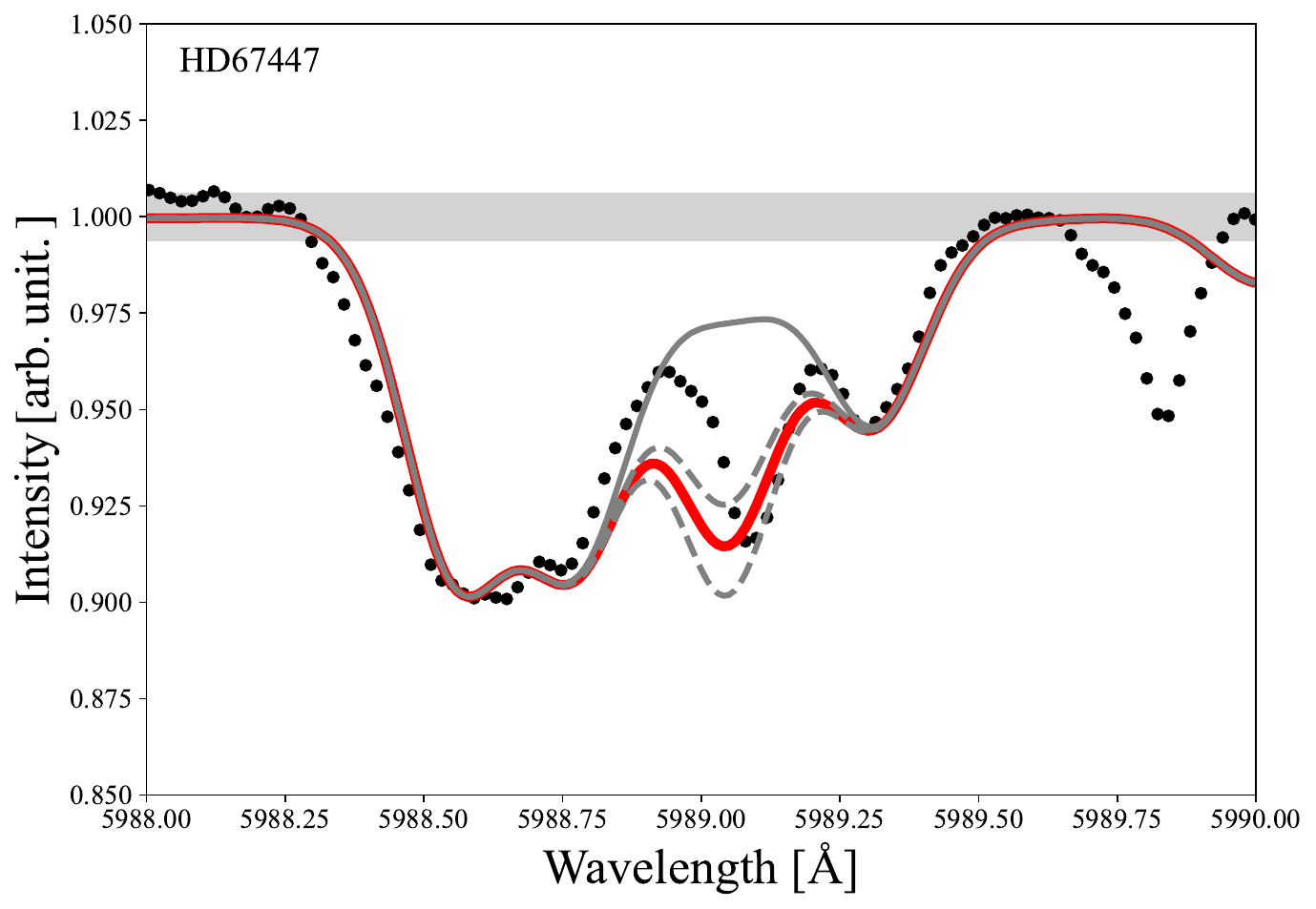}
\\[2mm]

\includegraphics[width=0.24\textwidth]{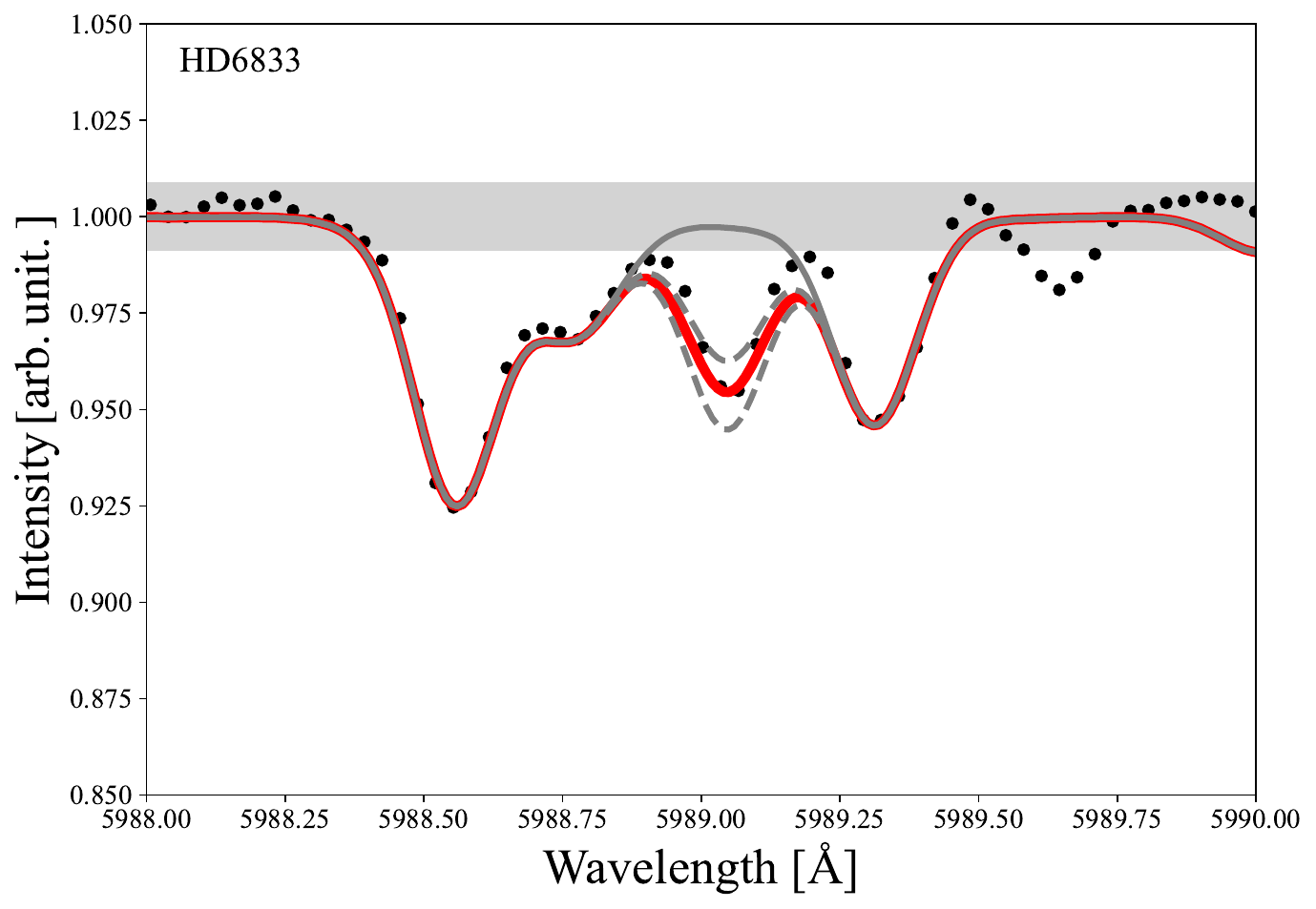}
\includegraphics[width=0.24\textwidth]{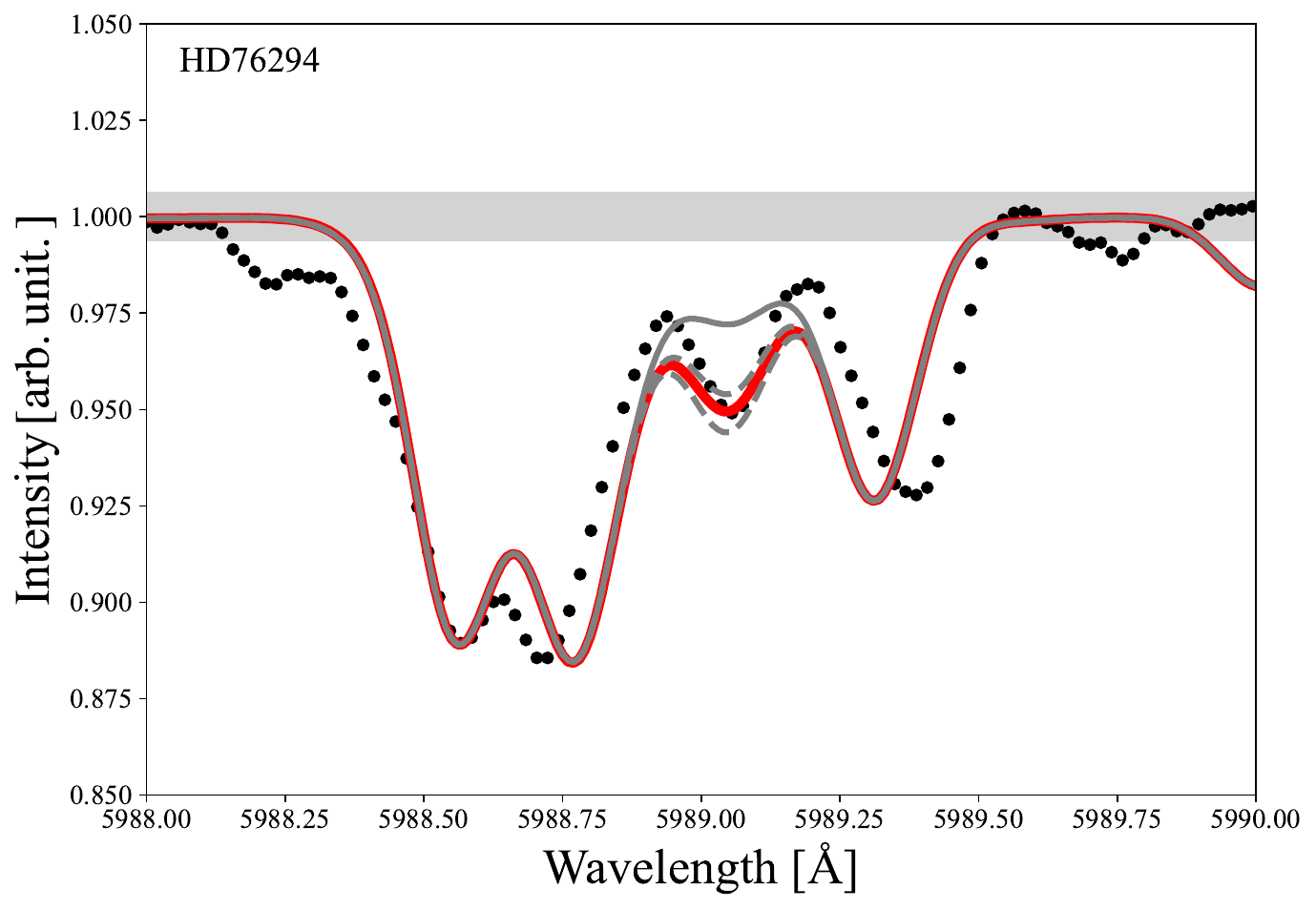}
\includegraphics[width=0.24\textwidth]{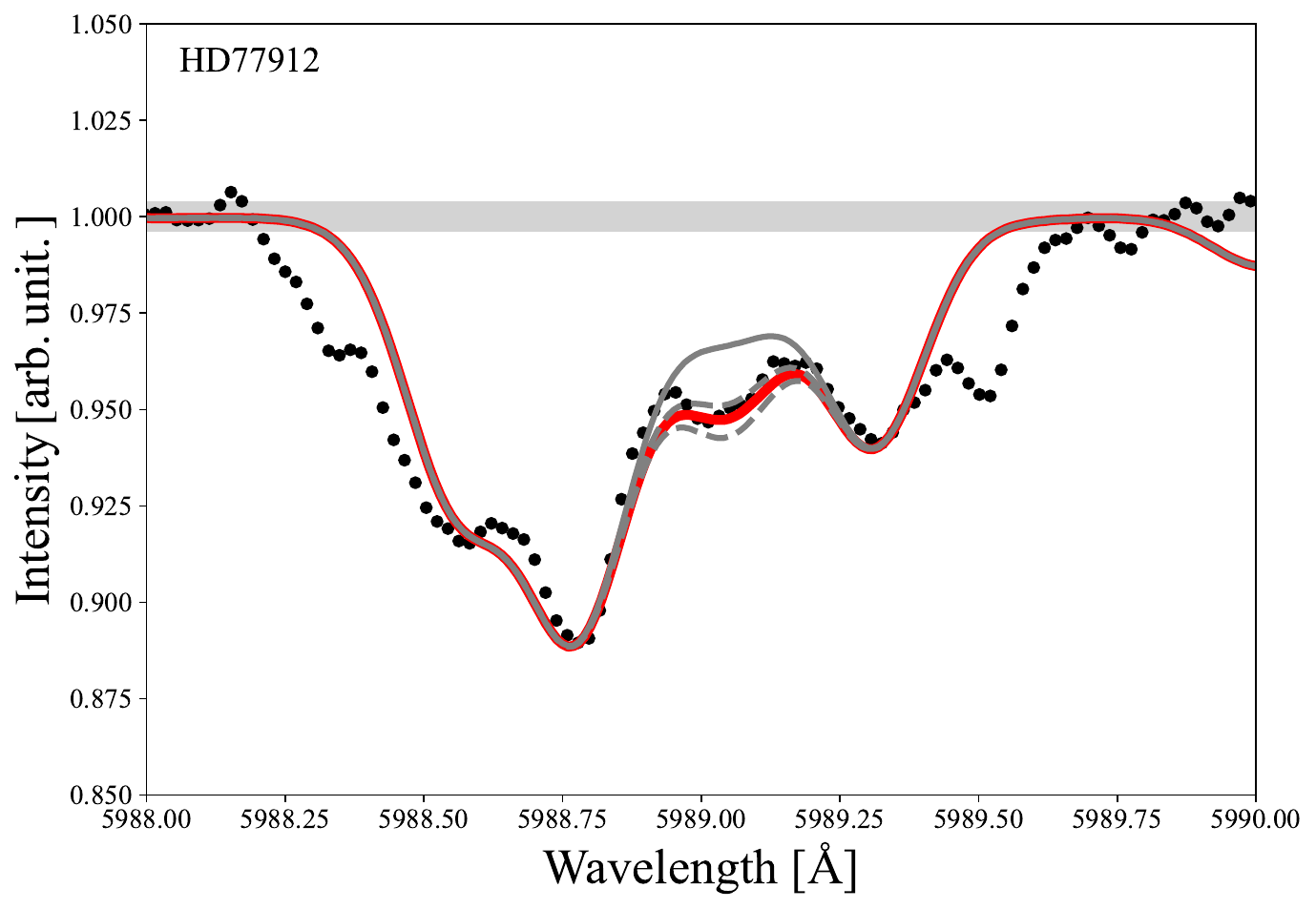}
\includegraphics[width=0.24\textwidth]{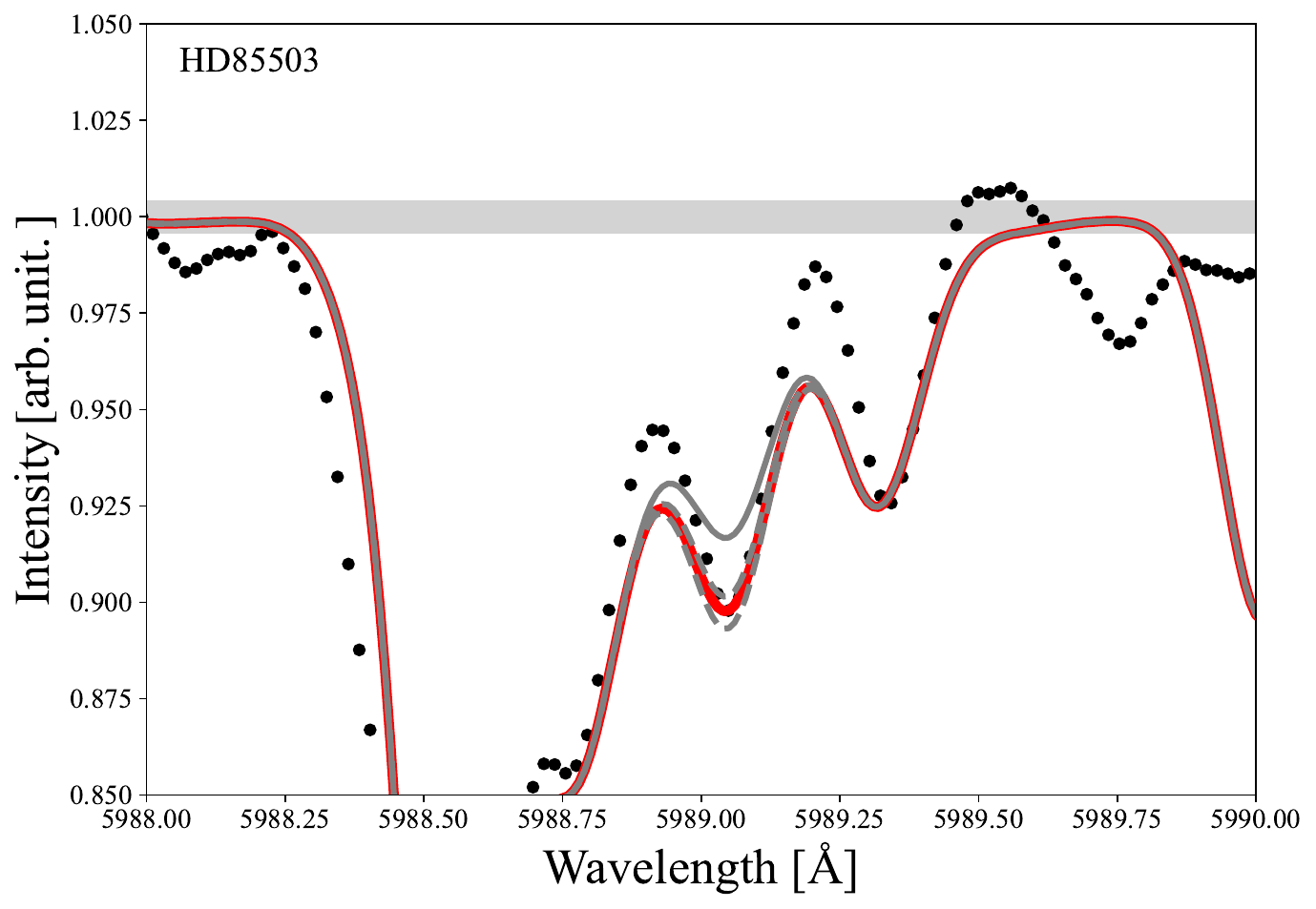}
\\[2mm]

\caption{All spectra of Th 5989 \AA ~line for objects that have Th detection. Black points show observations, red solid lines show the best-fit synthetic spectra, dashed lines show Th abundance changed by 0.1 dex, and the gray solid line shows no Th. The gray shade shows 1 $\upvarsigma$ error from S/N noise.
 {Alt text: All spectra of Thorium 5989 angstrom line. } 
 }
 \label{fig:allspe_Thdetct1}
\end{figure*}

\addtocounter{figure}{-1}

\begin{figure*}[t]

\centering
\includegraphics[width=0.24\textwidth]{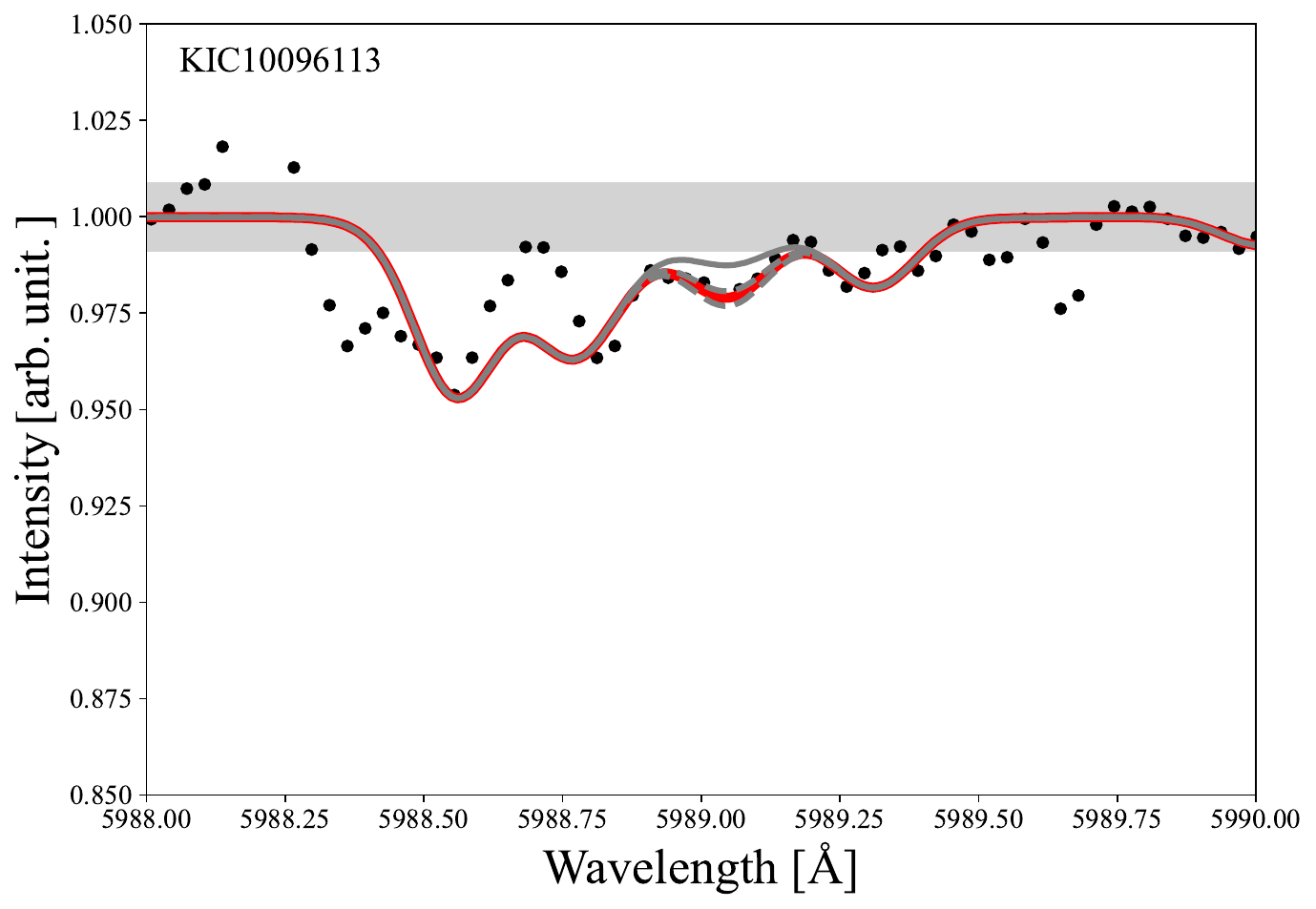}
\includegraphics[width=0.24\textwidth]{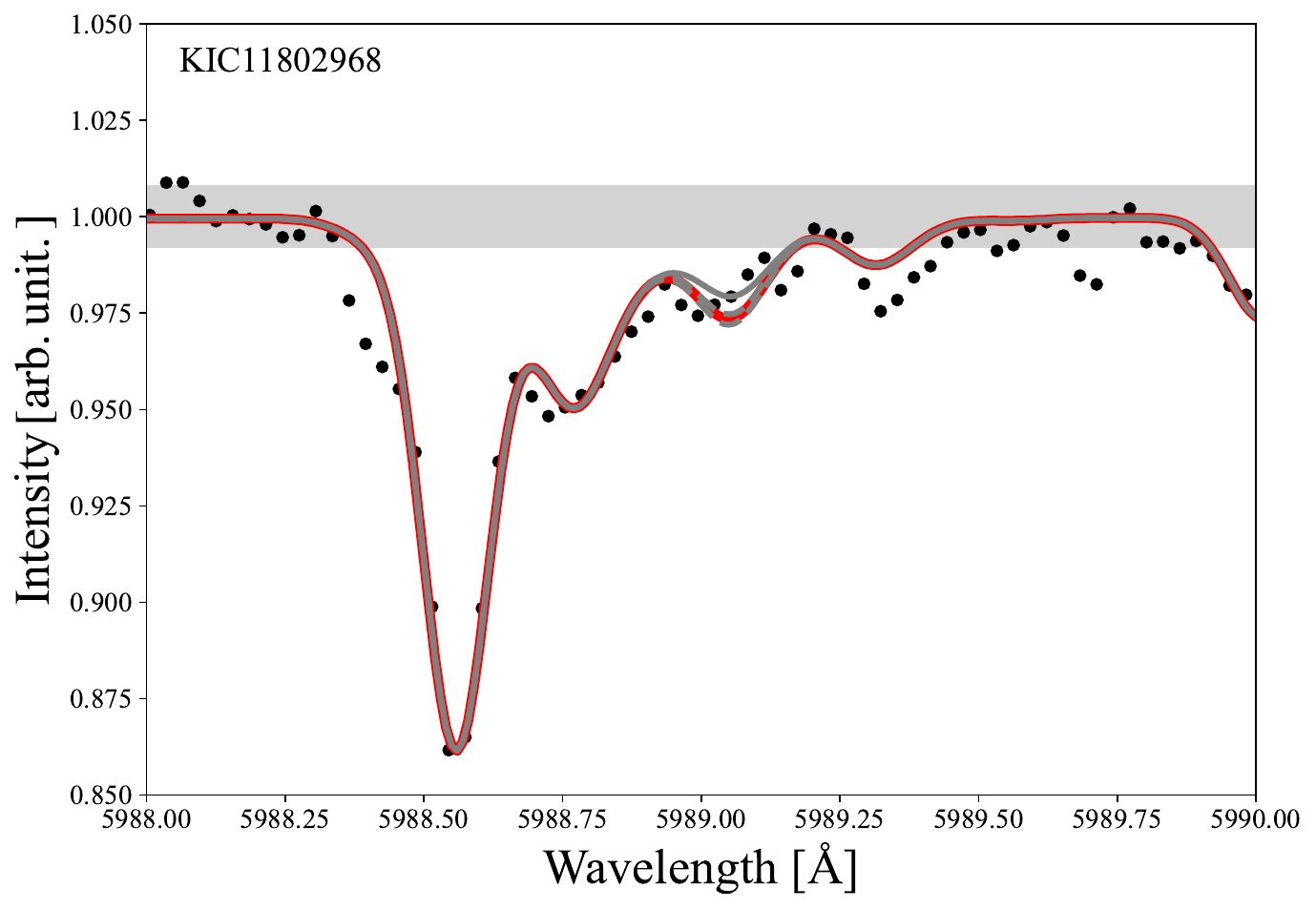}
\includegraphics[width=0.24\textwidth]{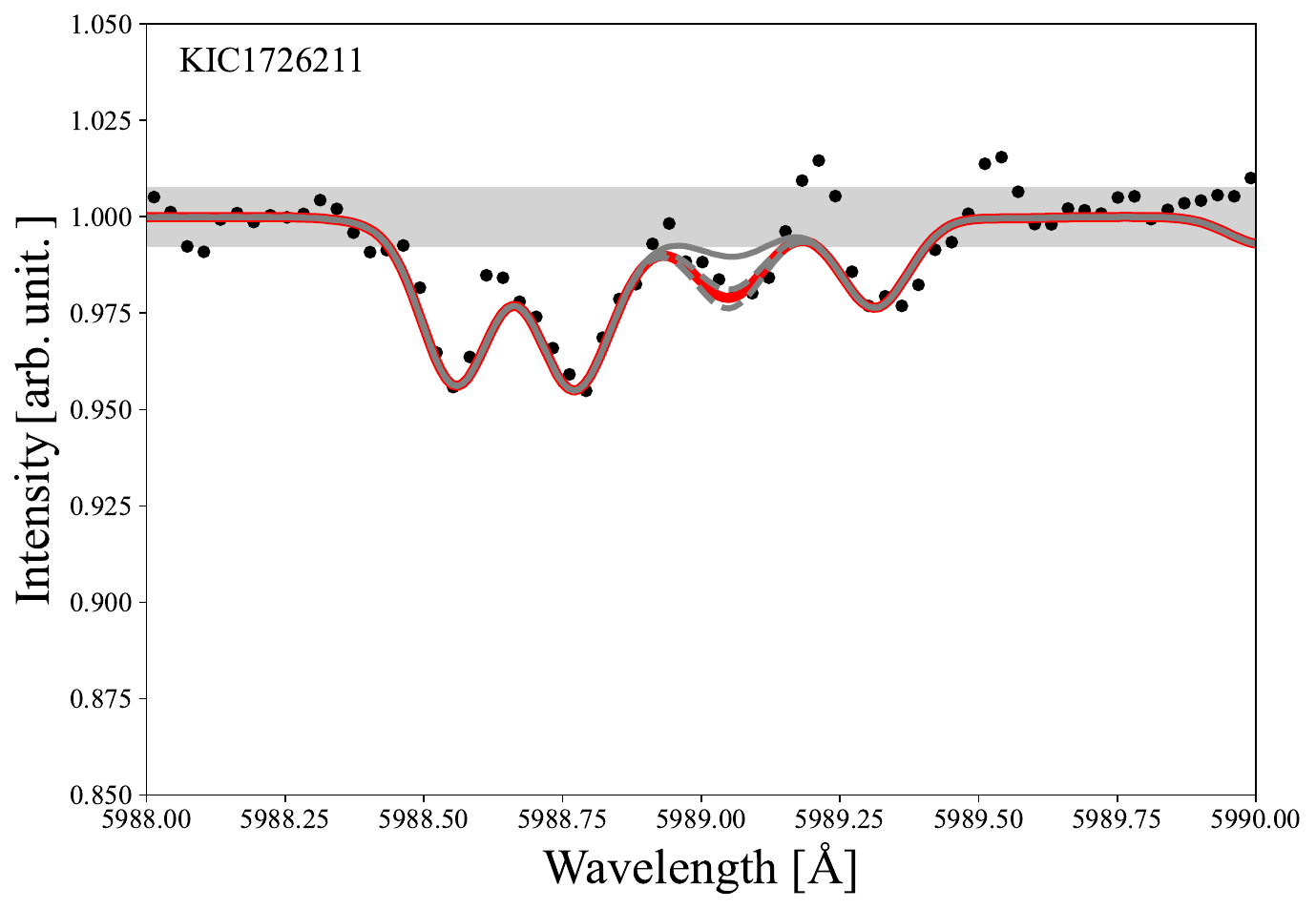}
\includegraphics[width=0.24\textwidth]{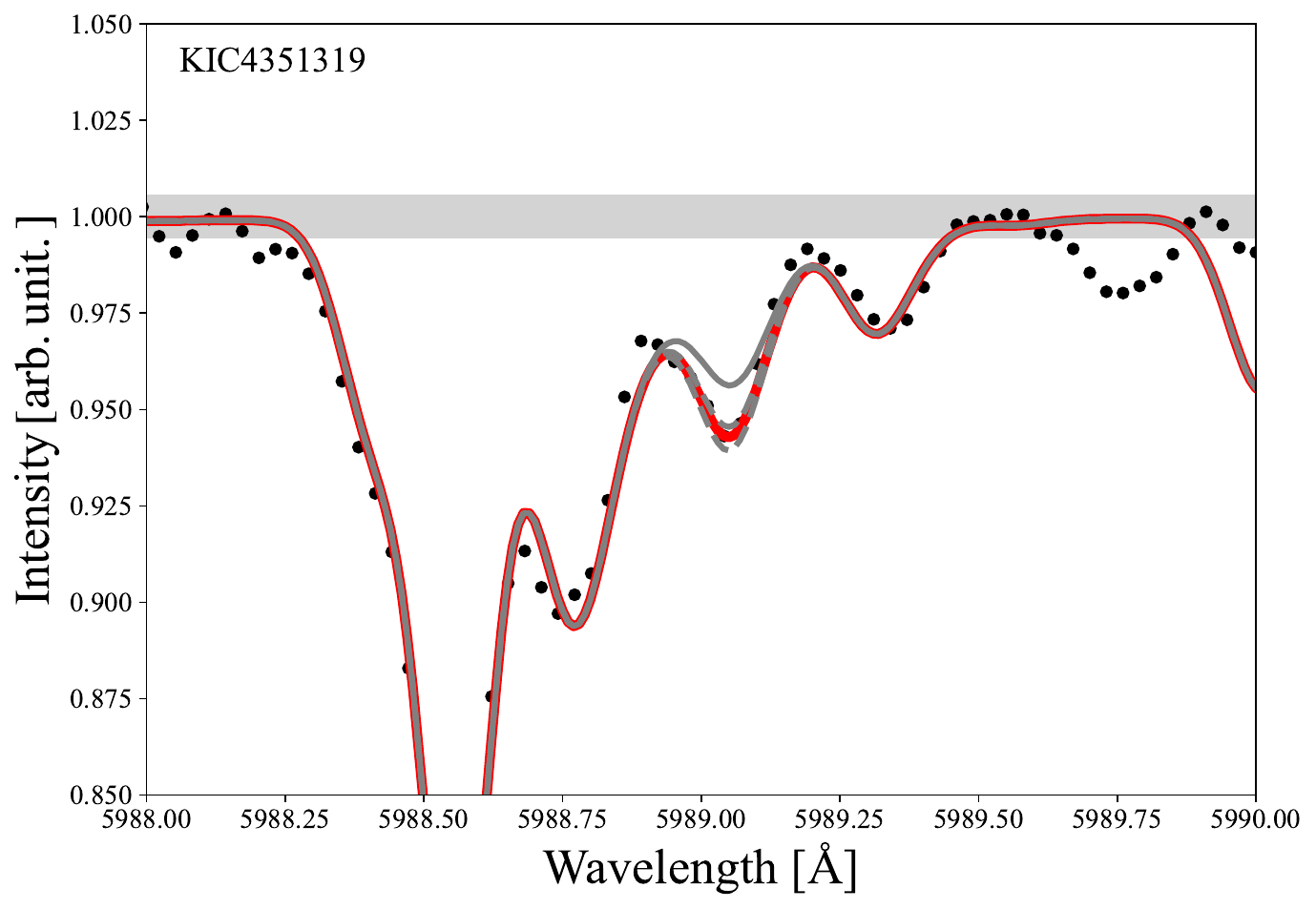}
\\[2mm]

\includegraphics[width=0.24\textwidth]{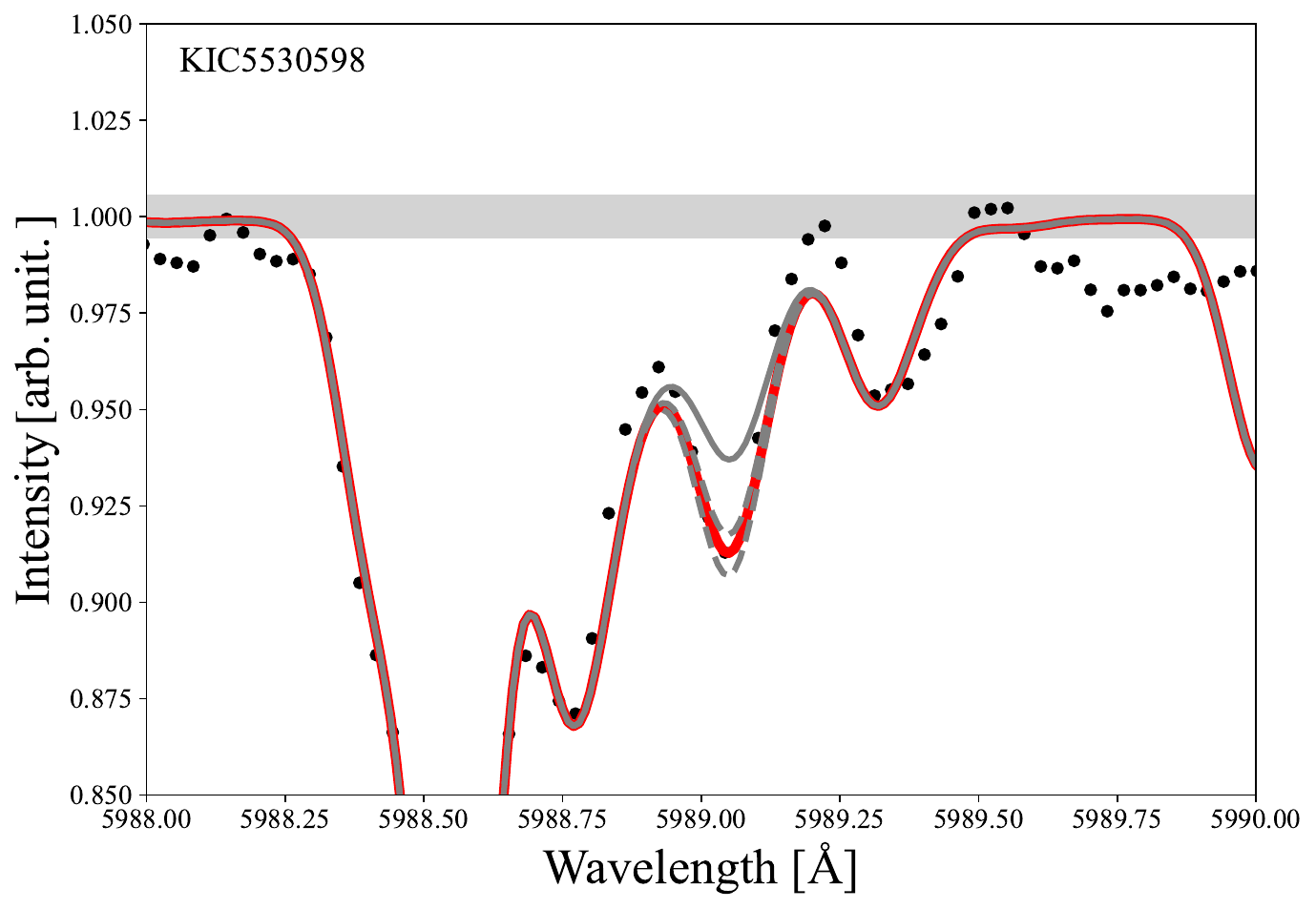}
\includegraphics[width=0.24\textwidth]{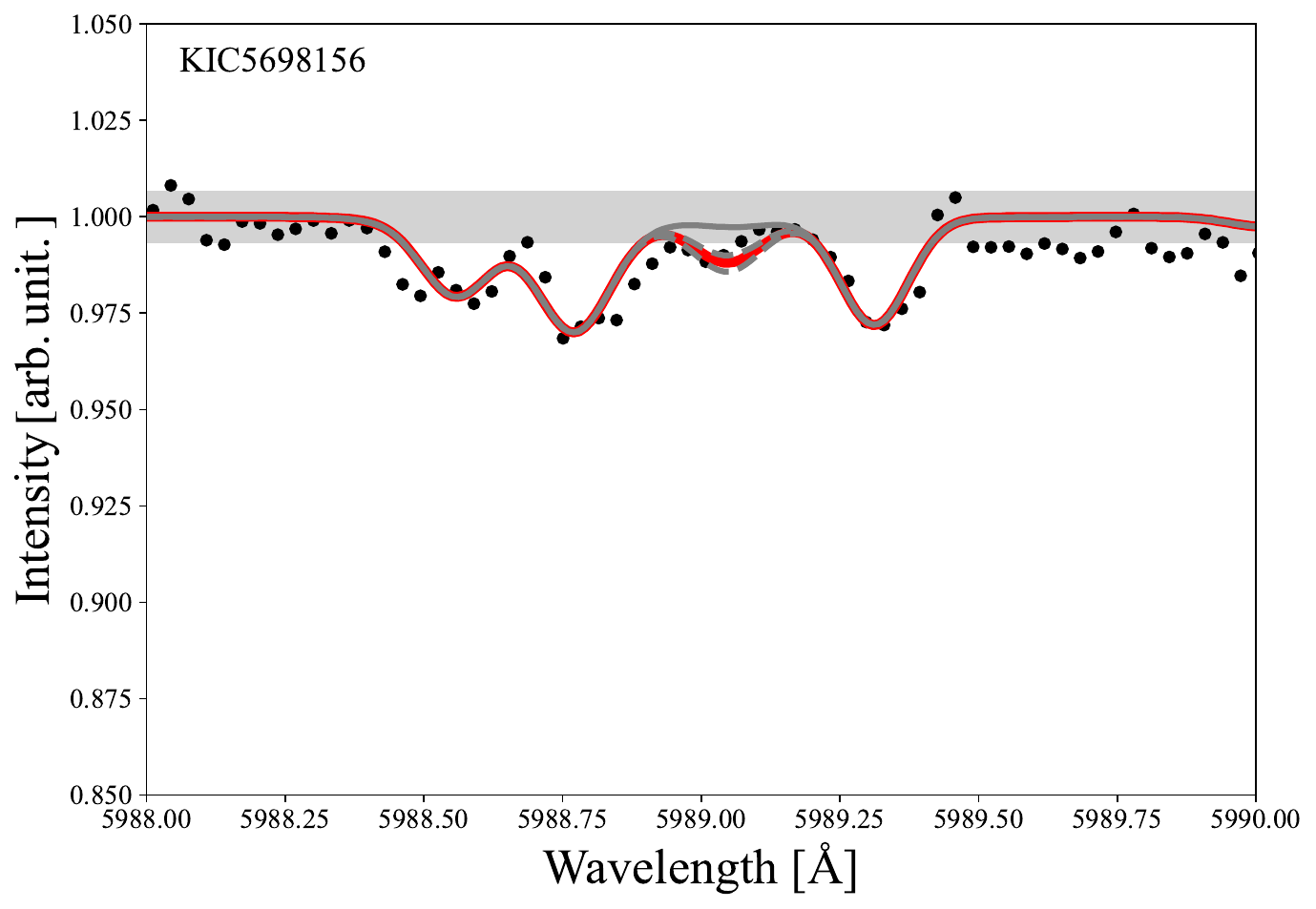}
\includegraphics[width=0.24\textwidth]{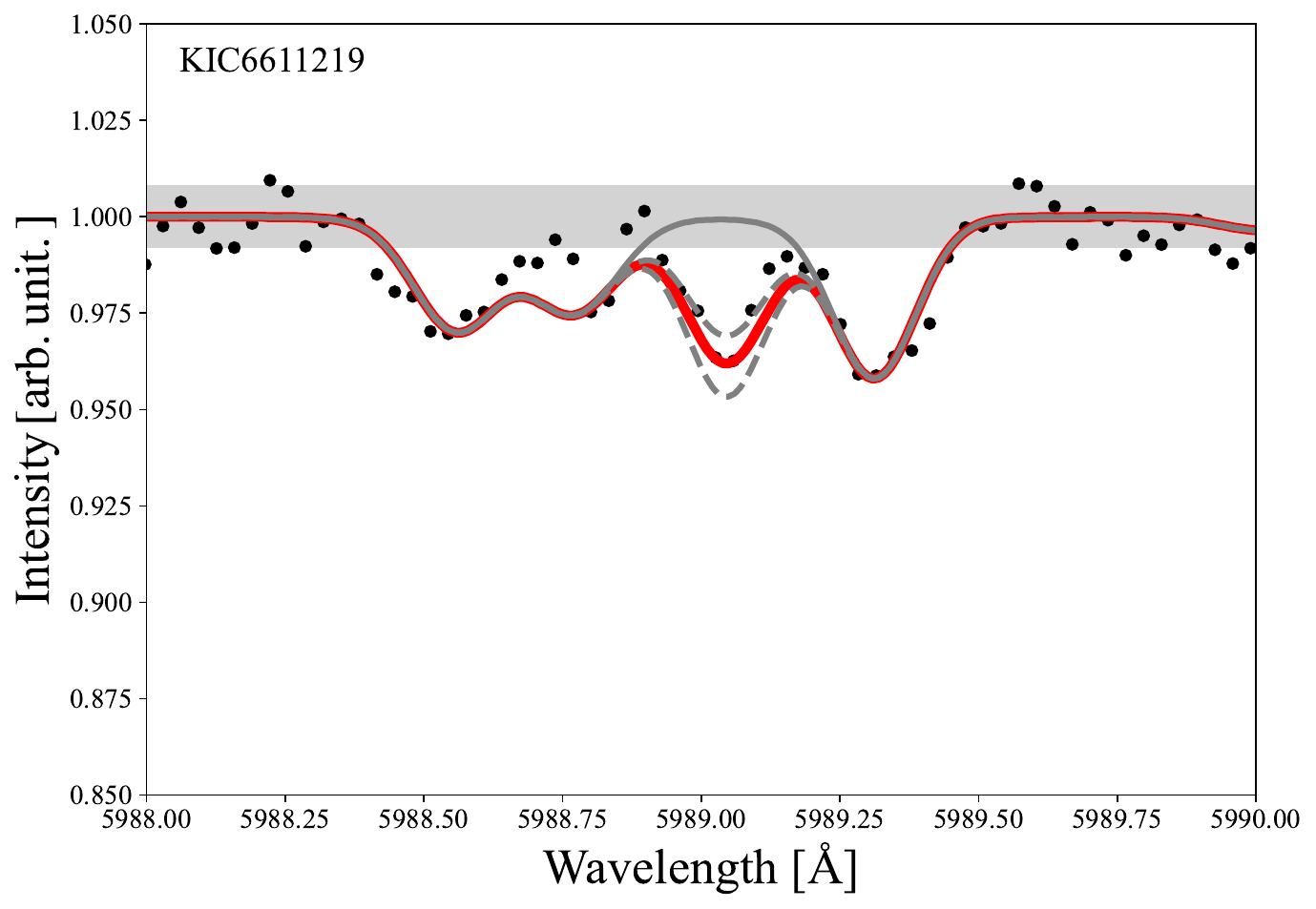}
\includegraphics[width=0.24\textwidth]{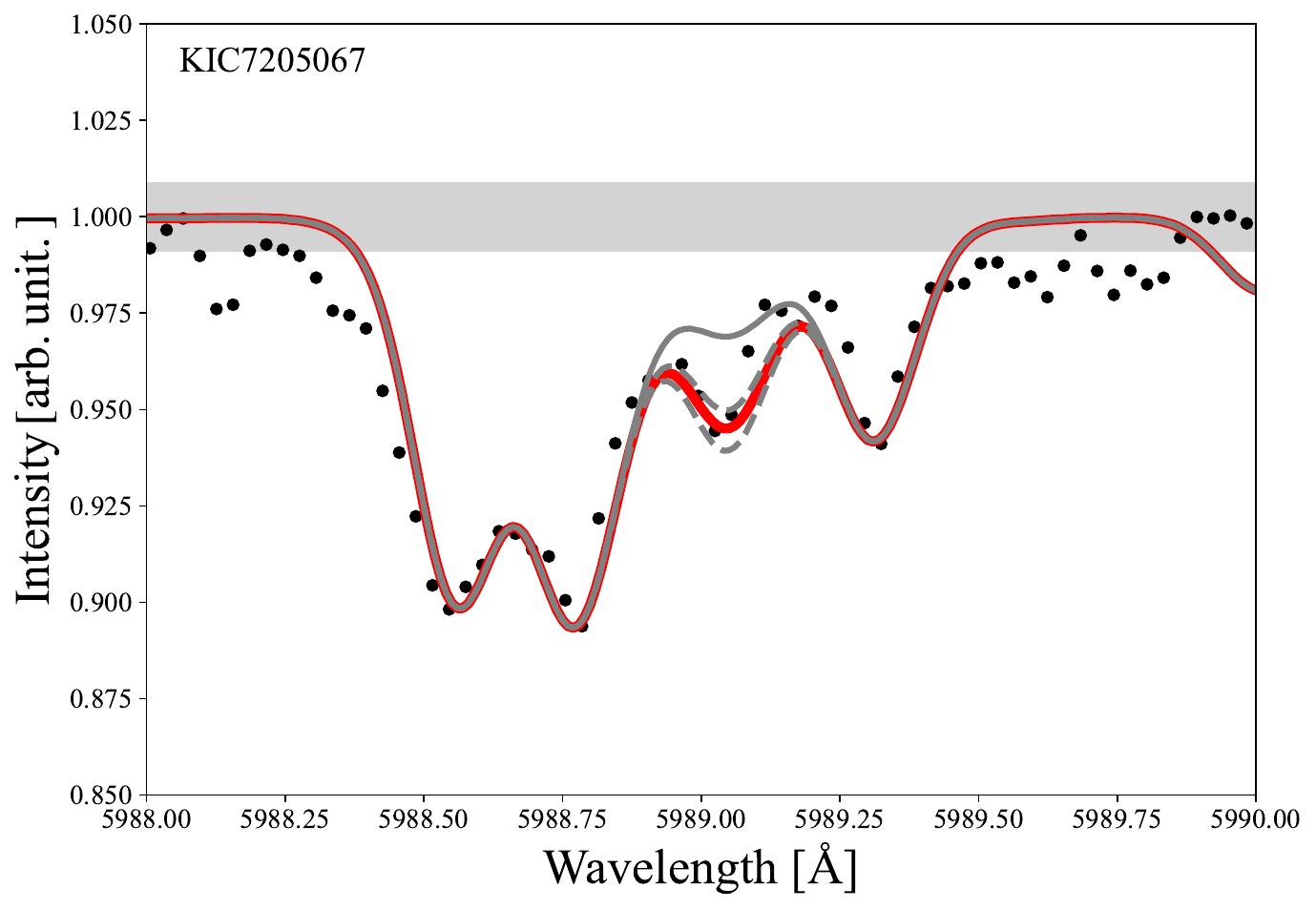}
\\[2mm]

\includegraphics[width=0.24\textwidth]{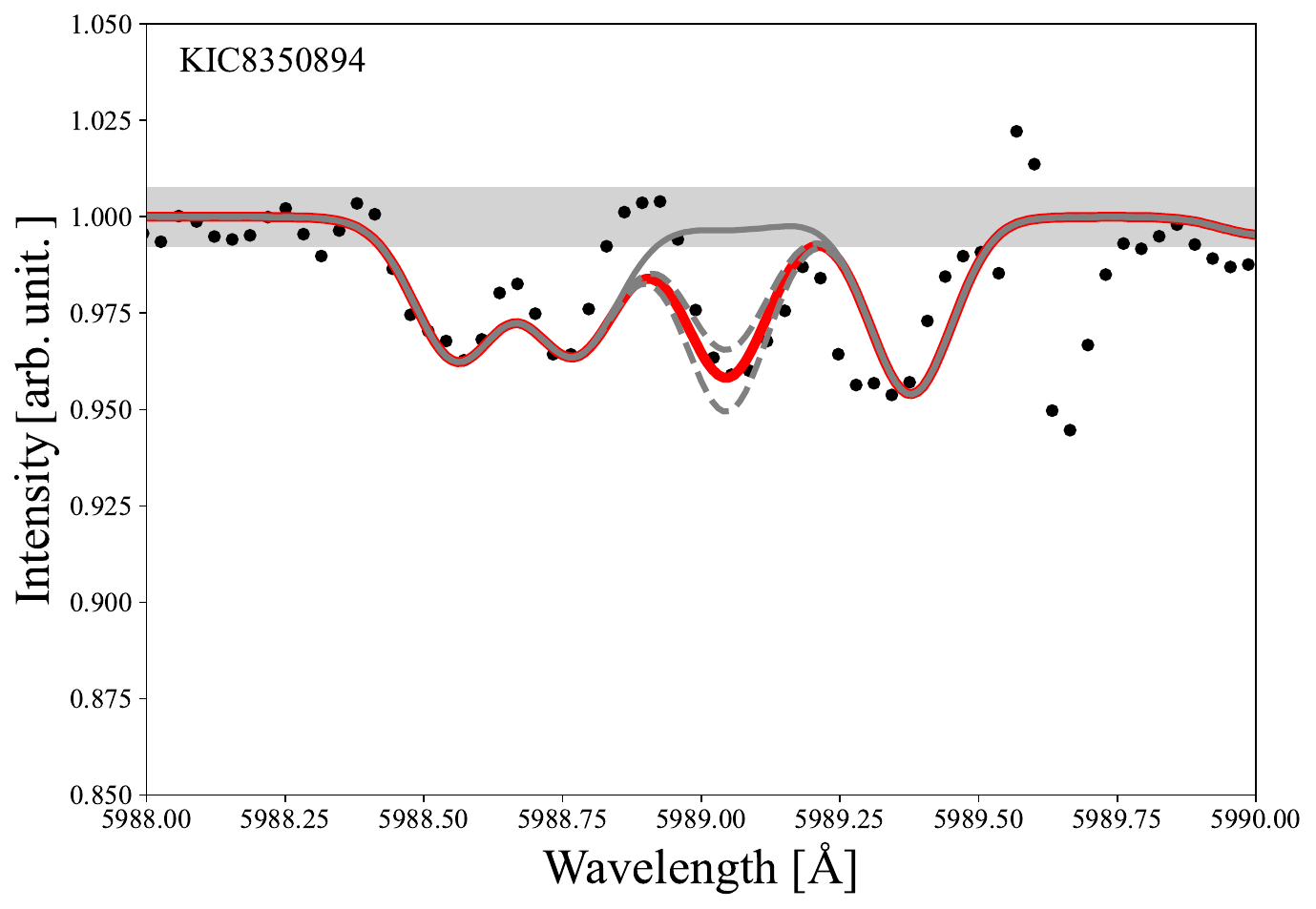}

\caption{(Continued)
 {Alt text: All spectra of Thorium 5989 angstrom line (Continued). } 
 }
 \label{fig:allspe_Thdetct2}
\end{figure*}

\begin{figure*}[t]
\centering

\includegraphics[width=0.24\textwidth]{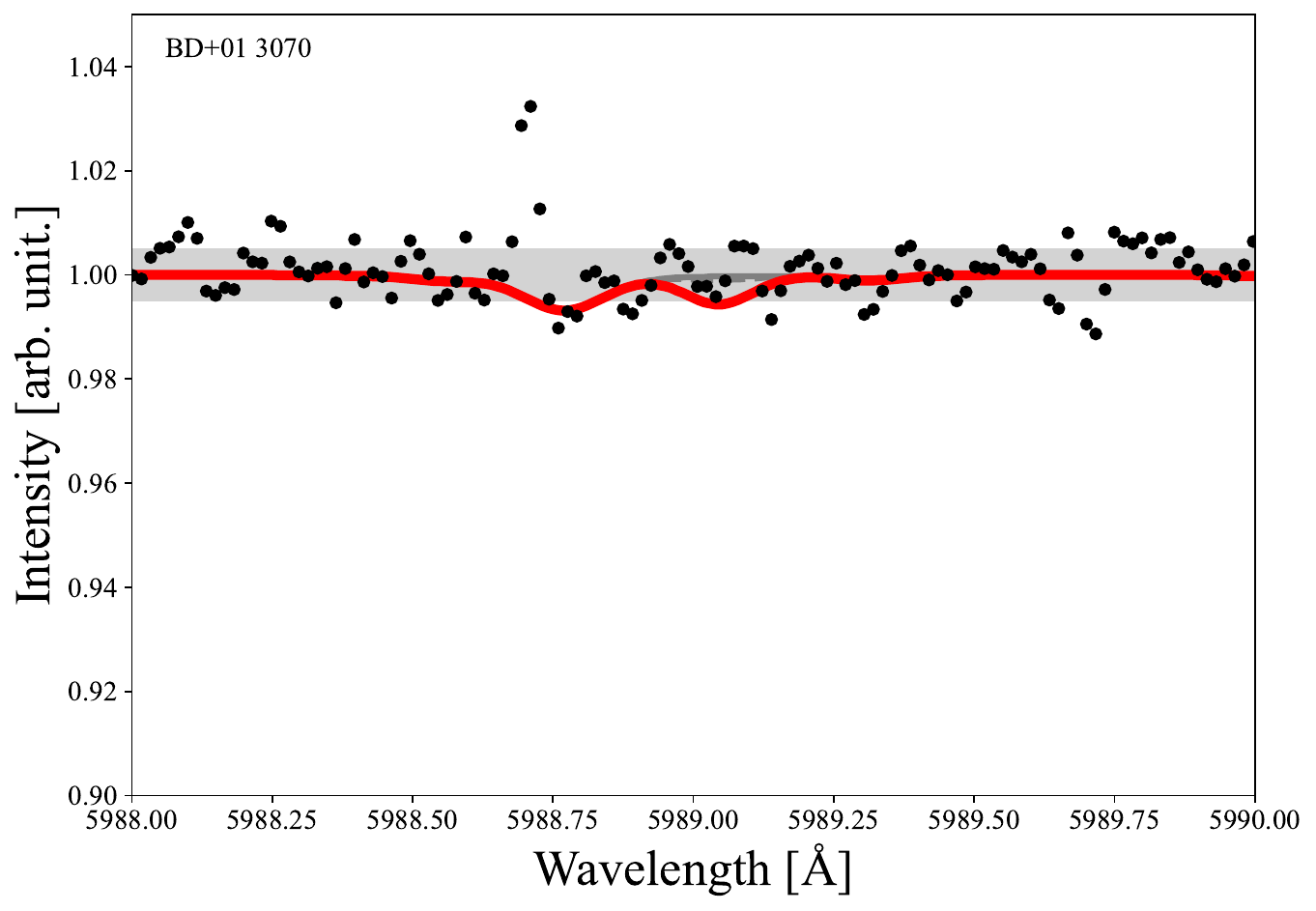}
\includegraphics[width=0.24\textwidth]{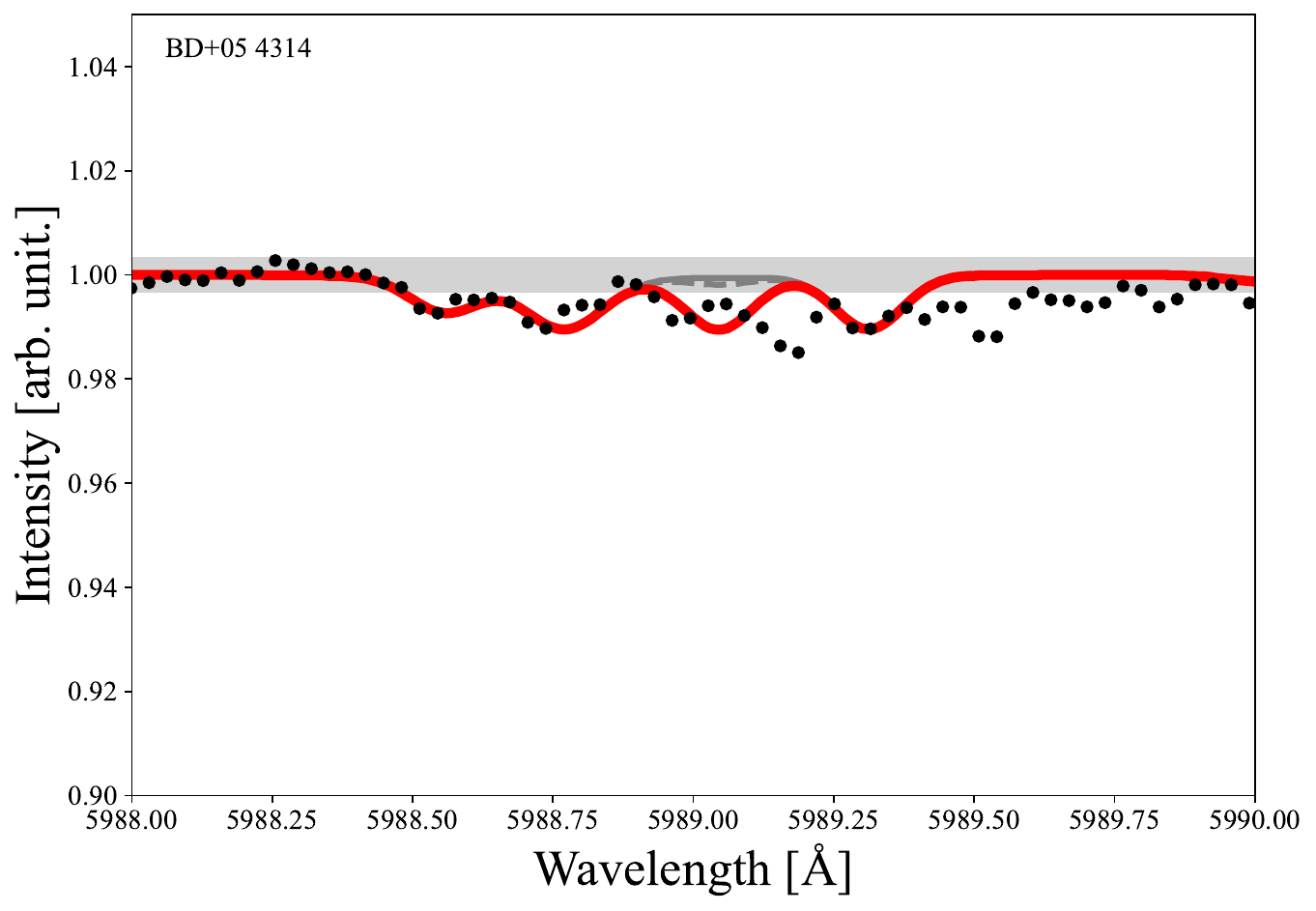}
\includegraphics[width=0.24\textwidth]{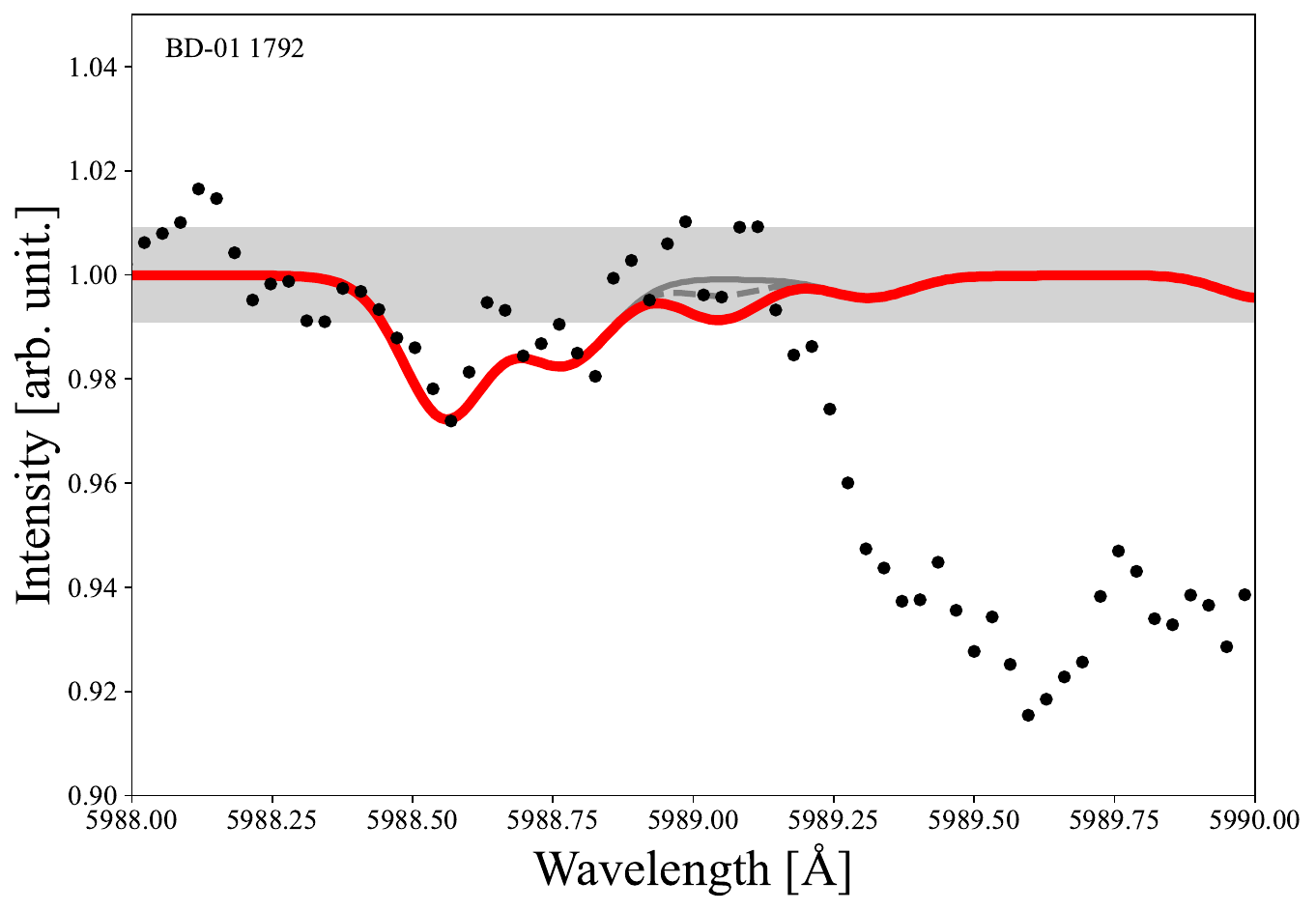}
\includegraphics[width=0.24\textwidth]{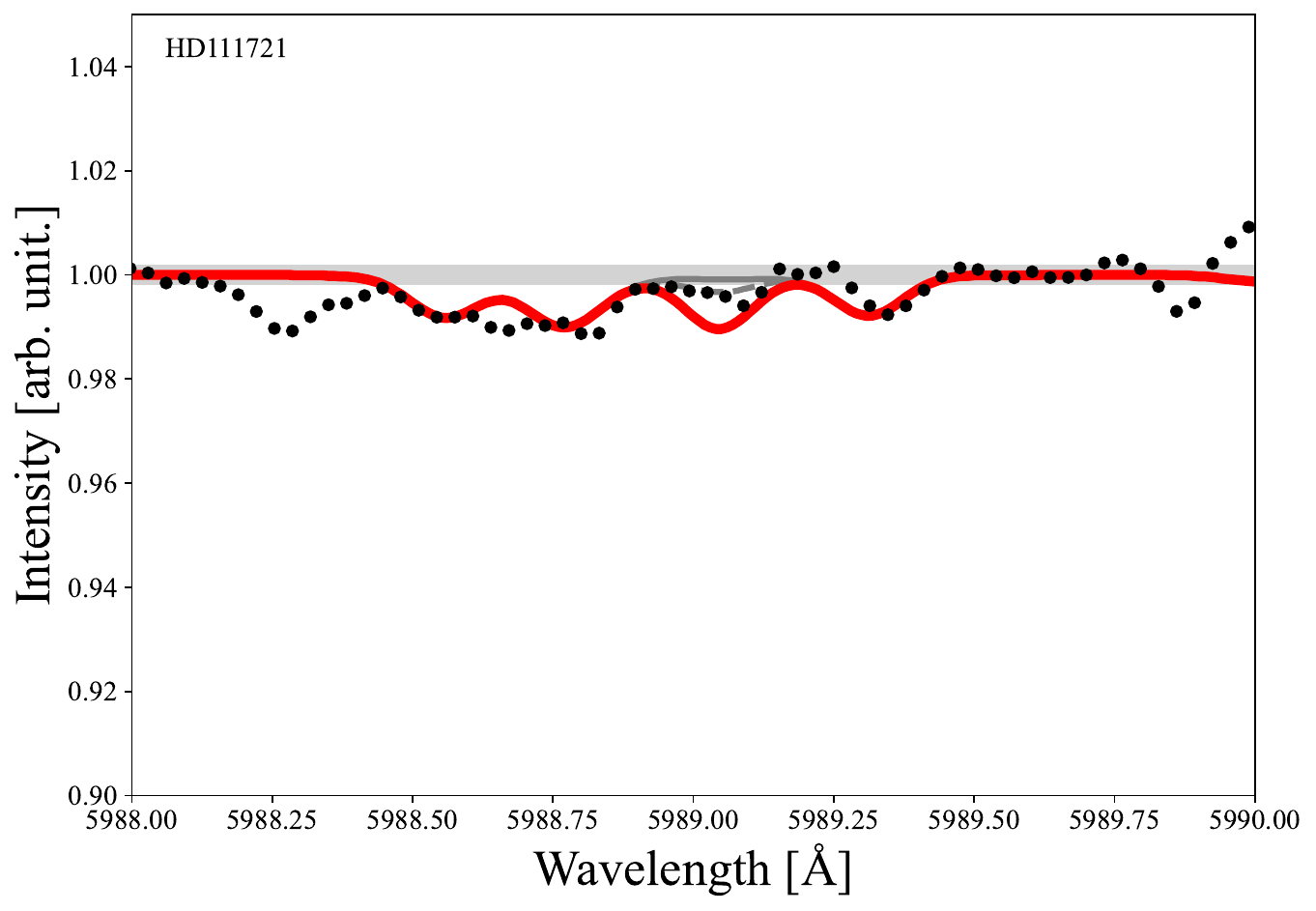}
\\[2mm]

\includegraphics[width=0.24\textwidth]{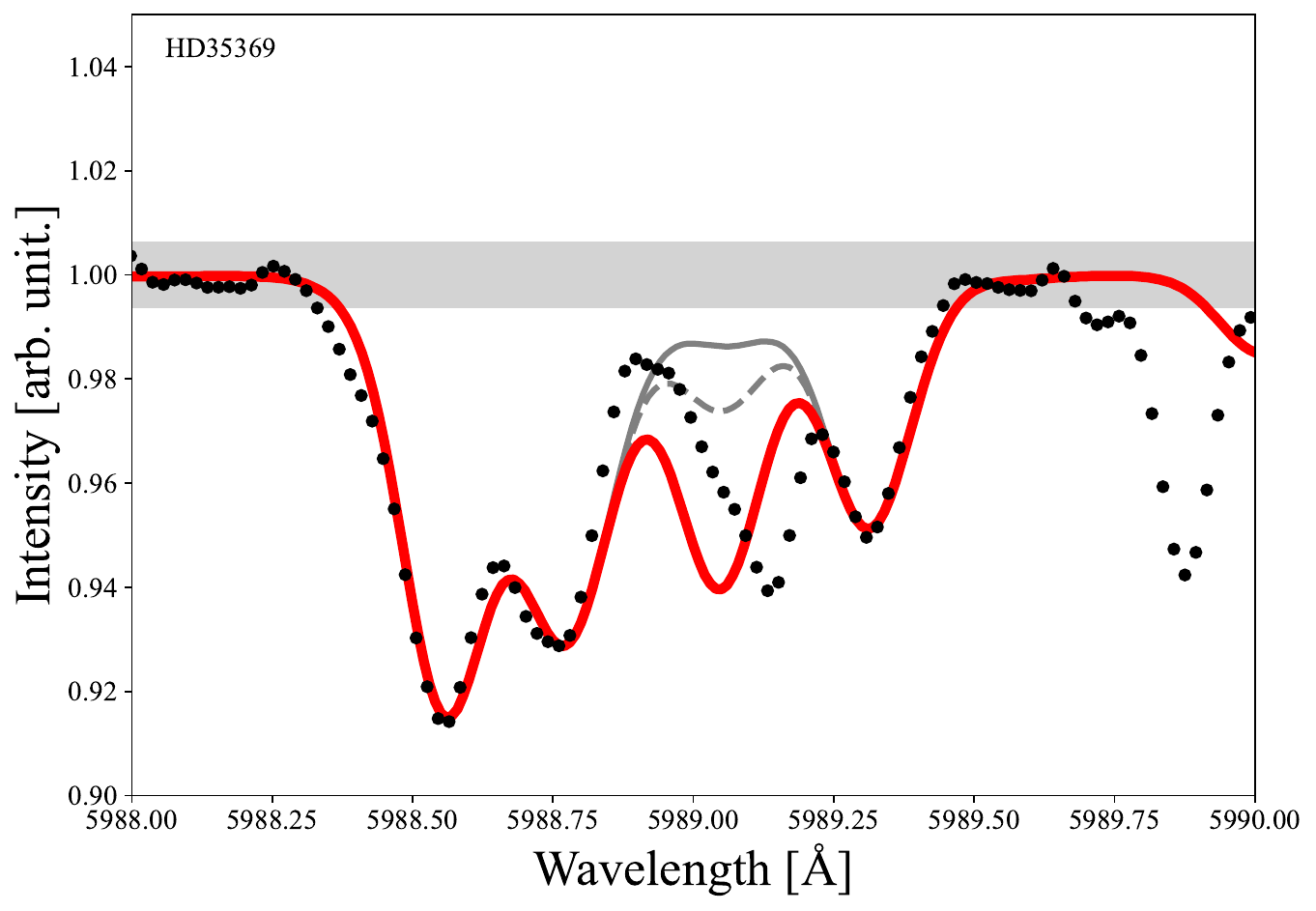}
\includegraphics[width=0.24\textwidth]{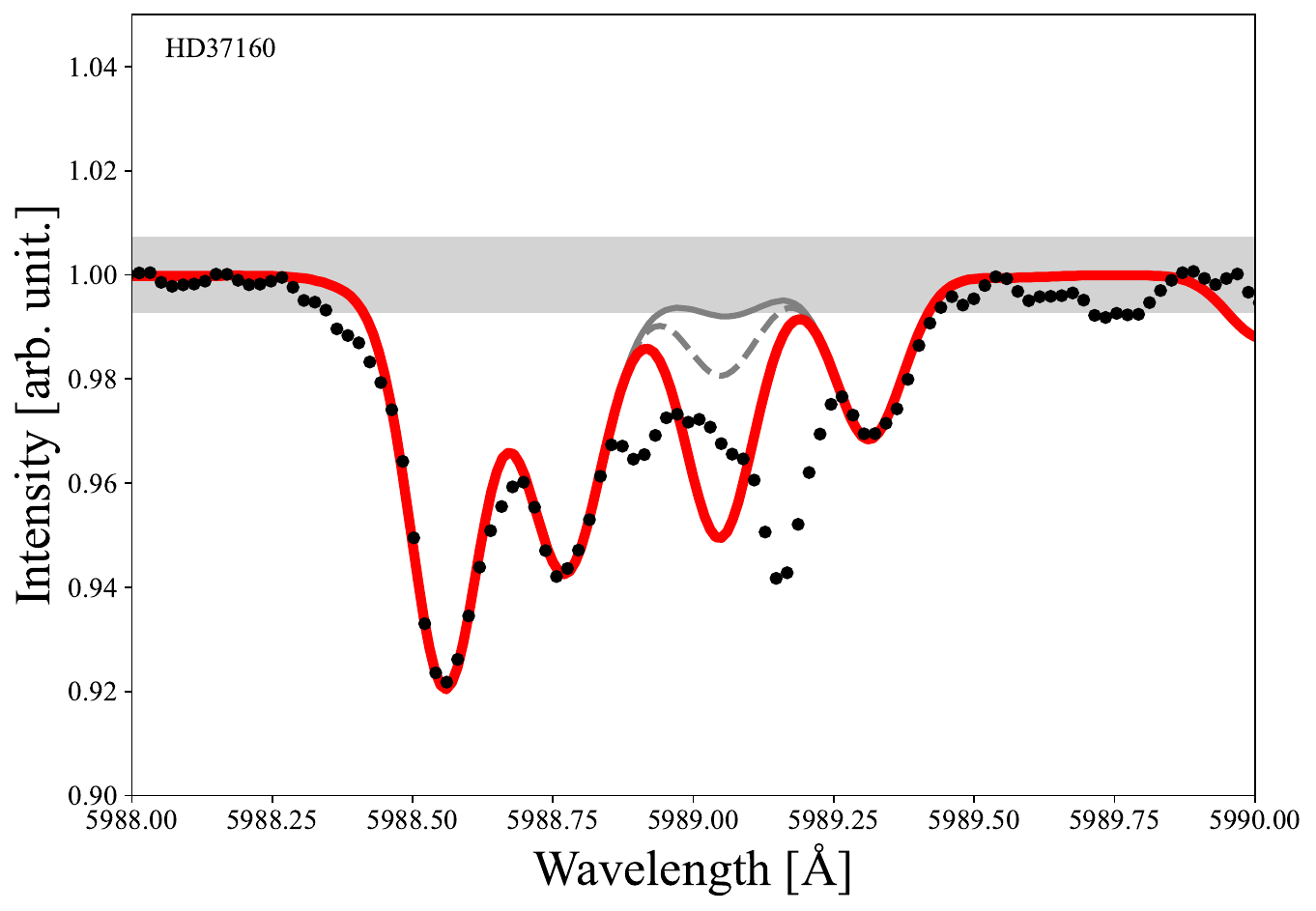}
\includegraphics[width=0.24\textwidth]{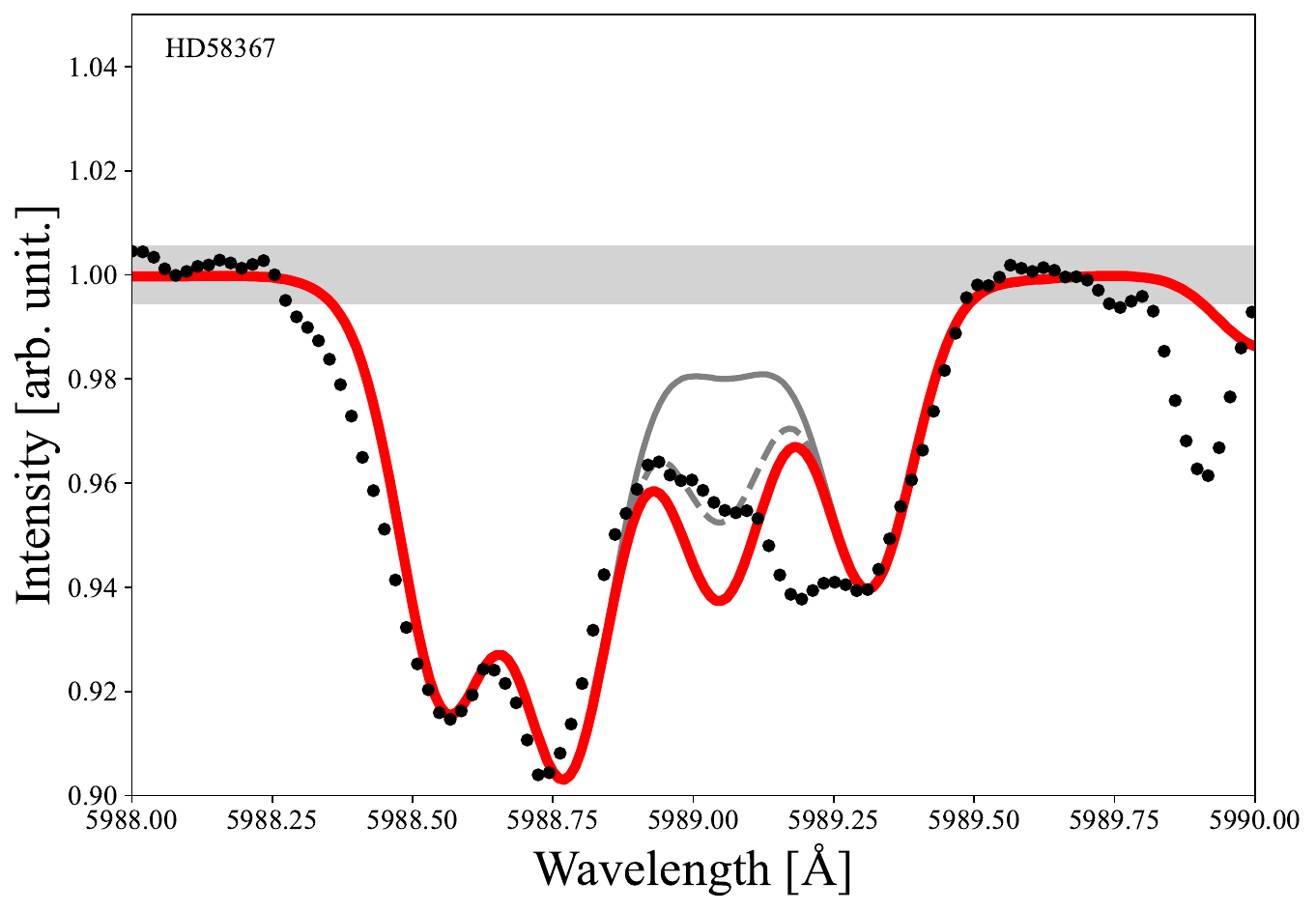}
\includegraphics[width=0.24\textwidth]{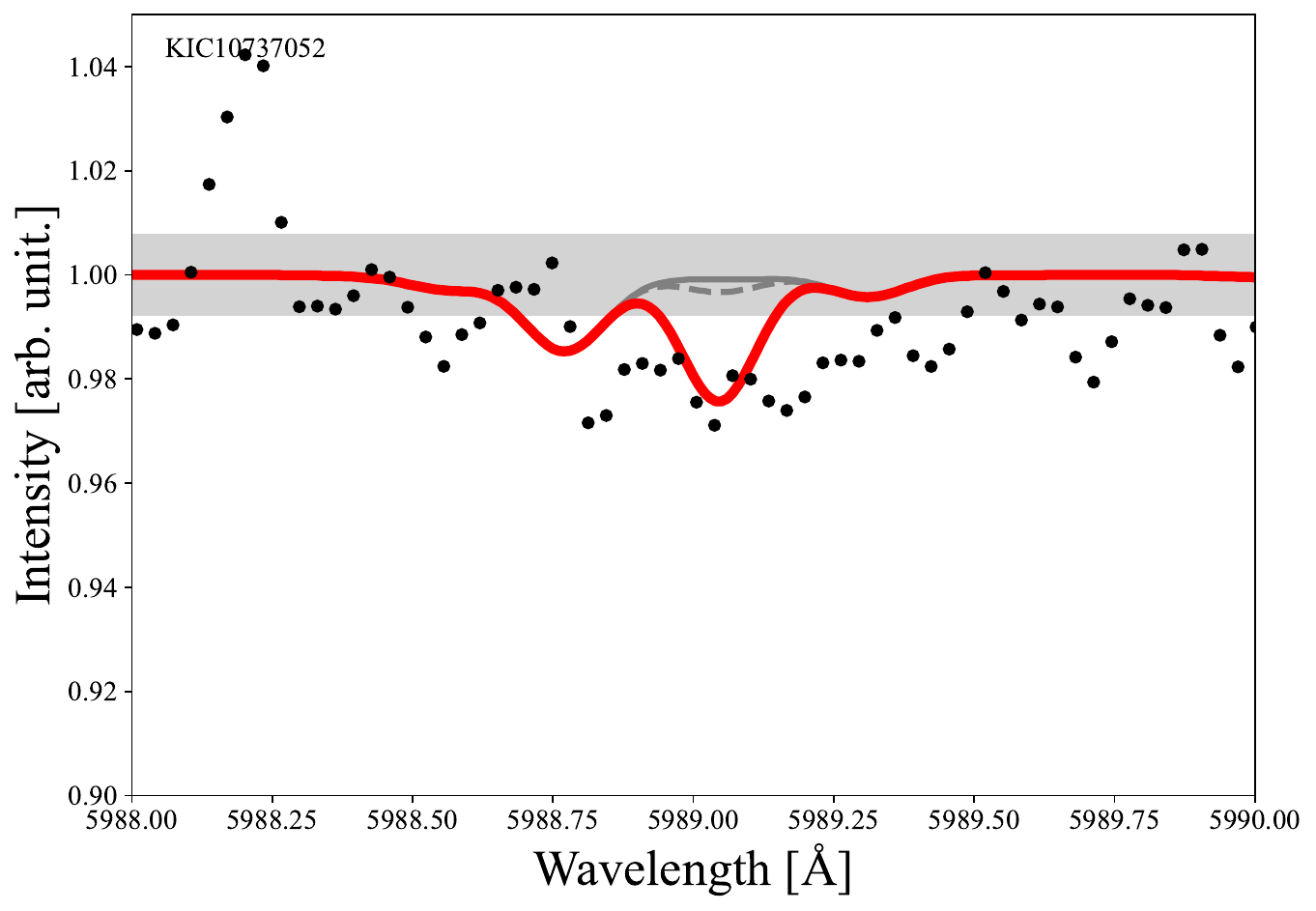}
\\[2mm]

\includegraphics[width=0.24\textwidth]{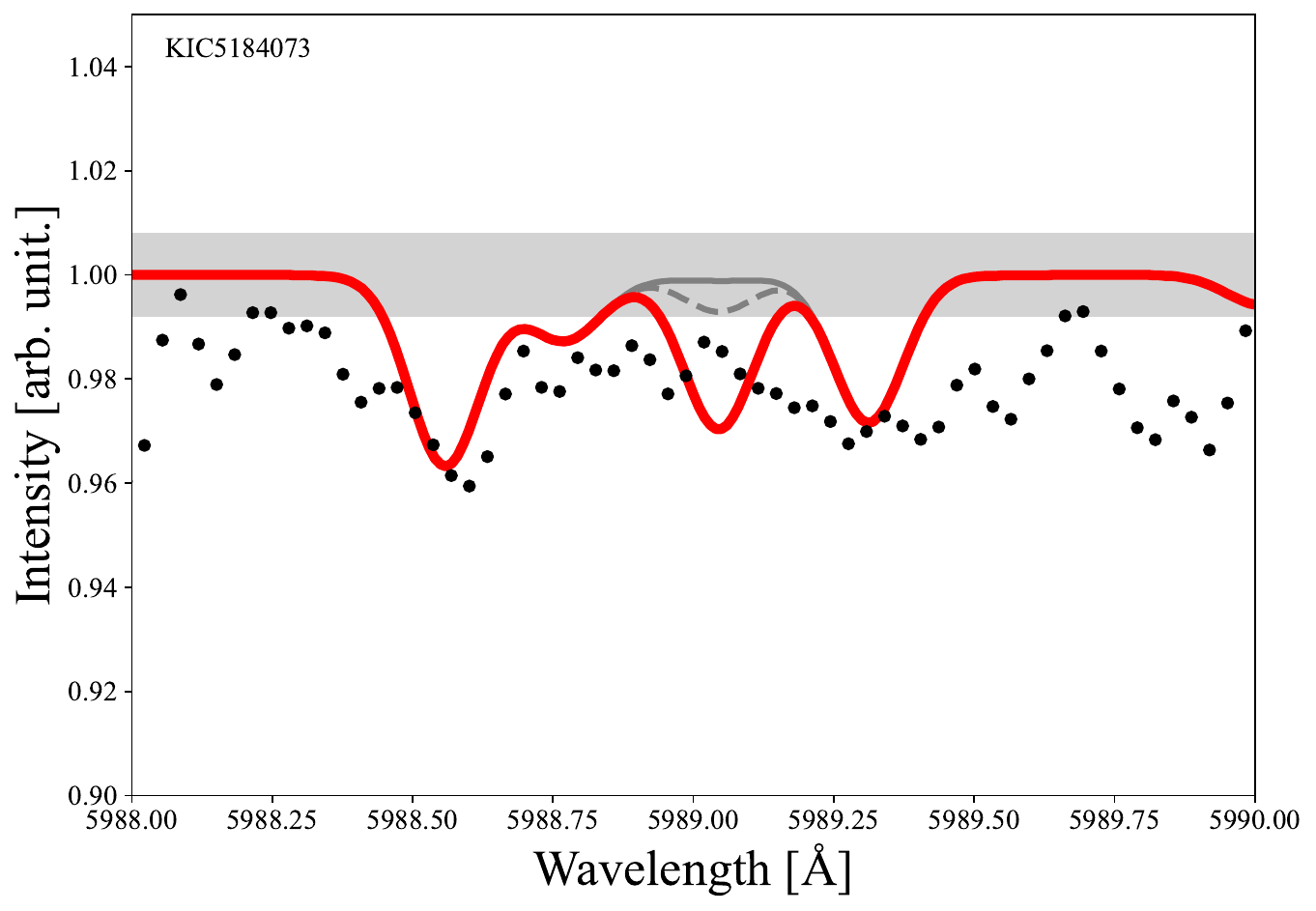}
\includegraphics[width=0.24\textwidth]{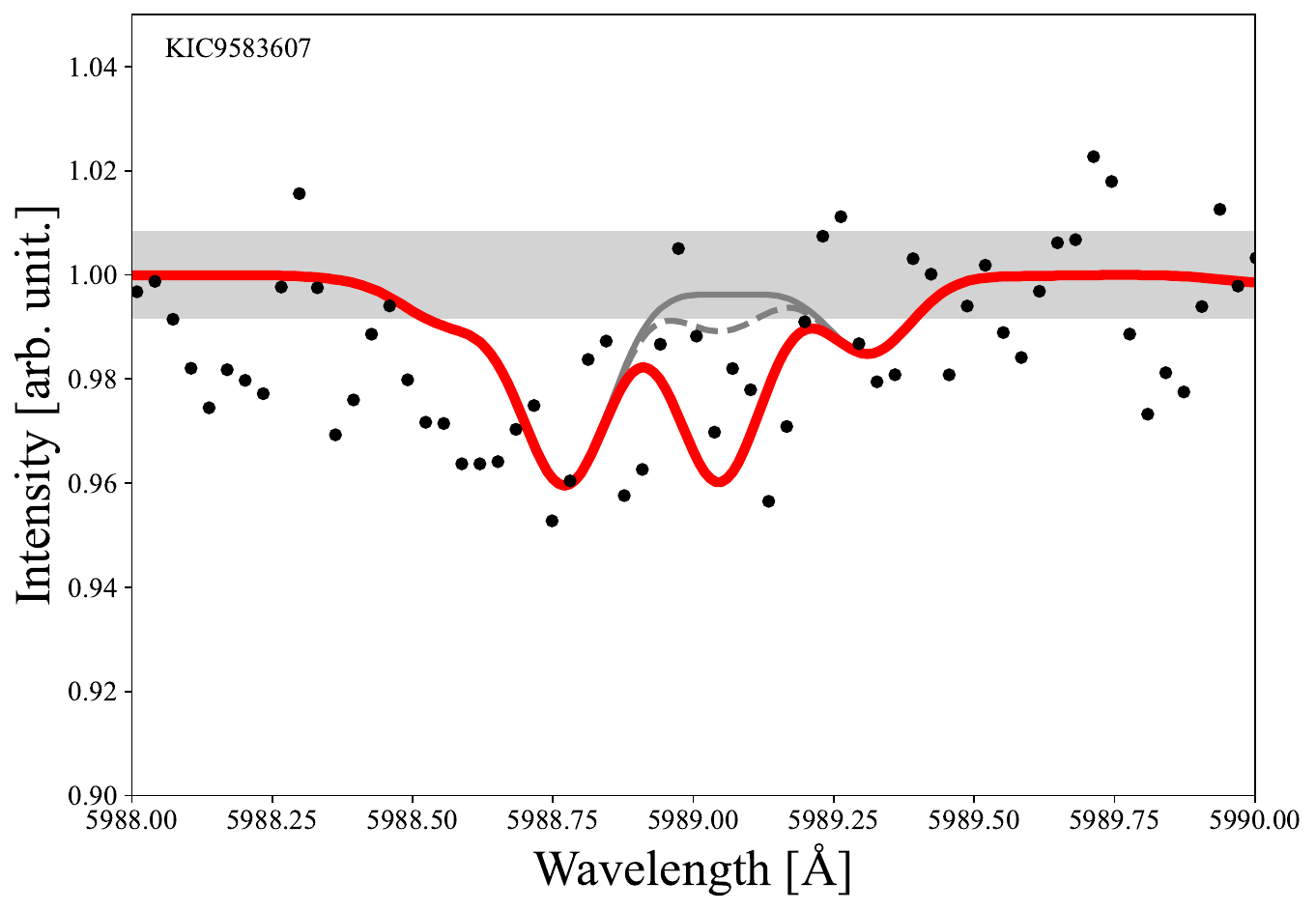}

 \caption{All spectra of Th 5989 \AA~ line for objects that have Th upper limit. Black points show observations, red solid lines show the upper limit of synthetic spectra, dashed lines show [Th/Fe] = 0 dex, and gray solid line shows no Th. Gray shade shows 1 $\upvarsigma$ error from S/N noise.
  {Alt text: All spectra of Thorium 5989 angstrom line for the upper limit. } 
 } 
 \label{fig:allspe_Thuplim}
 \end{figure*}

\clearpage

\textbf{\section{Line list}}
    Atomic line list used for the Th and Eu abundance analyses.
    Table \ref{tab:linelist} lists the wavelengths, excitation potentials, log \textit{gf}.
    
\begin{table*}[]
\centering

\caption{Th, Eu, Ti, Si, Nd, and CN absorption line list. }
\footnotesize
\begin{tabular}{llllrl}
\hline\noalign{\vskip3pt}
Z  & Element & Wavelength (\AA) & E.P. (eV) & log \textit{gf} & Reference$^\ast$    \\
\hline\noalign{\vskip3pt}
-  & CN    & 5972.985 & 0.76    & $-$2.05   &  1 \\
14 & Si~{\sc i}      & 5988.770   & 5.94                & $-$1.45 & 2  \\
22 & Ti~{\sc i}      & 5988.725   & 3.45                & $-$2.15 & 2 \\
-  & CN     & 5988.855 & 0.75   & $-$3.52   &          1               \\
-  & CN     & 5988.875 & 2.09   & $-$9.76   &          1               \\
-  & CN     & 5988.898 & 2.38   & $-$3.21   &          1              \\
-  & CN     & 5988.930  & 0.66   & $-$3.07   &        1                 \\
-  & CN     & 5988.987 & 2.34   & $-$2.87   &         1                \\
-  & CN     & 5989.000  & 2.34   & $-$2.85   &        1                 \\
90 & Th~{\sc ii}     & 5989.045 & 0.18 & $-$1.41 &  3 \\
-  & CN     & 5989.054 & 0.89   & $-$2.23   &       1                  \\
-  & CN     & 5989.083 & 3.04   & $-$3.68   &       1                  \\
-  & CN     & 5989.106 & 2.47   & $-$2.23   &      1                  \\
-  & CN     & 5989.197 & 1.50   & $-$3.11   &      1                   \\
-  & CN     & 5989.240  & 3.04   & $-$3.02   &     1                    \\
-  & CN     & 5989.249 & 2.33   & $-$3.40    &     1                    \\
-  & CN     & 5989.252 & 2.33   & $-$3.70    &     1                    \\
-  & CN     & 5989.261 & 2.45    & $-$2.08   &      1                   \\
-  & CN    & 5989.287 & 3.04   & $-$3.09   &        1                 \\
-  & CN    & 5989.294 & 2.45    & $-$3.04   &       1                  \\
-  & CN    & 5989.300 & 1.86   & $-$5.16   &        1                  \\
-  & CN    & 6644.304 & 0.81 & $-$2.05   &         1                 \\
60 & Nd~{\sc ii}      & 5989.312   & 0.74             & $-$1.48 & 4 \\
-  & CN    & 6644.352 & 0.81 & $-$2.66   &           1               \\
-  & CN    & 6644.922 & 1.07 & $-$1.97   &          1                \\
-  & CN    & 6644.965 & 1.66 & $-$1.95   &          1                \\
63 & Eu~{\sc ii}     & 6645.064 & 1.38 & 0.12  &  4 \\
63 & Eu~{\sc ii}     & 6645.057 & 1.38 & $-$0.84 &  5 \\
63 & Eu~{\sc ii}     & 6645.060 & 1.38 & $-$0.78 &  5 \\
63 & Eu~{\sc ii}     & 6645.068 & 1.38 & $-$2.13 &  5 \\ 
63 & Eu~{\sc ii}     & 6645.074 & 1.38 & $-$0.84 &  5 \\ 
63 & Eu~{\sc ii}     & 6645.083 & 1.38 & $-$0.91 &  5 \\ 
63 & Eu~{\sc ii}     & 6645.086 & 1.38 & $-$0.90  &  5 \\ 
63 & Eu~{\sc ii}     & 6645.098 & 1.38 & $-$0.60  &  5 \\ 
63 & Eu~{\sc ii}     & 6645.101 & 1.38 & $-$0.95 &  5 \\ 
63 & Eu~{\sc ii}     & 6645.121 & 1.38 & $-$1.01 &  5 \\ 
63 & Eu~{\sc ii}     & 6645.137 & 1.38 & $-$1.09 &  5 \\ 
63 & Eu~{\sc ii}     & 6645.149 & 1.38 & $-$1.19 &  5 \\ 
\hline\noalign{\vskip3pt}
\label{tab:linelist}
\end{tabular}

\begin{tabnote}
$^\ast$ Reference 1 is \citet{2025A&A...699A.276A}, 2 is Kurucz online line data (http://kurucz.harvard.edu/), 3 is \citet{2007PASJ...59L..15A}, 4 is \citet{2024ApJ...971..158R}, 5 is \citet{2019A&A...631A.113F}.
\end{tabnote}

\end{table*}

\textbf{\section{CN abundance estimation}}
Table \ref{table:CNestim} shows CN abundance estimation to measure Th 5989 \AA~line for relatively.

\begin{table*}[h]
\footnotesize
\centering
\caption{Defference of the adopted N abundance on the Th abundance derived from the 5989 \AA~line.}
\label{table:CNestim}
\begin{tabular}{lrrrrrrr}
\hline
Object & [Fe/H]  & {[}N/Fe{]} & $\log\varepsilon\mathrm{N}$ & $\log\varepsilon\mathrm{N}\_{\mathrm{err}}$ &  $\log\varepsilon\mathrm{Th}_{\mathrm{no CN}}$ &  $\log\varepsilon\mathrm{Th}_{\mathrm{CN}}$ & $\Delta\log\varepsilon\mathrm{Th}~({\mathrm{noCN-CN}})$ \\
\hline
GAOES            &       &          &                     &                &                   &              & \\
\hline
HD85503	& 0.44	& 0.35	& 8.84	& 0.20	& 1.11	& 0.52	& 0.59	\\
HD10761	& 0.03	& 0.15	& 8.23	& 0.20	& 0.51	& 0.20	& 0.31	\\
HD14770	& 0.01	& 0.33	& 8.39	& 0.20	& 0.68	& 0.39	& 0.29	\\
HD67447	& 0.00	& 0.28	& 8.33	& 0.18	& 0.85	& 0.75	& 0.10	\\
HD18970	& $-$0.06	& 0.00	& 7.99	& 0.20	& 0.64	& 0.50	& 0.14	\\
HD77912	& $-$0.07	& 0.42	& 8.40	& 0.18	& 0.53	& 0.21	& 0.32	\\
HD76294	& $-$0.12	& 0.23	& 8.16	& 0.20	& 0.48	& 0.15	& 0.33	\\
HD35369	& $-$0.12	& $-$0.05	& 7.88	& 0.20	& 0.71	& 0.61	& 0.10	\\
HD58367	& $-$0.13	& 0.03	& 7.95	& 0.18	& 0.34	& 0.19	& 0.15	\\
HD54131	& $-$0.18	& $-$0.02	& 7.85	& 0.20	& 0.28	& 0.04	& 0.24	\\
HD54810	& $-$0.26	& 0.10	& 7.89	& 0.20	& 0.34	& 0.14	& 0.20	\\
HD6186	& $-$0.30	& $-$0.10	& 7.65	& 0.20	& 0.22	& 0.10	& 0.12	\\
HD37160	& $-$0.56	& $-$0.05	& 7.44	& 0.20	& 0.32	& 0.26	& 0.06	\\
HD41597	& $-$0.59	& 0.04	& 7.50	& 0.19	& $-$0.17	& $-$0.31	& 0.14	\\
HD124897	& $-$0.62	& 0.00	& 7.43	& 0.19 & $-$0.27	& $-$0.38	& 0.11	\\
\hline								
HDS &	&	&       &		&	&		&	\\
\hline								
KIC5530598	& 0.34	& 0.34	& 8.73	& 0.20	& 1.06	& 0.63	& 0.43	\\
KIC4351319	& 0.29	& 0.31	& 8.65	& 0.20	& 1.11	& 0.61	& 0.50	\\
KIC7205067	& $-$0.02	& 0.40	& 8.43	& 0.20	& 0.74	& 0.40	& 0.34	\\
KIC11802968	& $-$0.11	& 0.30	& 8.24	& 0.20	& 0.61	& 0.11	& 0.50	\\
HD124897	& $-$0.59	& 0.00	& 7.46	& 0.19	& $-$0.24	& $-$0.37	& 0.13	\\
KIC1726211	& $-$0.66	& 0.30	& 7.69	& 0.20	& $-$0.22	& $-$0.44	& 0.22	\\
HD171496	& $-$0.67	& 0.20	& 7.58	& 0.20	& $-$0.19	& $-$0.30	& 0.11	\\
HD6833	& $-$0.69	& $-$1.00	& 6.36	& 0.19	& $-$0.15	& $-$0.15	& 0.00	\\
KIC10096113	& $-$0.74	& 0.50	& 7.81	& 0.20	& $-$0.14	& $-$0.47	& 0.33	\\
\hline
\end{tabular}
\begin{tabnote}
\raggedright
The units for all abundance-related columns are dex.
\end{tabnote}
\end{table*}

\clearpage

\bibliographystyle{aasjournal}
\bibliography{citation}

@ARTICLE{2017PASJ...69..102T,
       author = {{Tanaka}, Masaomi and {Utsumi}, Yousuke and {Mazzali}, Paolo A. and {Tominaga}, Nozomu and {Yoshida}, Michitoshi and {Sekiguchi}, Yuichiro and {Morokuma}, Tomoki and {Motohara}, Kentaro and {Ohta}, Kouji and {Kawabata}, Koji S. and {Abe}, Fumio and {Aoki}, Kentaro and {Asakura}, Yuichiro and {Baar}, Stefan and {Barway}, Sudhanshu and {Bond}, Ian A. and {Doi}, Mamoru and {Fujiyoshi}, Takuya and {Furusawa}, Hisanori and {Honda}, Satoshi and {Itoh}, Yoichi and {Kawabata}, Miho and {Kawai}, Nobuyuki and {Kim}, Ji Hoon and {Lee}, Chien-Hsiu and {Miyazaki}, Shota and {Morihana}, Kumiko and {Nagashima}, Hiroki and {Nagayama}, Takahiro and {Nakaoka}, Tatsuya and {Nakata}, Fumiaki and {Ohsawa}, Ryou and {Ohshima}, Tomohito and {Okita}, Hirofumi and {Saito}, Tomoki and {Sumi}, Takahiro and {Tajitsu}, Akito and {Takahashi}, Jun and {Takayama}, Masaki and {Tamura}, Yoichi and {Tanaka}, Ichi and {Terai}, Tsuyoshi and {Tristram}, Paul J. and {Yasuda}, Naoki and {Zenko}, Tetsuya},
        title = "{Kilonova from post-merger ejecta as an optical and near-Infrared counterpart of GW170817}",
      journal = {\pasj},
         year = 2017,
        month = dec,
       volume = {69},
       number = {6},
          eid = {102},
        pages = {102},
          doi = {10.1093/pasj/psx121},
archivePrefix = {arXiv},
       eprint = {1710.05850},
 primaryClass = {astro-ph.HE},
       adsurl = {https://ui.adsabs.harvard.edu/abs/2017PASJ...69..102T}
}

@ARTICLE{2002A&A...387..560H,
       author = {{Hill}, V. and {Plez}, B. and {Cayrel}, R. and {Beers}, T.~C. and {Nordstr{\"o}m}, B. and {Andersen}, J. and {Spite}, M. and {Spite}, F. and {Barbuy}, B. and {Bonifacio}, P. and {Depagne}, E. and {Fran{\c{c}}ois}, P. and {Primas}, F.},
        title = "{First stars. I. The extreme r-element rich, iron-poor halo giant CS 31082-001. Implications for the r-process site(s) and radioactive cosmochronology}",
      journal = {\aap},
         year = 2002,
        month = may,
       volume = {387},
        pages = {560-579},
          doi = {10.1051/0004-6361:20020434},
archivePrefix = {arXiv},
       eprint = {astro-ph/0203462},
 primaryClass = {astro-ph},
       adsurl = {https://ui.adsabs.harvard.edu/abs/2002A&A...387..560H}
}

@ARTICLE{2018ApJ...859L..24H,
       author = {{Holmbeck}, Erika M. and {Beers}, Timothy C. and {Roederer}, Ian U. and {Placco}, Vinicius M. and {Hansen}, Terese T. and {Sakari}, Charli M. and {Sneden}, Christopher and {Liu}, Chao and {Lee}, Young Sun and {Cowan}, John J. and {Frebel}, Anna},
        title = "{The R-Process Alliance: 2MASS J09544277+5246414, the Most Actinide-enhanced R-II Star Known}",
      journal = {\apjl},
         year = 2018,
        month = jun,
       volume = {859},
       number = {2},
          eid = {L24},
        pages = {L24},
          doi = {10.3847/2041-8213/aac722},
archivePrefix = {arXiv},
       eprint = {1805.11925},
 primaryClass = {astro-ph.SR},
       adsurl = {https://ui.adsabs.harvard.edu/abs/2018ApJ...859L..24H}
}

@ARTICLE{2022MNRAS.516.3786M,
       author = {{Mishenina}, T. and {Pignatari}, M. and {Gorbaneva}, T. and {C{\^o}t{\'e}}, B. and {Yag{\"u}e L{\'o}pez}, A. and {Thielemann}, F.-K. and {Soubiran}, C.},
        title = "{Enrichment of the Galactic disc with neutron-capture elements: Gd, Dy, and Th}",
      journal = {\mnras},
         year = 2022,
        month = nov,
       volume = {516},
       number = {3},
        pages = {3786-3801},
          doi = {10.1093/mnras/stac2361},
archivePrefix = {arXiv},
       eprint = {2208.11779},
 primaryClass = {astro-ph.GA},
       adsurl = {https://ui.adsabs.harvard.edu/abs/2022MNRAS.516.3786M}
}

@ARTICLE{2025A&A...699A.276A,
       author = {{Azhari}, Ainun and {Matsuno}, Tadafumi and {Aoki}, Wako and {Ishigaki}, Miho N. and {Tolstoy}, Eline},
        title = "{Th/Eu abundance ratio of red giants in the Kepler field}",
      journal = {\aap},
         year = 2025,
        month = jul,
       volume = {699},
          eid = {A276},
        pages = {A276},
          doi = {10.1051/0004-6361/202555281},
archivePrefix = {arXiv},
       eprint = {2505.11223},
 primaryClass = {astro-ph.SR},
       adsurl = {https://ui.adsabs.harvard.edu/abs/2025A&A...699A.276A}
}

@ARTICLE{2005PASJ...57...27T,
       author = {{Takeda}, Yoichi and {Ohkubo}, Michiko and {Sato}, Bun'ei and {Kambe}, Eiji and {Sadakane}, Kozo},
        title = "{Spectroscopic Study on the Atmospheric Parameters of Nearby F--K Dwarfs and Subgiants}",
      journal = {\pasj},
         year = 2005,
        month = feb,
       volume = {57},
        pages = {27-43},
          doi = {10.1093/pasj/57.1.27},
       adsurl = {https://ui.adsabs.harvard.edu/abs/2005PASJ...57...27T}
}

@ARTICLE{2002PASJ...54..451T,
       author = {{Takeda}, Yoichi and {Ohkubo}, Michiko and {Sadakane}, Kozo},
        title = "{Spectroscopic Determination of Atmospheric Parameters of Solar-Type Stars: Description of the Method and Application to the Sun}",
      journal = {\pasj},
         year = 2002,
        month = jun,
       volume = {54},
        pages = {451-462},
          doi = {10.1093/pasj/54.3.451},
       adsurl = {https://ui.adsabs.harvard.edu/abs/2002PASJ...54..451T}
}

@ARTICLE{1993KurCD..13.....K,
       author = {{Kurucz}, Robert},
        title = "{ATLAS9 Stellar Atmosphere Programs and 2 km/s grid.}",
      journal = {Robert Kurucz CD-ROM},
         year = 1993,
        month = jan,
       volume = {13},
       adsurl = {https://ui.adsabs.harvard.edu/abs/1993KurCD..13.....K}
}

@ARTICLE{2024ApJ...971..158R,
       author = {{Roederer}, Ian U. and {Beers}, Timothy C. and {Hattori}, Kohei and {Placco}, Vinicius M. and {Hansen}, Terese T. and {Ezzeddine}, Rana and {Frebel}, Anna and {Holmbeck}, Erika M. and {Sakari}, Charli M.},
        title = "{The R-Process Alliance: 2MASS J22132050─5137385, the Star with the Highest-known r-process Enhancement at [Eu/Fe] = +2.45}",
      journal = {\apj},
         year = 2024,
        month = aug,
       volume = {971},
       number = {2},
          eid = {158},
        pages = {158},
          doi = {10.3847/1538-4357/ad57bf},
archivePrefix = {arXiv},
       eprint = {2406.02691},
 primaryClass = {astro-ph.SR},
       adsurl = {https://ui.adsabs.harvard.edu/abs/2024ApJ...971..158R}
}

@ARTICLE{2007PASJ...59L..15A,
       author = {{Aoki}, Wako and {Honda}, Satoshi and {Sadakane}, Kozo and {Arimoto}, Nobuo},
        title = "{First Determination of the Actinide Thorium Abundance for a Red Giant of the Ursa Minor Dwarf Galaxy}",
      journal = {\pasj},
         year = 2007,
        month = jun,
       volume = {59},
        pages = {L15-L19},
          doi = {10.1093/pasj/59.3.L15},
archivePrefix = {arXiv},
       eprint = {0704.3104},
 primaryClass = {astro-ph},
       adsurl = {https://ui.adsabs.harvard.edu/abs/2007PASJ...59L..15A}
}

@ARTICLE{2008PASJ...60..781T,
       author = {{Takeda}, Yoichi and {Sato}, Bun'ei and {Murata}, Daisuke},
        title = "{Stellar Parameters and Elemental Abundances of Late-G Giants}",
      journal = {\pasj},
         year = 2008,
        month = aug,
       volume = {60},
        pages = {781},
          doi = {10.1093/pasj/60.4.781},
archivePrefix = {arXiv},
       eprint = {0805.2434},
 primaryClass = {astro-ph},
       adsurl = {https://ui.adsabs.harvard.edu/abs/2008PASJ...60..781T}
}

@ARTICLE{2026arXiv260412892S,
       author = {{Shah}, Shivani P. and {Ezzeddine}, Rana and {Holmbeck}, Erika M. and {Ji}, Alexander P. and {Placco}, Vinicius M. and {Roederer}, Ian U. and {Mardini}, Mohammad K. and {Usman}, Sam A. and {Bandyopadhyay}, Avrajit and {Beers}, Timothy C. and {Frebel}, Anna and {Hansen}, Terese T. and {Sakari}, Charli M. and {Sneden}, Chris},
        title = "{The $R$-Process Alliance: Actinide Abundances, Variation, and Evolution in Metal-Poor Stars}",
      journal = {arXiv e-prints},
         year = 2026,
        month = apr,
          eid = {arXiv:2604.12892},
        pages = {arXiv:2604.12892},
          doi = {10.48550/arXiv.2604.12892},
archivePrefix = {arXiv},
       eprint = {2604.12892},
 primaryClass = {astro-ph.SR},
       adsurl = {https://ui.adsabs.harvard.edu/abs/2026arXiv260412892S}
}

@ARTICLE{2006ApJ...645..613I,
       author = {{Ivans}, Inese I. and {Simmerer}, Jennifer and {Sneden}, Christopher and {Lawler}, James E. and {Cowan}, John J. and {Gallino}, Roberto and {Bisterzo}, Sara},
        title = "{Near-Ultraviolet Observations of HD 221170: New Insights into the Nature of r-Process-rich Stars}",
      journal = {\apj},
         year = 2006,
        month = jul,
       volume = {645},
       number = {1},
        pages = {613-633},
          doi = {10.1086/504069},
archivePrefix = {arXiv},
       eprint = {astro-ph/0604180},
 primaryClass = {astro-ph},
       adsurl = {https://ui.adsabs.harvard.edu/abs/2006ApJ...645..613I}
}

@ARTICLE{2012A&A...540A..98M,
       author = {{Mashonkina}, L. and {Ryabtsev}, A. and {Frebel}, A.},
        title = "{Non-LTE effects on the lead and thorium abundance determinations for cool stars}",
      journal = {\aap},
         year = 2012,
        month = apr,
       volume = {540},
          eid = {A98},
        pages = {A98},
          doi = {10.1051/0004-6361/201218790},
archivePrefix = {arXiv},
       eprint = {1202.2630},
 primaryClass = {astro-ph.GA},
       adsurl = {https://ui.adsabs.harvard.edu/abs/2012A&A...540A..98M}
}

@ARTICLE{2025A&A...693A.211G,
       author = {{Guo}, Yanjun and {Storm}, Nicholas and {Bergemann}, Maria and {Lian}, Jianhui and {Alexeeva}, Sofya and {Li}, Yangyang and {Ezzeddine}, Rana and {Jeffrey}, Gerber and {Chen}, XueFei},
        title = "{Non-local thermodynamic equilibrium (NLTE) abundances of europium (Eu) for a sample of metal-poor stars in the galactic halo and metal-poor disk with 1D and <3D> models}",
      journal = {\aap},
         year = 2025,
        month = jan,
       volume = {693},
          eid = {A211},
        pages = {A211},
          doi = {10.1051/0004-6361/202451536},
archivePrefix = {arXiv},
       eprint = {2412.06277},
 primaryClass = {astro-ph.SR},
       adsurl = {https://ui.adsabs.harvard.edu/abs/2025A&A...693A.211G}
}

@ARTICLE{2014A&A...569A..43M,
       author = {{Mashonkina}, L. and {Christlieb}, N. and {Eriksson}, K.},
        title = "{The Hamburg/ESO R-process Enhanced Star survey (HERES). X. HE 2252-4225, one more r-process enhanced and actinide-boost halo star}",
      journal = {\aap},
         year = 2014,
        month = sep,
       volume = {569},
          eid = {A43},
        pages = {A43},
          doi = {10.1051/0004-6361/201424017},
archivePrefix = {arXiv},
       eprint = {1407.5379},
 primaryClass = {astro-ph.SR},
       adsurl = {https://ui.adsabs.harvard.edu/abs/2014A&A...569A..43M}
}

@ARTICLE{2018ApJ...868..110S,
       author = {{Sakari}, Charli M. and {Placco}, Vinicius M. and {Farrell}, Elizabeth M. and {Roederer}, Ian U. and {Wallerstein}, George and {Beers}, Timothy C. and {Ezzeddine}, Rana and {Frebel}, Anna and {Hansen}, Terese and {Holmbeck}, Erika M. and {Sneden}, Christopher and {Cowan}, John J. and {Venn}, Kim A. and {Davis}, Christopher Evan and {Matijevi{\v{c}}}, Gal and {Wyse}, Rosemary F.~G. and {Bland-Hawthorn}, Joss and {Chiappini}, Cristina and {Freeman}, Kenneth C. and {Gibson}, Brad K. and {Grebel}, Eva K. and {Helmi}, Amina and {Kordopatis}, Georges and {Kunder}, Andrea and {Navarro}, Julio and {Reid}, Warren and {Seabroke}, George and {Steinmetz}, Matthias and {Watson}, Fred},
        title = "{The R-Process Alliance: First Release from the Northern Search for r-process-enhanced Metal-poor Stars in the Galactic Halo}",
      journal = {\apj},
         year = 2018,
        month = dec,
       volume = {868},
       number = {2},
          eid = {110},
        pages = {110},
          doi = {10.3847/1538-4357/aae9df},
archivePrefix = {arXiv},
       eprint = {1809.09156},
 primaryClass = {astro-ph.SR},
       adsurl = {https://ui.adsabs.harvard.edu/abs/2018ApJ...868..110S}
}

@ARTICLE{2019ApJ...881....5H,
       author = {{Holmbeck}, Erika M. and {Frebel}, Anna and {McLaughlin}, G.~C. and {Mumpower}, Matthew R. and {Sprouse}, Trevor M. and {Surman}, Rebecca},
        title = "{Actinide-rich and Actinide-poor r-process-enhanced Metal-poor Stars Do Not Require Separate r-process Progenitors}",
      journal = {\apj},
         year = 2019,
        month = aug,
       volume = {881},
       number = {1},
          eid = {5},
        pages = {5},
          doi = {10.3847/1538-4357/ab2a01},
archivePrefix = {arXiv},
       eprint = {1904.02139},
 primaryClass = {astro-ph.HE},
       adsurl = {https://ui.adsabs.harvard.edu/abs/2019ApJ...881....5H}
}

@ARTICLE{2008PASJ...60.1159S,
       author = {{Suda}, Takuma and {Katsuta}, Yutaka and {Yamada}, Shimako and {Suwa}, Tamon and {Ishizuka}, Chikako and {Komiya}, Yutaka and {Sorai}, Kazuo and {Aikawa}, Masayuki and {Fujimoto}, Masayuki Y.},
        title = "{Stellar Abundances for the Galactic Archeology (SAGA) Database --- Compilation of the Characteristics of Known Extremely Metal-Poor Stars}",
      journal = {\pasj},
         year = 2008,
        month = oct,
       volume = {60},
        pages = {1159},
          doi = {10.1093/pasj/60.5.1159},
archivePrefix = {arXiv},
       eprint = {0806.3697},
 primaryClass = {astro-ph},
       adsurl = {https://ui.adsabs.harvard.edu/abs/2008PASJ...60.1159S}
}

@ARTICLE{2023ApJ...942...39F,
       author = {{Fujibayashi}, Sho and {Kiuchi}, Kenta and {Wanajo}, Shinya and {Kyutoku}, Koutarou and {Sekiguchi}, Yuichiro and {Shibata}, Masaru},
        title = "{Comprehensive Study of Mass Ejection and Nucleosynthesis in Binary Neutron Star Mergers Leaving Short-lived Massive Neutron Stars}",
      journal = {\apj},
         year = 2023,
        month = jan,
       volume = {942},
       number = {1},
          eid = {39},
        pages = {39},
          doi = {10.3847/1538-4357/ac9ce0},
archivePrefix = {arXiv},
       eprint = {2205.05557},
 primaryClass = {astro-ph.HE},
       adsurl = {https://ui.adsabs.harvard.edu/abs/2023ApJ...942...39F}
}

@INPROCEEDINGS{2002ASPC..281..298B,
       author = {{Baba}, Hajime and {Yasuda}, Naoki and {Ichikawa}, Shin-Ichi and {Yagi}, Masafumi and {Iwamoto}, Nobuyuki and {Takata}, Tadafumi and {Horaguchi}, Toshihiro and {Taga}, Masatoshi and {Watanabe}, Masaru and {Ozawa}, Tomohiko and {Hamabe}, Masaru},
        title = "{Development of the Subaru-Mitaka-Okayama-Kiso Archive System}",
    booktitle = {Astronomical Data Analysis Software and Systems XI},
         year = 2002,
       editor = {{Bohlender}, David A. and {Durand}, Daniel and {Handley}, Thomas H.},
       series = {Astronomical Society of the Pacific Conference Series},
       volume = {281},
        month = jan,
        pages = {298},
       adsurl = {https://ui.adsabs.harvard.edu/abs/2002ASPC..281..298B}
}

@ARTICLE{2000A&AS..143....9W,
       author = {{Wenger}, M. and {Ochsenbein}, F. and {Egret}, D. and {Dubois}, P. and {Bonnarel}, F. and {Borde}, S. and {Genova}, F. and {Jasniewicz}, G. and {Lalo{\"e}}, S. and {Lesteven}, S. and {Monier}, R.},
        title = "{The SIMBAD astronomical database. The CDS reference database for astronomical objects}",
      journal = {\aaps},
         year = 2000,
        month = apr,
       volume = {143},
        pages = {9-22},
          doi = {10.1051/aas:2000332},
archivePrefix = {arXiv},
       eprint = {astro-ph/0002110},
 primaryClass = {astro-ph},
       adsurl = {https://ui.adsabs.harvard.edu/abs/2000A&AS..143....9W}
}

@ARTICLE{2000AJ....120.1841F,
       author = {{Fulbright}, Jon P.},
        title = "{Abundances and Kinematics of Field Halo and Disk Stars. I. Observational Data and Abundance Analysis}",
      journal = {\aj},
         year = 2000,
        month = oct,
       volume = {120},
       number = {4},
        pages = {1841-1852},
          doi = {10.1086/301548},
archivePrefix = {arXiv},
       eprint = {astro-ph/0006260},
 primaryClass = {astro-ph},
       adsurl = {https://ui.adsabs.harvard.edu/abs/2000AJ....120.1841F}
}

@ARTICLE{2012ApJ...753...64I,
       author = {{Ishigaki}, Miho N. and {Chiba}, Masashi and {Aoki}, Wako},
        title = "{Chemical Abundances of the Milky Way Thick Disk and Stellar Halo. I. Implications of [{\ensuremath{\alpha}}/Fe] for Star Formation Histories in Their Progenitors}",
      journal = {\apj},
         year = 2012,
        month = jul,
       volume = {753},
       number = {1},
          eid = {64},
        pages = {64},
          doi = {10.1088/0004-637X/753/1/64},
archivePrefix = {arXiv},
       eprint = {1205.2406},
 primaryClass = {astro-ph.GA},
       adsurl = {https://ui.adsabs.harvard.edu/abs/2012ApJ...753...64I}
}

@ARTICLE{2021ApJ...912...72M,
       author = {{Matsuno}, Tadafumi and {Aoki}, Wako and {Casagrande}, Luca and {Ishigaki}, Miho N. and {Shi}, Jianrong and {Takata}, Masao and {Xiang}, Maosheng and {Yong}, David and {Li}, Haining and {Suda}, Takuma and {Xing}, Qianfan and {Zhao}, Jingkun},
        title = "{Star Formation Timescales of the Halo Populations from Asteroseismology and Chemical Abundances}",
      journal = {\apj},
         year = 2021,
        month = may,
       volume = {912},
       number = {1},
          eid = {72},
        pages = {72},
          doi = {10.3847/1538-4357/abeab2},
archivePrefix = {arXiv},
       eprint = {2006.03619},
 primaryClass = {astro-ph.SR},
       adsurl = {https://ui.adsabs.harvard.edu/abs/2021ApJ...912...72M}
}

@ARTICLE{2009ARA&A..47..481A,
       author = {{Asplund}, Martin and {Grevesse}, Nicolas and {Sauval}, A. Jacques and {Scott}, Pat},
        title = "{The Chemical Composition of the Sun}",
      journal = {\araa},
         year = 2009,
        month = sep,
       volume = {47},
       number = {1},
        pages = {481-522},
          doi = {10.1146/annurev.astro.46.060407.145222},
archivePrefix = {arXiv},
       eprint = {0909.0948},
 primaryClass = {astro-ph.SR},
       adsurl = {https://ui.adsabs.harvard.edu/abs/2009ARA&A..47..481A}
}

@ARTICLE{2002PASJ...54..855N,
       author = {{Noguchi}, Kunio and {Aoki}, Wako and {Kawanomoto}, Satoshi and {Ando}, Hiroyasu and {Honda}, Satoshi and {Izumiura}, Hideyuki and {Kambe}, Eiji and {Okita}, Kiichi and {Sadakane}, Kozo and {Sato}, Bun'ei and {Tajitsu}, Akito and {Takada-Hidai}, Tasahide and {Tanaka}, Wataru and {Watanabe}, Etsuji and {Yoshida}, Michitoshi},
        title = "{High Dispersion Spectrograph (HDS) for the Subaru Telescope}",
      journal = {\pasj},
         year = 2002,
        month = dec,
       volume = {54},
        pages = {855-864},
          doi = {10.1093/pasj/54.6.855},
       adsurl = {https://ui.adsabs.harvard.edu/abs/2002PASJ...54..855N}
}

@ARTICLE{2019A&A...631A.113F,
       author = {{Forsberg}, R. and {J{\"o}nsson}, H. and {Ryde}, N. and {Matteucci}, F.},
        title = "{Abundances of disk and bulge giants from high-resolution optical spectra. IV. Zr, La, Ce, Eu}",
      journal = {\aap},
         year = 2019,
        month = nov,
       volume = {631},
          eid = {A113},
        pages = {A113},
          doi = {10.1051/0004-6361/201936343},
archivePrefix = {arXiv},
       eprint = {1909.10535},
 primaryClass = {astro-ph.GA},
       adsurl = {https://ui.adsabs.harvard.edu/abs/2019A&A...631A.113F}
}

@ARTICLE{1996ApJ...467..819S,
       author = {{Sneden}, Christopher and {McWilliam}, Andrew and {Preston}, George W. and {Cowan}, John J. and {Burris}, Debra L. and {Armosky}, Bradley J.},
        title = "{The Ultra--Metal-poor, Neutron-Capture--rich Giant Star CS 22892-052}",
      journal = {\apj},
         year = 1996,
        month = aug,
       volume = {467},
        pages = {819},
          doi = {10.1086/177656},
       adsurl = {https://ui.adsabs.harvard.edu/abs/1996ApJ...467..819S}
}

@ARTICLE{2021RvMP...93a5002C,
       author = {{Cowan}, John J. and {Sneden}, Christopher and {Lawler}, James E. and {Aprahamian}, Ani and {Wiescher}, Michael and {Langanke}, Karlheinz and {Mart{\'\i}nez-Pinedo}, Gabriel and {Thielemann}, Friedrich-Karl},
        title = "{Origin of the heaviest elements: The rapid neutron-capture process}",
      journal = {Reviews of Modern Physics},
         year = 2021,
        month = jan,
       volume = {93},
       number = {1},
          eid = {015002},
        pages = {015002},
          doi = {10.1103/RevModPhys.93.015002},
archivePrefix = {arXiv},
       eprint = {1901.01410},
 primaryClass = {astro-ph.HE},
       adsurl = {https://ui.adsabs.harvard.edu/abs/2021RvMP...93a5002C}
}

@ARTICLE{2018Natur.563...85H,
       author = {{Helmi}, Amina and {Babusiaux}, Carine and {Koppelman}, Helmer H. and {Massari}, Davide and {Veljanoski}, Jovan and {Brown}, Anthony G.~A.},
        title = "{The merger that led to the formation of the Milky Way's inner stellar halo and thick disk}",
      journal = {\nat},
         year = 2018,
        month = oct,
       volume = {563},
       number = {7729},
        pages = {85-88},
          doi = {10.1038/s41586-018-0625-x},
archivePrefix = {arXiv},
       eprint = {1806.06038},
 primaryClass = {astro-ph.GA},
       adsurl = {https://ui.adsabs.harvard.edu/abs/2018Natur.563...85H}
}

@ARTICLE{2022A&A...661A.103M,
       author = {{Matsuno}, Tadafumi and {Koppelman}, Helmer H. and {Helmi}, Amina and {Aoki}, Wako and {Ishigaki}, Miho N. and {Suda}, Takuma and {Yuan}, Zhen and {Hattori}, Kohei},
        title = "{High-precision chemical abundances of Galactic building blocks. The distinct chemical abundance sequence of Sequoia}",
      journal = {\aap},
         year = 2022,
        month = may,
       volume = {661},
          eid = {A103},
        pages = {A103},
          doi = {10.1051/0004-6361/202142752},
archivePrefix = {arXiv},
       eprint = {2111.15423},
 primaryClass = {astro-ph.GA},
       adsurl = {https://ui.adsabs.harvard.edu/abs/2022A&A...661A.103M}
}

@ARTICLE{2008AJ....135..209M,
       author = {{Massarotti}, Alessandro and {Latham}, David W. and {Stefanik}, Robert P. and {Fogel}, Jeffrey},
        title = "{Rotational and Radial Velocities for a Sample of 761 HIPPARCOS Giants and the Role of Binarity}",
      journal = {\aj},
         year = 2008,
        month = jan,
       volume = {135},
       number = {1},
        pages = {209-231},
          doi = {10.1088/0004-6256/135/1/209},
       adsurl = {https://ui.adsabs.harvard.edu/abs/2008AJ....135..209M}
}

@ARTICLE{2022A&A...668A.168C,
       author = {{Cescutti}, G. and {Bonifacio}, P. and {Caffau}, E. and {Monaco}, L. and {Franchini}, M. and {Lombardo}, L. and {Matas Pinto}, A.~M. and {Lucertini}, F. and {Fran{\c{c}}ois}, P. and {Spitoni}, E. and {Lallement}, R. and {Sbordone}, L. and {Mucciarelli}, A. and {Spite}, M. and {Hansen}, C.~J. and {Di Marcantonio}, P. and {Ku{\v{c}}inskas}, A. and {Dobrovolskas}, V. and {Korn}, A.~J. and {Valentini}, M. and {Magrini}, L. and {Cristallo}, S. and {Matteucci}, F.},
        title = "{MINCE. I. Presentation of the project and of the first year sample}",
      journal = {\aap},
         year = 2022,
        month = dec,
       volume = {668},
          eid = {A168},
        pages = {A168},
          doi = {10.1051/0004-6361/202244515},
archivePrefix = {arXiv},
       eprint = {2211.06086},
 primaryClass = {astro-ph.SR},
       adsurl = {https://ui.adsabs.harvard.edu/abs/2022A&A...668A.168C}
}

@ARTICLE{2025MNRAS.538.3284S,
       author = {{Storm}, Nicholas and {Bergemann}, Maria and {Eitner}, Philipp and {Hoppe}, Richard and {Kemp}, Alex J. and {Ruiter}, Ashley J. and {Janka}, Hans-Thomas and {Sieverding}, Andre and {de Mink}, Selma E. and {Seitenzahl}, Ivo R. and {Owusu}, Evans K.},
        title = "{Observational constraints on the origin of the elements. IX. 3D NLTE abundances of metals in the context of Galactic Chemical Evolution models and 4MOST}",
      journal = {\mnras},
         year = 2025,
        month = apr,
       volume = {538},
       number = {4},
        pages = {3284-3313},
          doi = {10.1093/mnras/staf472},
archivePrefix = {arXiv},
       eprint = {2503.16946},
 primaryClass = {astro-ph.SR},
       adsurl = {https://ui.adsabs.harvard.edu/abs/2025MNRAS.538.3284S}
}

@ARTICLE{2019ApJ...875..106C,
       author = {{C{\^o}t{\'e}}, Benoit and {Eichler}, Marius and {Arcones}, Almudena and {Hansen}, Camilla J. and {Simonetti}, Paolo and {Frebel}, Anna and {Fryer}, Chris L. and {Pignatari}, Marco and {Reichert}, Moritz and {Belczynski}, Krzysztof and {Matteucci}, Francesca},
        title = "{Neutron Star Mergers Might Not Be the Only Source of r-process Elements in the Milky Way}",
      journal = {\apj},
         year = 2019,
        month = apr,
       volume = {875},
       number = {2},
          eid = {106},
        pages = {106},
          doi = {10.3847/1538-4357/ab10db},
archivePrefix = {arXiv},
       eprint = {1809.03525},
 primaryClass = {astro-ph.HE},
       adsurl = {https://ui.adsabs.harvard.edu/abs/2019ApJ...875..106C}
}

@ARTICLE{2021MNRAS.506.5410I,
       author = {{Ishigaki}, Miho N. and {Hartwig}, Tilman and {Tarumi}, Yuta and {Leung}, Shing-Chi and {Tominaga}, Nozomu and {Kobayashi}, Chiaki and {Magg}, Mattis and {Simionescu}, Aurora and {Nomoto}, Ken'ichi},
        title = "{Origin of metals in old Milky Way halo stars based on GALAH and Gaia}",
      journal = {\mnras},
         year = 2021,
        month = oct,
       volume = {506},
       number = {4},
        pages = {5410-5429},
          doi = {10.1093/mnras/stab1982},
archivePrefix = {arXiv},
       eprint = {2107.04194},
 primaryClass = {astro-ph.GA},
       adsurl = {https://ui.adsabs.harvard.edu/abs/2021MNRAS.506.5410I}
}

\end{document}